\documentclass[journal,onecolumn]{IEEEtran}
\ifCLASSINFOpdf
\else
\fi
\ifCLASSOPTIONcompsoc
  \usepackage[caption=false,font=normalsize,labelfont=sf,textfont=sf]{subfig}
\else
  \usepackage[caption=false,font=footnotesize]{subfig}
\fi

\usepackage{amsthm}

\usepackage{amssymb}
\usepackage{mathtools}
\usepackage{cases}
\usepackage{autobreak}

\usepackage{booktabs}
\usepackage{multirow}

\usepackage{color}
\usepackage{hyperref}
\usepackage{cite}

\usepackage{amsmath}
\usepackage{extarrows}
\usepackage{bm}

\usepackage{comment}

\usepackage{algorithmic}
\usepackage{algorithm}
\usepackage{caption}
\usepackage{graphicx}

\newcommand{\Var}{\mathrm{Var}}
\renewcommand\Re{\operatorname{Re}}
\renewcommand\Im{\operatorname{Im}}

\newcommand{\diff}{\mathrm{d}}

\newcommand*{\rom}[1]{\expandafter\@slowromancap\romannumeral #1@}
\newcommand{\QED}{\hfill \ensuremath{\blacksquare}}

\DeclareMathOperator{\Tr}{Tr}
\DeclareMathOperator{\dist}{dist}

\DeclarePairedDelimiter\abs{\lvert}{\rvert}%
\DeclarePairedDelimiter\norm{\lVert}{\rVert}%
\makeatletter
\let\oldabs\abs
\def\abs{\@ifstar{\oldabs}{\oldabs*}}
\let\oldnorm\norm
\def\norm{\@ifstar{\oldnorm}{\oldnorm*}}
\makeatother

\newcommand{\RNum}[1]{\uppercase\expandafter{\romannumeral #1\relax}}

\newtheorem{remark}{Remark}
\newtheorem{theorem}{Theorem}
\newtheorem{lemma}{Lemma}
\newtheorem{proposition}{Proposition}
\newtheorem{assumption}{Assumption}
\newtheorem{corollary}{Corollary}

\allowdisplaybreaks[4]
\newtheorem{assumptionalt}{Assumption}[assumption]
\newenvironment{assumptionp}[1]{
  \renewcommand\theassumptionalt{#1}
  \assumptionalt
}{\endassumptionalt}

\title{Fundamental Limits of Two-Hop MIMO
Channels: An Asymptotic Approach}
\author{\IEEEauthorblockN {Zeyan Zhuang, Xin Zhang, Dongfang Xu, and Shenghui Song}

Dept. of ECE, The Hong Kong University of Science and Technology, Hong Kong\\

}

\begin{document}
\maketitle

\begin{abstract}
Multi-antenna relays and intelligent reflecting surfaces (IRSs) have been utilized to construct favorable channels to improve the performance of wireless systems. A common feature between relay systems and IRS-aided systems is the two-hop multiple-input multiple-output (MIMO) channel. 
As a result, the mutual information (MI)
of two-hop MIMO channels has been widely investigated with very engaging results. However, 
a rigorous investigation on the fundamental limits of two-hop MIMO channels, i.e., the first and second-order analysis, is not yet available in the literature, due to the difficulties caused by the two-hop (product) channel and the noise introduced by the relay (active IRS).
In this  paper,
we employ large-scale random matrix theory (RMT), specifically Gaussian tools, 
to derive the closed-form deterministic approximation for the mean and variance of the MI.
Additionally, we determine the convergence rate for the mean, 
variance and the characteristic function of the MI, and prove the asymptotic Gaussianity.
Furthermore,  we also investigate the analytical properties of the
 fundamental equations that describe the closed-form approximation and 
 prove the existence and uniqueness of the solution. An iterative algorithm is then proposed to obtain the solution for the fundamental equations.
 Numerical results validate the accuracy of the theoretical analysis.
\end{abstract}
\section{Introduction}
\subsection{Two-Hop MIMO Channels}
Relays \cite{van1971three} were proposed as a form of active scattering that can effectively enhance wireless channels and improve communication quality. To this end, multiple-input multiple-out (MIMO) relays have been widely investigated to improve the spectral efficiency and system reliability of two-hop channels.  
There are three types of relay protocols: decode-and-forward (DF), compress-and-forward (CF), and amplify-and-forward (AF), among which the AF protocol is simplest in implementation and has drawn immense research interests \cite{RelaybasedDeployment}.
\par
However, relay systems normally have high energy consumption and hardware cost.
To this end, intelligent reflecting surfaces (IRSs) \cite{IRSWu2021} were proposed to dynamically control the radio propagation channel in an energy-efficient way.
 However, one critical issue for passive IRSs is the ``multiplicative fading" effect  \cite{Najafi2020}  
due to the two-hop transmission. 
As a solution, 
active IRSs were proposed to reflect and amplify incoming signals with moderate energy consumption, tackling the multiplicative fading effect.
It is worth mentioning that, unlike AF relays,
active IRSs do not require expensive and power-consuming radio frequency (RF) chains to reflect incident signals\cite{wuqingqing2019, kang2023active}.
\par
MIMO relay and IRS-aided MIMO systems have been recognized as effective ways to improve the spectral efficiency and
reliability of wireless communications.
The common feature between them is the two-hop MIMO channel. 
In fact, such a two-hop structure can also be utilized to model other widely adopted
 wireless channels, including double-scattering and Rayleigh-product channels \cite{Gesbert2002Double}.
However, the product structure of the channel and the noise introduced at the relay node make the corresponding 
theoretical analysis very challenging.
To the best of the authors' knowledge, the fundamental limits of two-hop MIMO channels, including the first-order and second-order analysis, 
  have not been rigorously investigated yet.
\subsection{Existing Works}
The performance of two-hop MIMO channels has been studied using various methods, leading to very innovative results.
In \cite{Jin2010DualHop}, Jin \textit{et al.} established the upper 
and lower bounds of the ergodic capacity for dual-hop AF MIMO systems 
by utilizing finite-dimensional random matrix theory (RMT).
Finite-dimensional RMT can help obtain exact results for arbitrary dimensions, but the computational complexity is in general very high.
In contrast, the use of asymptotic RMT normally leads to strikingly simple approximations,
 which have been shown accurate even at low dimensions.
In \cite{AFMIMO2006Replica}, the replica method was used to derive the
central limit theorem (CLT) for the MI of AF
MIMO relay channels with correlated Rayleigh fading. 
In \cite{Wen2011}, Wen \textit{et al.} 
further considered the direct link,
and derived the deterministic approximation of the ergodic mutual information (EMI) by replica method.
The  EMI for MIMO relay channels with direct link under Rician fading was studied in \cite{Wen2012Robust}, 
which generalized the results in \cite{Wen2011}. The asymptotic MI of the 
IRS-aided MIMO Rician channel was
analyzed in \cite{Zhang2021Large} where Zhang \textit{et al.} highlighted the difference between AF relays and IRSs.
\par
The replica method came from statistical
physics \cite{edwards1975theory} and was subsequently used in communication systems \cite{Dong2005, Muller2003, Wen2007}. 
Although replica method can efficiently derive the asymptotic mean and variance of the
MI by finding saddle points, a rigorous mathematical 
derivation is missing, as pointed out by some works \cite{Wen2007,Hachem2008ANewApproach}.
There are also works that used other methods to study the two-hop MIMO channels. By utilizing free probability theory\cite{mingo2017free}, Zheng \textit{et al.} investigated the asymptotic mean and variance for the MI of Rayleigh product channels when
the number of transmitter antennas is equal to that of reveiver antennas and established the CLT in \cite{Zheng2017DoubleScattering}. 
In \cite{ZhangXin2022JSTSP}, the CLT for the MI of IRS-assisted MIMO communication over correlated channels 
was derived based on the Bai-Silverstein method \cite{bai2010spectral}, but the result can not be applied to general relay channels, because
Bai-Silverstein method requires one of the correlation matrices to be
diagonal \cite{Zhangjun2013Large}.
In summary, a general and rigorous investigation for the fundamental limits of general MIMO relay channels is not yet available in the literature. 
\par
In this work, we will fill this research gap by utilizing  another asymptotic approach: 
Gaussian tools\cite{ALytova2009AOP, pastur2011eigenvalue}. 
Compared to the replica method,
Gaussian tools are more rigorous \cite{Hachem2008ANewApproach}. 
Compared to the Bai-Silverstein method and free probability theory, Gaussian tools can handle arbitrary correlations 
as well as different antenna dimensions.
In fact, Gaussian tools have been utilized for the analysis of  
single-hop MIMO channels, e.g.,  CLT for the MI of Rayleigh channels \cite{Hachem2008ANewApproach} and EMI approximation for Rician channels 
\cite{Dumount2010Rician}.  In \cite{Xin2023Double-Scattering}, the CLT for the
 MI of double-scattering channels and 
the $\mathcal{O}(\frac{1}{N})$ convergence rate of the mean and variance were proved by Gaussian tools, where $N$ is the number of receive antennas. In this paper, 
 we will utilize Gaussian tools to analyze a general two-hop channel model, 
which takes the AF relay and active IRS-aided MIMO channel as special cases. 
The result can also be degenerated to the double-scattering channel \cite{Xin2023Double-Scattering} and 
single-hop Rayleigh channel \cite{Hachem2008ANewApproach}. 
Besides the rigorousness of the derivation, we also provide the necessary assumptions and investigate the properties of the fundamental equations of the system.  Additionally, we propose an iterative algorithm for solving the  fundamental equations
and establish the CLT, which can degenerate to the univariate case in \cite{AFMIMO2006Replica}.
\subsection{Challenges}
Due to the noise at the relay node, 
the MI of MIMO relay channels is very complex.
 In particular, it involves the subtraction of two MI terms. One is
  the MI between the received signal and the transmitted signal as well as the relay noise, 
  while the other is the conditional MI between the received signal and the relay noise. The two MI terms share the same random 
 channel and therefore are correlated. 
 The first MI term contains a general structure of random matrices, 
 where  the sample covariance matrix of the second hop channel is added to the sample covariance matrix of the two-hop channel. This structure is different  from that of  the double-scattering channel \cite{Xin2023Double-Scattering} which
 only contains the two-hop sample covariance matrix.
 For the second-order analysis, we
need to study the joint distribution of these two MI terms, where their correlation and the two-hop structure make the evaluation of their covariance matrix and the proof of the
 asymptotic Gaussianity very challenging.
 Additionally, the presence of the relay noise and the general two-hop structure prevent simultaneous diagonalization of the antenna correlation matrices for both hops. 
 Finally, closed-form approximation of the fundamental limits is described by a set of coupled fixed-point equations. 
Due to the coupling between the equations, obtaining the existence, uniqueness, and analytical properties of the solutions to this equation system is not straightforward.

 \par
\subsection{Contributions}
The main contributions of this paper are summarized from communication and RMT perspectives as follows.
\par
\textbf{1. Communication perspective:} We derive the 
deterministic approximation for the two MIs of the 
 two-hop MIMO systems, and prove the convergence rate of the mean, covariance matrix, and characteristic function of the two MIs to be of order $\mathcal{O}(\frac{1}{N})$. 
Based on the convergence of the characteristic function, we build up the joint CLT. Additionally, we analyze the impact of the number of reflecting elements in the active IRS-aided system and reveal the relationship between 
the single-hop and two-hop MIMO channels.
Numerical results 
validate the accuracy of the derived theoretical results as well as the tightness of the  $\mathcal{O}(\frac{1}{N})$ bound.
\par
\textbf{2. RMT perspective:}  
We study the resolvent of a general random matrix model that involves the sum of two sample covariance matrices, one of which includes the product of two random matrices.
We reveal the analytical properties of the fundamental equations describing the concerned system and prove the existence and uniqueness of the solution. 
Different from the signle-hop channels \cite{Hachem2008ANewApproach, Dumount2010Rician}, 
the two-hop system considered in this paper increases the order of the fundamental equations
and may cause divergence of conventional iterative algorithms. 
To this end, we propose an iterative algorithm to calculate the concerned solutions. The results in this paper can be regarded as a generalization of those in \cite{hachem2007deterministic}
and the idea can be further generalized to equations of higher order and coupling degree.
\subsection{Paper Organization and Notations}
The paper is organized as follows. In Section \ref{Sec_System_Model_and_Problem_Formulation}, 
we introduce the system model, formulate the problem, and give the important assumptions. 
In section \ref{Sec_Useful_Notations_and_Statement_of_the_Main_Results},
we present the main results including the deterministic approximation of EMI and the joint CLT analysis with convergence rate. 
 In Section \ref{Sec_First-order} and \ref{Sec_Second-order}, we rigorously prove the first-order and second-order results, respectively. Numerical results are provided in Section \ref{Sec_Numerical_Experiments} and Section \ref{Sec_Conclusion}  concludes the paper.
 The introduction of mathematical tools and some preliminary results are given in Appendix \ref{App_Lemma_DE_Q1_1}.
 \subsubsection*{Notations} 
 Throughout the paper, we use bold, 
 upper-case letters and bold lower-case letters to represent matrices and vectors, respectively.
$\mathbb{C}^N$ and $\mathbb{C}^{M \times N}$ denote the space of $N$-dimensional complex vectors and the $M$ by $N$ 
complex matrix space. The $(i, j)$-th element of matrix $\mathbf{A}$ is denoted as $[\mathbf{A}]_{ij}$. The $i$-th element of vector $\mathbf{a}$ is denoted $[\mathbf{a}]_i$. $\mathbf{A}^T$ and $\mathbf{A}^H$ represent the transpose or conjugate transpose of $\mathbf{A}$, 
respectively.  $\norm{\cdot}$ denotes the spectral norm of a matrix or the Euclidean norm of a vector. 
 $\mathrm{Tr} \mathbf{A}$ refers to the trace of $\mathbf{A}$ and $\mathbf{I}_N$ denotes the $N$ by $N$ identity matrix. The notations $\mathbf{0}_N$ and $\mathbf{0}_{N \times M}$ denote the column vector of zeros of size $N$ and $N$ by $M$ matrix with all zero entries, respectively.
 $\mathbb{P}(\cdot)$ denotes the probability measure and $\mathbb{E}(\cdot)$ represents the expectation operator. 
 $\Im(a)$, $\Re(a)$ denote the imaginary and real parts of a complex number respectively. 
 $(\cdot)^*$ denotes the conjugate of a complex number. 
$\mathbb{R}^+$ and $\mathbb{R}_-$ represent the sets  $\{x \in \mathbb{R}: x \geq 0\}$ and $\{x \in \mathbb{R}: x \leq 0\}$, respectively. $\mathbb{C}^+$ and $\mathbb{C}_-$ denote the half complex planes $\{z \in \mathbb{C}: \Im(z) > 0 \}$ and $\{z \in \mathbb{C}: \Im(z) < 0 \}$, respectively. $\delta(\cdot)$ denotes the Kronecker function, i.e., $\delta(0) = 1$ otherwise $\delta(x) = 0, x\neq 0$. The indicator function is represented by $\mathbb{I}_{\{\cdot\}}$. The notations $\xrightarrow[\mathcal{N}]{a.s.}$ and $\xrightarrow[\mathcal{N}]{d}$ indicate the convergence almost surely and in distribution, respectively, under the process $\mathcal{N}$. Lastly, $\mathcal{O}(\cdot)$ and $o(\cdot)$ denote the big-O and small-o notations, respectively.
\section{System Model and Problem Formulation}
\label{Sec_System_Model_and_Problem_Formulation}
\subsection{System Model}
Consider a two-hop point-to-point MIMO system. The transmitter and the receiver are 
equipped with $M$  and $N$ antennas, respectively. The middle node can be an AF relay or an active IRS, equipped with $L$ antennas (AF relay) or reflecting elements (active IRS). The received signal $\mathbf{y} \in \mathbb{C}^N$ is denoted as
\begin{equation}
    \mathbf{y} = \mathbf{G}_1 \mathbf{\Phi} (\mathbf{G}_2 \mathbf{x} + \mathbf{n}_1) + \mathbf{n}_2,
\end{equation}
where $\mathbf{x} \in \mathbb{C}^M$ is the zero mean transmitted signal with 
covariance $\mathbb{E}\mathbf{x}\mathbf{x}^H = \mathbf{P}$. Here $\mathbf{n}_1 \in \mathbb{C}^L$ denotes the noise vector at the relay 
\cite{Bowang2005} or the amplification noise introduced by the active component of the active IRS \cite{You2021,Dongfang2021}
 and $\mathbf{n}_2 \in \mathbb{C}^N$ represents the static noise vector at the receiver. Both $\mathbf{n}_1$ and $\mathbf{n}_2$ 
 are modeled as additive white Gaussian noise (AWGN), i.e.,
  $\mathbf{n}_1 \sim \mathcal{CN}(0, \sigma_1^2\mathbf{I}_L)$, 
  $\mathbf{n}_2 \sim \mathcal{CN}(0, \sigma_2^2\mathbf{I}_N)$. 
The channel matrices of the second and first hop are denoted by $\mathbf{G}_1 \in \mathbb{C}^{N \times L}$ and $\mathbf{G}_2 \in \mathbb{C}^{L \times M}$, respectively.
$\mathbf{\Phi}$ represents the amplification matrix of the AF relay or the reflection matrix of active IRS \cite{RezaeiAFAIRS2023}. 
\par
We consider the Kronecker correlated Rayleigh 
model \cite{DaShan2000} with
\begin{equation}
    \mathbf{G}_1 = \mathbf{A}_{1}\mathbf{X}_1\mathbf{B}_{1}, ~~  \mathbf{G}_2 = \mathbf{A}_{2}\mathbf{X}_2\mathbf{B}_{2}, \label{Channel_Model}
\end{equation}
where $\mathbf{A}_{1} \in \mathbb{C}^{N \times N}$, $\mathbf{B}_{1} \in \mathbb{C}^{L \times L}$ , $\mathbf{A}_{2} \in \mathbb{C}^{L \times L}$ 
and $\mathbf{B}_{2}\in \mathbb{C}^{M \times M}$ are deterministic matrices.
 $\mathbf{X}_1 \in \mathbb{C}^{N \times L}$ and $\mathbf{X}_2 \in \mathbb{C}^{L \times M}$  
 are matrices whose entries are independent and identically distributed (i.i.d.), circularly symmetric, zero-mean, complex Gaussian random variables
 (r.v.s) with variances $\frac{1}{L}$ and $\frac{1}{M}$, respectively. 
\par
In this work, we assume that the receiver knows 
the perfect channel state information (CSI), i.e. the 
instantaneous channel matrices $\mathbf{G}_i, i=1,2$, but 
the transmitter and relay only know the statistics of channel matrices. 
The  MI of the two-hop system is given by \cite{Lazano2002}
\begin{equation}
    I(\sigma_1^2, \sigma_2^2)  = 
    \log \left(\frac{\det\left(\mathbf{G}_1 \mathbf{\Phi} \mathbf{G}_2 \mathbf{P} \mathbf{G}_2^H \mathbf{\Phi}^H\mathbf{G}_1^H + \sigma_1^2 \mathbf{G}_1 \mathbf{\Phi}\mathbf{\Phi}^H\mathbf{G}_1^H  + \sigma_2^2 \mathbf{I}_N \right)}{\det \left(\sigma_1^2 \mathbf{G}_1 \mathbf{\Phi}\mathbf{\Phi}^H\mathbf{G}_1^H + \sigma_2^2 \mathbf{I}_N\right)}\right).
\label{MI_total}
\end{equation}
Due to the underlying randomness of the channel matrices, $I$ is a r.v. 
The focus of this paper is to characterize the first and second-order statistical information of $I$. 
The first-order information can be used to analyze the ergodic capacity of the concerned system, 
while the second-order information can be used to analyze the outage probability.
\subsection{Problem Formulation}
\label{Sec_Problem_formulation}
Directly analyzing \eqref{MI_total} is very challenging. To tackle this issue, 
we first express $I$ as $I(\sigma_1^2, \sigma_2^2) =  I_1(\sigma_1^2, \sigma_2^2) -  I_2(\sigma_1^2, \sigma_2^2)$, where
\begin{subequations}
\label{MI} %
\begin{align}
    I_1(\sigma_1^2, \sigma_2^2) &= \log \det \left(\frac{1}{\sigma_2^2}\left(\mathbf{G}_1 \mathbf{\Phi} \mathbf{G}_2 \mathbf{P} \mathbf{G}_2^H \mathbf{\Phi}^H\mathbf{G}_1^H + \sigma_1^2 \mathbf{G}_1 \mathbf{\Phi}\mathbf{\Phi}^H\mathbf{G}_1^H \right) +  \mathbf{I}_N \right), \label{MI_1} \\
    I_2(\sigma_1^2, \sigma_2^2) &=
     \log\det \left(\frac{\sigma_1^2}{\sigma_2^2} \mathbf{G}_1 \mathbf{\Phi}\mathbf{\Phi}^H\mathbf{G}_1^H + \mathbf{I}_N\right). \label{MI_2} %
\end{align}
\end{subequations} 
In fact, $I_1$ is the MI between the received signal $\mathbf{y}$ and the transmitted signal $\mathbf{x}$ as well as the AWGN at the receiver $\mathbf{n}_1$, i.e.,
$\mathcal{I}(\mathbf{y}; \mathbf{n}_1, \mathbf{x})$ and $I_2$ is the MI 
between the received signal $\mathbf{y}$ and the AWGN at the receiver $\mathbf{n}_1$ conditioned on the transmitted signal $\mathbf{x}$, i.e., $\mathcal{I}(\mathbf{y}, \mathbf{n}_1| \mathbf{x})$. Thus we call them the two MI terms. Next, we simplify $I_1$ and $I_2$ in \eqref{MI}
to obtain an equivalent channel model. Denote the singular value decomposition (SVD) of the concerned matrices as  
$\mathbf{A}_{i} = \mathbf{U}_{A_i} \mathbf{\Sigma}_{A_i} \mathbf{V}_{A_i}^H, i=1,2$, $\mathbf{B}_{1}\mathbf{\Phi} 
= \mathbf{U}_{B_1} \mathbf{\Sigma}_{B_1} \mathbf{V}_{B_1}^H$
 and $\mathbf{B}_{2}\mathbf{P}^{\frac{1}{2}} = \mathbf{U}_{B_2} \mathbf{\Sigma}_{B_2} \mathbf{V}_{B_2}^H$, and define $\mathbf{W}_{A_i} = \mathbf{V}_{A_i}\mathbf{U}_{A_i}^H$ and $\mathbf{W}_{B_i} = \mathbf{V}_{B_i}\mathbf{U}_{B_i}^H$ 
 for $i=1,2$. By using the identity $\det(\mathbf{I} + \mathbf{X}\mathbf{Y}) = \det(\mathbf{I} + \mathbf{Y}\mathbf{X})$, we can obtain
\begin{equation}
\begin{split}
     I_1(\sigma_1^2, \sigma_2^2) &= \log \det \Big(\frac{1}{\sigma_2^2}\mathbf{A}_{1}\mathbf{X}_1\mathbf{B}_{1} \mathbf{\Phi} \mathbf{A}_{2}\mathbf{X}_2\mathbf{B}_{2} \mathbf{P} \mathbf{B}_{2}^{H}\mathbf{X}_2^H\mathbf{A}_{2}^{H} \mathbf{\Phi}^H\mathbf{B}_{1}^{H}\mathbf{X}_1^H\mathbf{A}_{1}^{H}  + \frac{\sigma_1^2}{ \sigma_2^2} \mathbf{A}_{1}\mathbf{X}_1\mathbf{B}_{1} \mathbf{\Phi}\mathbf{\Phi}^H\mathbf{B}_{1}^H\mathbf{X}_1^H\mathbf{A}_{1}^{H}  +  \mathbf{I}_N \Big) \\
    & =  \log \det \Big(\frac{1}{\sigma_2^2}\mathbf{R}_1^{\frac{1}{2}}\widetilde{\mathbf{X}}_1 \mathbf{T}_1^{\frac{1}{2}} \mathbf{R}_{2}^{\frac{1}{2}}\widetilde{\mathbf{X}}_2\mathbf{T}_{2}\widetilde{\mathbf{X}}_2^H \mathbf{R}_{2}^{\frac{H}{2}}\mathbf{T}_1^{\frac{H}{2}} \widetilde{\mathbf{X}}_1^H\mathbf{R}_{1}^{\frac{H}{2}}  + \frac{\sigma_1^2}{ \sigma_2^2} \mathbf{R}_1^{\frac{1}{2}}\widetilde{\mathbf{X}}_1\mathbf{T}_1\widetilde{\mathbf{X}}_1^H\mathbf{R}_{1}^{\frac{H}{2}}  +  \mathbf{I}_N \Big),
\end{split}
\end{equation}
where $\mathbf{R}_i = \mathbf{U}_{A_i} \mathbf{\Sigma}_{A_i}^2 \mathbf{U}_{A_i}^H$, $\mathbf{T}_i = \mathbf{V}_{B_i} \mathbf{\Sigma}_{B_i}^2 \mathbf{V}_{B_i}^H$, and $\widetilde{\mathbf{X}}_i = \mathbf{W}_{A_i}^H\mathbf{X}_i\mathbf{W}_{B_i}^H, i=1,2$. 
Note that $\mathbf{W}_{A_i}$ and $\mathbf{W}_{B_i}, i=1,2$ are unitary matrices, and 
$\widetilde{\mathbf{X}}_i$ and $\mathbf{X}_i$ have the same distribution for $i=1,2$ 
due to the rotational invariance of Gaussian distribution. With the same argument, we can get
\begin{equation}
     I_2(\sigma_1^2, \sigma_2^2) = \log \det \Big(\frac{\sigma_1^2}{\sigma_2^2} \mathbf{R}_1^{\frac{1}{2}}\widetilde{\mathbf{X}}_1\mathbf{T}_1\widetilde{\mathbf{X}}_1^H\mathbf{R}_{1}^{\frac{H}{2}} + \mathbf{I}_N \Big).
\end{equation}
Next, we define two random matrices 
\begin{equation}
\mathbf{H}_1 = \mathbf{R}_{1}^{\frac{1}{2}}\mathbf{X}_1\mathbf{T}_{1}^{\frac{1}{2}}, ~~ \mathbf{H}_2 = \mathbf{R}_{2}^{\frac{1}{2}}\mathbf{X}_2\mathbf{T}_{2}^{\frac{1}{2}}. \label{Eq_Channel_Model}
\end{equation}
Then \eqref{MI_1} and \eqref{MI_2} are statistically equivalent to
\begin{subequations}
\label{MI_seq}
    \begin{align}
        I_1(\sigma_1^2, \sigma_2^2) &= \log \det \left(\frac{1}{\sigma_2^2}(\mathbf{H}_1  \mathbf{H}_2 \mathbf{H}_2^H \mathbf{H}_1^H + {\sigma}_1^2 \mathbf{H}_1\mathbf{H}_1^H ) +  \mathbf{I}_N  \right), \label{MI_1_seq} \\
        I_2(\sigma_1^2, \sigma_2^2) &=
         \log\det \left(\frac{\sigma_1^2}{\sigma_2^2} \mathbf{H}_1\mathbf{H}_1^H + \mathbf{I}_N\right). \label{MI_2_seq} %
    \end{align}
    \end{subequations} 
In this paper, we will consider a more general case and investigate the asymptotic distribution of the following random vector, 
\begin{equation}
\mathbf{m}(\overline{\sigma}_1^2, \underline{\sigma}_1^2, \sigma_2^2) 
= \begin{bmatrix}
    I_1(\overline{\sigma}_1^2,\sigma_2^2)\\
    I_2(\underline{\sigma}_1^2, \sigma_2^2)
\end{bmatrix}, \label{m_Joint_I_1_I_2}
\end{equation}
where $\overline{\sigma}_1^2 \geq 0$ and $\underline{\sigma}_1^2 \geq 0$ are not necessary equal.

\subsection{Assumptions}
To continue the study, we make the following mild assumptions
\begin{assumptionp}{A.1} \label{A-1}
    $0  \leq  \underset{N \rightarrow + \infty}{\lim\inf} \frac{M}{N}  \leq   \underset{N \rightarrow + \infty}{\lim\sup} \frac{M}{N} < + \infty $, and  $0 <  \underset{N \rightarrow + \infty}{\liminf} \frac{L}{N} 
     \leq  \underset{N \rightarrow + \infty}{\limsup} \frac{L}{N} < + \infty $.
\end{assumptionp}
\begin{assumptionp}{A.2} \label{A-2}
    $ \underset{N \geq 1}{\sup} \max \big\{ \norm{ \mathbf{R}_1 }, \norm{ \mathbf{R}_2 }, \norm{ \mathbf{T}_1 }, \norm{ \mathbf{T}_2} \big\} \leq r $, where $r < + \infty$ is a positive constant.
\end{assumptionp}
\begin{assumptionp}{A.3} \label{A-3}
    $ \underset{N \geq 1}{\inf}\min\big\{ \frac{1}{L}\mathrm{Tr}(\mathbf{R}_2\mathbf{T}_1), \frac{1}{N}\mathrm{Tr}\mathbf{R}_1 , \frac{1}{L}\mathrm{Tr}\mathbf{R}_2, \frac{1}{L}\mathrm{Tr}\mathbf{T}_1, \frac{1}{M}\Tr\mathbf{T}_2\big\} \geq l$, where $l > 0$ is a positive constant.
\end{assumptionp}
\textbf{\ref{A-1}} assumes that the system dimensions $L$, $M$, and $N$ are comparable. More specifically, $L = L(N)$ and $M = M(N)$ form integer sequences that 
increase to infinity in the same order with $N$, which is a common assumption in large system analysis. 
\textbf{\ref{A-2}} implies that all the eigenvalues of $\mathbf{R}_i$ and $\mathbf{T}_i$ are uniformly 
bounded in $N$. We can observe that if the spectral norm of the amplification or reflection matrix $\mathbf{\Phi}$ and that of the 
transmit covariance matrix $\mathbf{P}$ is uniformly bounded, and the singular values of the antenna correlation matrices 
$\mathbf{A}_i$ and $\mathbf{B}_i$, $i=1,2$, do not diverge to infinity, then assumption \textbf{\ref{A-2}} will be satisfied. 
\textbf{\ref{A-3}} guarantees that the extremely low-rank case of $\mathbf{R}_2\mathbf{T}_1$, $\mathbf{R}_i$, and $\mathbf{T}_i$ for $i=1,2$, i.e., 
the rank of these matrices does not increase with system dimensions, will not occur. 
\section{Useful Notations and the Main Results}
\label{Sec_Useful_Notations_and_Statement_of_the_Main_Results}
\subsection{Useful Notations}
Before presenting the main results, we first introduce several useful notations that will 
be utilized throughout the rest of this paper, along with some related properties.
\subsubsection{Fundamental System of Equations}
First, we introduce the key fixed-point system of equations 
that describe the statistical properties of the concerned two MI terms. 
The first-order and the second-order results are related to the following system of equations
\begin{subequations}
\label{DE_system_12}
\begin{align}
&\left\{ \begin{array}{lr}
        \delta(\overline{s}, z) =  \frac{1}{L} \mathrm{Tr}\big[ \mathbf{R}_1(z\mathbf{I}_N + (\overline{s} \overline{\omega}\left(\overline{s}, z\right) + \gamma\left(\overline{s}, z\right)\underline{\omega}(\overline{s}, z))\mathbf{R}_1)^{-1} \big] ,&\\
        \overline{\omega}(\overline{s}, z) =  \frac{1}{L} \mathrm{Tr} \big[\mathbf{T}_1(\mathbf{I}_L + \overline{s} \delta(\overline{s}, z) \mathbf{T}_1 + \delta(\overline{s}, z) \gamma(\overline{s}, z) \mathbf{R}_2 \mathbf{T}_1 )^{-1}\big] ,&\\
        \underline{\omega}(\overline{s}, z)  = \frac{1}{L} \mathrm{Tr} \big[\mathbf{R}_2 \mathbf{T}_1(\mathbf{I}_L + \overline{s}  \delta(\overline{s}, z) \mathbf{T}_1 + \delta(\overline{s}, z) \gamma(\overline{s}, z) \mathbf{R}_2 \mathbf{T}_1 )^{-1}\big], & \\
        \gamma(\overline{s}, z) = \frac{1}{M} \mathrm{Tr}\big[\mathbf{T}_2(\mathbf{I}_M + \frac{L}{M}\delta(\overline{s}, z)\underline{\omega}(\overline{s}, z)  \mathbf{T}_2)^{-1}\big],
    \end{array} \label{DE_system_1}%
\right. \\ 
& \left\{ \begin{array}{lr}
        \tau(\underline{s},z) =  \frac{1}{L} \mathrm{Tr}\big[ \mathbf{R}_1(z\mathbf{I}_N + \underline{s} \overline{\tau}(\underline{s},z) \mathbf{R}_1)^{-1} \big] ,&\\
        \overline{\tau}(\underline{s},z) =  \frac{1}{L} \mathrm{Tr} \big[\mathbf{T}_1(\mathbf{I}_L + \underline{s} \tau(\underline{s},z) \mathbf{T}_1 )^{-1}\big], &\\
    \end{array} \label{DE_system_2}
\right.
\end{align} 
\end{subequations}
where \eqref{DE_system_1} and \eqref{DE_system_2} are related to $I_1(\overline{s}, z)$ and $I_2(\underline{s}, z)$, respectively. 
In fact, \eqref{DE_system_2} also occurs for the MI analysis of single-hop MIMO channels \cite{Hachem2008ANewApproach}. In particular, by setting 
$\delta^{\mathrm{NEW}}(t) = \frac{1}{t} \tau(1, \frac{1}{t})$ and $\widetilde{\delta}^{\mathrm{NEW}}(t) = \overline{\tau}(1, \frac{1}{t})$, 
\eqref{DE_system_2} can be transformed to \cite[Eq. (2)]{Hachem2008ANewApproach}. 
For ease of illustration, we will omit the variables $\overline{s}, \underline{s}, z$  
in functions $(\delta(\overline{s},z), \overline{\omega}(\overline{s},z), 
\underline{\omega}(\overline{s},z), \gamma(\overline{s},z))$ and $(\tau(\underline{s},z), \overline{\tau}(\underline{s},z))$ when appropriate 
 in the following. 
The existence and uniqueness of the positive solutions to 
system \eqref{DE_system_2} can be obtained by 
following the methods in \cite[Proposition 1]{Hachem2008ANewApproach}.
However, the existence and uniqueness of the solutions to system \eqref{DE_system_1}
have not been proved in the literature, and no algorithm is available for their numerical calculation. 
The following proposition guarantees the uniqueness of the solutions. 
\begin{proposition} \label{Prop_Stieltjes}
    Assuming that assumptions \textbf{\ref{A-1}}-\textbf{\ref{A-3}} hold, 
    the system of equations in \eqref{DE_system_1}  admit unique positive solutions $(\delta, \overline{\omega}, \underline{\omega}, \gamma)$ for $z > 0$, $\overline{s} \geq 0$.
\end{proposition}
\textit{Proof}: The proof is given in Appendix \ref{App_Prop_Stiltjes}. 
\begin{remark}
    The existence and analytical properties of the solutions to the fundamental fixed point equations \eqref{DE_system_1},
     as well as the numerical iteration method for computing the positive solutions, are also provided in Appendix \ref{App_Prop_Stiltjes}.
\end{remark}
\subsubsection{The Resolvent} 
Resolvent has wide applications 
in the analysis of eigenvalue distributions and linear spectral statistics (LSS) of large random matrices \cite{bai2010spectral,pastur2011eigenvalue}, and 
MI is an important type of LSS. To investigate the statistical properties of the MI terms in this paper,  we define
two resolvent matrices
\begin{subequations}
\begin{align}
    \mathbf{Q}_1(\overline{s}, z) &= \left(\mathbf{H}_1 \mathbf{H}_2  \mathbf{H}_2^H \mathbf{H}_1^H + \overline{s} \mathbf{H}_1\mathbf{H}_1^H+ z \mathbf{I}_N \right)^{-1}, \label{Res_Q1}\\
    \mathbf{Q}_2(\underline{s}, z) &= \left(\underline{s} \mathbf{H}_1\mathbf{H}_1^H + z \mathbf{I}_N \right)^{-1}. \label{Res_Q2}
\end{align} \label{Res_matrices}%
\end{subequations}
In the sequel, we will omit the variables $\overline{s}$, $\underline{s}$ and $z$ for ease of presentation. 
The following resolvent identities will also be utilized 
\begin{subequations}
\begin{align}
    & \mathbf{Q}_1\mathbf{H}_1 \mathbf{H}_2\mathbf{H}_2^H\mathbf{H}_1^H  + \overline{s}\mathbf{Q}_1\mathbf{H}_1\mathbf{H}_1^H + z \mathbf{Q}_1 = \mathbf{I}_N, \label{Resolvent_identity_Q1}\\
    & \underline{s} \mathbf{Q}_2 \mathbf{H}_1\mathbf{H}_1^H + z \mathbf{Q}_2 = \mathbf{I}_N. \label{Resolvent_identity_Q2}
\end{align}
\end{subequations}
\subsubsection{Deterministic Quantities}
 Denote the positive solutions for 
 \eqref{DE_system_1} and \eqref{DE_system_2} by 
 $(\delta, \overline{\omega}, \underline{\omega}, \gamma)$ and $(\tau, \overline{\tau})$, respectively. Define deterministic matrices
\begin{align}
    & \mathbf{F}_{\delta} = (z\mathbf{I}_N + (\overline{s} \overline{\omega} + \gamma\underline{\omega})\mathbf{R}_1)^{-1},\mathbf{F}_{\omega} = (\mathbf{I}_L + \overline{s}\delta \mathbf{T}_1 + \delta \gamma \mathbf{R}_2 \mathbf{T}_1 )^{-1}, \mathbf{F}_{\gamma} = (\mathbf{I}_M + \frac{L}{M}\delta\underline{\omega}  \mathbf{T}_2)^{-1}, \notag \\
    & \mathbf{G}_{\tau} = (z\mathbf{I}_N + \underline{s} \overline{\tau} \mathbf{R}_1)^{-1}, \mathbf{G}_{\overline{\tau}} = (\mathbf{I}_L + \underline{s}\tau \mathbf{T}_1 )^{-1}.
\end{align}
We will show that $\mathbf{F}_{\delta}$ and $ \mathbf{G}_{\tau}$ are good approximations of $\mathbb{E}\mathbf{Q}_1$ and $\mathbb{E} \mathbf{Q}_2$, respectively.
For ease of reading, we gather the notations used in the derivation in Table \ref{tabel_of_notations}.
\begin{table}[t]
\centering
\caption{Table of Main Notations}
\label{tabel_of_notations}
 \begin{tabular}{|c c| c c| c c |} 
\toprule[1pt]
\midrule
 Notation & Expression & Notation & Expression & Notation & Expression\\ [0.5ex] 
\midrule
 $\delta_k$ & $\frac{1}{L} \mathrm{Tr} [(\mathbf{R}_1 \mathbf{F}_\delta)^k]$ & $\delta_{k,I}$ & $\frac{1}{L} \mathrm{Tr} [(\mathbf{R}_1 \mathbf{F}_\delta)^{k-1}\mathbf{F}_\delta ]$ &$\overline{\omega}_k$ & $\frac{1}{L}  \mathrm{Tr} [(\mathbf{T}_1 \mathbf{F}_\omega)^k]$ \\ 
 [1ex]
 $\underline{\omega}_k$ & $\frac{1}{L}  \mathrm{Tr} [(\mathbf{R}_2 \mathbf{T}_1 \mathbf{F}_\omega)^k]$ &$\underline{\omega}_{k, I}$ & $\frac{1}{L}  \mathrm{Tr} [(\mathbf{R}_2 \mathbf{T}_1 \mathbf{F}_\omega)^{k-1}\mathbf{F}_\omega]$& $\underline{\overline{\omega}}_{k, l}$ & $\frac{1}{L}\mathrm{Tr} [(\mathbf{T}_1 \mathbf{F}_\omega)^k (\mathbf{R}_2 \mathbf{T}_1 \mathbf{F}_\omega)^l ]$\\
 [1ex]
$\overline{\underline{\omega}}_{k,l,I}$ &$\frac{1}{L}  \mathrm{Tr} [(\mathbf{T}_1 \mathbf{F}_\omega)^k (\mathbf{R}_2 \mathbf{T}_1 \mathbf{F}_\omega)^{l-1}\mathbf{F}_\omega ]$ &  $\gamma_k$ & $\frac{1}{M} \mathrm{Tr} [(\mathbf{T}_2 \mathbf{F}_\gamma)^k]$ & $\varsigma$ & $2 \overline{s} \gamma \overline{\underline{\omega}}_{1,1} + \gamma^2 \underline{\omega}_2 + \overline{s}^2 \overline{\omega}_2$  \\
[1ex]
 $\Delta$ & $1 - \frac{L}{M} \gamma_2 \underline{\omega}_2 \delta^2 $ & $\Delta_{V_1}$ & $(1 - \varsigma \delta_2) \Delta - \frac{L}{M} \gamma_2 \underline{\omega}_{2, I}^2 \delta_2$ && \\
[1ex]
\midrule
$\tau_k$ & $\frac{1}{L} \mathrm{Tr}[(\mathbf{R}_1 \mathbf{G}_{\tau})^k]$ & $\tau_{k, I}$ & $\frac{1}{L} \mathrm{Tr}[(\mathbf{R}_1 \mathbf{G}_{\tau})^{k-1}\mathbf{G}_{\tau}]$ &$\overline{\tau}_k$ &$\frac{1}{L} \mathrm{Tr}[(\mathbf{T}_1 \mathbf{G}_{\overline{\tau}})^k]$ \\
[1ex]
$\Delta_{V_2}$& $1 - \underline{s}^2 \tau_2 \overline{\tau}_2$ &&&& \\
 [1ex]
 \midrule
$\vartheta$ & $\frac{1}{L} \mathrm{Tr} [\mathbf{R}_1 \mathbf{G}_\tau \mathbf{R}_1 \mathbf{F}_\delta ]$ & $\overline{\phi}$ & $\frac{1}{L}\mathrm{Tr} [\mathbf{T}_1 \mathbf{G}_{\overline{\tau}} \mathbf{T}_1 \mathbf{F}_\omega]$ & $\underline{\phi}$ & $\frac{1}{L}\mathrm{Tr} [ \mathbf{T}_1 \mathbf{G}_{\overline{\tau}} \mathbf{T}_1 \mathbf{F}_\omega \mathbf{R}_2]$ \\
[1ex]
$\vartheta_{k,l}$& $\frac{1}{L}\mathrm{Tr} [(\mathbf{R}_1 \mathbf{G}_{{\tau}})^k (\mathbf{R}_1 \mathbf{F}_\delta)^l]$ & $\vartheta_{k,l,I}$ & $\frac{1}{L}\mathrm{Tr} [(\mathbf{R}_1 \mathbf{G}_{\overline{\tau}})^k(\mathbf{R}_1 \mathbf{F}_\delta)^{l-1}\mathbf{F}_\delta]$ &$\vartheta_{k,I,l}$ & $\frac{1}{L}\mathrm{Tr} [(\mathbf{R}_1 \mathbf{G}_{{\tau}})^{k-1}\mathbf{G}_{{\tau}} (\mathbf{R}_1 \mathbf{F}_\delta)^l]$ \\
[1ex]
$\overline{\phi}_{k, l}$ &$\frac{1}{L}\mathrm{Tr} [(\mathbf{T}_1 \mathbf{G}_{\overline{\tau}})^k (\mathbf{T}_1 \mathbf{F}_\omega)^l]$
 &$\underline{\phi}_{k,l}$ & $\frac{1}{L}\mathrm{Tr} [(\mathbf{T}_1 \mathbf{G}_{\overline{\tau}})^k (\mathbf{T}_1 \mathbf{F}_\omega \mathbf{R}_2)^l]$ & $\underline{\overline{\phi}}_{1,2}$  & $\frac{1}{L}\mathrm{Tr} [\mathbf{T}_1 \mathbf{G}_{\overline{\tau}} \mathbf{T}_1 \mathbf{F}_\omega \mathbf{R}_2 \mathbf{T}_1 \mathbf{F}_\omega]$ \\
[1ex]
$\widetilde{\overline{\underline{\phi}}}_{1,2}$ &$\frac{1}{L}\mathrm{Tr} [ \mathbf{T}_1 \mathbf{G}_{\overline{\tau}} \mathbf{T}_1 \mathbf{F}_\omega  \mathbf{T}_1 \mathbf{F}_\omega\mathbf{R}_2]$& $\Delta_C$ & $1 - \underline{s} \vartheta (\gamma \underline{\phi} + \overline{s}\overline{\phi})$ && \\
[1ex]
\bottomrule[1pt]
 \end{tabular}
\end{table}
\par
In Table \ref{tabel_of_notations}, $k$ and $l$ are positive integers. 
 The notations in the table are divided into three blocks. 
 The top block is defined by 
 matrices $\mathbf{F}_{\delta}, \mathbf{F}_{\omega}$, and $\mathbf{F}_{\gamma}$. 
 The term $\Delta_{V_1}$  is related to the variance of $I_1$ and the details are presented in 
 Theorem \ref{Thm_Second-order}. 
 Similarly, the terms $\Delta_{V_2}$ in the middle block and $\Delta_{C}$ in the bottom block are related to the variance of $I_2$ and the covariance between $I_1$ and $I_2$, respectively. In the following proposition, we provide the regularity conditions for these terms.
\begin{proposition} \label{Prop_tightness}
    Given that assumptions \textbf{\ref{A-1}}-\textbf{\ref{A-3}} hold true, if we define 
\begin{equation}
    \mathbf{V}(\overline{s}, \underline{s}, z) = \begin{bmatrix}
        -\log(\Delta_{V_1})& -\log(\Delta_{C}) \\
        -\log(\Delta_C)& -\log(\Delta_{V_2})
    \end{bmatrix}, \hspace*{1mm}\overline{s},\hspace*{1mm} \underline{s} \geq 0,\hspace*{1mm} z > 0, \label{Cov_matrix}
\end{equation}
    then there exist two positive numbers $m_V(\overline{s}, \underline{s}, z)$ and $M_V(\overline{s}, \underline{s}, z)$ such that
    \begin{equation}
       0 < m_V(\overline{s}, \underline{s}, z) \leq \inf_{N \geq 1} \lambda_{min}(\mathbf{V}(\overline{s}, \underline{s}, z)) \leq \sup_{N \geq 1} \lambda_{max}(\mathbf{V}(\overline{s}, \underline{s}, z)) \leq M_V(\overline{s}, \underline{s}, z) < + \infty , 
       \label{regularity_cond}
    \end{equation}
    where  $\lambda_{min}(\mathbf{V}(\overline{s}, \underline{s}, z))$ and $\lambda_{max}(\mathbf{V}(\overline{s}, \underline{s}, z))$ 
    represent the minimum and maximum eigenvalue of $\mathbf{V}(\overline{s}, \underline{s}, z)$, respectively.
\end{proposition}
\textit{Proof}: The proof of Proposition \ref{Prop_tightness} is given in Appendix \ref{App_Prop_tightness}.
\begin{remark}
    Proposition \ref{Prop_tightness} implies that  $\mathbf{V}$
     is positive definite and hence invertible uniformly in $N$. Different from previous works 
     \cite[Proposition 2]{Hachem2008ANewApproach}, \cite[Proposition 1]{Xin2023Double-Scattering} that only discussed the scalar cases,
      the covariance matrix of the joint distribution is given in this paper. Specifically, the matrix $\mathbf{V}$ is a good approximation for the covariance matrix of
      $\mathbf{m}$, and more details will be given in Theorem \ref{Thm_Second-order}. The boundedness property of the eigenvalues for $\mathbf{V}$ indicates
      $\{ \mathbf{m} \}_{N \geq 1}$ is tight.
      Note that in \cite{AFMIMO2006Replica}, the authors used \cite[Theorem 1]{Normal1988} to prove the Gaussianity, without validating the conditions of \cite[Theorem 1]{Normal1988}, i.e., the convergence of the mean and variance of the MI to a constant. 
      From the proof of Proposition \ref{Prop_tightness}, we note that
      the properties of $\Delta_{V1}$, $\Delta_{V_2}$, and $\Delta_C$ are only related to the correlation matrices $\mathbf{R}_i, \mathbf{T}_i$ and these terms may not necessarily converge as $N$ goes to infinity.
\end{remark}
\subsection{Main Results}
\subsubsection{First-order Analysis}We give the approximation of EMI and the convergence order of the approximation result
in the following Lemma \ref{Lemma_EMI_2} and Theorem \ref{Th_First_Order}.
\begin{lemma} \label{Lemm_DE_I_2} 
\label{Lemma_EMI_2}
{\cite[Theorem 1]{Hachem2008ANewApproach}}
    Given that assumptions \textbf{\ref{A-1}} - \textbf{\ref{A-3}} hold true and 
     $(\tau, \overline{\tau})$ are the positive solutions of \eqref{DE_system_2} 
     when $\underline{s} = \underline{\sigma}_1^2$ and $z = \sigma_2^2$, the expected value of  $I_2$ can be approximated by
\begin{equation}
    \mathbb{E} I_2(\underline{\sigma}_1^2, \sigma_2^2) = \overline{I}_2(\underline{\sigma}_1^2, \sigma_2^2) + \mathcal{O}(\frac{1}{N}),
\end{equation}
where
\begin{equation}
     \overline{I}_2(\underline{\sigma}_1^2, \sigma_2^2) = \log\det(\mathbf{I}_N + \frac{\underline{\sigma}_1^2\overline{\tau}}{\sigma_2^2}  \mathbf{R}_1) + \log \det(\mathbf{I}_L + \underline{\sigma}_1^2\tau \mathbf{T}_1 ) - L \underline{\sigma}_1^2 \tau \overline{\tau}.  \label{I_2}
\end{equation}
\end{lemma}
\begin{theorem} \label{Th_First_Order}
Given that assumptions \textbf{\ref{A-1}} - \textbf{\ref{A-3}} hold true 
and $(\delta, \overline{\omega}, \underline{\omega}, \gamma)$ are the positive 
solutions of \eqref{DE_system_1} when $\overline{s} = \overline{\sigma}_1^2$ and $z = \sigma_2^2$ , 
the expected value of $I_1$ can be approximated by
\begin{equation}
    \mathbb{E}I_1(\overline{\sigma}_1^2, \sigma_2^2) = \overline{I}_1(\overline{\sigma}_1^2, \sigma_2^2) + \mathcal{O}(\frac{1}{N}),
\label{Thm_Eq_DE_I_1}
\end{equation}
where
\begin{equation}
\begin{split}
\overline{I}_1(\overline{\sigma}_1^2, \sigma_2^2) &= \log\det(\mathbf{I}_N + \frac{\overline{\sigma}_1^2  \overline{\omega} + \gamma\underline{\omega}}{\sigma_2^2}\mathbf{R}_1) + \log \det (\mathbf{I}_L + \overline{\sigma}_1^2\delta \mathbf{T}_1 + \delta \gamma \mathbf{R}_2 \mathbf{T}_1 )  \\
&+ \log \det (\mathbf{I}_M + \frac{L}{M}\delta\underline{\omega}  \mathbf{T}_2)- \overline{\sigma}_1^2 L\delta \overline{\omega} - 2L\delta \underline{\omega}\gamma. \label{I_1}
\end{split}
\end{equation}
\end{theorem}
\textit{Proof}: The proof of Theorem \ref{Th_First_Order} is given in Section \ref{Sec_First-order}. \QED
\begin{remark} \label{remark_first_order}
    Combining Lemma \ref{Lemma_EMI_2} and Theorem \ref{Th_First_Order}, 
    we can obtain an approximation of the EMI for the two-hop system as 
     $\mathbb{E} I = \overline{I}_1 - \overline{I}_2 + \mathcal{O}(\frac{1}{N})$. 
     This result can degenerate to some special scenarios. 
     When both $\overline{\sigma}_1^2$ and $\underline{\sigma}_1^2$ are 
     set to $0$, Theorem \ref{Th_First_Order} degenerates to the EMI of 
     double-scattering channels\cite{Xin2023Double-Scattering, Hoydis2011double-scattering} or
      passive IRS-aided system over Rayleigh fading \cite{Zhangjun2020IRS}. 
      When  $\overline{\sigma}_1^2 = \underline{\sigma}_1^2$, the result is identical to that in \cite[Theorem 1]{AFMIMO2006Replica}.
\end{remark}
\subsubsection{Second-order Analysis} The CLT for the joint distribution of $I_1$ and $I_2$ is given by the following theorem.
\begin{theorem}{(Joint CLT)} \label{Thm_Second-order}
    Given that assumptions \textbf{\ref{A-1}}-\textbf{\ref{A-3}} hold true and $(\delta, \overline{\omega}, \underline{\omega}, \gamma)$,  
    $(\tau, \overline{\tau})$ are the positive solutions of \eqref{DE_system_12}, when $\overline{s} = \overline{\sigma}_1^2$, 
    $\underline{s} =  \overline{\sigma}_1^2$ and $z = \sigma_2^2$, the joint distribution of $I_1$ and $I_2$ will converge in distribution to Gaussian, 
\begin{equation}
    \mathbf{V}^{-\frac{1}{2}}(\overline{\sigma}_1^2, \underline{\sigma}_1^2, \sigma_2^2)\begin{bmatrix}
        I_1(\overline{\sigma}_1^2, \sigma_2^2) - \overline{I}_1(\overline{\sigma}_1^2, \sigma_2^2) \\
        I_2(\underline{\sigma}_1^2, \sigma_2^2) - \overline{I}_2(\underline{\sigma}_1^2, \sigma_2^2)
    \end{bmatrix} \xrightarrow[N \rightarrow + \infty ]{d} \mathcal{N}(\mathbf{0}_2, \mathbf{I}_2),
\end{equation}
where $\overline{I}_1(\overline{\sigma}_1^2, \sigma_2^2)$ and $\overline{I}_2(\underline{\sigma}_1^2, \sigma_2^2)$ are given in 
\eqref{I_1} and \eqref{I_2}, respectively. 
The covariance matrix $\mathbf{V}(\overline{\sigma}_1^2, \underline{\sigma}_1^2, \sigma_2^2)$ is given in \eqref{Cov_matrix}.
Furthermore, the convergence of the covariance matrix is given by
\begin{equation}
\norm{\mathbf{V}(\overline{\sigma}_1^2, \underline{\sigma}_1^2, \sigma_2^2) - \mathrm{Cov}\left(
\begin{bmatrix}
    I_1(\overline{\sigma}_1^2, \sigma_2^2)\\
    I_2(\underline{\sigma}_1^2, \sigma_2^2)
\end{bmatrix}
\right)} = \mathcal{O}(\frac{1}{N}).
\end{equation}
\end{theorem}
\textit{Proof}: The proof of Theorem \ref{Thm_Second-order} is given in Section \ref{Sec_Second-order}.
\begin{remark} Similar to Remark \ref{remark_first_order}, the CLT 
    results can degenerate to Theorem 2 in \cite{Xin2023Double-Scattering} 
and Theorem 1 in \cite{AFMIMO2006Replica}. 
In the numerical experiments, we will show that the convergence rate $\mathcal{O}(\frac{1}{N})$ is tight.
\end{remark}
By using Theorem \ref{Thm_Second-order}, we can obtain the asymptotic distribution of $I$. In fact, when $\overline{\sigma}_1^2 = \underline{\sigma}_1^2 = {\sigma}_1^2$, 
 we have 
$I \xrightarrow[]{d} \mathcal{N}(0, [\mathbf{V}]_{11} + [\mathbf{V}]_{22} - 2[\mathbf{V}]_{12}) $.
Therefore, we can obtain an approximation of the outage probability $p_{out}$ for a given transmission 
rate threshold $R$ and conversely the outage rate for a given outage probability $p_{out}$ as, respectively,
\begin{subequations}
\begin{align}
    p_{out} &= \mathbb{P}\left( I \leq R \right) \approx 1 - Q\big( \frac{R - \overline{I}_1 + \overline{I}_2}{\sqrt{[\mathbf{V}]_{11} + [\mathbf{V}]_{22} - 2[\mathbf{V}]_{12}}} \big), \label{outage_prob}\\
    C_{out} &= \sup_{R\geq 0} \left\{ \mathbb{P}\left( I \leq R \right) \leq p_{out} \right\} \approx \overline{I}_1 - \overline{I}_2 + \sqrt{[\mathbf{V}]_{11} + [\mathbf{V}]_{22} - 2[\mathbf{V}]_{12}}Q^{-1}(1 - p_{out}),
\end{align}
\end{subequations}
where $Q(x)$ is defined as $Q(x) = \frac{1}{\sqrt{2\pi}}\int_{x}^{\infty} \exp\{-\frac{t^2}{2}\} \mathrm{d} t$. 
\begin{figure}[t]
    \centering
    \includegraphics[width=5.6in]{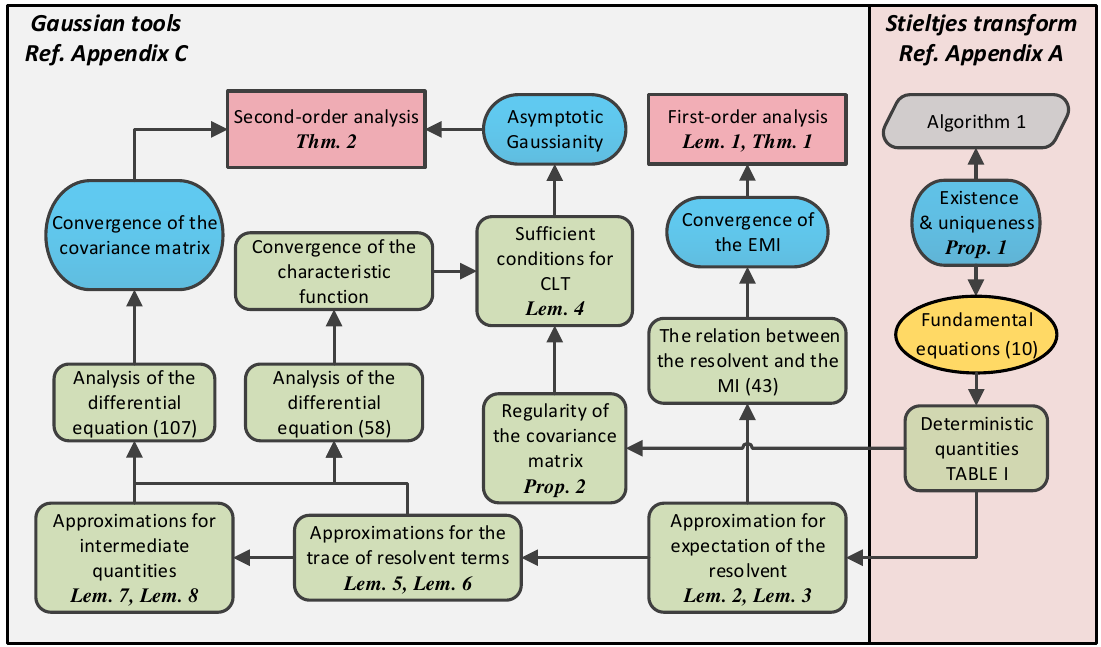}
    \caption{Diagram for the logic flow of this paper}
    \label{Logic_glow}
\end{figure}
\par
\subsubsection{Logic Flow}
Due to the complexity of the concerned problem, a much involved research is required. 
In order to clearly illustrate the relationships between the key results, we have summarized the logic flow in Fig. \ref{Logic_glow}. This paper consists of two major technical contributions, i.e., 
the first and second-order analysis for the MI and the existence and uniqueness proof for the solutions of the fixed point systems, as shown in the two blocks of Fig. \ref{Logic_glow}. The former relies mainly on Gaussian tools and the latter involves Stieltjes transform and its corresponding properties. These mathematical tools will be introduced later in this paper. The red square boxes in Fig. \ref{Logic_glow} represent the main results, the blue boxes denote the corresponding properties, and the green boxes illustrate the derivation steps. The yellow elliptical box and the gray parallelogram shaped box represent the fundamental equations \eqref{DE_system_12} and Algorithm \ref{DE_algorithm}, respectively.
\subsection{Case study: i.i.d.}
\label{Sec_CaseIID}
To get more physical insights of the two-hop system, in this subsection we consider the special case where 
all the matrices $\mathbf{A}_i$, $\mathbf{B}_i$, $\mathbf{P}$, and 
$\mathbf{\Phi}$ in \eqref{Channel_Model} and \eqref{MI_total} are identity matrices with appropriate dimensions.
As a result, $\mathbf{R}_i$ and $\mathbf{T}_i$ in \eqref{Eq_Channel_Model} are identity matrices.
For ease of illustration, we first define two ratios related to the number of antennas $c_1 = \frac{N}{L}, c_2 = \frac{L}{M}$ and two quantities 
$m_{F} = \frac{1}{N} \Tr \mathbf{F}_{\delta}$, $m_{G} = \frac{1}{N} \Tr \mathbf{G}_{\tau}$. Plugging $m_F$ and $m_G$ into \eqref{DE_system_12}, we get
two fundamental equations
\begin{subequations}
\begin{align}
     \mathcal{L}_F(m_F) &= c_1c_2 m_F (c_1 z m_F + 1 - c_1)\left( zm_F - 1 + \overline{s} m_F \left( c_1 z m_F  + 1 - c_1 \right)\right) 
     \notag \\
     &+ (\overline{s} + 1) m_F \left(c_1 z m_F + 1 - c_1 \right) + z m_F - 1  = 0, \label{DE_system_iid_1} \\  
     \mathcal{L}_G(m_G) &= \underline{s} m_G(c_1zm_G + 1 - c_1) + z m_G - 1 = 0, \label{DE_system_iid_2}
\end{align}
\label{DE_system_iid_2_eq}%
\end{subequations}
where \eqref{DE_system_iid_1} is a quartic equation obtained from \eqref{DE_system_1} by using $\delta = c_1 m_F$ to
express other terms  as $\overline{\omega} = \underline{\omega} = c_1zm_F +1 - c_1$ and $\gamma = 1 + c_1c_2 (zm_F-1+\overline{s}m_F(zc_1m_F+1-c_1)) $.
Similarly, \eqref{DE_system_iid_2} is a quadratic equation derived from \eqref{DE_system_2}. Similar to Proposition \ref{Prop_Stieltjes},
 we have the following properties for the above equations.
\begin{proposition}
    For $\overline{s}, \underline{s} \geq 0 $ and $ z > 0$, the equations $\mathcal{L}_F(x) = 0 $
    and $\mathcal{L}_F(y) = 0$ have unique solutions in the interval $(\max(0, \frac{1 - \frac{1}{c_1}}{z}), \frac{1}{z})$.
\end{proposition}
\textit{Proof:} For a quadratic equation \eqref{DE_system_iid_2}, 
one just need to check its roots. As for \eqref{DE_system_iid_1}, we can
check $\mathcal{L}_F( \frac{1 - \frac{1}{c_1}}{z}) = -\frac{1}{c_1} < 0$, $\mathcal{L}_F( 0) = -1 < 0$, and  $\mathcal{L}_F( \frac{1}{z}) = \frac{c_1c_2 \overline{s}}{z^2} + \frac{\overline{s} + 1}{z} > 0$, 
which implies that there exist roots in the interval $(\max(0, \frac{1 - \frac{1}{c_1}}{z}), \frac{1}{z})$. Assuming \eqref{DE_system_iid_1}
has two roots in this interval, denoted by $m_{F,1} < m_{F, 2}$, we can rewrite \eqref{DE_system_iid_1} as 
\begin{equation}
 \frac{1 - zm_{F, i}}{m_{F, i}(c_1 z m_{F, i} + 1 - c_1)} = \overline{s} + 1 + c_1c_2(zm_{F, i} - 1 + \overline{s}m_{F, i}(c_1 z m_{F, i} + 1 - c_1) ), i = 1, 2.
\label{EQ_re_iid}
\end{equation}
By defining the left-hand side (LHS) and right-hand side (RHS) of \eqref{EQ_re_iid} as $A_i$ and $B_i$, respectively, 
we have $0 < A_2 < A_1 = B_1 < B_2$. Therefore, $m_{F, 1} = m_{F,2}$ must hold. \QED
\begin{remark}
 For the i.i.d. case, the system is described by a quartic equation, 
 which is different from that for double-scattering channels \cite{Xin2023Double-Scattering}, 
 where the fundamenal equation is cubic.
Additionally, to obtain the positive solution for the i.i.d. case,
we can use Algorithm \ref{DE_algorithm} or apply classical root-finding algorithms on the interval
 $(\max(0, \frac{1 - \frac{1}{c_1}}{z}), \frac{1}{z})$ to get $m_F$.
 Both methods yield the same results.
\end{remark}
To simplify the discussion, we set $\overline{s} = \underline{s}$.
The following proposition presents the joint CLT for the i.i.d. case. 
\begin{corollary} (i.i.d. Joint CLT)
    Given  assumptions \textbf{\ref{A-1}}-\textbf{\ref{A-3}} hold and the matrices $\mathbf{R}_i$, $\mathbf{T}_i$, $i=1,2$,
     are identity matrices with appropriate  dimensions,
    we have
\begin{equation}
    \mathbf{V}_{\mathrm{iid}}^{-\frac{1}{2}}( \sigma_1^2, \sigma_2^2)\begin{bmatrix}
        I_{1, \mathrm{iid}}({\sigma}_1^2, \sigma_2^2) - \overline{I}_{1,\mathrm{iid}}({\sigma}_1^2, \sigma_2^2) \\
        I_{2, \mathrm{iid}}({\sigma}_1^2, \sigma_2^2) - \overline{I}_{2,\mathrm{iid}}({\sigma}_1^2, \sigma_2^2)
    \end{bmatrix} \xrightarrow[N \rightarrow + \infty ]{d} \mathcal{N}(\mathbf{0}_2, \mathbf{I}_2),
\end{equation}
where the mean is given by
\begin{subequations}
\begin{align}
    \overline{I}_{1,\mathrm{iid}}({\sigma}_1^2, \sigma_2^2)  &= - N \log(\sigma_2^2m_F) - L \log(c_1\sigma_2^2m_F + 1-c_1) 
    - M\log(1 + c_1c_2(\sigma_2^2 m_F - 1 + {\sigma}_1^2 m_F (c_1\sigma_2^2 m_F + 1 - c_1)) \notag\\
    &+N[2\sigma_2^2 m_F -2 + {\sigma}_1^2 m_F(c_1\sigma_2^2m_F + 1 -c_1)], \\
    \overline{I}_{2,\mathrm{iid}}({\sigma}_1^2, \sigma_2^2) &= -N \log(\sigma_2^2 m_G) 
    -  L \log(c_1\sigma_2^2m_G + 1 - c_1 ) + N(\sigma_2^2m_G - 1),
\end{align}
\label{iid_I}%
\end{subequations}
and the covariance matrix is given by
\begin{equation}
    \mathbf{V}_{\mathrm{iid}}(\sigma_1^2, \sigma_2^2) = \begin{bmatrix}
        -\log(\Delta_{V_1, \mathrm{iid}}) & -\log(\Delta_{C, \mathrm{iid}})\\
        -\log(\Delta_{V_C, \mathrm{iid}}) & -\log(\Delta_{V_2, \mathrm{iid}})
    \end{bmatrix},
    \label{iid_V}
\end{equation}
with
\begin{subequations}
    \begin{align}
        \Delta_{V_1, \mathrm{iid}} &= 1 + (2\sigma_2^2 + (c_1 - 1)(\sigma_1^2 + c_2 + 1))(\sigma_2^2 m_F - 1 + {\sigma}_1^2 m_F (c_1\sigma_2^2 m_F + 1 - c_1)) \notag\\
        & - c_1(\sigma_2^2 m_F -1)^2 + \frac{c_1-1}{c_1} + \frac{c_1+1}{c_1}(c_1\sigma_2^2 m_F + 1 - c_1) \notag\\
        &+ (2(\sigma_1^2+1) + c_2(c_1+1))[\sigma_2^2 m_F - 1 + {\sigma}_1^2 m_F (c_1\sigma_2^2 m_F + 1 - c_1)](c_1\sigma_2^2 m_F + 1 - c_1), \\ 
        \Delta_{V_2, \mathrm{iid}} &= (\frac{\sigma_2^2}{\sigma_1^2} + 1 + c_1) \sigma_2^2 m_G + 1 - c_1 - \frac{\sigma_2^2}{\sigma_1^2}, \\
        \Delta_{C, \mathrm{iid}} &= 1 - c_1(1 - \sigma_2^2 m_F)(1 - \sigma_2^2 m_G).
    \end{align}
\end{subequations}
\label{Prop_iid}
\end{corollary}
\textit{Proof:} The proof can be obtained directly by setting $\mathbf{R}_1 = \mathbf{I}_N, 
\mathbf{R}_2 = \mathbf{T}_1 = \mathbf{I}_L$ and $ \mathbf{T}_2 = \mathbf{I}_M$ in Theorem \ref{Thm_Second-order} and is omitted here. \QED
\par
In the following, we will derive some asymptotic results to reveal the physical insights.
In particular, we consider the case with $c_1 \rightarrow  0$, $c_2 \rightarrow  +\infty$ and $c_1c_2 \rightarrow  c$ where constant
$c \in (0, +\infty)$. 
Note that this asymptotic regime violates assumption \textbf{\ref{A-1}}, but the convergence of the mean and variance, as well as the CLT, could still hold. 
In particular, when $N=M$ and $L=N^{1+\epsilon}$ for sufficiently small $\epsilon$, it can be proved that as $N \rightarrow +\infty$, the convergence rate of the mean, covariance matrix, and characteristic function will become 
$\mathcal{O}(N^{K\epsilon - 1})$ for some constant $K$. By using the bounded properties of $m_F$, \eqref{DE_system_iid_1} becomes
\begin{equation}
    \mathcal{L}_F(m_F) \approx c (\sigma_2^2 + \sigma_1^2) m_F^2 +
    (\sigma_2^2 + \sigma_1^2 + 1 - c)m_F - 1 = 0,
\end{equation}
and the deterministic approximation for the mean of ${I}_{1, \mathrm{iid}}$ becomes
\begin{equation}
\begin{split}
    \overline{I}_{1,\mathrm{iid}}({\sigma}_1^2, \sigma_2^2)& \approx - N \log(\sigma_2^2m_F) 
    - M\log(1 + c(\sigma_2^2 + {\sigma}_1^2) m_F - c) +N[(\sigma_2^2 + {\sigma}_1^2) m_F - 1] \\
    & + \{ N  [ {\sigma}_1^2 m_F  - 1] - L \log(1 + c_1(\sigma_2^2m_F- 1)) \} \\
&\overset{(a)}{=} - N \log(\sigma_2^2m_F)  - M\log(1 + c(\sigma_2^2 + {\sigma}_1^2) m_F - c) +N[(\sigma_2^2 + {\sigma}_1^2) m_F - 1]\\
& + \{ N  [ {\sigma}_1^2 m_F  - 1] - L ( \frac{N}{L}(\sigma_2^2m_F- 1) - \frac{ N^2(\sigma_2^2m_F- 1)^2}{L^22}  + \cdots ) \} \\
& \approx - N \log(\sigma_2^2m_F) - M\log(1 + c(\sigma_2^2 + {\sigma}_1^2) m_F - c) +N[(\sigma_2^2 + {\sigma}_1^2) m_F - 1], \label{Approx_iid_I1}
\end{split}
\end{equation}
where $(a)$ follows by Taylor's formula. Similarly, we can get $\overline{I}_{2,\mathrm{iid}}({\sigma}_1^2, \sigma_2^2) \approx -N \log(\frac{\sigma_2^2}{\sigma_1^2 + \sigma_2^2})$, which when combined  with \eqref{Approx_iid_I1} gives
\begin{equation}
    \overline{I}_{iid} =  \overline{I}_{1,\mathrm{iid}} - \overline{I}_{2,\mathrm{iid}} \approx 
    - N \log((\sigma_2^2 + {\sigma}_1^2)m_F)  - M\log(1 + c(\sigma_2^2 + {\sigma}_1^2) m_F - c) + N[(\sigma_2^2 + {\sigma}_1^2) m_F - 1].
\label{inf_L_iid_I}
\end{equation}
We can also observe that in this asymptotic regime, $\Delta_{V_2, \mathrm{iid}} \approx 1$ and $\Delta_{C,\mathrm{iid}} \approx 1$, which means that $I_{2, \mathrm{iid}}$ becomes deterministic
and $I_{1, \mathrm{iid}}$ and $I_{2, \mathrm{iid}}$ are asymptotically uncorrelated. The term $\Delta_{V_1,\mathrm{iid}}$ can be approximated as
\begin{equation}
    \Delta_{V_1,\mathrm{iid}} \approx (\sigma_2^2 + \sigma_1^2 + 1 + c)(\sigma_2^2 + \sigma_1^2)m_F + 1 - c - (\sigma_2^2 + \sigma_1^2).
\end{equation} 
Thus when $N$, $L$, and $M$ go to infinity with $L$ growing at a suitably faster rate, $[-\log(\Delta_{V_1, \mathrm{iid}})]^{-\frac{1}{2}}({I}_{1, \mathrm{iid}} - \overline{I}_{1, \mathrm{iid}}) $ converges to standard Gaussian in distribution.
Moreover, by employing Markov’s inequality, for any $\varepsilon > 0$, there holds
\begin{equation}
    \begin{split}
    \mathbb{P}(\abs{{I}_{2, \mathrm{iid}} - \overline{I}_{2, \mathrm{iid}}} &\geq \varepsilon) \leq \frac{\mathbb{E}\abs{{I}_{2, \mathrm{iid}} - \overline{I}_{2, \mathrm{iid}}}^2 }{\varepsilon^2} = \frac{\Var(I_2) +\abs{ \mathbb{E} I_{2, \mathrm{iid}} - \overline{I}_{2, \mathrm{iid}}}^2 }{\varepsilon^2}\\
    & = \frac{-\log(\Delta_{V_2,\mathrm{iid}}) + o(1)}{\varepsilon^2} \xrightarrow{N, c_2 \rightarrow + \infty, c_1c_2 \rightarrow c} 0,
    \end{split}
\end{equation}
which implies ${I}_{2, \mathrm{iid}} - \overline{I}_{2, \mathrm{iid}}$ converges to 0 in probability. According to Slutsky’s theorem, the following holds
\begin{equation}
    \frac{I_{\mathrm{iid}} - \overline{I}_{\mathrm{iid}}}{\sqrt{-\log(\Delta_{V_1, \mathrm{iid}})}} \xrightarrow[N, c_2 \rightarrow + \infty, c_1c_2 \rightarrow c]{d} \mathcal{N}(0, 1). 
    \label{inf_L_iid_CLT}
\end{equation}
We give some physical insights in the following two remarks.
\begin{remark}(Comparison with single-hop Rayleigh MIMO Channels)
As $L$ approaches infinity at a rate faster than $M$ and $N$, 
    the first-order \eqref{inf_L_iid_I} and second-order \eqref{inf_L_iid_CLT} results 
    become equivalent to those of single-hop Rayleigh MIMO channels
    with AWGN power $\sigma_1^2 + \sigma_2^2$ at the receiver \cite{Kamath2005iidRayleigh}.
\end{remark}
\begin{remark}
(Large number of reflecting elements)
Here, we consider the active IRS-aided MIMO systems where the
number reflecting elements is very large compared to the number of transmitter and receiver antennas. With i.i.d. channels, the reflection matrix is given by $\mathbf{\Phi} = a\mathbf{I}_L$, where $a$ is determined by the power constraint for the active IRS 
\cite{zhuang2023active}. Thus $\sigma_{1}^2$ and $\sigma_{1}^2$ are related to $L$. Moreover, 
from \cite{zhuang2023active} we have ${\lim}_{L \rightarrow + \infty} a(L) = 1$. Hence, $\sigma_{1}^2$ and $\sigma_{2}^2$ will converge and the asymptotic results \eqref{inf_L_iid_I} and \eqref{inf_L_iid_CLT} still hold. \eqref{inf_L_iid_I} indicates that when the number of reflecting elements increases while the total energy for the transmitted signal and amplification remains constant, the performance will eventually degrade. This is because each reflecting element of the active IRS will introduce noise. As the number of active IRS elements increases, the noise power increases, reducing signal-to-noise ratio (SNR) and throughput.
\end{remark}
\section{Proof of the First-order Result: Approximation for the Mean of $I_1$} \label{Sec_First-order}
In this section, we will focus on proving Theorem \ref{Th_First_Order}, which is based on the estimation of $\mathbb{E} \mathbf{Q}_1$. This will rely on an asymptotic RMT 
 method: Gaussian Tools, a brief introduction of which is provided in Appendix \ref{App_Lemma_DE_Q1_1}. We will divide the proof into the two steps:
\begin{itemize}
    \item [1)] Approximate the mean of the resolvent $\mathbf{Q}_1$, and give the
     convergence rate for elements and trace of the resolvent.
    \item [2)] Build up the relation between $I_1$ and $\mathbf{Q}_1$, and give an estimation of the convergence rate of $\mathbb{E} I_1$.
\end{itemize}
\subsection{Approximation of $\mathbb{E}\mathbf{Q}_1$}
\label{Approximation_of_EQ1}
\par 
In this section, we will show that $\mathbf{F}_{\delta}$ is a good approximation of $\mathbb{E} \mathbf{Q}_1$.
 It is difficult to directly transform $\mathbb{E} \mathbf{Q}_1$ to $\mathbf{F}_{\delta}$ since $\mathbf{F}_{\delta}$ 
 contains solutions of fixed points. In order to connect $\mathbf{F}_{\delta}$ and $\mathbb{E} \mathbf{Q}_1$, 
 we need to introduce some intermediate variables, including random and deterministic quantities.
 To this end, we first define two random quantities
\begin{equation}
    \widehat{\alpha}_{\delta} = \frac{1}{L}\mathrm{Tr}[\mathbf{R}_1 \mathbf{Q}_1] , ~~ \widehat{\kappa} = \mathrm{Tr} [ \frac{1}{M}\mathbf{R}_2\mathbf{H}_1^H \mathbf{Q}_1\mathbf{H}_1], \label{Random_quantities}
\end{equation}
and denote $\alpha_{\delta} = \mathbb{E}\widehat{\alpha}_{\delta}$, $\mathring\alpha_{\delta} =  \widehat{\alpha}_{\delta} - {\alpha}_{\delta}$, and $\kappa = \mathbb{E}\widehat{\kappa}$, $\mathring \kappa = \widehat{\kappa} - {\kappa}$.
Then, based on \eqref{Random_quantities}, we define the following deterministic quantities
\begin{align}
     &\mathbf{\Theta}_{\gamma} = (\mathbf{I}_M + \kappa \mathbf{T}_2)^{-1}, ~~  \alpha_{\gamma} = \frac{1}{M}\mathrm{Tr}[\mathbf{T}_2 \mathbf{\Theta}_{\gamma}], \notag \\
     &\mathbf{\Theta}_{\omega} = (\mathbf{I}_L + \overline{s}\alpha_{\delta} \mathbf{T}_1 + \alpha_{\delta} \alpha_{\gamma} \mathbf{R}_2 \mathbf{T}_1 )^{-1}, ~~ \alpha_{\overline{\omega}} = \frac{1}{L}\mathrm{Tr}[\mathbf{T}_1 \mathbf{\Theta}_{\omega}], ~~ \alpha_{\underline{\omega}} = \frac{1}{L}\mathrm{Tr}[\mathbf{R}_2\mathbf{T}_1 \mathbf{\Theta}_{\omega}],  \notag \\
      &\mathbf{\Theta}_{\delta} = (z\mathbf{I}_{N} + (\overline{s} \alpha_{\overline{\omega}} + \alpha_\gamma \alpha_{\underline{\omega}} )\mathbf{R}_1)^{-1}. \label{Deterministic_quantities}
\end{align}
Next, we will show that $\mathbf{\Theta}_{\delta}$ is a direct approximation 
of $\mathbb{E} \mathbf{Q}_1$ in the entry-wise and trace sense by Gaussian tools. 
Then, the term $\mathbf{\Theta}_{\delta}$ is connected to $\mathbf{F}_{\delta}$ by normal family theorem. 
To facilitate a better discussion on the convergence order, we introduce the notation 
$\mathcal{O}_{z}^s(A) = \mathcal{O}(Az^{-2}\mathcal{G}_s(z^{-1}) )$.
Here, function $\mathcal{G}_s$ is defined as
\begin{equation} 
\mathcal{G}_s(\cdot) = \Big\{ \sum_{i=1}^T [\mathcal{P}_{s, i}(\cdot)]^{\lambda_i} \Big\}^{\lambda},
\end{equation}
where $i$ is the index and $\mathcal{P}_{s, i}(\cdot)$ is a polynomial whose positive coefficients depending on $\overline{s}$ and $\underline{s}$. 
$T$ is a finite positive integer, and $\lambda$ and $\lambda_i$ are positive rational numbers. 
$\mathcal{G}(x)$ is monotonically increasing on $\mathbb{R}^+$.
Next, we give two lemmas to demonstrate the relationship between $\mathbb{E}\mathbf{Q}_1$, $\mathbf{\Theta}_{\delta}$,
and $\mathbf{F}_{\delta}$. 
\begin{lemma} \label{Lemma_DE_Q1_1}
Given that assumptions \textbf{\ref{A-1}}-\textbf{\ref{A-3}} hold true and $\mathbf{A}$ is a deterministic matrix with uniformly bounded norm, i.e., $\sup_{N \geq 1} \norm{\mathbf{A}} < + \infty$, $\mathbb{E}\mathbf{Q}_1$ can be approximated as 
\begin{equation}
   \mathbb{E}\mathbf{Q}_1 = \mathbf{\Theta}_{\delta} + \boldsymbol{\mathcal{E}}_{\delta}, \label{Approx_of_Q_1_eq1}
\end{equation}
where the deterministic matrix  $\boldsymbol{\mathcal{E}}_{\delta}$ satisfies $\sup_{i,j} |[\boldsymbol{\mathcal{E}}_{\delta}]_{ij}| = \mathcal{O}_{z}^s(\frac{1}{N^{\frac{3}{2}}z})$ and  $\Tr [\mathbf{A}\boldsymbol{\mathcal{E}}_{\delta} ] = \mathcal{O}_{z}^s(\frac{1}{Nz})$.
\end{lemma}
\textit{Proof}: The proof of Lemma \ref{Lemma_DE_Q1_1} is given in Appendix \ref{App_Lemma_DE_Q1_1}. \QED
\begin{lemma} \label{Lemma_DE_Q1_2}
With the same conditions as in Lemma \ref{Lemma_DE_Q1_1}, the following relation holds true
\begin{equation}
   \mathbf{\Theta}_{\delta} =  \mathbf{F}_{\delta} + \widetilde{\boldsymbol{\mathcal{E}}}_{\delta}, \label{Lemma_DE_Q1_2_Eq}
\end{equation}
where the deterministic matrix  $\widetilde{\boldsymbol{\mathcal{E}}}_{\delta}$ satisfies $\sup_{i,j} |[\widetilde{\boldsymbol{\mathcal{E}}}_{\delta}]_{ij}| = \mathcal{O}_{z}^s(\frac{1}{N^{2}z^3})$ and  $\Tr [\mathbf{A}\widetilde{\boldsymbol{\mathcal{E}}}_{\delta} ] = \mathcal{O}_{z}^s(\frac{1}{Nz^3})$.
\end{lemma}
\textit{Proof}: The proof of Lemma \ref{Lemma_DE_Q1_2} is given in Appendix \ref{App_Lemma_DE_Q1_2}. \QED
\par
\begin{remark}
    The entry-wise convergence results of Lemma \ref{Lemma_DE_Q1_1} and \ref{Lemma_DE_Q1_2} indicate that $\sup_{i,j} \abs{[\mathbf{F}_{\delta} - \mathbb{E}\mathbf{Q}_1]_{ij}} = \mathcal{O}_z^s(\frac{1}{N^{\frac{3}{2}}z})$ and $\norm{\mathbf{F}_{\delta} - \mathbb{E}\mathbf{Q}_1} = \mathcal{O}_z^s(\frac{1}{\sqrt{N}z})$, which we provide here for completeness but will not be used in the paper. The entry-wise approximation results can be applied to various problems, such as the EMI evaluation for MIMO systems equipped with MMSE receivers \cite{Artigue2011MMSE},  eigenvalue and eigenvector estimation for large sample covariance matrices \cite{Mestre2008} or asymptotic behavior analysis for subspace-based Direction-of-Arrival (DoA) estimation methods \cite{Schenck2022}.
\end{remark}
To demonstrate the usage of the resolvent and its connection to eigenvalues, we define the empirical spectral distribution (ESD) of a Hermitian matrix $\mathbf{M} \in \mathbb{C}^{N \times N}$ as
$
    F^{\mathbf{M}}(x) = \frac{1}{N} \sum_{k=1}^N \mathbb{I}_{\{\lambda_{k} \leq x\}},
$
where $\lambda_{k}$, $k = 1, \ldots, N$, represent the eigenvalues of $\mathbf{M}$. Then, there holds 
\begin{equation}
    \frac{1}{N} \Tr \mathbf{Q}_1(\overline{s} , -z) = \frac{1}{N} \sum_{i=1}^N \frac{1}{\lambda_i(\mathbf{B}) - z} = \int \frac{F^{\mathbf{B}, N}(\mathrm{d} \lambda)}{\lambda - z} := m^{\mathbf{B}, N}(z),
\end{equation}
where $\mathbf{B} = \mathbf{H}_1 \mathbf{H}_2  \mathbf{H}_2^H \mathbf{H}_1^H + \overline{s} \mathbf{H}_1\mathbf{H}_1^H$ 
and $m^{\mathbf{B}, N}(z)$ is the Stieltjes transform (see Appendix \ref{App_Prop_Stiltjes}) of ESD $F^{\mathbf{B}, N}$. 
By taking $\mathbf{A} = \mathbf{I}_N$  in the variance control result \eqref{Prop_variance_3_Trace_1} and Markov inequality, we have
\begin{equation}
    \sum_{N \geq 1}\mathbb{P}\left({N^{-1}}\abs{\Tr (\mathbf{Q}_1 - \mathbb{E}\mathbf{Q}_1)} \geq \epsilon \right) \leq \sum_{N \geq 1} \frac{\Var(\Tr \mathbf{Q}_1)}{\epsilon^2N^{2}} < + \infty.
    \label{SUM_AS}
\end{equation}
According to Borel-Cantelli lemma and Lemma \ref{Lemma_DE_Q1_2}, we know that  the difference
$
    \frac{1}{N} \Tr \mathbf{F}_{\delta}(\overline{s}, -z) - m^{\mathbf{B}, N}(z) 
$ almost surely converges to $0$. Following the same method in Appendix \ref{App_Prop_Stiltjes}, we can show that $\frac{1}{N} \Tr \mathbf{F}_{\delta}(\overline{s}, -z) $ is the Stieltjes transform of some deterministic probability measure $\overline{F}^N$.
Moreover, if $\frac{1}{N} \Tr \mathbf{F}_{\delta}(\overline{s}, -z)$ converges to a Stieltjes transform with a probability measure $\overline{F}$ as $N$ approaches infinity, Vitali’s convergence theorem \cite[Lemma 2.14]{bai2010spectral} and \cite[Theorem B.9]{bai2010spectral} together imply that 
\begin{equation}
    \mathbb{P}\left(F^{\mathbf{B}, N} \xrightarrow[N \rightarrow + \infty]{d} \overline{F} \right) = 1.
    \label{Eq_Conv_d_as}
\end{equation}
\begin{remark}
    As discussed in Section \ref{Sec_Convergence_rate_of_covariance}, we can show that each order of the central moment for $\Tr \mathbf{Q}_1$ is uniformly bounded. Consequently, for any $\iota > 0$, there holds ${N^{-\iota}} \Tr \mathbf{A}(\mathbf{Q}_1 - \mathbf{F}_{\delta})$ converges to 0 almost surely.
\end{remark}
\begin{remark}
We can use the inverse transform formula \cite[Eq. (1.4), (1.5)]{gesztesy2000matrix}
to characterize the ESD of $F^{\mathbf{B}, N}$ by $\mathbf{F}_{\delta}$, as will be shown in Section \ref{Sec_Numerical_Experiments}. 
For the i.i.d. case studied in Section \ref{Sec_CaseIID}, the conditions for \eqref{Eq_Conv_d_as} are satisfied. The solution $m_F$ in \eqref{DE_system_iid_2_eq} is the Stieltjes transform of the limiting spectral distribution (LSD) of $\mathbf{B}$, while $m_G$ corresponds to the Stieltjes transform of Marčenko-Pastur law \cite{Marčenko_1967}. By solving a quartic equation and using \cite[Eq. (1.5)]{gesztesy2000matrix}, an explicit expression for the LSD of $\mathbf{B}$ could be obtained, which facilitates the estimation of LSS. This will be left for future works.
\end{remark}
\subsection{Approximation of $I_1$}
\label{Sec_Approximation_of_I_1}
In the following, we will establish the relationship between $\mathbf{Q}_1$ and $I_1$
as well as that between $\mathbf{F}_{\delta}$ and $\overline{I}_1$. Then, we will determine
the convergence rate of $\mathbb{E}I_1 - \overline{I}_1$. Based on the definition of 
$I(\overline{s}, z)$ in Section \ref{Sec_Problem_formulation}, we can obtain $\lim_{z \rightarrow + \infty} I(\overline{s} , z) = 0$. By taking the derivative with respect to $z$ and using $\partial \log\det(\mathbf{X}) = \Tr(\mathbf{X}^{-1} \partial \mathbf{X})$, we have
\begin{equation}
    I_1(\overline{\sigma}_1^2, \sigma_2^2) = \int_{\sigma_2^2}^{+ \infty} -\frac{\partial I_1(\overline{\sigma}_1^2, z)}{\partial z} \diff{z} =  \int_{\sigma_2^2}^{+ \infty} \left( \frac{N}{z} - \Tr \mathbf{Q}_1(\overline{\sigma}_1^2, z) \right) \diff{z}.
\end{equation}
To prove \eqref{Thm_Eq_DE_I_1}, we will now take the derivative 
of $\overline{I}_1$ defined in \eqref{I_1} using the same method. For ease of calculation, we will change the form of $\overline{I}_1 = f_{1}(\overline{s}, z, \delta, \overline{\omega}, \underline{\omega}, \gamma)$. By observing that $\frac{ f_1}{ \partial \delta} = \frac{ \partial f_1}{ \partial \overline{\omega}} = \frac{ \partial f_1}{ \partial \underline{\omega}} = \frac{ \partial f_1}{ \partial \gamma} = 0$, we have  
\begin{equation}
   \frac{\partial \overline{I}_1}{\partial z} = \frac{\partial f_1}{\partial z} + \frac{ \partial f_1}{ \partial \delta} \frac{ \partial \delta}{ \partial z} + \frac{ \partial f_1}{ \partial \overline{\omega}} \frac{ \partial \overline{\omega}}{\partial z} + \frac{ \partial f_1}{ \partial \underline{\omega}} \frac{ \partial \underline{\omega}}{\partial z} + \frac{ \partial f_1}{ \partial \gamma}  \frac{ \partial \gamma}{\partial z} = - \frac{N}{z} + \mathrm{Tr}\mathbf{F}_{\delta}. \label{derivite_of_I_1_}
\end{equation}
By the fundamental equations \eqref{DE_system_1} and Proposition \ref{Prop_Stieltjes}, we have $\underset{z \rightarrow + \infty}{\lim} \overline{I}_1(\overline{s}, z)= 0$, and thus
\begin{equation}
    \overline{I}_1(\overline{\sigma}_1^2, \sigma_2^2) = \int_{\sigma_2^2}^{+ \infty} \Big( \frac{N}{z} - \mathrm{Tr}\mathbf{F}_{\delta}(\overline{\sigma}_1^2, z) \Big) \mathrm{d}z.
\end{equation}
According to Fubini's theorem, 
the term 
$\mathbb{E} I_1(\overline{\sigma}_1^2, \sigma_2^2) - \overline{I}_1(\overline{\sigma}_1^2, \sigma_2^2)$ 
can be calculated as
\begin{equation}
    \mathbb{E} [I_1(\overline{\sigma}_1^2, \sigma_2^2)]- \overline{I}_1(\overline{\sigma}_1^2, \sigma_2^2) = 
    \int_{\sigma_2^2}^{+ \infty} \left( \Tr\mathbf{F}_{\delta}(\overline{\sigma}_1^2, z)  - 
    \Tr \mathbb{E}\mathbf{Q}_1(\overline{\sigma}_1^2, z)\right) \mathrm{d}z \overset{(a)}{=} \int_{\sigma_2^2}^{+ \infty} 
    \mathcal{O}(\frac{\mathcal{G}_{\sigma_1^2}(z^{-1})\diff z}{N z^3}) \overset{(b)}{\leq} \mathcal{O}(\frac{\mathcal{G}_{\sigma_1^2}(\sigma_2^{-2})}{N\sigma_2^4}),
\end{equation}
where $ (a)$ follows from the setting  $\mathbf{A} = \mathbf{I}_N$ in Lemma \ref{Lemma_DE_Q1_1} and Lemma \ref{Lemma_DE_Q1_2}, and $(b)$ is due to
 the integrability of $z^{-3}$ over $[\sigma_2^2, + \infty)$ and the monotonicity of $\mathcal{G}_{\sigma_1^2}(z^{-1})$. Thus, we proved Theorem \ref{Th_First_Order}. \QED
\section{Proof of the Second-order Result: CLT for the Joint Distribution of MIs} \label{Sec_Second-order}
This section is devoted to the detailed proof of Theorem \ref{Thm_Second-order}. The proof  mainly uses the technique of characteristic functions, so we first define the characteristic function of centralized MI 
\begin{equation}
    \varphi(\mathbf{t}, \overline{s}, \underline{s}, z) = \mathbb{E}\exp[\jmath \mathbf{t}^T (\mathbf{m} - \overline{\mathbf{m}})], \hspace*{1mm}\mathbf{t} \in \mathbb{R}^2, \label{Chara_func_MI}
\end{equation} 
where $\mathbf{m}$ is defined in \eqref{m_Joint_I_1_I_2} and 
$\overline{\mathbf{m}} = (\overline{I}_1, \overline{I}_2)^T$. 
For brevity, we introduce the notation
 $\widehat{\psi} = \exp[\jmath \mathbf{t}^T{\mathbf{m}}]$ and ${\psi} = \mathbb{E}\widehat{\psi}$. 
 The proof uses Gaussian tools (see Appendix \ref{App_Lemma_DE_Q1_1}) 
 to estimate the characteristic function and involves a large amount of computation.
  For the sake of convenience, we first present some preliminary results.
\subsection{Preliminaries of the Proof}
\subsubsection{The Condition for the Asymptotic Gaussianity of Random Vector Sequence} In the proof, we will need the following lemma to justify the asymptotic Gaussianity of $\mathbf{m} - \overline{\mathbf{m}}$ to establish the CLT.
\begin{lemma} \label{Lemma_tight_CLT}
Assume that $\{\mathbf{x}_{n}\}_{n=1}^{+ \infty}$ are $p$-dimensional random vector sequence
 with characteristic functions $\varphi_n(\mathbf{t}) = \mathbb{E} \exp[\jmath \mathbf{t}^T \mathbf{x}_n]$.
  If, for any $\mathbf{t} \in \mathbb{R}^p$,
\begin{equation}
    \lim_{n \rightarrow +\infty} \left\{ \varphi_n(\mathbf{t}) - \exp[- \frac{1}{2}\mathbf{t}^T \mathbf{V}_n\mathbf{t}] \right\} = 0, \label{CLT_Lemma_Cond1}
\end{equation}
where $\mathbf{V}_n$ is a positive definite matrix whose eigenvalues are uniformly contained in
a compact set bounded away from $0$, i.e., 
\begin{equation}
    0 < m_V =  \inf_{n \geq 1} \lambda_{min}(\mathbf{V}_n) \leq \sup_{n \geq 1} \lambda_{max}(\mathbf{V}_n) = M_V < + \infty, \label{Regu_lemma_Gaussian}
\end{equation}
then 
\begin{equation}
    \mathbf{V}_n^{-\frac{1}{2}} \mathbf{x}_n  \xrightarrow[n \rightarrow + \infty]{d} \mathcal{N}(\mathbf{0}_p, \mathbf{I}_p). 
\end{equation}
\end{lemma}
\textit{Proof:} The proof of Lemma \ref{Lemma_tight_CLT} is given in Appendix \ref{App_Lemma_tight_CLT}. \QED
\begin{remark}
    This lemma can be viewed as a multivariate generalization of Proposition 6 in \cite{Hachem2008ANewApproach}. 
    The main purpose of condition \eqref{Regu_lemma_Gaussian} is to ensure the tightness. 
    This lemma can be used in proving the joint CLT for multiple LSS of random matrices.
     For example, it can be applied to the analysis of multi-user 
     secure communication systems \cite{Xin2023Secrecy}.
\end{remark}
\par
\subsubsection{Approximations of Necessary Quantities}  We define the trace of resolvent functions in Table \ref{Table_trace_of_resolvent_terms}, where $\mathbf{A}, \mathbf{B}, \mathbf{C}$, and $\mathbf{D}$ 
denote deterministic matrices with appropriate dimensions and bounded norms.
\begin{table}[t]
\centering
\caption{Trace of Resolvent Terms}
\label{Table_trace_of_resolvent_terms}
 \begin{tabular}{| cc | cc |} 
\toprule[1pt]
\midrule
 Function & Expression & Function & Expression \\ [0.5ex] 
\midrule
 $\widehat{\Psi}_1(\mathbf{A}, \mathbf{B})$ & $\mathrm{Tr}[\mathbf{A} \mathbf{Q}_1 \mathbf{H}_1 \mathbf{B} \mathbf{H}_1^H]$ & $\widehat{\Phi}_1(\mathbf{A}, \mathbf{B})$ & $\mathrm{Tr}[\mathbf{A} \mathbf{Q}_2 \mathbf{H}_1 \mathbf{B} \mathbf{H}_1^H]$ \\[1ex] 
 $\widehat{\Psi}_2(\mathbf{A}, \mathbf{B}, \mathbf{C})$ & $\mathrm{Tr}[\mathbf{A} \mathbf{Q}_1 \mathbf{H}_1 \mathbf{H}_2 \mathbf{B} \mathbf{H}_2^H \mathbf{C}\mathbf{H}_1^H]$ & $\widehat{\Phi}_2(\mathbf{A}, \mathbf{B})$ & $\mathrm{Tr}[\mathbf{A} \mathbf{Q}_2 \mathbf{R}_1 \mathbf{Q}_2 \mathbf{H}_1 \mathbf{B} \mathbf{H}_1^H]$ \\[1ex] 
$\widehat{\Psi}_3(\mathbf{A}, \mathbf{B}, \mathbf{C})$ & $\mathrm{Tr}[\mathbf{A} \mathbf{Q}_1 \mathbf{B} \mathbf{Q}_1 \mathbf{H}_1 \mathbf{C}\mathbf{H}_1^H]$ & $\widehat{\Phi}_3(\mathbf{A}, \mathbf{B})$ & $\mathrm{Tr}[\mathbf{A} \mathbf{Q}_2 \mathbf{R}_1 \mathbf{Q}_1 \mathbf{H}_1 \mathbf{B} \mathbf{H}_1^H]$ \\[1ex] 
 $\widehat{\Psi}_4(\mathbf{A}, \mathbf{B}, \mathbf{C}, \mathbf{D})$ & $\mathrm{Tr}[\mathbf{A} \mathbf{Q}_1 \mathbf{B} \mathbf{Q}_1  \mathbf{H}_1 \mathbf{H}_2  \mathbf{C} \mathbf{H}_2^H \mathbf{D}\mathbf{H}_1^H]$ & $\widehat{\Phi}_4(\mathbf{A}, \mathbf{B}, \mathbf{C})$ & $\mathrm{Tr}[\mathbf{A} \mathbf{Q}_2 \mathbf{R}_1 \mathbf{Q}_1 \mathbf{H}_1 \mathbf{H}_2\mathbf{B} \mathbf{H}_2^H  \mathbf{C} \mathbf{H}_1^H]$\\[1ex] 
$\widehat{\Psi}_5(\mathbf{A}, \mathbf{B})$ & $\mathrm{Tr}[\mathbf{A} \mathbf{Q}_1 \mathbf{H}_1 \mathbf{R}_2 \mathbf{H}_1^H \mathbf{Q}_1  \mathbf{H}_1 \mathbf{B}\mathbf{H}_1^H]$ && \\[1ex] 
$\widehat{\Psi}_6(\mathbf{A}, \mathbf{B}, \mathbf{C})$ & $\mathrm{Tr}[\mathbf{A} \mathbf{Q}_1 \mathbf{H}_1 \mathbf{R}_2 \mathbf{H}_1^H \mathbf{Q}_1  \mathbf{H}_1 \mathbf{H}_2 \mathbf{B} \mathbf{H}_2^H \mathbf{C} \mathbf{H}_1^H]$ && \\[1ex] 
$\widehat{\Psi}_{7}(\mathbf{A}, \mathbf{B})$ & $\mathrm{Tr}[\mathbf{A} \mathbf{Q}_1 \mathbf{R}_1 \mathbf{Q}_2 \mathbf{H}_1 \mathbf{B}\mathbf{H}_1^H]$ && \\ [1ex] 
\bottomrule[1pt]
 \end{tabular}
\end{table}
As discussed in Section \ref{Approximation_of_EQ1}, the mean of $\widehat{x}$ is denoted by $x = \mathbb{E}\widehat{x}$
and its centralized version is defined as $\mathring{x} = \widehat{x} - x$, 
where $\widehat{x}$ can be any functions in Table 
\ref{Table_trace_of_resolvent_terms} or quantities in \eqref{Random_quantities}.
Here, $\widehat{\Psi}_k, k = 1,\ldots, 6$ , $\widehat{\Phi}_1$, and $\widehat{\Phi}_2$ 
are functions only contain  
the resolvent matrix $\mathbf{Q}_1$ and $\mathbf{Q}_2$. 
The terms $\widehat{\Psi}_7, \widehat{\Phi}_3$, and $\widehat{\Phi}_4$ are related to 
both $\mathbf{Q}_1$ and $\mathbf{Q}_2$. 
Similar to Proposition \ref{Prop_variance}, we can show the variance control results by Poincaré-Nash inequality \eqref{Poincare_Nash_ineq}
\begin{equation}
    \mathrm{Var}(\widehat{\Psi}_i) = \mathcal{O}_{z}^s(\frac{1}{z^{2}}), i=3,\ldots,7, \ \mathrm{Var}(\widehat{\Phi}_1) =  \mathcal{O}_{z}^s(1), \ \mathrm{Var}(\widehat{\Phi}_j) =  \mathcal{O}_{z}^s(\frac{1}{z^{2}}), j=2,3,4. \label{Var_control}
\end{equation}
The approximation of the mean of the above terms is presented in the following two lemmas.
\begin{lemma} \label{Lemma_DE_terms_Approx_Psi}
    Given assumptions \textbf{\ref{A-1}}-\textbf{\ref{A-3}} hold true, the approximations of functions $\Psi_i, i=1,2,..,7$, are given by, respectively, 
\begin{subequations}
\begin{align}
    \Psi_1(\mathbf{A}, \mathbf{B}) &= \frac{1}{L}\mathrm{Tr}[\mathbf{B}\mathbf{T}_1 \mathbf{F}_{\omega}]\mathrm{Tr}[\mathbf{A}\mathbf{F}_{\delta} \mathbf{R}_1 ] + \mathcal{O}_{z}^s(\frac{1}{N}), \label{DE_Psi_1} \\
    \Psi_2(\mathbf{A}, \mathbf{B}, \mathbf{C}) &= \frac{1}{M}\mathrm{Tr}[\mathbf{B}\mathbf{T}_2 \mathbf{F}_{\gamma}]\frac{1}{L}\mathrm{Tr}[\mathbf{R}_2 \mathbf{C} \mathbf{T}_1 \mathbf{F}_{\omega}]\mathrm{Tr}[\mathbf{A}\mathbf{F}_{\delta} \mathbf{R}_1 ] + \mathcal{O}_{z}^s(\frac{1}{N}), \label{DE_Psi_2} \\
    \Psi_3(\mathbf{A}, \mathbf{B}, \mathbf{C})
    &= -\frac{\Delta}{\Delta_{V_1}} \frac{1}{L}\mathrm{Tr}[\mathbf{B}\mathbf{F}_{\delta} \mathbf{R}_1 \mathbf{F}_\delta]\frac{1}{L} \mathrm{Tr}[(\overline{s} \mathbf{I}_L + \gamma\mathbf{R}_2)\mathbf{T}_1\mathbf{F}_{\omega}\mathbf{C}\mathbf{T}_1 \mathbf{F}_{\omega}]\mathrm{Tr}[\mathbf{A} \mathbf{F}_{\delta} \mathbf{R}_1] \notag \\
    &+ \frac{\frac{L}{M}  \underline{\omega}_{2, I} \gamma_{2}\delta}{\Delta_{V_1}}\frac{1}{L}\mathrm{Tr}[\mathbf{B}\mathbf{F}_{\delta} \mathbf{R}_1 \mathbf{F}_\delta]\frac{1}{L} \mathrm{Tr}[\mathbf{R}_2\mathbf{T}_1\mathbf{F}_{\omega}\mathbf{C}\mathbf{T}_1 \mathbf{F}_{\omega}]\mathrm{Tr}[\mathbf{A} \mathbf{F}_{\delta} \mathbf{R}_1] \notag \\
    &+ \frac{\Delta \varsigma + \frac{L}{M} \underline{\omega}_{2, I}^2 \gamma_2}{\Delta_{V_1}} \frac{1}{L}\mathrm{Tr}[\mathbf{C}\mathbf{T}_{1}\mathbf{F}_{\omega}]\frac{1}{L} \mathrm{Tr}[\mathbf{B}\mathbf{F}_{\delta}\mathbf{R}_1 \mathbf{F}_{\delta}] \mathrm{Tr}[\mathbf{A}\mathbf{F}_{\delta}\mathbf{R}_1 \mathbf{F}_{\delta}\mathbf{R}_1] \notag \\
    &+ \frac{1}{L}\mathrm{Tr}[\mathbf{C}\mathbf{T}_{1}\mathbf{F}_{\omega}]\mathrm{Tr}[\mathbf{A}\mathbf{F}_{\delta}\mathbf{B} \mathbf{F}_{\delta}\mathbf{R}_1] + \mathcal{O}_{z}^s(\frac{1}{N}), \label{DE_Psi_3} \\
    \Psi_4(\mathbf{A}, \mathbf{B}, \mathbf{C}, \mathbf{D})&= -\frac{ \underline{\omega}_{2, I}}{\Delta_{V_1}} \frac{1}{M} \mathrm{Tr}[\mathbf{B} \mathbf{F}_{\delta} \mathbf{R}_1 \mathbf{F}_{\delta}]\frac{1}{M} \mathrm{Tr}[\mathbf{C} \mathbf{T}_2 \mathbf{F}_{\gamma} \mathbf{T}_2 \mathbf{F}_{\gamma}] \frac{1}{L} \mathrm{Tr}[\mathbf{R}_2 \mathbf{D} \mathbf{T}_1 \mathbf{F}_{\omega}] \mathrm{Tr}[\mathbf{A} \mathbf{F}_{\delta} \mathbf{R}_1]\notag \\
    &+ \frac{1}{M}\mathrm{Tr} [\mathbf{C} \mathbf{T}_{2}\mathbf{F}_{\gamma}] \Psi_3(\mathbf{A}, \mathbf{B}, \mathbf{R}_2 \mathbf{D})  + \mathcal{O}_{z}^s(\frac{1}{N}), \label{DE_Psi_4} \\
     \Psi_5(\mathbf{A}, \mathbf{B}) 
  &= -\frac{\delta_2 \underline{\omega}_{2, I}}{\Delta_{V_1}} \frac{1}{L} \mathrm{Tr}[(\overline{s} \mathbf{I}_L + \gamma \mathbf{R}_2) \mathbf{T}_1 \mathbf{F}_{\omega} \mathbf{B} \mathbf{T}_1 \mathbf{F}_{\omega}] \mathrm{Tr}[\mathbf{A} \mathbf{F}_{\delta} \mathbf{R}_1] \notag \\
    &+ \frac{\delta(1 - \varsigma \delta_2)}{\Delta_{V_1}} \frac{1}{L} \mathrm{Tr}[\mathbf{R}_2 \mathbf{T}_1 \mathbf{F}_{\omega} \mathbf{B} \mathbf{T}_1 \mathbf{F}_{\omega}]\mathrm{Tr}[\mathbf{A} \mathbf{F}_{\delta} \mathbf{R}_1] \notag \\
     &+ \frac{\underline{\omega}_{2, I}}{\Delta_{V_1}}\frac{1}{L}\mathrm{Tr}[\mathbf{T}_{1} \mathbf{F}_{\omega}\mathbf{B}]\mathrm{Tr}[\mathbf{A}\mathbf{F}_{\delta} \mathbf{R}_1 \mathbf{F}_{\delta} \mathbf{R}_1 ] + \mathcal{O}_{z}^s(\frac{1}{N}),  \label{DE_Psi_5} \\
     \Psi_6(\mathbf{A}, \mathbf{B}, \mathbf{C}) &=  -(\frac{\frac{L}{M} \underline{\omega}_{2, I}^2 \delta_2}{\Delta_{V_1} \Delta} + \frac{ \frac{L}{M}\delta^2 \underline{\omega}_2}{\Delta}) \frac{1}{M} \mathrm{Tr}[ \mathbf{B} \mathbf{T}_2 \mathbf{F}_{\gamma}\mathbf{T}_2 \mathbf{F}_{\gamma}] \frac{1}{L} \mathrm{Tr}[\mathbf{R}_2 \mathbf{C} \mathbf{T}_1 \mathbf{F}_{\omega}]\mathrm{Tr}[\mathbf{A}\mathbf{F}_{\delta}\mathbf{R}_1] \notag \\
    &+ \frac{1}{M} \mathrm{Tr}[\mathbf{B}\mathbf{T}_{2}\mathbf{F}_{\gamma}] \Psi_5(\mathbf{A}, \mathbf{R}_2 \mathbf{C}) + \mathcal{O}_{z}^s(\frac{1}{N}), \label{DE_Psi_6}\\
    \Psi_7(\mathbf{A}, \mathbf{B}) &=  - \frac{\vartheta}{\Delta_C}   \frac{1}{L} \mathrm{Tr}[(\overline{s} \mathbf{I}_L + \gamma \mathbf{R}_2) \mathbf{T}_1 \mathbf{G}_{\overline{\tau}} \mathbf{B} \mathbf{T}_1 \mathbf{F}_{\omega}] \mathrm{Tr}[\mathbf{A} \mathbf{F}_{\delta} \mathbf{R}_1] \notag \\
    &+ \frac{1}{\Delta_C}\frac{1}{L}\mathrm{Tr} [\mathbf{B}\mathbf{T}_{1}\mathbf{G}_{\overline{\tau}}] \mathrm{Tr}[\mathbf{R}_1 \mathbf{A} \mathbf{F}_{\delta} \mathbf{R}_1 \mathbf{G}_{\tau}] + \mathcal{O}_{z}^s(\frac{1}{N}).\label{DE_Psi_7}%
\end{align}
\end{subequations}
\end{lemma}
\textit{Proof:} The proof of Lemma \ref{Lemma_DE_terms_Approx_Psi} is given in Appendix \ref{App_Lemma_DE_terms_Approx_Psi}. \QED
\begin{lemma} \label{Lemma_DE_terms_Approx_Phi}
    Given assumptions \textbf{\ref{A-1}}-\textbf{\ref{A-3}} hold true, the approximations of functions $\Phi_j, j=1,2,3,4$, are given by, respectively,
\begin{subequations}
\begin{align}
    \Phi_1(\mathbf{A}, \mathbf{B})
    &= \frac{1}{L} \mathrm{Tr} [\mathbf{T}_1\mathbf{G}_{\overline{\tau}} \mathbf{B}]  \mathrm{Tr}[\mathbf{A}\mathbf{G}_{\tau} \mathbf{R}_1]  + \mathcal{O}_{z}^s(\frac{1}{N}), \label{DE_Phi_1} \\
    \Phi_2(\mathbf{A}, \mathbf{B})
    &= -\frac{\underline{s} \tau_2}{\Delta_{V_2}} \frac{1}{L} \mathrm{Tr}[\mathbf{T}_1 \mathbf{G}_{\overline{\tau}}\mathbf{T}_1 \mathbf{G}_{\overline{\tau}}\mathbf{B}] \mathrm{Tr}[\mathbf{A} \mathbf{G}_{\tau} \mathbf{R}_1] + \frac{1}{\Delta_{V_2}}\mathrm{Tr}[\mathbf{T}_1 \mathbf{G}_{\overline{\tau}}\mathbf{B}] \mathrm{Tr}[\mathbf{A} \mathbf{G}_{\tau} \mathbf{R}_1\mathbf{G}_{\tau} \mathbf{R}_1] + \mathcal{O}_{z}^s(\frac{1}{N}), \label{DE_Phi_2} \\
    \Phi_3(\mathbf{A}, \mathbf{B}) &= -\frac{\underline{s} \vartheta}{\Delta_C}\frac{1}{L} \mathrm{Tr}[\mathbf{B} \mathbf{T}_1 \mathbf{G}_{\overline{\tau}} \mathbf{T}_1 \mathbf{F}_{\omega} ] \mathrm{Tr}[\mathbf{A}\mathbf{G}_{\tau} \mathbf{R}_1] + \frac{1}{\Delta_C}\frac{1}{L}\mathrm{Tr}[\mathbf{T}_1 \mathbf{F}_{\omega}\mathbf{B}] \mathrm{Tr}[\mathbf{R}_1 \mathbf{F}_{\delta}\mathbf{R}_1\mathbf{A} \mathbf{G}_{\tau} ] + \mathcal{O}_{z}^s(\frac{1}{N}), \label{DE_Phi_3} \\
    \Phi_4(\mathbf{A}, \mathbf{B}, \mathbf{C}) &=  \frac{1}{M}\mathrm{Tr}[\mathbf{T}_{2} \mathbf{F}_{\gamma}\mathbf{B}] \Phi_3(\mathbf{A}, \mathbf{R}_2\mathbf{C})  + \mathcal{O}_{z}^s(\frac{1}{N}). \label{DE_Phi_4} 
\end{align}
\end{subequations}
\end{lemma}
\textit{Proof:} The proof of Lemma \ref{Lemma_DE_terms_Approx_Phi} is given in Appendix \ref{App_Lemma_DE_terms_Approx_Phi}. \QED
\subsection{Proof of Theorem \ref{Thm_Second-order}}
The outline for the proof of Theorem \ref{Thm_Second-order} is described as follows:
\begin{itemize}
 \item [1)] In order to evaluate the characteristic function \eqref{Chara_func_MI}, 
we first take the derivative of $\frac{\partial \varphi(\mathbf{t}, \overline{s}, \underline{s}, z)}{\partial z}$ and show that 
\begin{equation}
    \frac{\partial \varphi(\mathbf{t}, \overline{s}, \underline{s}, z) }{\partial z} = \left[- \frac{1}{2} \mathbf{t}^T \boldsymbol{\mathcal{V}}(\overline{s}, \underline{s}, z) \mathbf{t} \right] \varphi(\mathbf{t}, \overline{s}, \underline{s}, z) + \mathcal{O}_{z}^s(\frac{1}{zN}), \label{Diff_eq}
\end{equation}
where $\boldsymbol{\mathcal{V}}(s, z)$ is a matrix whose elements are linear combinations of the terms in Table \ref{Table_trace_of_resolvent_terms}.
 \item [2)] Next, we show that  the following differential equation holds
\begin{equation}
    \frac{\partial \varphi(\mathbf{t}, \overline{s}, \underline{s}, z) }{\partial z} = \left[- \frac{1}{2} \mathbf{t}^T \frac{\partial \mathbf{V}(\overline{s}, \underline{s}, z)}{\partial z}\mathbf{t} \right] \varphi(\mathbf{t}, \overline{s}, \underline{s}, z) + \mathcal{O}_{z}^{s}(\frac{1}{z N}), 
    \label{Diff_eqq}
\end{equation}
where $ \mathbf{V}$ is defined in \eqref{Cov_matrix}.
\item [3)] We analyze the differential equation \eqref{Diff_eqq} and show that for $\overline{\sigma}_1^2, \underline{\sigma}_1^2 \geq 0, \sigma_2^2 > 0$, the convergence of the characteristic function holds with
\begin{equation}
    \varphi(\mathbf{t}, \overline{\sigma}_1^2, \underline{\sigma}_1^2, \sigma_2^2) - \exp[-\frac{1}{2}\mathbf{t}^T \mathbf{V}(\overline{\sigma}_1^2, \underline{\sigma}_1^2, \sigma_2^2) \mathbf{t}] \xrightarrow[N\rightarrow +\infty]{} 0, \label{Conv_cha_Eq}
\end{equation}
and the convergence rate is of order $\mathcal{O}(\frac{1}{N})$. Thus, by using Lemma \ref{Lemma_tight_CLT}, we can establish Theorem \ref{Thm_Second-order}.
\item  [4)] Finally, we determine the convergence rate of the covariance matrix of $\mathbf{m}$ by
\begin{equation}
    \norm{\mathbf{V}(\overline{\sigma}_1^2, \underline{\sigma}_1^2,\sigma_2^2) - \mathrm{Cov}[\mathbf{m}(\overline{\sigma}_1^2, \underline{\sigma}_1^2, \sigma_2^2)]} = \mathcal{O}(\frac{1}{N}).
\end{equation}
\end{itemize}
Now, we start from the first step. 
\subsubsection{Decomposition of $\frac{\partial\varphi}{\partial z}$}
By the definition in \eqref{Chara_func_MI} and the dominated convergence theorem, 
we can get
\begin{equation}
\begin{split}
    \frac{\partial \varphi }{\partial z} &= \frac{\partial (\psi \exp[-\jmath \mathbf{t}^T \overline{\mathbf{m}}])}{\partial z} = \mathbb{E}[j(t_1\frac{\partial I_1(\overline{s}, z)}{\partial z} + t_2\frac{\partial I_2(\underline{s}, z)}{\partial z} )\widehat{\psi} ] \exp[-\jmath \mathbf{t}^T \overline{\mathbf{m}}] - \jmath (t_1\frac{\partial \overline{I}_1(\overline{s}, z)}{\partial z} + t_2\frac{\partial \overline{I}_2(\underline{s}, z)}{\partial z} ) \varphi \\
    & {=} -\frac{\jmath}{z}(t_1 \mathcal{X} + t_2 \mathcal{Y})\exp[-\jmath \mathbf{t}^T \overline{\mathbf{m}}] + \frac{\jmath}{z}(t_1L ( \overline{s} \overline{\omega} + \gamma \underline{\omega}) \delta +  t_2 L \underline{s} \overline{\tau} \tau) \varphi, \label{Diff_Eq_CLT}
\end{split}
\end{equation}
where $\mathcal{X} = \mathbb{E}[\overline{s}\widehat{\Psi}_1(\mathbf{I}_N, \mathbf{I}_L) + \widehat{\Psi}_2(\mathbf{I}_N, \mathbf{I}_M, \mathbf{I}_L)]\widehat{\psi}$
 and $\mathcal{Y} = \underline{s}\mathbb{E} \widehat{\Phi}_2(\mathbf{I}_N,\mathbf{I}_L)\widehat{\psi}$. 
 The derivative $\frac{\partial \overline{I}_1}{\partial z}$ can be calculated using \eqref{derivite_of_I_1_}.
Furthermore, we have $\Tr \mathbf{F}_{\delta}-\frac{N}{z} = \Tr( \mathbf{F}_{\delta}-(z{\mathbf{I}_N})^{-1}) = \Tr \left[ \mathbf{F}_{\delta}(-(\overline{s}\overline{\omega} + \gamma \underline{\omega})\mathbf{R}_1)(z{\mathbf{I}_N})^{-1}\right] = -\frac{L}{z}(\overline{s \omega} + \gamma \underline{\omega})\delta$. 
Using the same method, the derivative $\frac{\partial \overline{I}_2}{\partial z}$ can be obtained. In order to establish the  equation \eqref{Diff_eq},
we need to use the integration by parts formula \eqref{Integration_by_parts} to decompose $\mathcal{X}$ and $\mathcal{Y}$. 
The result of the decomposition involves the expectation for the product of some quantities in Table \ref{Table_trace_of_resolvent_terms}. 
Using the Poincaré-Nash inequality \eqref{Poincare_Nash_ineq}, we can prove that these terms are asymptotically uncorrelated, i.e., the expectation of their product is approximately equal to the product of their expectations.
\par
\paragraph{Decomposition of $\mathcal{X}$}
By taking the integration by parts formula \eqref{Integration_by_parts} with respect to $[\mathbf{H}_2]_{kl}$ and using $\partial \log\det(\mathbf{X}) = \Tr \mathbf{X}^{-1} \partial \mathbf{X}$, we have
\begin{equation}
\begin{split}
    &\sum_{k} \mathbb{E} [\mathbf{Q}_1\mathbf{H}_1]_{ik} [\mathbf{H}_2]_{kl} [\mathbf{H}_2]_{ms}^* [\mathbf{H}_1]_{jp}^* \widehat{\psi} = \sum_{k, a, b}  \frac{[\mathbf{R}_2]_{ka} [\mathbf{T}_2]_{bl}}{M} \mathbb{E} \frac{\partial}{\partial [\mathbf{H}_2]_{ab}^*} \left([\mathbf{Q}_1\mathbf{H}_1]_{ik} [\mathbf{H}_2]_{ms}^* [\mathbf{H}_1]_{jp}^* \widehat{\psi} \right) \\
    &= \sum_{k, a, b}  \frac{[\mathbf{R}_2]_{ka} [\mathbf{T}_2]_{bl}}{M} \mathbb{E} \Big\{ -[\mathbf{Q}_1\mathbf{H}_1 \mathbf{H}_2]_{ib} [\mathbf{H}_1^H \mathbf{Q}_1 \mathbf{H}_1]_{ak} [\mathbf{H}_2]_{ms}^* [\mathbf{H}_1]_{jp}^* + [\mathbf{Q}_1\mathbf{H}_1]_{ik} \delta(m-a) \delta(s-b) [\mathbf{H}_1]_{jp}^* \\
    &+ \jmath t_1  [\mathbf{H}_1^H \mathbf{Q}_1 \mathbf{H}_1 \mathbf{H}_2]_{ab}[\mathbf{Q}_1\mathbf{H}_1]_{ik} [\mathbf{H}_2]_{ms}^* [\mathbf{H}_1]_{jp}^* \Big\}\widehat{\psi}. \label{CLT_Eq_Inte_By_Parts_Q_1}
\end{split} 
\end{equation}
Thus, by  writing $\widehat{\kappa} = \mathring{\kappa} + \kappa$ we can get
\begin{equation}
\begin{split}
    \mathbb{E} [\mathbf{Q}_1\mathbf{H}_1 \mathbf{H}_2 \mathbf{\Theta}_{\gamma}^{-1}]_{il} [\mathbf{H}_2]_{ms}^* [\mathbf{H}_1]_{jp}^* \widehat{\psi} &= \mathbb{E} \Big\{ -\mathring{\kappa}   [\mathbf{Q}_1\mathbf{H}_1 \mathbf{H}_2 \mathbf{T}_2]_{il}  [\mathbf{H}_2]_{ms}^* [\mathbf{H}_1]_{jp}^*  +  \frac{[\mathbf{T}_2]_{sl}}{M}[\mathbf{Q}_1\mathbf{H}_1 \mathbf{R}_2]_{im}  [\mathbf{H}_1]_{jp}^* \\
    &+  \frac{\jmath t_1}{M} [\mathbf{Q}_1\mathbf{H}_1\mathbf{R}_2 \mathbf{H}_1^H \mathbf{Q}_1 \mathbf{H}_1 \mathbf{H}_2 \mathbf{T}_2]_{il} [\mathbf{H}_2]_{ms}^* [\mathbf{H}_1]_{jp}^*  \Big\}\widehat{\psi}. \label{CLT_1}
\end{split}
\end{equation}
By using the same method in \eqref{CLT_Eq_Inte_By_Parts_Q_1} and \eqref{CLT_1}, we have
\begin{equation}
\begin{split}
     & \mathbb{E} [\mathbf{Q}_1\mathbf{H}_1]_{iq}  [\mathbf{H}_1]_{jp}^* \widehat{\psi} = \mathbb{E} \Big\{ - \widehat{\alpha}_{\delta} [\mathbf{Q}_1 \mathbf{H}_1 \mathbf{H}_2 \mathbf{H}_2^H \mathbf{T}_1]_{iq}   [\mathbf{H}_1]_{jp}^* - \overline{s} \widehat{\alpha}_{\delta} [\mathbf{Q}_1 \mathbf{H}_1\mathbf{T}_1]_{iq} [\mathbf{H}_1]_{jp}^* + \frac{[\mathbf{T}_1]_{pq}}{L}[\mathbf{Q}_{1} \mathbf{R}_1]_{ij} \\
     & + \frac{\jmath t_1}{L}[ \mathbf{Q}_1\mathbf{R}_1 \mathbf{Q}_1 \mathbf{H}_1 \mathbf{H}_2 \mathbf{H}_2^H \mathbf{T}_1]_{iq} [\mathbf{H}_1]_{jp}^*  
     + \frac{\jmath \overline{s} t_1}{L} [\mathbf{Q}_1 \mathbf{R}_1 \mathbf{Q}_1\mathbf{H}_1\mathbf{T}_1]_{iq} [\mathbf{H}_1]_{jp}^*+ 
     \frac{\jmath \underline{s} t_2}{L}[\mathbf{Q}_1 \mathbf{R}_1 \mathbf{Q}_2\mathbf{H}_1 \mathbf{T}_1]_{iq} [\mathbf{H}_1]_{jp}^*\Big\} \widehat{\psi}. \label{Eq_CLT_QH1}
\end{split}
\end{equation}
Multiplying both sides of equation \eqref{CLT_1} by $[\mathbf{\mathbf{\Theta}}_{\gamma}]_{ls}$, $[\mathbf{T}_1]_{mq}$ and summing over subscripts $l$, $s$, and $m$, we have
\begin{equation}
\begin{split}
    \mathbb{E} [\mathbf{Q}_1\mathbf{H}_1 \mathbf{H}_2\mathbf{H}_2^H \mathbf{T}_1]_{iq} [\mathbf{H}_1]_{jp}^* \widehat{\psi} &= \mathbb{E} \Big\{ -\mathring{\kappa}   [\mathbf{Q}_1\mathbf{H}_1 \mathbf{H}_2 \mathbf{T}_2\mathbf{\mathbf{\Theta}}_{\gamma}\mathbf{H}_2^H \mathbf{T}_1]_{iq}^* [\mathbf{H}_1]_{jp}^*  +  \alpha_{\gamma} [\mathbf{Q}_1\mathbf{H}_1 \mathbf{R}_2 \mathbf{T}_1]_{iq}  [\mathbf{H}_1]_{jp}^* \\
    &+  \frac{\jmath t_1}{M} [\mathbf{Q}_1\mathbf{H}_1\mathbf{R}_2 \mathbf{H}_1^H \mathbf{Q}_1 \mathbf{H}_1 \mathbf{H}_2 \mathbf{T}_2\mathbf{\mathbf{\Theta}}_{\gamma}\mathbf{H}_2^H \mathbf{T}_1]_{iq} [\mathbf{H}_1]_{jp}^*  \Big\}\widehat{\psi}. \label{Eq_CLT_QH3T}
\end{split}
\end{equation}
Multiplying both sides of \eqref{Eq_CLT_QH3T} by $\alpha_{\delta}$ and subtracting the result from \eqref{Eq_CLT_QH1}, the following can be obtained by expressing all the r.v.s $\widehat{x}$ as $x + \mathring{x}$
\begin{equation}
\begin{split}
     & \mathbb{E} [\mathbf{Q}_1\mathbf{H}_1 \mathbf{\Theta_{\omega}}^{-1}]_{iq}  [\mathbf{H}_1]_{jp}^* \widehat{\psi} = \mathbb{E}\Big\{  \alpha_{\delta}  \mathring \kappa[\mathbf{Q}_1\mathbf{H}_1 \mathbf{H}_2 \mathbf{T}_2\mathbf{\Theta}_{\gamma} \mathbf{H}_2^H \mathbf{T}_1]_{iq} [\mathbf{H}_1]_{jp}^*  
     -\mathring{\alpha}_{\delta} [\mathbf{Q}_1 \mathbf{H}_1 \mathbf{H}_2 \mathbf{H}_2^H \mathbf{T}_1]_{iq}   [\mathbf{H}_1]_{jp}^* \\
     & - \frac{\jmath t_1 \alpha_{\delta} }{M} [\mathbf{Q}_1\mathbf{H}_1\mathbf{R}_2 \mathbf{H}_1^H \mathbf{Q}_1 \mathbf{H}_1 \mathbf{H}_2 \mathbf{T}_2\mathbf{\Theta}_{\gamma} \mathbf{H}_2^H \mathbf{T}_1]_{iq} [\mathbf{H}_1]_{jp}^*   - \overline{s} \mathring{\alpha}_{\delta} [\mathbf{Q}_1 \mathbf{H}_1\mathbf{T}_1]_{iq} [\mathbf{H}_1]_{jp}^* + \frac{[\mathbf{T}_1]_{pq}}{L}[\mathbf{Q}_{1} \mathbf{R}_1]_{ij}   \\
     &+ \frac{\jmath t_1}{L} [ \mathbf{Q}_1\mathbf{R}_1 \mathbf{Q}_1 \mathbf{H}_1 \mathbf{H}_2 \mathbf{H}_2^H \mathbf{T}_1]_{iq} [\mathbf{H}_1]_{jp}^*  + \frac{\jmath \overline{s} t_1}{L}\mathbb{E}  [\mathbf{Q}_1 \mathbf{R}_1 \mathbf{Q}_1\mathbf{H}_1\mathbf{T}_1]_{iq} [\mathbf{H}_1]_{jp}^* + \frac{\jmath \underline{s} t_2}{L}[\mathbf{Q}_1 \mathbf{R}_1 \mathbf{Q}_2\mathbf{H}_1 \mathbf{T}_1]_{iq} [\mathbf{H}_1]_{jp}^* \Big\}\widehat{\psi}. \label{CLT_2}
\end{split}
\end{equation}
By multiplying $[\mathbf{\mathbf{\Theta}_{\omega}}(\overline{s} \mathbf{I}_L + \alpha_{\gamma} \mathbf{R}_2)]_{qp}$ on both sides of \eqref{CLT_2} then summing over the subscript $q, p$, 
and multiplying $[\mathbf{\mathbf{\Theta}_{\gamma}}]_{ls}$ and $\delta(m - p)$ on both sides of \eqref{CLT_1} then summing over $l, s, m$ and $p$, we can get two equations. 
By adding these two equations together, we have
\begin{equation}
\begin{split}
    &\mathbb{E} \{ [\mathbf{Q}_1\mathbf{H}_1 \mathbf{H}_2 \mathbf{H}_2^H \mathbf{H}_1^H]_{ij} \widehat{\psi}+ \overline{s} [\mathbf{Q}_1\mathbf{H}_1 \mathbf{H}_1^H]_{ij} \widehat{\psi}\} 
    = \mathbb{E} \Big\{ (\overline{s} \alpha_{\overline{\omega}} + \alpha_{\gamma} \alpha_{\underline{\omega}}) [\mathbf{Q}_{1} \mathbf{R}_1]_{ij} - \mathring \alpha_{\delta} [\mathbf{Q}_1 \mathbf{H}_1 \mathbf{H}_2 \mathbf{H}_2^H  \overline{\mathbf{\Theta}}_{\omega} \mathbf{H}_1^H]_{ij}  \\
    & + \frac{\jmath \underline{s} t_2}{L} [\mathbf{Q}_1 \mathbf{R}_1 \mathbf{Q}_2\mathbf{H}_1\overline{\mathbf{\Theta}}_{\omega} \mathbf{H}_1^H]_{ij} + \frac{\jmath t_1}{L}   [ \mathbf{Q}_1\mathbf{R}_1 \mathbf{Q}_1 \mathbf{H}_1 \mathbf{H}_2 \mathbf{H}_2^H \overline{\mathbf{\Theta}}_{\omega} \mathbf{H}_1^H]_{ij} + \frac{\jmath \overline{s} t_1}{L}  [\mathbf{Q}_1 \mathbf{R}_1 \mathbf{Q}_1\mathbf{H}_1\overline{\mathbf{\Theta}}_{\omega} \mathbf{H}_1^H]_{ij}  \\
     &- \overline{s}  \mathring \alpha_{\delta} [\mathbf{Q}_1 \mathbf{H}_1\overline{\mathbf{\Theta}}_{\omega} \mathbf{H}_1^H]_{ij}  -  \mathring \kappa[\mathbf{Q}_1\mathbf{H}_1 \mathbf{H}_2 \mathbf{T}_2\mathbf{\Theta}_{\gamma} \mathbf{H}_2^H\mathbf{\Theta}_{\omega}^H \mathbf{H}_1^H]_{ij} +  \frac{\jmath t_1}{M} [\mathbf{Q}_1\mathbf{H}_1\mathbf{R}_2 \mathbf{H}_1^H \mathbf{Q}_1 \mathbf{H}_1 \mathbf{H}_2 \mathbf{T}_2\mathbf{\Theta}_{\gamma} \mathbf{H}_2^H \mathbf{\Theta}_{\omega}^H\mathbf{H}_1^H]_{ij}  \Big\}\widehat{\psi} , \label{CLT_Hole}
\end{split}   
\end{equation}
where the deterministic matrix $\overline{\mathbf{\Theta}}_{\omega}$ follows the definition in \eqref{overline_Theta_omega}. 
By applying the resolvent identity of $\mathbf{Q}_1$ \eqref{Resolvent_identity_Q1} to 
the term $[\mathbf{Q}_{1} \mathbf{R}_1]_{ij}$ in the first line of \eqref{CLT_Hole}, multiplying $[\mathbf{\Theta}_{\delta}]_{ji}$ on both sides of \eqref{CLT_Hole} and summing over $i$, $j$, the following holds
\begin{equation}
\begin{split}
    \mathcal{X} 
    &= \mathbb{E}\Big[ -z \overline{s}\mathring \alpha_{\delta} \widehat{\Psi}_1(\mathbf{\Theta}_{\delta}, \overline{\mathbf{\Theta}}_{\omega})   -z\mathring \alpha_{\delta} \widehat{\Psi}_2(\mathbf{\Theta}_{\delta}, \mathbf{I}_M, \overline{\mathbf{\Theta}}_{\omega})  -z\mathring \kappa \widehat{\Psi}_2(\mathbf{\Theta}_\delta, \mathbf{T}_2 \mathbf{\Theta}_{\gamma}, \mathbf{\Theta}_{\omega}^H)   \\
    &+ \frac{\jmath t_1z \overline{s} \widehat{\Psi}_3(\mathbf{\Theta}_{\delta}, \mathbf{R}_1, \overline{\mathbf{\Theta}}_{\omega})}{L}  + \frac{\jmath t_1 z \widehat{\Psi}_4(\mathbf{\Theta}_{\delta}, \mathbf{R}_1, \mathbf{I}_M, \overline{\mathbf{\Theta}}_{\omega})}{L}  + \frac{\jmath zt_1 \widehat{\Psi}_6(\mathbf{\Theta}_{\delta}, \mathbf{T}_2 \mathbf{\Theta}_{\gamma}, \mathbf{\Theta}_{\omega}^H)}{M}  \\
    &+ \frac{\jmath t_2z \underline{s} \widehat{\Psi}_7(\mathbf{\Theta}_{\delta}, \overline{\mathbf{\Theta}}_{\omega})}{L} \Big]\widehat{\psi} + L (s \alpha_{\overline{\omega}} + \alpha_{\gamma} \alpha_{\underline{\omega}}) \widetilde{\alpha}_{\delta}\psi  \\
    & \overset{(a)}{=} -z \overline{s} {\Psi}_1(\mathbf{\Theta}_{\delta}, \overline{\mathbf{\Theta}}_{\omega}) \mathcal{Z}_{\alpha_\delta} - z {\Psi}_2(\mathbf{\Theta}_{\delta}, \mathbf{I}_M, \overline{\mathbf{\Theta}}_{\omega}) \mathcal{Z}_{\alpha_\delta} -z{\Psi}_2(\mathbf{\Theta}_\delta, \mathbf{T}_2 \mathbf{\Theta}_{\gamma}, \mathbf{\Theta}_{\omega}^H) \mathcal{Z}_{\kappa}  \\
    &+ \frac{\jmath t_1z \overline{s} {\Psi}_3(\mathbf{\Theta}_{\delta}, \mathbf{R}_1, \overline{\mathbf{\Theta}}_{\omega})}{L}{\psi} + \frac{\jmath t_1 z {\Psi}_4(\mathbf{\Theta}_{\delta}, \mathbf{R}_1, \mathbf{I}_M, \overline{\mathbf{\Theta}}_{\omega})}{L}{\psi} + \frac{\jmath zt_1 {\Psi}_6(\mathbf{\Theta}_{\delta}, \mathbf{T}_2 \mathbf{\Theta}_{\gamma}, \mathbf{\Theta}_{\omega}^H)}{M} {\psi} \\
    &+ \frac{\jmath t_2z \underline{s} {\Psi}_7(\mathbf{\Theta}_{\delta}, \overline{\mathbf{\Theta}}_{\omega}) }{L}{\psi} + L (\overline{s} \alpha_{\overline{\omega}} + \alpha_{\gamma} \alpha_{\underline{\omega}}) \widetilde{\alpha}_{\delta}\psi + \varepsilon_{\mathcal{X}} {=} \sum_{i=1}^8 \mathcal{X}_i + \varepsilon_{\mathcal{X}}, \label{X_to_Psi_terms}
\end{split}  
\end{equation}
where $\widetilde{\alpha}_{\delta} $ follows the definition in Table \ref{Temp_Notations}.
Here, we define $\mathcal{Z}_{\alpha_\delta} = \mathbb{E}\mathring{\alpha}_{\delta} \widehat{\psi}$ and 
  $\mathcal{Z}_{\kappa} = \mathbb{E}\mathring{\kappa}  \widehat{\psi}$. Step $(a)$ follows by 
expressing $\widehat{\Psi}_i$ in the form of $\Psi_i + \mathring{\Psi}_i$. The term $\varepsilon_{\mathcal{X}}$ can be proved to be $\mathcal{O}_{z}^s(\frac{1}{N})$ by using variance control in Proposition \ref{Prop_variance} and \eqref{Var_control}, and  Cauchy–Schwarz inequality as follows 
\begin{equation}
\begin{split}
\abs{\varepsilon_\mathcal{X}} &= \abs{ \mathbb{E}\Big( -z\overline{s}\mathring \alpha_{\delta} \mathring{\Psi}_1   -z\mathring \alpha_{\delta} \mathring{\Psi}_2  -z\mathring \kappa \mathring{\Psi}_2  
    + \frac{\jmath t_1z \overline{s} \mathring{\Psi}_3}{L}  + \frac{\jmath t_1 z \mathring{\Psi}_4}{L}  + \frac{\jmath zt_1 \mathring{\Psi}_6}{M} 
    + \frac{\jmath t_2z \underline{s} \mathring{\Psi}_7}{L} \Big) \mathring{\psi}} \\
   &\leq \mathbb{E}^{\frac{1}{2}} \abs{\mathring \alpha_{\delta}}^2 \left(z \overline{s}  \mathbb{E}^{\frac{1}{2}} \abs{\mathring{\Psi}_1}^2   +  z  \mathbb{E}^{\frac{1}{2}} \abs{\mathring{\Psi}_2}^2 \right)  + z \mathbb{E}^{\frac{1}{2}} \abs{\mathring \kappa}^2 \mathbb{E}^{\frac{1}{2}} \abs{\mathring{\Psi}_2}^2 
    + \frac{\abs{t_1} z \overline{s} \mathbb{E}^{\frac{1}{2}}\abs{\mathring{\Psi}_3}^2}{L}  + \frac{|t_1| z \mathbb{E}^{\frac{1}{2}}\abs{\mathring {\Psi}_4}^2}{L} \\
    &+ \frac{ z|t_1| \mathbb{E}^{\frac{1}{2}}\abs{\mathring{\Psi}_6}^2}{M} + \frac{|t_2| z \underline{s} \mathbb{E}^{\frac{1}{2}}\abs{\mathring{\Psi}_7}^2}{L} \overset{(a)}{=} \mathcal{O}_{z}^s(\frac{1}{N}). \label{CLT_res_X}
\end{split}
\end{equation}
Here, we omit the variables of $\Psi_i$ for clarity. Step $(a)$ follows by the bound of the norm $\norm{\mathbf{\Theta}_{\delta}} \leq \frac{1}{z}$. According to Lemma \ref{Lemma_DE_Q1_1} and Lemma \ref{Lemma_DE_Q1_2},
 we can obtain $\mathcal{X}_8 = L (\overline{s} \alpha_{\overline{\omega}} + \alpha_{\gamma} \alpha_{\underline{\omega}}) \widetilde{\alpha}_{\delta}  
 \psi = L( \overline{s} \overline{\omega} + \gamma \underline{\omega}) \delta \psi + \mathcal{O}_{z}^s(\frac{1}{N})$. 
Next, we utilize Lemma \ref{Lemma_DE_terms_Approx_Psi} and 
 the approximation rules \eqref{Approx_omega_gamma} to derive the following result. By
 replacing $\mathbf{\Theta}_{sym}$ with $\mathbf{F}_{sym}$, where $sym$ can be any symbol of
$\delta$, $\omega$, or $\gamma$, we obtain
\begin{subequations}
\label{CLT_X_terms}
\begin{align}
    \mathcal{X}_1 &= -z \overline{s} L(\overline{s} \overline{\omega}_2 + \gamma \overline{\underline{\omega}}_{1,1}) \delta_{2,I} \mathcal{Z}_{\alpha_\delta}  + \mathcal{O}_{z}^s(\frac{1}{N^2}), \label{X_1} \\
    \mathcal{X}_2 &= -z L \gamma (\overline{s} \overline{\underline{\omega}}_{1,1} + \gamma \underline{\omega}_2) \delta_{2,I} \mathcal{Z}_{\alpha_\delta} + \mathcal{O}_{z}^s(\frac{1}{N^2}),  \\
    \mathcal{X}_3 &= -z L\gamma_2 \underline{\omega}_{2, I} \delta_{2, I} \mathcal{Z}_{\kappa} + \mathcal{O}_{z}^s(\frac{1}{N^2}), \\
    \mathcal{X}_4 &= \frac{\jmath t_1 z }{\Delta_{V_1}} \Big[ \underbrace{- \overline{s}{\Delta \delta_2 \delta_{2,I}}(\overline{s}^2 \overline{\omega}_3 + 
    2 \overline{s} \gamma  \overline{\underline{\omega}}_{2,1}  + \gamma^2  \overline{\underline{\omega}}_{1,2} ) }_{\mathcal{X}_{4,1}} 
    + \underbrace{{\frac{L}{M}\overline{s}\underline{\omega}_{2, I} \gamma_2 \delta \delta_2\delta_{2,I}}(\overline{s} \overline{\underline{\omega}}_{2,1} 
    + \gamma \overline{\underline{\omega}}_{1,2} ) }_{\mathcal{X}_{4,2}} \notag \\
    &+ \underbrace{ \overline{s}{\Delta \delta_{3,I}}(\overline{s}\overline{\omega}_2 + \gamma \underline{\overline{\omega} }_{1,1})}_{\mathcal{X}_{4,3}} \Big] \psi 
    +  \mathcal{O}_{z}^s(\frac{1}{N^2}), \\
    \mathcal{X}_5 &=  \frac{\jmath t_1 z}{\Delta_{V_1}}\Big[ \underbrace{- {\frac{L}{M}\underline{\omega}_{2, I} \delta_2 \gamma_2 \delta_{2, I}}( \overline{s} \overline{\underline{\omega}}_{1,1}
     + \gamma \underline{\omega}_2)}_{\mathcal{X}_{5,1}} +  \underbrace{[- \Delta \delta_2 \gamma \delta_{2, I}(\overline{s}^2
      \overline{\underline{\omega}}_{2, 1} + 2 \overline{s} \gamma \overline{\underline{\omega}}_{1,2} + \gamma^2 \underline{\omega}_3)]}_{\mathcal{X}_{5,2}} \notag \\
    &+  \underbrace{\frac{L}{M}\underline{\omega}_{2, I} \gamma_2 \delta \delta_2\gamma \delta_{2,I} (\overline{s} \underline{\overline{\omega}}_{1,2} 
    + \gamma \underline{\omega}_3)}_{\mathcal{X}_{5,3}} + \underbrace{\Delta \gamma \delta_{3,I}( \overline{s} \overline{\underline{\omega}}_{1,1} + \gamma \underline{\omega}_2)}_{\mathcal{X}_{5, 4}}\Big]\psi + \mathcal{O}_{z}^s(\frac{1}{N^2}), \\
    \mathcal{X}_6 &= \frac{\jmath z t_1 }{\Delta_{V_1}} \Big[ \underbrace{[- (\frac{\frac{L}{M} \underline{\omega}_{2,I}^2 \delta_2}{\Delta} +  \frac{\frac{L}{M} \delta^2 \underline{\omega}_2\Delta_{V_1}} {\Delta})\frac{L}{M}\gamma_3\underline{\omega}_{2,I}\delta_{2,I}]}_{\mathcal{X}_{6,1}} 
    + \underbrace{[- \frac{L}{M}\gamma_2 \delta_2 \underline{\omega}_{2,I} \delta_{2,I}( \overline{s} \overline{\underline{\omega}}_{1,2,I} + \gamma \underline{\omega}_{3,I})]}_{\mathcal{X}_{6,2}} \notag \\
    &+ \underbrace{\frac{L}{M}\gamma_2\delta(1-\varsigma \delta_2) \delta_{2,I} \underline{\omega}_{3,I}}_{\mathcal{X}_{6,3}}  + \underbrace{\frac{L}{M}\gamma_2 \underline{\omega}_{2,I}^2 \delta_{3,I}}_{\mathcal{X}_{6,4}}  \Big] \psi + \mathcal{O}_{z}^s(\frac{1}{N^2}), \\
    \mathcal{X}_7  &= \frac{\jmath t_2 z }{\Delta_{C}} \Big[\underbrace{-(\overline{s}^2 \overline{\phi}_{1,2} +  \overline{s}\gamma \widetilde{\overline{\underline{\phi}}}_{1,2}
     + \overline{s} \gamma \overline{\underline{\phi}}_{1,2} +  \gamma^2 \underline{\phi}_{1,2})\underline{s}\vartheta \delta_{2,I}}_{\mathcal{X}_{7,1}} + \underbrace{(\overline{s} \overline{\phi} + \gamma \underline{\phi}) \underline{s} \vartheta_{1,2,I}}_{\mathcal{X}_{7,2}} \Big] \psi  + \mathcal{O}_{z}^s(\frac{1}{N^2}), 
\end{align}
\end{subequations}
In the above equations, $\mathcal{X}_i$, $i=4,5, 6,7$, has been decomposed into the product of $\psi$ and some deterministic terms defined in
 Table \ref{tabel_of_notations}. 
 To further analyze the term $\mathcal{X}$, we need to decompose and estimate $\mathcal{Z}_{\alpha_\delta}$ and $\mathcal{Z}_{\kappa}$ in the same way, and the results are given in the following lemma.
\begin{lemma} \label{Lemma_Z_alpha_kappa}
Assume that assumptions \textbf{\ref{A-1}} - \textbf{\ref{A-3}} hold, $\mathcal{Z}_{\alpha_\delta}$ and $\mathcal{Z}_{\kappa}$ are defined in \eqref{X_to_Psi_terms}, and $\mathcal{X}_{i,j}$ is defined in \eqref{CLT_X_terms}. Then the following holds true
\begin{equation}
     \begin{bmatrix}
        \mathcal{Z}_{\alpha_\delta} \\
        \mathcal{Z}_{\kappa}
    \end{bmatrix} = \frac{1}{L \Delta_{V_1}}\begin{bmatrix}
         \Delta & \frac{L}{M} \gamma_2 \underline{\omega}_{2,I} \delta_2 \\
        \frac{L}{M} \underline{\omega}_{2,I} & \frac{L}{M}(1 - \varsigma \delta_2)
    \end{bmatrix}\begin{bmatrix}
            X_{1,V_1} + X_{1, C} \\
            X_{2,V_1} + X_{2, C}
    \end{bmatrix} + \begin{bmatrix}
        \mathcal{O}_{z}^s(\frac{1}{N^2}) \\
        \mathcal{O}_{z}^s(\frac{1}{N^2})
    \end{bmatrix} 
\end{equation}
where
\begin{subequations}
\begin{align}
    X_{1, V_1}  &=   - \frac{\jmath t_1}{\Delta_{V_1}}\Big[\frac{\delta_2}{\delta_{2,I}} (\mathcal{X}_{4,1} + \mathcal{X}_{4,2} + \mathcal{X}_{5,1} + \mathcal{X}_{5,2} + \mathcal{X}_{5,3}+ \mathcal{X}_{6,1} + \mathcal{X}_{6,2} + \mathcal{X}_{6,3}) + \frac{\delta_3}{\delta_{3,I}} (\mathcal{X}_{4,3} +\mathcal{X}_{5,4} +\mathcal{X}_{6,4})\Big]\psi,  \\
   X_{1, C}
   &= -\frac{\jmath  t_2}{\Delta_{C}}( \mathcal{X}_{7,1} \frac{\delta_2}{\delta_{2,I}}  + \mathcal{X}_{7,2} \frac{\vartheta_{1,2}}{\vartheta_{1,2,I}} )\psi,  \\
   X_{2, V_1} &=  -\frac{\jmath t_1}{\Delta_{V_1}} \Big[\underbrace{[-(\frac{\frac{L}{M} \underline{\omega}_{2,I}^2 \delta_2}{\Delta} +  \frac{\frac{L}{M} \delta^2 \underline{\omega}_2\Delta_{V_1}} {\Delta})\frac{L}{M}\gamma_3\underline{\omega}_{2}\delta^2]}_{\widetilde{\mathcal{X}}_1} + \underbrace{\frac{L}{M}\gamma_2(1-\varsigma \delta_2) \delta^3 \underline{\omega}_{3}}_{\widetilde{\mathcal{X}}_2} \notag \\ 
   &+ \underbrace{2\frac{L}{M} \gamma_2 \underline{\omega}_{2,I}  \delta_{2} \delta \underline{\omega}_{3,I}}_{\widetilde{\mathcal{X}}_3} + \underbrace{[- \Delta  \delta_{2}(\gamma\underline{\omega}_{3,I} + \overline{s}\overline{\underline{\omega}}_{1,2,I})]}_{\widetilde{\mathcal{X}}_4}\Big]\psi, \\
    X_{2, C} &= -{\frac{\jmath  t_2}{\Delta_C}\Big[\underbrace{(\overline{s} \overline{\underline{\phi}}_{1,2} + \gamma \underline{\phi}_{1,2}) \underline{s}\delta \vartheta}_{\widetilde{\mathcal{X}}_5} 
    + \underbrace{(- \underline{s}\underline{\phi} \vartheta)}_{\widetilde{\mathcal{X}}_6} \Big]}\psi.
\end{align}
\label{Lemma_Z_alpha_kappa_X_terms}
\end{subequations}
\end{lemma}
\textit{Proof:} The proof is quite similar to the evaluation from \eqref{CLT_Eq_Inte_By_Parts_Q_1} to \eqref{CLT_X_terms} and given in Appendix \ref{App_Lemma_Z_alpha_kappa}. \QED
\par
Plugging $\mathcal{Z}_{\alpha_{\delta}}$ and $\mathcal{Z}_{\kappa}$ into $\mathcal{X}_1$, $\mathcal{X}_2$, and $\mathcal{X}_3$, the following holds
\begin{equation}
    \sum_{i=1}^8\mathcal{X}_i = \frac{-z \delta_{2,I}}{\Delta_{V_1}}\begin{bmatrix}
        \Delta \varsigma + \frac{L}{M} \gamma_2 \underline{\omega}_{2,I}^2 & \frac{L}{M} \gamma_2 \underline{\omega}_{2,I}
    \end{bmatrix}\begin{bmatrix}
            X_{1,V_1} + X_{1, C} \\
            X_{2,V_1} + X_{2, C}
    \end{bmatrix} + \sum_{i=4}^8 \mathcal{X}_i + \mathcal{O}_{z}^s(\frac{1}{N})
    = \mathcal{X}_{V_1} + \mathcal{X}_{C} + \mathcal{X}_8 + \mathcal{O}_{z}^s(\frac{1}{N}),
\end{equation}
where 
\begin{subequations}
\begin{align}
    \mathcal{X}_{V_1} &= \frac{-z \delta_{2,I}}{\Delta_{V_1}}\begin{bmatrix}
        \Delta \varsigma + \frac{L}{M} \gamma_2 \underline{\omega}_{2,I}^2 & \frac{L}{M} \gamma_2 \underline{\omega}_{2,I}
    \end{bmatrix}\begin{bmatrix}
            X_{1,V_1} \\
            X_{2,V_1}
    \end{bmatrix} + \sum_{i=4}^6 \mathcal{X}_i,  \label{X_V_1_terms}\\
    \mathcal{X}_{C} &= \frac{-z \delta_{2,I}}{\Delta_{V_1}}\begin{bmatrix}
        \Delta \varsigma + \frac{L}{M} \gamma_2 \underline{\omega}_{2,I}^2 & \frac{L}{M} \gamma_2 \underline{\omega}_{2,I}
    \end{bmatrix}\begin{bmatrix}
            X_{1,C} \\
            X_{2,C}
    \end{bmatrix} + \mathcal{X}_7.
\end{align}
\end{subequations}
Thus, we have completed the decomposition of $\mathcal{X}$.
\paragraph{Decomposition of $\mathcal{Y}$} Next, we will use the same technique to tackle the term $\mathcal{Y}$. 
Applying the integration by parts formula \eqref{Integration_by_parts} on $[\mathbf{H}_1]_{kl}$ yields
\begin{equation}
    \begin{split}
    &\sum_{k} \mathbb{E}  [\mathbf{Q}_2]_{ik} [\mathbf{H}_1]_{kl} [\mathbf{H}_1]_{jm}^*\widehat{\psi} 
    = \sum_{k, a,b} \frac{1}{L} [\mathbf{R}_1]_{ka}[\mathbf{T}_1]_{bl} \mathbb{E} \Big\{ - \underline{s} [\mathbf{Q}_2 \mathbf{H}_1]_{ib} [\mathbf{Q}_2]_{ak}[\mathbf{H}_1]_{jm}^* \\
    & + [\mathbf{Q}_2]_{ik} \delta(j-a) \delta(m-b)  +  \jmath [\mathbf{Q}_2]_{ik} [\mathbf{H}_1]_{jm}^*(t_1 [\mathbf{Q}_1 \mathbf{H}_1 \mathbf{H}_2 \mathbf{H}_2^H]_{ab} 
    +  \overline{s} t_1[\mathbf{Q}_1 \mathbf{H}_1]_{ab}+ \underline{s} t_2[\mathbf{Q}_2 \mathbf{H}_1]_{ab} ) \Big\}\widehat{\psi}.  \label{CLT_Eq_Inte_by_Parts_Q2} 
    \end{split}
\end{equation}
By writing $\widehat{\beta}_{\tau} = \mathring{\beta}_{\tau} + \beta_{\tau}$, multiplying $[\mathbf{\Omega}_{\overline{\tau}}]_{lm}$ on both sides, and summing the subscripts $l$ and $m$, we have
\begin{equation}
    \begin{split}
    & \mathbb{E} [\mathbf{Q}_2 \mathbf{H}_1 \mathbf{H}_1^H]_{ij}\widehat{\psi} = \mathbb{E} \Big\{ -\underline{s} \mathring{\beta}_{\tau} [\mathbf{Q}_2 \mathbf{H}_1 \mathbf{T}_1\mathbf{\Omega}_{\overline{\tau}} \mathbf{H}_1^H]_{ij} 
    + \frac{\jmath t_1 }{L}  [\mathbf{Q}_2 \mathbf{R}_1 \mathbf{Q}_1 \mathbf{H}_1 \mathbf{H}_2 \mathbf{H}_2^H\mathbf{T}_1\mathbf{\Omega}_{\overline{\tau}} \mathbf{H}_1^H]_{ij} \\
    &+  \beta_{\overline{\tau}}  [\mathbf{Q}_{2}\mathbf{R}_{1}]_{ij}    + \frac{\jmath \overline{s} t_1}{L} [\mathbf{Q}_2 \mathbf{R}_1\mathbf{Q}_1 \mathbf{H}_1 \mathbf{T}_1\mathbf{\Omega}_{\overline{\tau}} \mathbf{H}_1^H]_{ij} 
     + \frac{\jmath \underline{s} t_2}{L} [\mathbf{Q}_2 \mathbf{R}_1\mathbf{Q}_2 \mathbf{H}_1\mathbf{T}_1\mathbf{\Omega}_{\overline{\tau}} \mathbf{H}_1^H]_{ij} \Big\}\widehat{\psi}. \label{CLT_Q2_resol}
    \end{split}
\end{equation}
By using the resolvent identity of $\mathbf{Q}_2$ on the term $  [\mathbf{Q}_{2}\mathbf{R}_{1}]_{ij} $ in the above equation
and multiplying $[\mathbf{\Omega}_{\tau}]_{ji}$ on both sides, the following can be obtained by summing over the subscripts $i$ and $j$
\begin{equation}
    \begin{split}
     \mathcal{Y} & = \mathbb{E} \Big[- z\underline{s}^2\mathring{\beta}_{\tau} \widehat{\Phi}_1(\mathbf{\Omega}_{\tau}, \mathbf{T}_1\mathbf{\Omega}_{\overline{\tau}})  +  \frac{\jmath \underline{s}^2t_2z \widehat{\Phi}_2(\mathbf{\Omega}_{\tau},  \mathbf{T}_1\mathbf{\Omega}_{\overline{\tau}})}{L}  + \frac{\jmath \overline{s} \underline{s} t_1z \widehat{\Phi}_3(\mathbf{\Omega}_{\tau},  \mathbf{T}_1\mathbf{\Omega}_{\overline{\tau}})}{L}  + \frac{\jmath \underline{s} t_1z \widehat{\Phi}_4(\mathbf{\Omega}_{\tau}, \mathbf{I}_M, \mathbf{T}_1\mathbf{\Omega}_{\overline{\tau}})}{L} \Big]\widehat{\psi}\\
    &+  L\underline{s} \beta_{\overline{\tau}} \widetilde{\beta}_{{\tau}} {\psi} = - z\underline{s}^2{\Phi}_1(\mathbf{\Omega}_{\tau}, \mathbf{T}_1\mathbf{\Omega}_{\overline{\tau}}) \mathcal{Z}_{\beta_{\tau}}  +  \frac{\jmath \underline{s}^2t_2z {\Phi}_2(\mathbf{\Omega}_{\tau},  \mathbf{T}_1\mathbf{\Omega}_{\overline{\tau}})}{L} \psi  + \frac{\jmath \underline{s}\overline{s} t_1z {\Phi}_3(\mathbf{\Omega}_{\tau},  \mathbf{T}_1\mathbf{\Omega}_{\overline{\tau}})}{L} \psi \\
    &+ \frac{\jmath \underline{s} t_1z {\Phi}_4(\mathbf{\Omega}_{\tau}, \mathbf{I}_M, \mathbf{T}_1\mathbf{\Omega}_{\overline{\tau}})}{L} \psi + L\underline{s}\beta_{\overline{\tau}} \widetilde{\beta}_{{\tau}} {\psi}  + \varepsilon_{\mathcal{Y}} = \sum_{i=1}^5 \mathcal{Y}_i + \varepsilon_{\mathcal{Y}}, \label{CLT_Eq_Y_to_Phi}
    \end{split}
\end{equation}
where $\mathcal{Z}_{\beta_{\tau}}$ is defined as $\mathcal{Z}_{\beta_{\tau}} = \mathbb{E}\mathring{\beta}_{\tau} \widehat{\psi}$ and 
$
\varepsilon_\mathcal{Y} = \mathbb{E}(- z\underline{s}^2\mathring{\beta}_{\tau} \mathring{\Phi}_1  +  \frac{\jmath \underline{s}^2t_2z }{L}\mathring{\Phi}_2  + \frac{\jmath \underline{s}\overline{s} t_1z }{L}\mathring{\Phi}_3  + \frac{\jmath \underline{s} t_1z }{L}\mathring{\Phi}_4 ) \mathring \psi
$. The variables for $\Phi_i$ are omitted here without causing ambiguity. Following the same step in \eqref{CLT_res_X}, we have $\abs{\varepsilon_{\mathcal{Y}}} = \mathcal{O}_{z}^s(\frac{1}{N})$. Through the approximation rules \eqref{Approx_rules_Omega}, we know that $\mathcal{Y}_5 =  L\underline{s} \beta_{\overline{\tau}} \widetilde{\beta}_{{\tau}} {\psi} =  L\underline{s}\overline{\tau}\tau \psi + \mathcal{O}_{z}^s(\frac{1}{N})$. Using Lemma \ref{Lemma_DE_terms_Approx_Phi} and \eqref{Approx_rules_Omega}, we obtain the following 
\begin{subequations}
\begin{align}
    \mathcal{Y}_1 &= - z\underline{s}^2  L\overline{\tau}_2 \tau_{2,I} \mathcal{Z}_{\beta_{\tau}} + \mathcal{O}_{z}^s(\frac{1}{N^2}), \label{Y_1}\\
    \mathcal{Y}_2 &=  \frac{\jmath t_2z}{\Delta_{V_2}}\Big(\underbrace{-{\underline{s}^3\tau_2}\overline{\tau}_3 \tau_{2,I}}_{\mathcal{Y}_{2,1}} + \underbrace{ \underline{s}^2\overline{\tau}_2 \tau_{3,I}}_{\mathcal{Y}_{2,2}}\Big) \psi  + \mathcal{O}_{z}^s(\frac{1}{N^2}), \\
    \mathcal{Y}_3 &= \frac{\jmath t_1z}{\Delta_{C}}\Big(\underbrace{-\underline{s}^2\overline{s} \vartheta \overline{\phi}_{2,1} \tau_{2, I}}_{\mathcal{Y}_{3,1}} + \underbrace{\underline{s}\overline{s}\overline{\phi} \vartheta_{2,I, 1}}_{\mathcal{Y}_{3,2}} \Big)\psi + \mathcal{O}_{z}^s(\frac{1}{N^2}), \\
    \mathcal{Y}_4 
    &= \frac{\jmath t_1z }{\Delta_C}\Big(\underbrace{-\underline{s}^2 \vartheta \gamma \underline{\phi}_{2,1} \tau_{2,I}}_{\mathcal{Y}_{4,1}} + \underbrace{\underline{s} \gamma \underline{\phi} \vartheta_{2,I, 1}}_{\mathcal{Y}_{4,2}} \Big)\psi  + \mathcal{O}_{z}^s(\frac{1}{N^2}), \label{Y_4} %
\end{align}\label{CLT_Ys}%
\end{subequations}
where $\mathcal{Z}_{\beta_{\tau}} = \mathbb{E}\mathring{\beta}_{\tau} \widehat{\psi}$. To evaluate $\mathcal{Z}_{\beta_{\tau}}$ we need the following lemma.
\begin{lemma} \label{Lemma_beta_tau}
Assume that assumptions \textbf{\ref{A-1}} - \textbf{\ref{A-3}} hold, $\mathcal{Z}_{\beta_{\tau}}$ is defined in \eqref{CLT_Eq_Y_to_Phi}, and $\mathcal{Y}_{i,j}$ is defined in \eqref{CLT_Ys}. Then, the following holds true
\begin{equation}
     \mathcal{Z}_{\beta_{\tau}} = Y_{V_2} + Y_{C} + \mathcal{O}_{z}^s(\frac{1}{N^2}),
\end{equation}
where 
\begin{subequations}
\begin{align}
Y_{V_2} &= - \frac{\jmath  t_2}{L\Delta_{V_2}^2}(\mathcal{Y}_{2,1} \frac{\tau_2}{\tau_{2,I}} + \mathcal{Y}_{2,2} \frac{\tau_3}{\tau_{3,I}}) \psi,\\
  Y_C  
    &= -\frac{\jmath t_1}{L\Delta_C\Delta_{V_2}} \Big[(\mathcal{Y}_{3,1} + \mathcal{Y}_{4,1}) \frac{\tau_2}{\tau_{2,I}} + (\mathcal{Y}_{3,2} + \mathcal{Y}_{4,2}) \frac{\vartheta_{2,1}}{\vartheta_{2,I, 1}} \Big] \psi.
\end{align}
\end{subequations}
\end{lemma}
\textit{Proof:} The proof of Lemma \ref{Lemma_beta_tau} is similar to the
 evaluation of \eqref{CLT_Eq_Inte_by_Parts_Q2}-\eqref{CLT_Ys}. 
 Moreover, the proof of Lemma \ref{Lemma_beta_tau} is very similar to that of 
 Lemma \ref{Lemma_Z_alpha_kappa} and the key step is to use the resolvent identity of $\mathbf{Q}_2$ on the term $\mathbb{E}[\mathbf{Q}_2\mathbf{H}_1\mathbf{H}_1^H]_{ij}\widehat{\psi}$ in \eqref{CLT_Q2_resol}. Therefore, we omit the details here. \QED
\par
By applying Lemma \ref{Lemma_beta_tau} to \eqref{Y_1}, the following is obtained
\begin{align}
    \sum_{i=1}^5 \mathcal{Y}_i& = -z \underline{s}^2 L \overline{\tau}_2 \tau_{2,I}(Y_{V_2} + Y_C) + \sum_{i=2}^5 \mathcal{Y}_i + \mathcal{O}_{z}^s(\frac{1}{N}) = \mathcal{Y}_{V_2} + \mathcal{Y}_C + \mathcal{Y}_5 + \mathcal{O}_{z}^s(\frac{1}{N}),\\
    \mathcal{Y}_{V_2} &= -z\underline{s}^2  L \overline{\tau}_2 \tau_{2,I}Y_{V_2} + \mathcal{Y}_2,  \hspace{2mm} \mathcal{Y}_C = -z\underline{s}^2  L \overline{\tau}_2 \tau_{2,I} Y_C + \mathcal{Y}_3 + \mathcal{Y}_4. \label{Y_V_2_and_Y_C}
\end{align}
\par
Thus, we have completed the decomposition of $\mathcal{Y}$. Summarizing the decomposition of $\mathcal{X}$ and $\mathcal{Y}$ and substituting the results into equation \eqref{Diff_Eq_CLT}, we can obtain the following
\begin{equation}
\begin{split}
    \frac{\partial \varphi }{\partial z} &{=} -\frac{\jmath}{z}(t_1 (\mathcal{X}_{V_1} + \mathcal{X}_{C} ) + t_2(\mathcal{Y}_{V_2} + \mathcal{Y}_C )) \exp[-\jmath \mathbf{t}^T \overline{\mathbf{m}}] +  \mathcal{O}_{z}^s(\frac{1}{zN}) \\
    &= -\frac{\varphi}{2} \mathbf{t}^T \begin{bmatrix}
        \frac{2 \jmath }{z t_1 \psi}\mathcal{X}_{V_1} & \frac{ \jmath }{z t_1t_2 \psi}(t_1 \mathcal{X}_C + t_2\mathcal{Y}_C)\\
       \frac{ \jmath }{z t_1t_2 \psi}(t_1 \mathcal{X}_C + t_2\mathcal{Y}_C) & \frac{2 \jmath }{z t_2 \psi} \mathcal{X}_{V_2}
    \end{bmatrix} \mathbf{t} + \mathcal{O}_{z}^s(\frac{1}{zN}). \label{Pre_CLT_Inge_Eq}
\end{split}
\end{equation}
\subsubsection{Construction of the Differential Equation} In this step, we will evaluate the terms $\mathcal{X}_{V_1}$, $\mathcal{Y}_{V_1}$ and $t_1 \mathcal{X}_{C} + t_2 \mathcal{Y}_C$, respectively, and show that $t_1 \mathcal{X}_{V_1} + t_2\mathcal{Y}_{V_1} + t_1 \mathcal{X}_{C} + t_2 \mathcal{Y}_C = \jmath z \psi [t_1^2 \log(\Delta_{V_1})' + t_2^2\log(\Delta_{V_2})' + 2 t_1 t_2 \log(\Delta_C)']$. We will start the proof from the LHS of the equation. Before entering into the specific calculations, we need the following derivative results, which can be obtained by implicit differentiation of equations \eqref{DE_system_1} and \eqref{DE_system_2} with respect to $z$. The differentiability of the positive solutions $(\delta, \overline{\omega}, \underline{\omega}, \gamma)$ is naturally obtained from the integration representations in the proof of Proposition \ref{Prop_Stieltjes}, and the differentiability of $(\tau, \overline{\tau})$ is quite similar. Thus, we have
\begin{equation}
    \mathbf{K}_1 \begin{bmatrix}
        \delta' \\
        \overline{\omega}' \\
        \underline{\omega}' \\
        \gamma' 
    \end{bmatrix} = \begin{bmatrix}
        1 & \overline{s} \delta_2 & \gamma \delta_2 & \underline{\omega} \delta_2 \\
        \overline{s} \overline{\omega}_2 + \gamma \overline{\underline{\omega}}_{1,1} &1&0& \delta \overline{\underline{\omega}}_{1,1} \\
        \overline{s}  \overline{\underline{\omega}}_{1,1} + \gamma \underline{\omega}_2& 0&1&\delta \underline
        {\omega}_2 \\
        \frac{L}{M}\underline{\omega} \gamma_2 & 0 & \frac{L}{M}\delta \gamma_2 & 1
    \end{bmatrix} \begin{bmatrix}
        \delta' \\
        \overline{\omega}' \\
        \underline{\omega}' \\
        \gamma' 
    \end{bmatrix} = \begin{bmatrix}
        -\delta_{2, I} \\
       0\\
        0 \\
        0 
    \end{bmatrix}. \label{Diff_funcamental_equations}
\end{equation}
Given $\det(\mathbf{K}_1) = \Delta_{V_1}$, the derivatives $(\delta', \overline{\omega}', \underline{\omega}', \gamma')$ can be calculated as, respectively,
\begin{align}
\delta' &= \frac{-\delta_{2, I}}{\Delta_{V_1}}\Delta, \hspace{2mm} \overline{\omega}' = \frac{\delta_{2, I}}{\Delta_{V_1}} [(\overline{s}  \overline{\omega}_2 + \gamma \overline{\underline{\omega}}_{1,1}) \Delta - \frac{L}{M}\gamma_2 \overline{\underline{\omega}}_{1,1} \underline{\omega}_{2, I} \delta], \notag \\
\underline{\omega}' &= \frac{\delta_{2, I}}{\Delta_{V_1}} [(\overline{s} \overline{\underline{\omega}}_{1,1} + \gamma \underline{\omega}_2) \Delta - \frac{L}{M}\gamma_2 \underline{\omega}_2 \underline{\omega}_{2, I} \delta], \hspace{2mm} \gamma' = \frac{\delta_{2,I}}{\Delta_{V_1}} \frac{L}{M}  \gamma_2 \underline{\omega}_{2, I}.\label{Diff_EQ_deltas}
\end{align} 
The other derivative equations are given by
\begin{equation}
    \mathbf{K}_2 \begin{bmatrix}
        \tau' \\
        \overline{\tau}' 
    \end{bmatrix} = \begin{bmatrix}
        1 & \underline{s} \tau_2 \\
        \underline{s} \overline{\tau}_2 & 1
    \end{bmatrix} \begin{bmatrix}
        \tau' \\
        \overline{\tau}' 
    \end{bmatrix} = \begin{bmatrix}
        -\tau_{2, I} \\
       0
    \end{bmatrix}.
\end{equation}
Given $\det(\mathbf{K}_2) = \Delta_{V_2}$, $(\tau', \overline{\tau}')$ can be expressed as, respectively, 
\begin{align}
\tau' = \frac{-\tau_{2, I}}{\Delta_{V_2}}, \hspace{2mm} \overline{\tau}' = \frac{\tau_{2, I}}{\Delta_{V_1}} \underline{s} \overline{\tau}_2. 
\end{align}
\paragraph{Evaluation of $\mathcal{X}_{V_1}$} Here, we further introduce some results, which can be obtained by using the definition and \eqref{Diff_EQ_deltas}
\begin{align}
    \delta_2' &= -2(\delta_{3,I}  + \frac{\delta_{2,I}\delta_3(\Delta \varsigma + \frac{L}{M} \gamma_2 \underline{\omega}_{2,I}^2)}{\Delta_{V_1}}),  \hspace*{2mm}\underline{\omega}_2' = -2( \overline{s} \delta' \overline{\underline{\omega}}_{1,2} + (\delta' \gamma + \delta \gamma')\underline{\omega}_3),~~
    \notag\\
\delta' \gamma + \delta \gamma' &= \frac{\delta_{2,I}}{\Delta_{V_1}}(- \gamma \Delta + \frac{L \gamma_2 \underline{\omega}_{2,I}\delta}{M}), ~~ \gamma_2' = \frac{\frac{L}{M}\gamma_3 \underline{\omega}_{2,I}\delta_{2,I}}{\Delta_{V_1}}.
\end{align}
These results will be used in the calculation of $\mathcal{X}_{V_1}$. Expanding the terms in \eqref{X_V_1_terms}, we obtain
\begin{equation}
    \begin{split}
    \mathcal{X}_{V_1} 
    &= \frac{\jmath t_1 z \psi}{\Delta_{V_1}} \Big\{\mathcal{X}_{4,1} + \mathcal{X}_{4,2} +  \mathcal{X}_{4,3} + \mathcal{X}_{5,1} + \mathcal{X}_{5,2} + \mathcal{X}_{5,3} + \mathcal{X}_{5,4} + \mathcal{X}_{6,1} + \mathcal{X}_{6,2} + \mathcal{X}_{6,3} + \mathcal{X}_{6,4}   \\
    &+ \frac{ \delta_{2}(\Delta \varsigma + \frac{L}{M} \gamma_2 \underline{\omega}_{2,I}^2)}{\Delta_{V_1}} (\mathcal{X}_{4,1} + \mathcal{X}_{4,2} + \mathcal{X}_{5,1} + \mathcal{X}_{5,2} + \mathcal{X}_{5,3} + \mathcal{X}_{6,1} + \mathcal{X}_{6,2} + \mathcal{X}_{6,3}) \\
    &+ \frac{ \delta_{2,I}(\Delta \varsigma + \frac{L}{M} \gamma_2 \underline{\omega}_{2,I}^2)}{\Delta_{V_1}}\frac{\delta_3}{\delta_{3,I}} (\mathcal{X}_{4,3} +\mathcal{X}_{5,4} +\mathcal{X}_{6,4}) + \frac{ \frac{L}{M} \gamma_2 \underline{\omega}_{2,I}\delta_{2,I}}{\Delta_{V_1}}  ( \widetilde{\mathcal{X}}_1 + \widetilde{\mathcal{X}}_2 + \widetilde{\mathcal{X}}_3 + \widetilde{\mathcal{X}}_4)\Big\}  + \mathcal{O}_{z}^s(\frac{1}{N^2}) \\
    &\overset{(a)}{=} \frac{\jmath t_1 z \psi}{\Delta_{V_1}} \Big\{(\frac{ \delta_{2,I}\delta_3(\Delta \varsigma + \frac{L}{M} \gamma_2 \underline{\omega}_{2,I}^2)}{\Delta_{V_1}\delta_{3,I}} + 1) (\mathcal{X}_{4,3}  + \mathcal{X}_{5,4} + \mathcal{X}_{6,4})  + \frac{\Delta}{\Delta_{V_1}} (\mathcal{X}_{4,1} + \mathcal{X}_{4,2} + \mathcal{X}_{5,1} + \mathcal{X}_{5,2} \\
    & + \mathcal{X}_{5,3} + \mathcal{X}_{6,1} + \mathcal{X}_{6,2} + \mathcal{X}_{6,3})  +  \frac{ \frac{L}{M} \gamma_2 \underline{\omega}_{2,I}\delta_{2,I}}{\Delta_{V_1}} ( \widetilde{\mathcal{X}}_1 + \widetilde{\mathcal{X}}_2 + \widetilde{\mathcal{X}}_3 + \widetilde{\mathcal{X}}_4) \Big\} + \mathcal{O}_{z}^s(\frac{1}{N^2}) \\
    &:= \frac{\jmath t_1 z\psi }{\Delta_{V_1}}(W_1 + W_2 + W_3 + W_4 + W_5) + \mathcal{O}_{z}^s(\frac{1}{N^2}),
    \end{split}
    \label{X_V1_terms_W}
\end{equation}
where $(a)$ follows by substituting
 $ \delta_{2}(\Delta \varsigma + \frac{L}{M} \gamma_2 \underline{\omega}_{2,I}^2) = \Delta - \Delta_{V_1}$ 
 into the second line of \eqref{X_V1_terms_W}. 
 The definition of $W_i$, $i=1,\ldots,5$, and their evaluation will be given below. We will show that they are equivalent 
 to the linear combinations of the derivatives of some terms in Table \ref{tabel_of_notations}. 
 In particular, $W_1$ is given by
\begin{equation}
\begin{split}
    W_1 &= \frac{\Delta}{\Delta_{V_1}} \left(\mathcal{X}_{4,1} + \mathcal{X}_{4,2} + \mathcal{X}_{5,1} + \mathcal{X}_{5,2} + \mathcal{X}_{5,3}\right) = \frac{\delta_{2,I}\delta_2 \Delta}{\Delta_{V_1}} \Big[-\overline{s}  {\Delta }(\overline{s}^2 \overline{\omega}_3 + 2 \overline{s} \gamma  \overline{\underline{\omega}}_{2,1} + \gamma^2 \overline{\underline{\omega}}_{1,2}) \\
    &+ (-\Delta \gamma + \frac{L}{M}\underline{\omega}_{2, I} \gamma_2 \delta) (\overline{s}^2  \overline{\underline{\omega}}_{2,1} + 2 \overline{s}  \gamma \overline{\underline{\omega}}_{1,2} + \gamma^2 \underline{\omega}_3) - {\frac{L}{M}\underline{\omega}_{2, I}\gamma_2}(\overline{s}  \overline{\underline{\omega}}_{1,1} + \gamma \underline{\omega}_2)\Big] \\
    &\overset{(a)}{=} \Delta \delta_2[ \delta' \overline{s} (\overline{s} ^2 \overline{\omega}_3 + 2 \overline{s}  \gamma  \overline{\underline{\omega}}_{2,1} + \gamma^2 \overline{\underline{\omega}}_{1,2}) + (\delta' \gamma + \delta \gamma') (\overline{s}^2  \overline{\underline{\omega}}_{2,1} + 2 \overline{s}  \gamma \overline{\underline{\omega}}_{1,2} + \gamma^2 \underline{\omega}_3) -\gamma' (\overline{s} \overline{\underline{\omega}}_{1,1} + \gamma \underline{\omega}_2)] \overset{(b)}{=} -\frac{\varsigma' \Delta \delta_2 }{2}, 
\label{W_1}
\end{split}
\end{equation}
where step $(a)$ follows by using the expression of $\delta'$ and $\gamma'$ in \eqref{Diff_EQ_deltas}. Step $(b)$ follows the definition in Table \ref{tabel_of_notations}. The term $W_2$ is given by
\begin{equation}
    \begin{split}
    W_2 &= \Big[ \frac{ \delta_{2,I}\delta_3(\Delta \varsigma + \frac{L}{M} \gamma_2 \underline{\omega}_{2,I}^2)}{\Delta_{V_1}\delta_{3,I}} + 1\Big](\mathcal{X}_{4,3} +\mathcal{X}_{5,4} +\mathcal{X}_{6,4}) \\
    & = \Big[ \frac{ \delta_{2,I}\delta_3(\Delta \varsigma + \frac{L}{M} \gamma_2 \underline{\omega}_{2,I}^2)}{\Delta_{V_1}} + \delta_{3,I}\Big]( \Delta \varsigma + \frac{L}{M} \gamma_2 \underline{\omega}_{2,I}^2) = -\frac{\delta_2'(\Delta \varsigma  + \frac{L}{M} \gamma_2 \underline{\omega}_{2,I}^2) }{2}.
\end{split}
\end{equation}
The term $W_3$ is given by
\begin{equation}
\begin{split}
    W_3 &= \frac{\Delta}{\Delta_{V_1}} \mathcal{X}_{6,1} + \frac{ \frac{L}{M} \gamma_2 \underline{\omega}_{2,I}\delta_{2,I}}{\Delta_{V_1}} \widetilde{\mathcal{X}}_1 \overset{(a)}{=} -\frac{\frac{L}{M}\gamma_3 \underline{\omega}_{2,I}\delta_{2,I}}{\Delta_{V_1}}\Big[\frac{L}{M} \underline{\omega}_{2,I}^2 \delta_2 +  \frac{L}{M} \delta^2 \underline{\omega}_2\Delta_{V_1} +  \frac{L}{M}\underline{\omega}_{2}\delta^2(1 - \varsigma\delta_2 - \Delta_{V_1}) \Big] \\
     &= -\frac{1}{2} \Big[\frac{L}{M} \underline{\omega}_{2,I}^2 \delta_2\gamma_2' +  \frac{L}{M}\underline{\omega}_{2}\delta^2\gamma_2' (1 - \varsigma\delta_2) \Big],
\end{split}
\end{equation}
where $(a)$ follows the relation $\gamma_2({\frac{L}{M} \underline{\omega}_{2,I}^2 \delta_2} + {\frac{L}{M} \delta^2 \underline{\omega}_2\Delta_{V_1}}) = (1 - \varsigma\delta_2 - \Delta_{V_1})\Delta$. The term $W_4$ is given by
\begin{equation}
    \begin{split}
    W_4 &= \frac{\Delta}{\Delta_{V_1}} \mathcal{X}_{6,3} +  \frac{ \frac{L}{M} \gamma_2 \underline{\omega}_{2,I}\delta_{2,I}}{\Delta_{V_1}} \widetilde{\mathcal{X}}_2 
    \overset{(a)}{=} \frac{\delta_{2,I}}{\Delta_{V_1}}(1-\varsigma \delta_2) \Big[\Delta\frac{L}{M}\gamma_2\delta (\underline{\omega}_2 - \overline{s} \delta \overline{\underline{\omega}}_{1,2} - \gamma \delta \underline{\omega}_3) + \frac{L}{M} \gamma_2 \underline{\omega}_{2,I}\frac{L}{M} \gamma_2 \delta^3 \underline{\omega}_{3}\Big] \\
    & = (1-\varsigma \delta_2) \Big[-\frac{L}{M}\delta'\gamma_2\delta \underline{\omega}_2 + \frac{L}{M} \gamma_2 \delta^2 (\overline{s} \delta' \overline{\underline{\omega}}_{1,2}+ (\gamma \delta' + \gamma' \delta) \underline{\omega}_{3})\Big] 
    = -\frac{1}{2} (1-\varsigma \delta_2) (\frac{L}{M}\gamma_2 \underline{\omega}_2(\delta^2)' + \frac{L}{M} \gamma_2 \underline{\omega}_2'\delta^2), 
\end{split}
\end{equation}
where in step $(a)$ we have used the relation $\underline{\omega}_2 = \underline{\omega}_{3,I} + \overline{s}  \delta \overline{\underline{\omega}}_{1,2} + \gamma \delta \underline{\omega}_3$. The term $W_5$ is given by
\begin{equation}
\begin{split}
    W_5 &=  \frac{\Delta}{\Delta_{V_1}} \mathcal{X}_{6,2} +  \frac{ \frac{L}{M} \gamma_2 \underline{\omega}_{2,I}\delta_{2,I}}{\Delta_{V_1}} (\widetilde{\mathcal{X}}_3 +\widetilde{\mathcal{X}}_4) \\
     &=  \frac{\delta_{2,I}}{\Delta_{V_1}}\frac{2L}{M}\gamma_2 \delta_2  \underline{\omega}_{2,I}\Big[-\Delta \overline{s} \overline{\underline{\omega}}_{1,2,I} + (-\Delta\gamma  + \frac{L}{M} \gamma_2 \underline{\omega}_{2,I} \delta) \underline{\omega}_{3,I}\Big] = -\frac{1}{2} \frac{L}{M}\gamma_2 \delta_2  (\underline{\omega}_{2,I}^2)' .
\label{W_5}
\end{split}
\end{equation}
Note that here we  have used  the result
\begin{equation}
    \underline{\omega}_{2,I}' = - \overline{s} \delta'(\frac{1}{L}  \mathrm{Tr} [\mathbf{T}_1 \mathbf{F}_\omega \mathbf{R}_2 \mathbf{T}_1 \mathbf{F}_\omega^2 ] + \frac{1}{L}  \mathrm{Tr} [\mathbf{T}_1 \mathbf{F}_\omega^2 \mathbf{R}_2 \mathbf{T}_1 \mathbf{F}_\omega]) - 2(\delta\gamma)' \underline{\omega}_{3,I}
     \overset{(a)}{=} -2 (\overline{s} \delta' \overline{\underline{\omega}}_{1,2,I} + (\delta\gamma)' \underline{\omega}_{3,I} ),
\end{equation}
where $(a)$ follows from the fact that
\begin{equation}
\begin{split}
     \overline{\underline{\omega}}_{1,2,I} &= \frac{1}{L} \Tr \mathbf{T}_1 \mathbf{F}_{\omega} \mathbf{R}_2 \mathbf{T}_1 \mathbf{F}_{\omega}^2 = \frac{1}{L} \Tr \mathbf{T}_1 \mathbf{F}_{\omega} \mathbf{R}_2 \mathbf{T}_1 \mathbf{F}_{\omega}\mathbf{F}_{\omega}^{-1}\mathbf{F}_{\omega}  -( \overline{s} \delta \overline{\underline{\omega}}_{2,1} + \delta \gamma \overline{\underline{\omega}}_{1,2}) \\
     &= \frac{1}{L} \Tr \mathbf{T}_1 \mathbf{F}_{\omega} \mathbf{F}_{\omega}^{-1}\mathbf{F}_{\omega}  \mathbf{R}_2 \mathbf{T}_1 \mathbf{F}_{\omega} -( \overline{s} \delta \overline{\underline{\omega}}_{2,1} + \delta \gamma \overline{\underline{\omega}}_{1,2}) = \frac{1}{L} \Tr \mathbf{T}_1 \mathbf{F}_{\omega}^2 \mathbf{R}_2 \mathbf{T}_1 \mathbf{F}_{\omega}.
\end{split}
\end{equation}
Combining the calculation results of $W_i$, $i=1, \ldots, 5$, in \eqref{W_1}-\eqref{W_5}, we immediately obtain
\begin{equation}
    \mathcal{X}_{V_1} = \frac{\jmath t_1 z \psi}{\Delta_{V_1}}\left(\sum_{i=1}^5 W_i\right) + \mathcal{O}_{z}^s(\frac{1}{N^2}) =  \frac{\jmath t_1 z \psi }{2 } \frac{\Delta_{V_1}'}{\Delta_{V_1}}  + \mathcal{O}_{z}^s(\frac{1}{N^2}). \label{Result_X_V1}
\end{equation}
\paragraph{Evaluation of $\mathcal{Y}_{V_2}$} We first introduce some useful derivatives
\begin{equation}
    \tau_2' = -2(\tau_{3,I} + \frac{\underline{s}^2 \overline{\tau}_2 \tau_{2,I}}{\Delta_{V_2}} \tau_3), \hspace*{2mm} \overline{\tau}_2' = \frac{2 \underline{s}\tau_{2,I} \overline{\tau}_3}{\Delta_{V_2}}.
\end{equation}
Based on the definition in \eqref{Y_V_2_and_Y_C}, we can directly obtain
\begin{equation}
\begin{split}
    \mathcal{Y}_{V_2} &= \mathcal{Y}_2 - z\underline{s}^2 L \overline{\tau}_2 \tau_{2,I} Y_{V_2} 
    = \frac{\jmath t_2z}{\Delta_{V_2}} \Big[ \frac{\mathcal{Y}_{2,1}}{\Delta_{V_2}}   + (1 + \frac{ \underline{s}^2  \overline{\tau}_2 \tau_{2,I}}{\Delta_{V_2}} \frac{\tau_3}{\tau_{3,I}}) \mathcal{Y}_{2,2}\Big] \psi + \mathcal{O}_{z}^s(\frac{1}{N^2}) \\
    &= \frac{\jmath t_2z}{\Delta_{V_2}} \Big[\frac{-\tau_{2,I}}{\Delta_{V_2}} \underline{s}^3\tau_{2}\overline{\tau}_3   + (\tau_{3,I} + \frac{ \underline{s}^2  \overline{\tau}_2 \tau_{2,I} \tau_3}{\Delta_{V_2}}) \underline{s}^2 \overline{\tau}_2 \Big] \psi + \mathcal{O}_{z}^s(\frac{1}{N^2}) \\
    &= -\frac{\jmath t_2z}{2\Delta_{V_2}} (  \underline{s}^2\overline{\tau}_2'\tau_{2}   + \underline{s}^2 \tau_2' \overline{\tau}_2 ) \psi + \mathcal{O}_{z}^s(\frac{1}{N^2}) = \frac{\jmath t_2z\psi}{2} \frac{\Delta_{V_2}'}{\Delta_{V_2}}+ \mathcal{O}_{z}^s(\frac{1}{N^2}). \label{Result_Y_V2}
\end{split}
\end{equation}
\paragraph{Evaluation of $t_1 \mathcal{X}_{C} + t_2 \mathcal{Y}_C$} Following the same steps as above, we list some derivative terms which will be used later
\begin{align}
    \underline{\phi}' &= -(\underline{s} \tau' \underline{\phi}_{2,1} + \overline{s} \delta' \widetilde{\overline{\underline{\phi}}}_{1,2} + (\delta \gamma)'\underline{\phi}_{1,2}), \hspace*{2mm} 
    \overline{\phi}' = -(\underline{s} \tau' \overline{\phi}_{2,1} + \overline{s} \delta' \overline{\phi}_{1,2} + (\delta \gamma)' \overline{\underline{\phi}}_{1,2}), \notag \\
    \vartheta' &= -(\vartheta_{2,I, 1} + \underline{s} \overline{\tau}' \vartheta_{2,1} + \vartheta_{1,2,I} + (\overline{s} \overline{\omega} + \gamma \underline{\omega})' \vartheta_{1,2}), \hspace*{2mm} 
    (\overline{s} \overline{\omega} + \gamma \underline{\omega})' = \frac{\delta_{2,I}(\Delta \varsigma + \frac{L}{M}\gamma_2 \underline{\omega}_{2,I}^2)}{ \Delta_{V_1}}.
\end{align}
To tidy up the form of the expression, we rewrite
\begin{equation}
\begin{split}
    t_1 \mathcal{X}_{C} + t_2 \mathcal{Y}_C &= \frac{j t_1 t_2 z \psi }{\Delta_C}\Big[\frac{\mathcal{Y}_{3,1} + \mathcal{Y}_{4,1}}{ \Delta_{V_2}} + 
     (\vartheta_{2,I, 1} + \frac{ \underline{s}^2 \overline{\tau}_2 \tau_{2,I}\vartheta_{2,1}}{ \Delta_{V_2}})\frac{\mathcal{Y}_{3,2} + \mathcal{Y}_{4,2}}{\vartheta_{2,I, 1}} \\
    &+ \frac{\Delta \mathcal{X}_{7,1}}{\Delta_{V_1}}  + (\vartheta_{1,2,I} + \frac{\delta_{2,I}(\Delta \varsigma + \frac{L}{M}\gamma_2 \underline{\omega}_{2,I}^2)\vartheta_{1,2}}{ \Delta_{V_1}}) \frac{\mathcal{X}_{7,2}}{\vartheta_{1,2,I}} + \frac{ \frac{L}{M} \gamma_2 \underline{\omega}_{2,I}\delta_{2,I}}{\Delta_{V_1}}({\widetilde{\mathcal{X}}_5 + \widetilde{\mathcal{X}}_6})\Big] +{\mathcal{O}_z^s(\frac{1}{N^2})} \\
    &= \frac{j t_1 t_2 z \psi}{\Delta_C} (V_1 + V_2 + V_3) + \mathcal{O}_{z}^s(\frac{1}{N^2}), \label{t_X_C_Y_C}
\end{split}
\end{equation}
where $V_1$, $V_2$, and $V_3$ will be defined in the subsequent calculations. The calculation of $V_1$ can be obtained as
\begin{equation}
\begin{split}
    V_1 &= (\vartheta_{2,I, 1} + \frac{ \underline{s}^2 \overline{\tau}_2 \tau_{2,I}\vartheta_{2,1}}{ \Delta_{V_2}})\frac{\mathcal{Y}_{3,2} + \mathcal{Y}_{4,2}}{\vartheta_{2,I, 1}} + (\vartheta_{1,2,I} + \frac{\delta_{2,I}(\Delta \varsigma + \frac{L}{M}\gamma_2 \underline{\omega}_{2,I}^2)\vartheta_{1,2}}{ \Delta_{V_1}}) \frac{\mathcal{X}_{7,2}}{\vartheta_{1,2,I}}  \\
    &= [\vartheta_{2,I, 1} + \underline{s} \overline{\tau}' \vartheta_{2,1} + \vartheta_{1,2,I} + (\overline{s} \overline{\omega} + \gamma \underline{\omega})' \vartheta_{1,2}]
     (\overline{s}\underline{s} \overline{\phi} + \underline{s} \gamma \underline{\phi}) = -  (\overline{s}\underline{s} \overline{\phi} + \underline{s} \gamma \underline{\phi})\vartheta'.
\end{split}
\end{equation}
Similarly, we define and calculate $V_2$ as follows 
\begin{equation}
\begin{split}
    V_2 &= \frac{\mathcal{Y}_{3,1} + \mathcal{Y}_{4,1}}{ \Delta_{V_2}} + \frac{\Delta}{\Delta_{V_1}} \mathcal{X}_{7,1} + \frac{ \frac{L}{M} \gamma_2 \underline{\omega}_{2,I}\delta_{2,I}}{\Delta_{V_1}}\widetilde{\mathcal{X}}_5 \\
    &= \underline{s} \vartheta \Big[ -(\overline{s}\underline{s} \overline{\phi}_{2,1} + \underline{s} \gamma \underline{\phi}_{2,1}) \frac{\tau_{2, I}}{\Delta_{V_2}} - (\overline{s}^2 \overline{\phi}_{1,2} + \overline{s} \gamma \widetilde{\overline{\underline{\phi}}}_{1,2}+ \overline{s} \gamma \overline{\underline{\phi}}_{1,2} + \gamma^2 \underline{\phi}_{1,2}) \frac{\Delta \delta_{2,I}}{\Delta_{V_1}} 
    + (\overline{s}\delta \overline{\underline{\phi}}_{1,2} + \delta\gamma \underline{\phi}_{1,2})\frac{  \frac{L}{M} \gamma_2 \underline{\omega}_{2,I}\delta_{2,I}}{\Delta_{V_1}} \Big] \\
    &= \underline{s} \vartheta \left[ (\overline{s}\underline{s}  \overline{\phi}_{2,1}  +\underline{s} \gamma \underline{\phi}_{2,1}) \tau' + (\overline{s}^2 \overline{\phi}_{1,2} +  \overline{s} \gamma \widetilde{\overline{\underline{\phi}}}_{1,2} +  \overline{s} \gamma \overline{\underline{\phi}}_{1,2} + \gamma^2 \underline{\phi}_{1,2}) \delta' 
    + (\overline{s} \delta \overline{\underline{\phi}}_{1,2} + \delta\gamma \underline{\phi}_{1,2}) \gamma' \right] = - \underline{s} \vartheta ( \gamma \underline{\phi}' + \overline{s} \overline{\phi}' ).
\end{split}
\end{equation}
In addition, $V_3$ is given by
\begin{equation}
\label{V_3}
    V_3 =  \frac{ \frac{L}{M} \gamma_2 \underline{\omega}_{2,I}\delta_{2,I}}{\Delta_{V_1}}\widetilde{\mathcal{X}}_6 = -\frac{ \frac{L}{M} \gamma_2 \underline{\omega}_{2,I}\delta_{2,I}\underline{s}\underline{\phi}\vartheta}{\Delta_{V_1}} =- \underline{s} \gamma' \underline{\phi} \vartheta. 
\end{equation}
Therefore, through the analysis of \eqref{t_X_C_Y_C}-\eqref{V_3}, we can obtain
\begin{equation}
    t_1 \mathcal{X}_{C} + t_2 \mathcal{Y}_C = j t_1 t_2 z \psi \frac{\Delta_C'}{\Delta_C}  + \mathcal{O}_{z}^s(\frac{1}{N^2}). \label{Result_XC_YC}
\end{equation}
\par
Finally, by substituting the results of  \eqref{Result_X_V1}, \eqref{Result_Y_V2}, and \eqref{Result_XC_YC} into \eqref{Pre_CLT_Inge_Eq}, we obtain 
 the differential equation as follows
\begin{equation}
        \frac{\partial \varphi }{\partial z} = -\frac{\varphi}{2} \mathbf{t}^T \begin{bmatrix}
            -\frac{\Delta_{V_1}'}{\Delta_{V_1} }& -\frac{\Delta_{C}'}{\Delta_{C} }\\
            -\frac{\Delta_{C}'}{\Delta_{C} } & -\frac{\Delta_{V_2}'}{\Delta_{V_2} }
        \end{bmatrix} \mathbf{t} + \mathcal{O}_{z}^s(\frac{1}{zN}) =  -\frac{1}{2} \mathbf{t}^T \frac{\partial \mathbf{V}}{\partial z}\mathbf{t} \varphi + \mathcal{O}_{z}^s(\frac{1}{zN}).
\label{Eq_O_DIff_Eq_V}
\end{equation}
\subsubsection{Convergence of the Characteristic Function} In order to obtain \eqref{Conv_cha_Eq}, we introduce the auxiliary quantity 
\begin{equation}
    \mathcal{K}(\mathbf{t}, \overline{s}, \underline{s}, z) = \exp\left[\frac{1}{2}\mathbf{t}^T \mathbf{V}(\overline{s}, \underline{s}, z)\mathbf{t}\right] \varphi(\overline{s}, \underline{s}, z).
\end{equation}
By taking the derivative with respect to $z$ and using \eqref{Eq_O_DIff_Eq_V}, we have
\begin{equation}
    \frac{\partial \mathcal{K}}{\partial z} = \left(\frac{1}{2} \mathbf{t}^T \frac{\partial \mathbf{V}}{\partial z} \mathbf{t} \right) \exp\left[\frac{1}{2}\mathbf{t}^T \mathbf{V} \mathbf{t}\right] \varphi
+  \exp\left[\frac{1}{2}\mathbf{t}^T \mathbf{V} \mathbf{t}\right] \frac{\partial \varphi }{\partial z} =  \exp\left[\frac{1}{2}\mathbf{t}^T \mathbf{V} \mathbf{t}\right] \mathcal{O}\left(\frac{ 
\mathcal{G}_s(z^{-1})}{z^3N}\right).
\end{equation}
By the properties of the positive solutions to the system of equations \eqref{DE_system_1} and \eqref{DE_system_2}, we can observe that $\lim_{z \rightarrow + \infty} \mathbf{V}
= \mathbf{0}_{2 \times 2}$. In conclusion, we have
\begin{equation}
    \begin{split}
   & \abs{1 - \mathcal{K}(\mathbf{t}, \overline{\sigma}_1^2, \underline{\sigma}_1^2, \sigma_2^2)} = \abs{\int_{\sigma_2^2}^{+ \infty}  
   \frac{\partial \mathcal{K}}{\partial z} \diff z} \leq
    \int_{\sigma_2^2}^{+ \infty} \diff z \exp\left[\frac{1}{2}\mathbf{t}^T \mathbf{V} \mathbf{t}\right] \mathcal{O}\left(\frac{ 
        \mathcal{G}_{\sigma_1^2}(z^{-1})}{z^3N}\right) \\
        &\leq  \int_{\sigma_2^2}^{+ \infty} \exp\left[\frac{\norm{\mathbf{t}}^2 }{2}  \norm{\mathbf{V}(\overline{\sigma}_1^2, \underline{\sigma}_1^2, z)} \right] \mathcal{O}\left(\frac{ 
        \mathcal{G}_{\sigma_1^2}(z^{-1})}{z^3N}\right)\diff z \overset{(a)}{\leq}  \int_{\sigma_2^2}^{+ \infty} \frac{\mathcal{O}\big(\frac{ 
        \mathcal{G}_{\sigma_1^2}(z^{-1})}{z^3N}\big)}{\left(\Delta_{V_1}\Delta_{V_2}\right)^\frac{\norm{\mathbf{t}}^2 }{2}} \diff z \\
&\overset{(b)}{\leq } C_1\int_{\sigma_2^2}^{+ \infty} \frac{\left(1 + \frac{C_2}{z}\right)^{4 \norm{\mathbf{t}}^2}\mathcal{G}_{\sigma_1^2}(z^{-1})}{z^3N} \diff z \leq 
\mathcal{O}\left(\frac{\left(1 + C_2 \sigma_2^{-2} \right)^{4 \norm{\mathbf{t}}^2} \mathcal{G}_{\sigma_1^2}(\sigma_2^{-2})}{N \sigma_2^{4}}\right).
\end{split}
\end{equation}
Here, step $(a)$ is due to the inequality $\norm{\mathbf{V}} \leq \Tr \mathbf{V}$ for positive definite $\mathbf{V}$. In step $(b)$, 
both $C_1$ and $C_2$ are positive constants that are independent of $z$. Step $(b)$ uses the uniform lower bound of $\Delta_{V_1}$ and $\Delta_{V_2}$,
 which are given in \eqref{lower_bounds_Delta_V1} and subsequent. From the definition of $\mathcal{K}$, we have
$
    \varphi = \exp\left[-\frac{1}{2}\mathbf{t}^T \mathbf{V}\mathbf{t}\right] +  \mathcal{O}(\frac{1}{N}).
$
Furthermore, we have proven that $\mathbf{V}$ satisfies the conditions of Lemma \ref{Lemma_tight_CLT} in Proposition \ref{Prop_tightness}. Therefore, we have proven the CLT.
\subsubsection{Convergence Rate of the Covariance Matrix}
\label{Sec_Convergence_rate_of_covariance}
For the sake of simplicity, we discuss the convergence rate for the variance of $I_1(\overline{\sigma_1}^2, \sigma_2^2)$. 
By taking the derivative of $\Var(I_1)$ with respective to $z$ , we obtain
\begin{equation}
    \frac{\partial \mathbb{E} \left( I_1(\overline{s}, z) -  \mathbb{E}I_1(\overline{s}, z) \right)^2}{\partial z} =  -\frac{2}{z} \left\{\mathbb{E}(\chi {I}_1)  - \mathbb{E} \chi \mathbb{E}I_1 \right\},
\label{var_diff_chi}
\end{equation}
where $\chi$ is defined as $\chi = \overline{s}\widehat{\Psi}_1(\mathbf{I}_N, \mathbf{I}_L) + \widehat{\Psi}_2(\mathbf{I}_N, \mathbf{I}_M, \mathbf{I}_L) $. When applying the integration by parts formula \eqref{Integration_by_parts} on the term $\mathbb{E} \chi I_1$, we need to evaluate $\partial I_1$. Since there holds $\partial(\exp[\jmath t I_1])
= \jmath (\partial I_1) \exp[\jmath t I_1]$, the estimation of the variance is very similar to the estimation of the 
characteristic function \eqref{CLT_Eq_Inte_By_Parts_Q_1}-\eqref{Eq_O_DIff_Eq_V}. In this process, the order of the approximation error is mainly dominated by expressions like $\mathbb{E}[\mathring{\alpha}_{\delta}\mathring{\Psi}_j \mathring{I}_1]$. To estimate the
error, we use the Hölder's inequality
\begin{equation}
    \mathbb{E}\left[\abs{\mathring{\alpha}_{\delta}\mathring{\Psi}_j \mathring{I}_1}\right] \leq \left(\mathbb{E}\abs{\mathring{\alpha}_{\delta}}^{4}\right)^{\frac{1}{4}}\left(\mathbb{E}
    \abs{\mathring{\Psi}_j}^{4}\right)^{\frac{1}{4}} \left(\mathbb{E}\abs{\mathring{I}_1}^2\right)^{\frac{1}{2}}.
\label{centralized_3_bounds}
\end{equation}
We set $j=1$ for example. By identity $\Var(|x|^2) + (\mathbb{E}|x|^2)^2 = \mathbb{E}|x|^4 $, the fourth central moment $\mathbb{E}|\mathring{\Psi}_1|^4$ can be written as $\mathbb{E}|\mathring{\Psi}_1|^4 =
\Var(|\mathring{\Psi}_1|^2) + (\mathbb{E}|\mathring{\Psi}_1|^2)^2$. Based on the variance control in Proposition \ref{Prop_variance},
we know that $(\mathbb{E}|\mathring{\Psi}_1|^2)^2$ is $\mathcal{O}_z^s(1)$. Using the Poincaré-Nash inequality \eqref{variance_f_Tr}, we have
\begin{equation}
     \mathrm{Var}(|\mathring{\Psi}_1|^2) \leq  U_{2,1} + U_{2,2} + U_{2,3} + U_{2,4}. 
\end{equation}
The form of $U_{2, j}$ is similar to \eqref{U_1_s} and we analyze $U_{2, 1}$ in the following
\begin{equation}
\begin{split}
    U_{2,1} &= \frac{6 Nr^2}{L}\mathbb{E}\abs{\mathring{\Psi}_1}^2 \left( \norm{\mathbf{H}_2 \mathbf{H}_2^H \mathbf{H}_1^H\mathbf{Q}_1\mathbf{H}_1 \mathbf{B} \mathbf{H}_1^H \mathbf{A} \mathbf{Q}_1}^2 + \norm{\mathbf{Q}_1\mathbf{H}_1 \mathbf{B}  
     \mathbf{H}_1^H\mathbf{A}\mathbf{Q}_1 \mathbf{H}_{1} \mathbf{H}_2 \mathbf{H}_2^H}^2 \right) \\
     & \leq \frac{12 N r^2 \norm{\mathbf{A}}^2 \norm{\mathbf{B}}^2}{Lz^4} \mathbb{E}\abs{\mathring{\Psi}_1}^2 \norm{\mathbf{H}_1}^6\norm{\mathbf{H}_2}^4 
     \leq  \frac{C_1}{z^4} \sqrt{\mathbb{E}\abs{\mathring{\Psi}_1}^4 \mathbb{E}\norm{\mathbf{H}_1}^{12}\norm{\mathbf{H}_2}^8} := z^{-4} C_2 \sqrt{M_{\Phi_1}},
    \end{split}
\end{equation}
where $C_1$ and $C_2$ are constants that are independent of $N$ and $z$ and $M_{\Phi_1} = \mathbb{E}|\mathring{\Psi}_1|^4$. Similar to the evaluation on $U_{2,2}$, $U_{2, 3}$, and $U_{2, 4}$, we obtain
\begin{equation}
    M_{\Phi_1} \leq \mathcal{O}_z^s(1)(1 + \sqrt{M_{\Phi_1}}),
\end{equation}
which implies that $M_{\Phi_1}$ is uniformly bounded in $N$, i.e. $\sup_{N \geq 1} M_{\Phi_1} < + \infty$. A more accurate analysis shows that 
\begin{equation}
\begin{split}
    & \mathbb{E} \abs{\mathring{\Psi}_1}^2 \norm{\mathbf{H}_1}^6\norm{\mathbf{H}_2}^4 = \mathbb{E} \abs{\mathring{\Psi}_1}^2 \norm{\mathbf{H}_1}^6\norm{\mathbf{H}_2}^4 \mathbb{I}_{A(\nu)} +  \mathbb{E} \abs{\mathring{\Psi}_1}^2 \norm{\mathbf{H}_1}^6\norm{\mathbf{H}_2}^4 \mathbb{I}_{A^c(\nu)} \\
    & \overset{(a)}{\leq }\nu \mathbb{E} \abs{\mathring{\Psi}_1}^2 + \sqrt{M_{\Phi_4}} \mathbb{E}^{\frac{1}{4}} \norm{\mathbf{H}_1}^{24}\norm{\mathbf{H}_2}^{16} 
    \mathbb{P}^{\frac{1}{4}}[A^c(\nu)] \leq  \mathcal{O}_s^z(1)\nu + C_3\mathbb{P}^{\frac{1}{4}}[A^c(\nu)],
\end{split}
\end{equation}
where $C_3$ is a constant. The set $A(\nu)= \{ \norm{\mathbf{H}_1}^6\norm{\mathbf{H}_2}^4 \leq \nu \}$ and  $A^c(\nu)$ represents its complement. 
Step $(a)$ follows by applying Hölder's inequality to the term 
$\mathbb{E} \abs{\mathring{\Psi}_1}^2 \norm{\mathbf{H}_1}^6\norm{\mathbf{H}_2}^4 \mathbb{I}_{A^c(\nu)}$. 
From \cite[Theorem II.13]{davidson2001local}, we know that there exists a large enough $\nu_{6,4} > 0$
such that $\sum_{N > 0}\mathbb{P}(A^c(\nu_{6,4})) < + \infty$. Taking $\nu = \nu_{6,4}$ and applying it to the above equation, 
we immediately get $\mathbb{E} |\mathring{\Psi}_1|^2 \norm{\mathbf{H}_1}^6\norm{\mathbf{H}_2}^4  = \mathcal{O}_z^s(1)$ and $U_{2,1} = \mathcal{O}_z^s(\frac{1}{z^4})$. 
With the same method, we can evaluate $U_{2,2}$, $U_{2,3}$, $U_{2, 4}$, and get $\mathbb{E} |\mathring{\Psi}_1|^4 = \mathcal{O}_z^s(z^{-2})$. 
Furthermore, we can also obtain the fourth-order central moment bounds for the other terms in Table \ref{Table_trace_of_resolvent_terms}.
 More generally, we can determine the order of $\mathbb{E}|\mathring{x}|^k$, where $x$ can be random quantities in Table \ref{Table_trace_of_resolvent_terms} or 
 defined in \eqref{Random_quantities} and \eqref{Q_2_quantities}.
The bounded property for the variance of $\mathring{I}_1$ can be proved straightforwardly by Poincaré-Nash inequality. At this stage, we have 
$\mathbb{E}[|\mathring{\alpha}_{\delta}\mathring{\Psi}_j \mathring{I}_1|] = \mathcal{O}_z^s(\frac{1}{N})$. 
\par
Returning to \eqref{var_diff_chi}, some tedious but straightforward calculations show that
\begin{equation}
    \Var(I_1(\overline{\sigma}_1, \sigma_2^2)) = \int_{\sigma_2^2}^{\infty} - \frac{\partial \Delta_{V_1}(\overline{\sigma}_1, z)}{\partial z} + \mathcal{O}_z^{\sigma_1^2}(\frac{1}{N}) \diff z
     = \Delta_{V_1}(\overline{\sigma}_1, \sigma_2^2) + \mathcal{O}(\frac{1}{N}).
\end{equation}
For each element in $\mathbf{V} - \mathrm{Cov}(\mathbf{m})$, we can obtain convergence of $\mathcal{O}(\frac{1}{N})$, 
and therefore, the norm is also of $\mathcal{O}(\frac{1}{N})$. \QED
 \section{Numerical Experiments}
\label{Sec_Numerical_Experiments}
In the experiments, we adopt the following model for the correlation matrices  \cite{ALMoustakas2003}
\begin{equation}
    \left[\boldsymbol{\mathcal{C}} (\eta, \delta,  d_s)\right]_{m, n} = \int_{-180}^{180} \frac{ 1}{\sqrt{2 \pi \delta^2_{c}}} \exp\left[
        {2\pi\jmath d_s (m-n) \sin(\frac{\pi \phi}{180}) - \frac{(\phi - \eta)^2}{2 \delta_c^2}}\right] \mathrm{d} \phi,
\end{equation}
where $\delta_c$ denotes the root-mean-square angle spread and $d_s$ 
is the relative antenna spacing in wavelength. $\phi$ and $\eta$ represent 
the angular spread of the signal and the mean angle, respectively. $\phi$ is randomly generated from $-\pi$ to $\pi$. 
The index for the  reflecting elements of active IRS/relay antennas
is denoted by $m$ and $n$. In the simulation, we set $\mathbf{A}_1 = \boldsymbol{\mathcal{C}}^{\frac{1}{2}}(60, 30, 1)  $, $\mathbf{B}_1 = \boldsymbol{\mathcal{C}}^{\frac{1}{2}}(0, 30, 1)$, 
$\mathbf{A}_2 = \boldsymbol{\mathcal{C}}^{\frac{1}{2}}(0, 5, 1)$, and $\mathbf{B}_2 = \boldsymbol{\mathcal{C}}^{\frac{1}{2}}(10, 5, 1)$. 
\par
\begin{figure}[t]
    \centering
    \subfloat[\label{ELSD_0}]{
        \centering
        \includegraphics[width=3.2in]{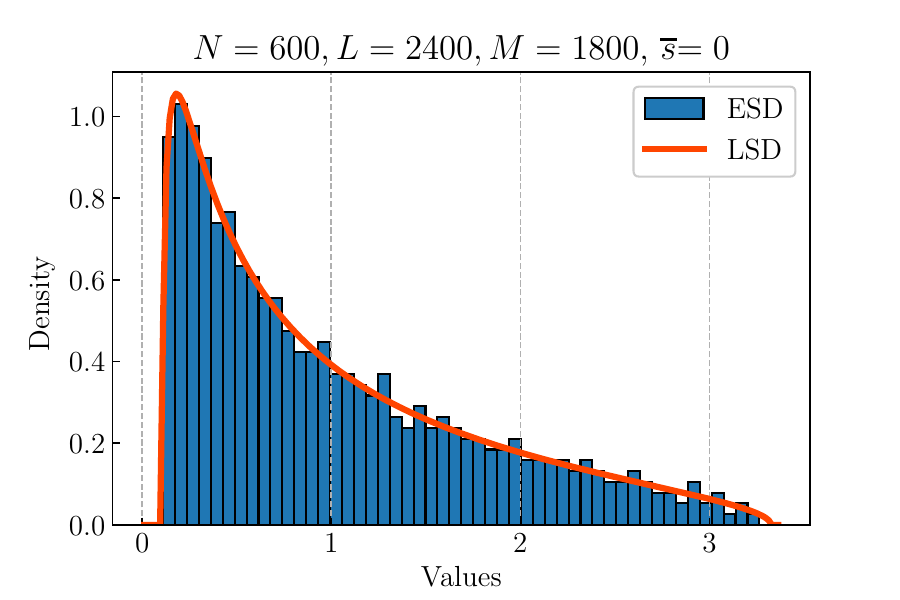}
    }
    \subfloat[\label{ELSD_2}]{
        \centering
        \includegraphics[width=3.2in]{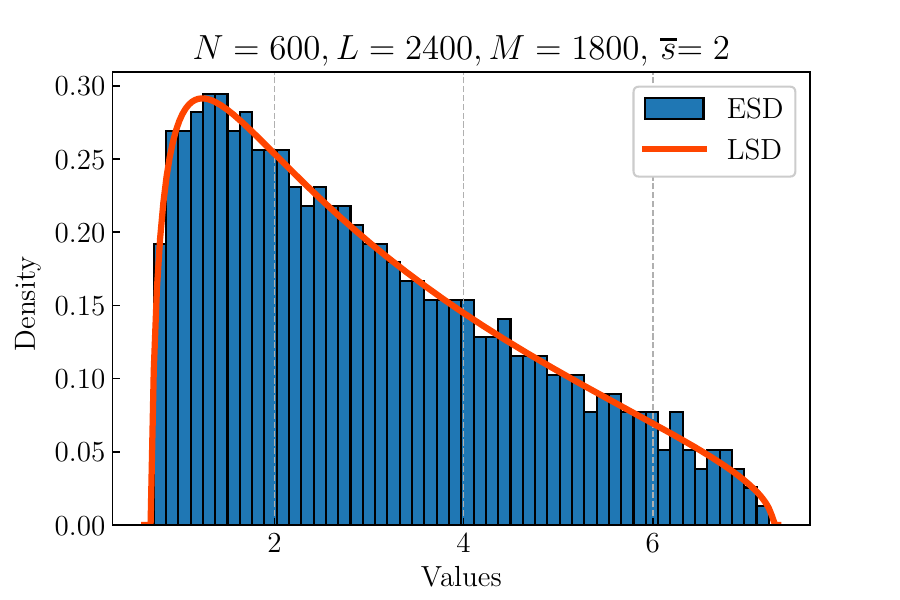}
    }
    \caption{ESD of $\mathbf{H}_1 \mathbf{H}_2  \mathbf{H}_2^H \mathbf{H}_1^H + \overline{s} \mathbf{H}_1\mathbf{H}_1^H$ versus LSD}
    \label{ESD_vs_LSD}
\end{figure}
\textbf{Accuracy of the deterministic approximation of the resolvent:} Fig. \ref{ESD_vs_LSD} shows the ESD of the random matrix $\mathbf{B} = \mathbf{H}_1 \mathbf{H}_2 \mathbf{H}_2^H \mathbf{H}_1^H + \overline{s} \mathbf{H}_1 \mathbf{H}_1^H$ and the 
LSD obtained through the 
inversion formula with $(N, L, M, \overline{s})$ = $(600, 2400, 1800, 0)$ for Fig. \ref{ELSD_0} and $(600, 2400, 1800, 2)$ for Fig. \ref{ELSD_2}. In the experiment, the correlation matrices $\mathbf{R}_i$ and $\mathbf{T}_i$, $i=1,2$, are both set to identity matrix.
It can be observed that $\frac{1}{N} \Tr \mathbf{F}_\delta$ admits an integral representation, and its corresponding
measure is very close to the ESD of $\mathbf{B}$. This indicates that Lemmas \ref{Lemma_DE_Q1_1} and \ref{Lemma_DE_Q1_2} are accurate.
\par
\begin{figure}[t]
    \centering
    \subfloat[\label{Fig_Error_Mean}]{
        \centering
        \includegraphics[width=3.2in]{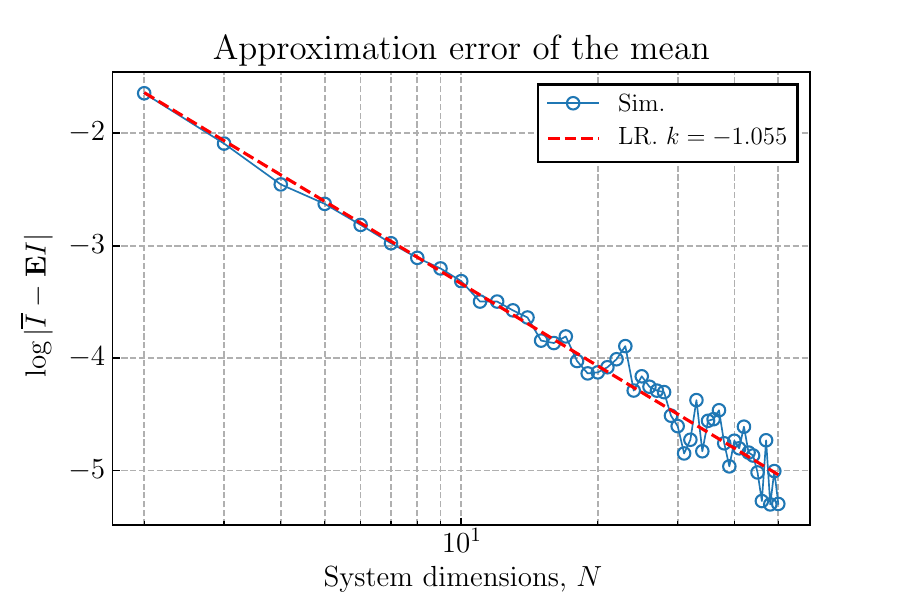}
    }
    \subfloat[\label{Fig_Error_I}]{
        \centering
        \includegraphics[width=3.2in]{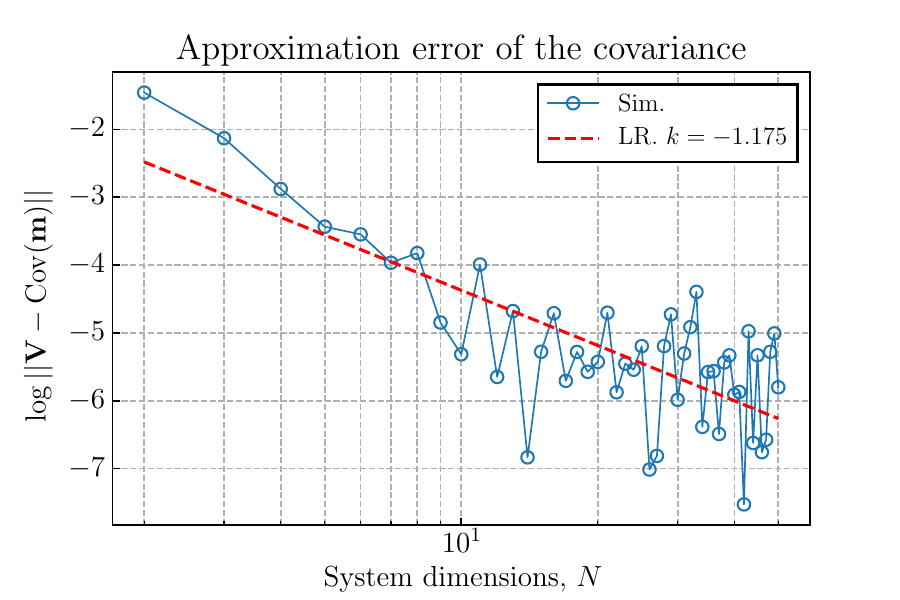}
    }
    \caption{Approximation Error versus $N$}
    \label{Fig_Approx_Error}
\end{figure}
\textbf{Tightness of convergence rate for mean and covariance matrix:} Fig. \ref{Fig_Approx_Error}
 shows the convergence rate for the mean and covariance. 
 In the experiment, we consider the i.i.d. case with $N=L=M$. 
 The horizontal axis represents the logarithmic coordinate of the system dimensions $N$, 
 while the vertical axis corresponds to the logarithm of the error for the deterministic approximations of the mean and covariance.
  The values of  $\mathbf{E}I$ and $\mathrm{Cov}(\mathbf{m})$ are obtained by Monte Carlo (MC) simulation for $5 \times 10^5$ realizations. 
  The red line represents the result of linear regression (LR).
  It can be observed that the slopes are slightly less than 1, indicating that $\mathcal{O}(\frac{1}{N})$ is tight.
\par
\begin{figure}[t]
    \centering
    \subfloat[\label{Fig_SNR}]{
        \centering
        \includegraphics[width=3.2in]{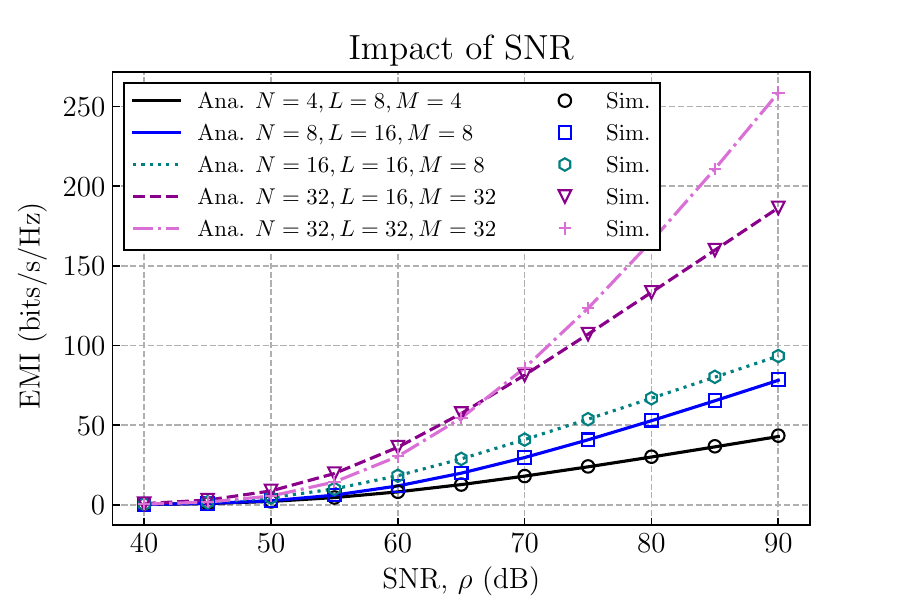}
    }
    \subfloat[\label{Fig_L}]{
        \centering
        \includegraphics[width=3.2in]{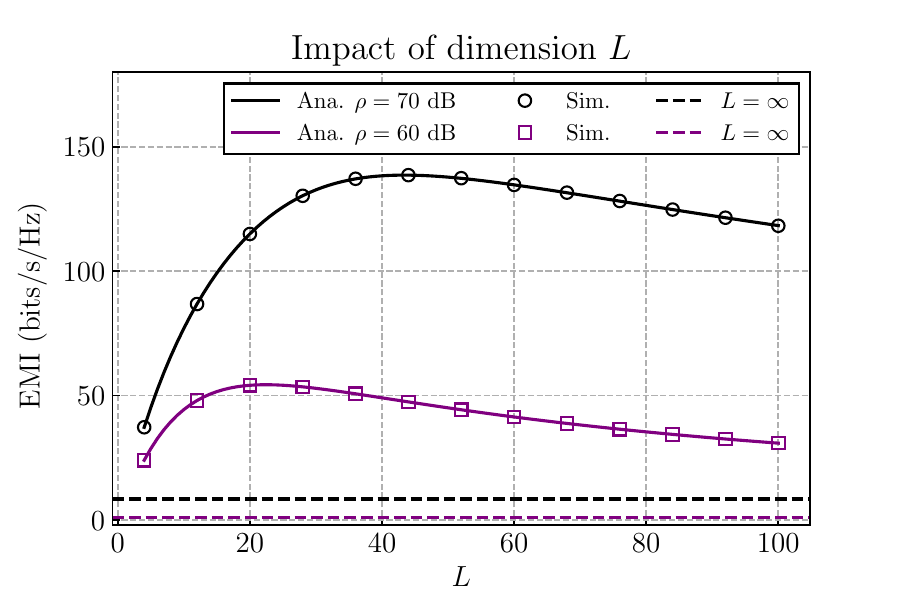}
    }
    \hfill
    \subfloat[\label{Fig_limit_L}]{
        \centering
        \includegraphics[width=3.2in]{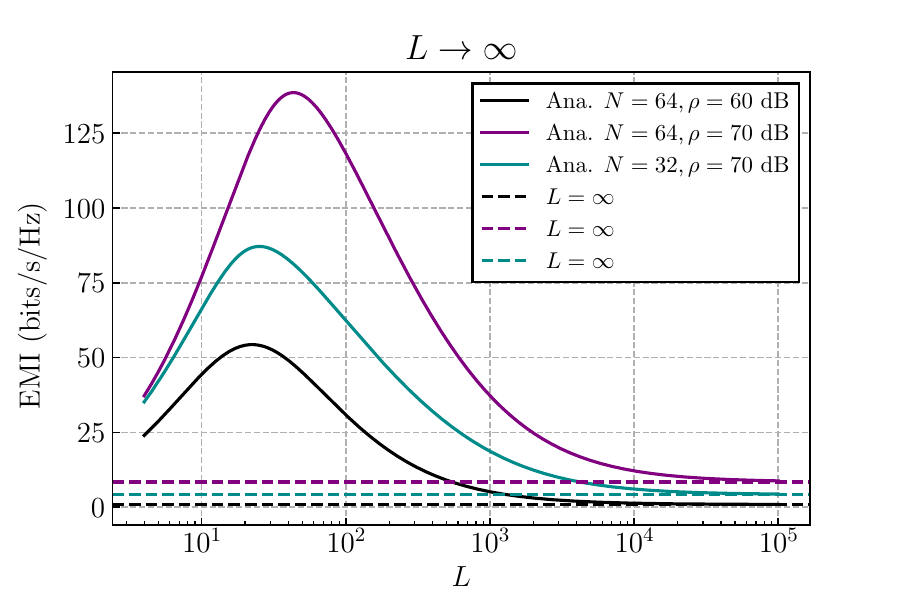}
    }
    \subfloat[\label{Fig_snr_inf_L}]{
        \centering
        \includegraphics[width=3.2in]{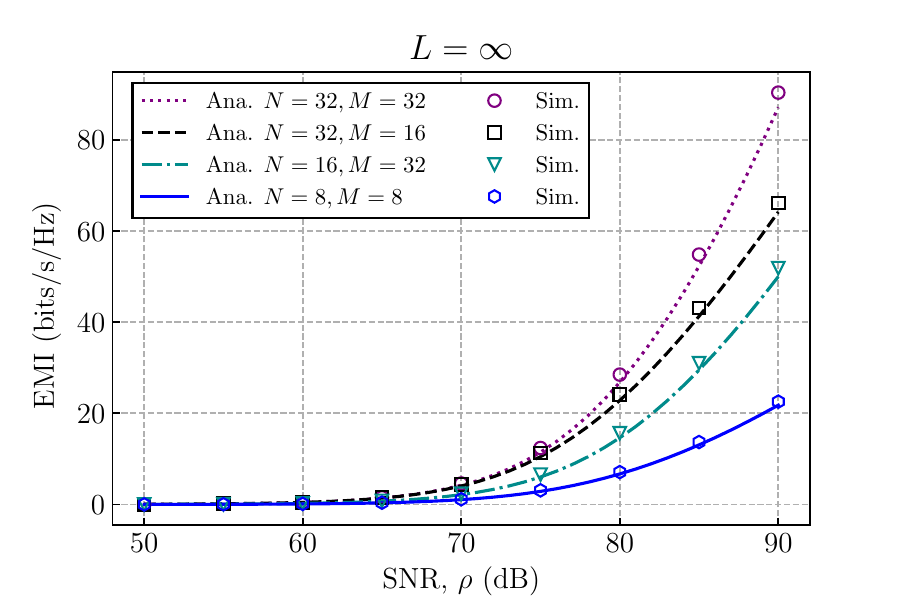}
    }
    \caption{i.i.d. Case}
    \label{Ergodic}
\end{figure}
\par
\textbf{The impact of number of active IRS elements:}  Fig. \ref{Ergodic} shows the impact of SNR and the number of reflecting elements of active IRSs under the i.i.d. setting. In the experiment, we set ${\sigma_1}^2 = \sigma_2^2 = \sigma^2$, i.e., the AWGN power at the active-IRS and the receiver are the same. The transmission power $P_T$ and the amplification power of the active-IRS $P_A$ are set to $P_T = 1$ W and $P_A = 0.5$ W, respectively. The path loss of each hop is set to $40$ dB and the SNR is defined as $\rho=\frac{P_T}{\sigma^2}$.  
MC simulations are represented by markers. 
\par
From Fig. \ref{Fig_SNR}, we can observe that with different settings of $N$, $ M$, and $L$, 
 the EMI increases with SNR and the first-order analysis is accurate. 
 In Fig. \ref{Fig_L}, we set $N=M=64$. It can be observed that as $L$ increases,
the EMI first increases and then decreases. Furthermore, at high SNR, the maximum EMI is achieved with a larger $L$. This is
because when the noise power at the IRSs decreases, the noise produced by more reflecting elements can be tolerated. 
We set $N=M$ for Fig. \ref{Fig_limit_L}. The solid line represents the analysis result in \eqref{iid_I},
while the dashed line corresponds to the case with $L \rightarrow \infty$ \eqref{inf_L_iid_I}.
 It can be observed that as $L$ increases, 
 the EMI approaches the limiting case. The experimental results are consistent with the theoretical analysis.
 In Fig. \ref{Fig_snr_inf_L}, 
 the lines represent the analysis result of \eqref{inf_L_iid_I}, i.e., the EMI of the single-hop MIMO Rayleigh channel.
The markers denote the MC simulation of active IRS-aided MIMO systems with $L = 5 \times 10^4$. It can be observed that when $L \rightarrow \infty$, the two-hop MIMO channel will degenerate to the single-hop channel.
\par
\begin{figure}[t]
    \centering
    \subfloat[]{
        \centering
        \includegraphics[width=3.2in]{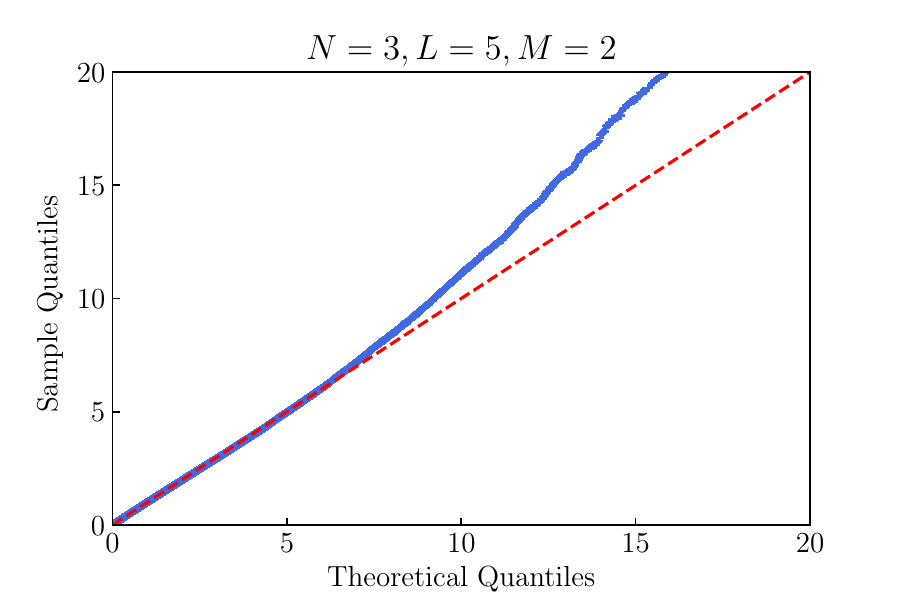}
    }
    \subfloat[]{
        \centering
        \includegraphics[width=3.2in]{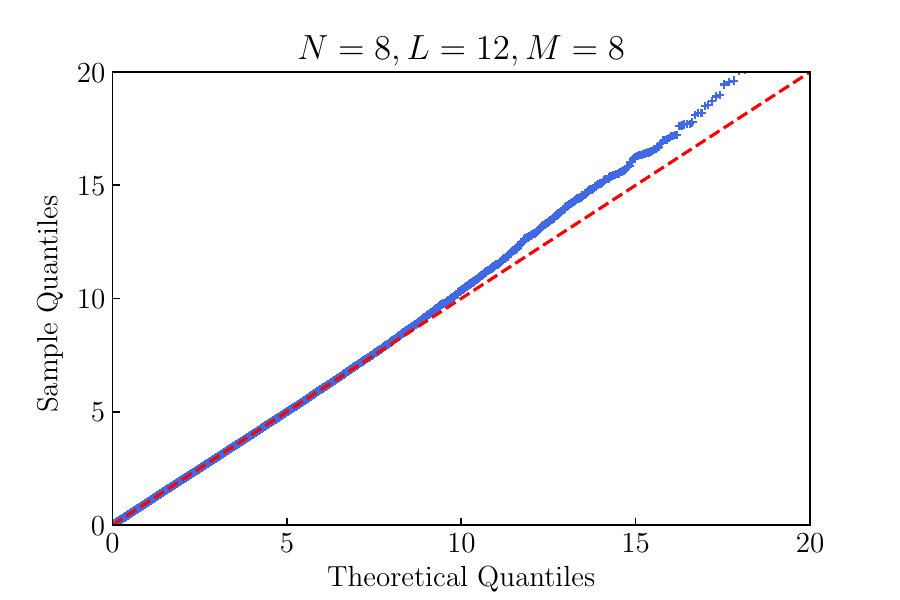}
    }
    \hfill
    \subfloat[]{
        \centering
        \includegraphics[width=3.2in]{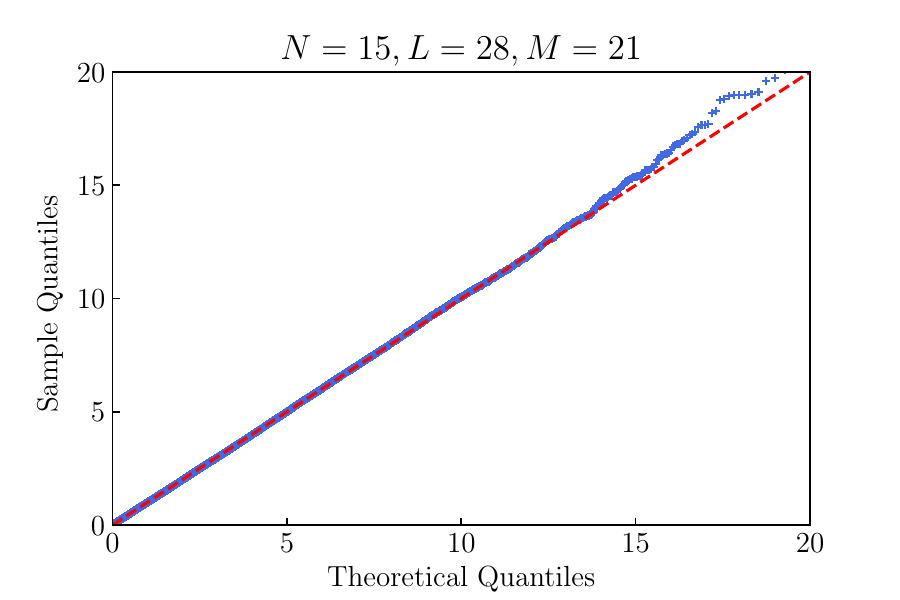}
    }
    \subfloat[]{
        \centering
        \includegraphics[width=3.2in]{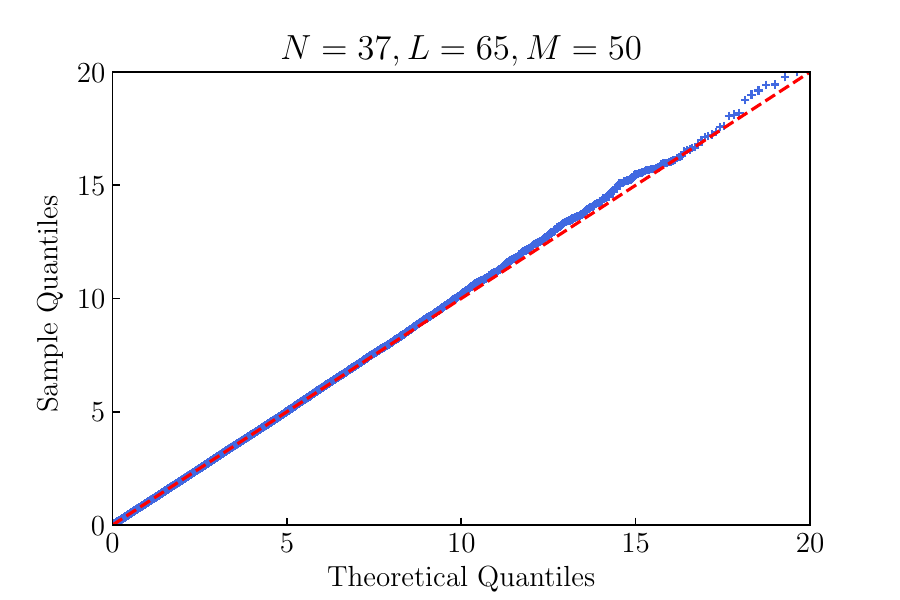}
    }
    \caption{Chi-square plot}
    \label{Chi_sq_plot}
\end{figure}
\textbf{Asymptotic Gaussianity:} Fig. \ref{Chi_sq_plot} shows the chi-square plot\cite{johnson2002applied} for the random vector $\mathbf{m}$. 
The horizontal axis represents the quantile of $\chi_2^2$, and the vertical axis is the order statistic of square Mahalanobis distance $d_j^2 = (\mathbf{m}_j - \overline{\mathbf{m}})^T \mathbf{V}^{-\frac{1}{2}}(\mathbf{m}_j - \overline{\mathbf{m}}) $ with $j=1, \ldots, 10^5$ number of samples. 
We have plotted the chi-square plot for different dimensions $(N, L, M)$, and the red line represents $y=x$. It can be observed that the CLT results are accurate even for small dimensions. 
As the dimension increases, the joint distribution approaches a Gaussian distribution.
\par
\begin{figure}[t]
    \centering
    \includegraphics[width=4.2in]{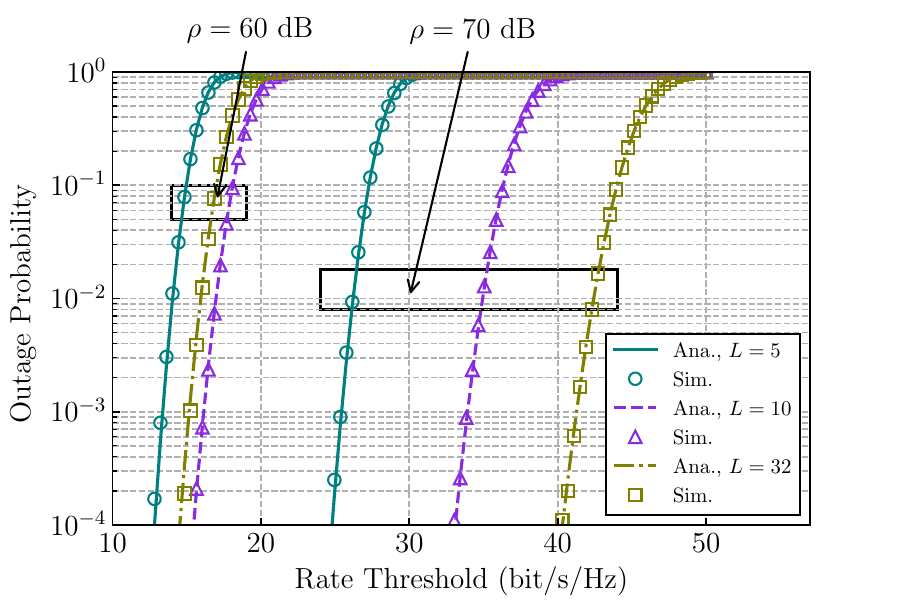}
    \caption{Outage probability approximation}
    \label{outage}
\end{figure}
\textbf{Accuracy of the second-order analysis:} Fig. \ref{outage} 
compares the outage probability calculated by \eqref{outage_prob} and the corresponding simulation results.
The number of transmit and receive antennas are set to be  $N=M=32$. The number of reflecting elements of the active IRS is set to $L=5$, $10$, and $32$. 
The two black-lined boxes indicate different SNRs. It can be observed that \eqref{outage_prob} is accurate. 


\section{Conclusion}
\label{Sec_Conclusion}
In this paper, we studied the statistical properties of the MI for two-hop MIMO channels by utilizing RMT. We derived the deterministic approximations 
for the mean and covariance matrix of the two MI terms required for the calculation of the MI, which were proved to be of order $\mathcal{O}(\frac{1}{N})$. 
By discussing the characteristic function, we further proved the asymptotic Gaussianity of the 
joint distribution for the two MI terms. 
To the best of the authors' knowledge, this is the first rigorous result regarding the fundamental limits of general top-hop MIMO channels. The results can be applied to different communication scenarios, such as active or passive IRS-assisted MIMO communications, double-scattering channels, AF relay systems, and so on. In addition, the analysis in this work can be extended to investigate the two-hop MIMO channels with the line-of-sight (LOS) component or direct links. From the perspective of RMT, the derived results regarding the existence and uniqueness of the solution for the fixed point systems can also be extended to cover the product model with an arbitrary number of Gaussian matrices.  Additionally, the discussions on the tightness of the random vector sequence can be extended to the convergence of the joint distribution for LSS of random matrices.
\appendices
\section{Proof of Proposition \ref{Prop_Stieltjes}} \label{App_Prop_Stiltjes}
For ease of presentation, we will introduce the concept of Stieltjes transform first. Let $\mu$ be a finite positive measure support over $\mathbb{R}$. Its Stieltjes transform $m(z)$ is defined as 
\begin{equation}
    m(z) = \int_{\mathbb{R}} \frac{\mu(\mathrm{d}\lambda)}{\lambda - z}, z \in \mathbb{C} - \mathbb{R}. \label{Integration_rep}
\end{equation}
In this sequel, the class of Stieltjes transforms of positive measure supported over $\mathbb{R}^+$ is denoted by $\mathcal{S}$. We now list some useful properties of the elements of $\mathcal{S}$.
\begin{proposition}\label{Prop_Sset}
    Let $f \in \mathcal{S}$ be the Stieltjes transform of positive measure $\mu$, then the following properties hold
    \begin{itemize}
        \item [1)] $f$ is holomorphic over $\mathbb{C}^+$.  For $z \in \mathbb{C}^+$, $f(z) \in \mathbb{C}^+$ and $zf(z) \in \mathbb{C}^+$. 
        \item [2)] $|f(z)| \leq \frac{\mu(\mathbb{R}^+)}{\dist(z, \mathbb{R}^+)}$ and  $\lim_{y \rightarrow + \infty} -\jmath y f(\jmath y) = \mu(\mathbb{R}^+)$.
        \item [3)] Conversely, if $g$ is holomorphic over $\mathbb{C}^+$ such that for $z \in \mathbb{C}^+$, then $g(z) \in \mathbb{C}^+$ and $zg(z) \in \mathbb{C}^+$. 
        Moreover, if $\lim_{y \rightarrow + \infty} -\jmath y g(\jmath y) = a \in (0, +\infty)$, then $g \in \mathcal{S}$ and the corresponding measure of $g$ satisfies $\mu(\mathbb{R}^+) = a$.
        \item [4)] If $\nu$ is a positive measure supported over $\mathbb{R}^+$, then
        \begin{equation}
            s(z) = \int_{\mathbb{R}^+}\frac{\nu(\mathrm{d}\lambda)}{\lambda + \frac{1}{f(z)}} \in \mathcal{S} \text{ and } -\frac{s(z)}{z f(z)} \in \mathcal{S}.
        \end{equation}
        \item [5)] If $g_1, g_2 \in \mathcal{S}$ and $a_1, a_2 > 0$ are constants, then $[\frac{a_1}{g_1(z)} - a_2 zg_2(z)]^{-1} \in \mathcal{S}$.
        \item [6)] (Matrix-valued Stieltjes transform) Let $\mathbf{F}(z)$ be an $N \times N$ matrix-valued function over $z \in \mathbb{C}^+$. 
        If $\mathbf{F}(z)$ is holomorphic over $\mathbb{C}^+$, with $\Im \mathbf{F}(z) \succ 0$, $\Im z\mathbf{F}(z) \succ 0$ and $\lim_{y \rightarrow + \infty} - \jmath y \mathbf{F}(\jmath y) = \mathbf{A} $,
        where $\mathbf{A}$ is a diagonal positive definite matrix, then there exists a positive matrix-valued measure  $\mu^{\mathbf{F}}$ (in the sense that $\mathbf{x}^H \mu^{\mathbf{F}}\mathbf{x} $ is a positive scalar measure for any $\mathbf{x} \neq \mathbf{0}_{N}$) supported over $\mathbb{R}^+$  such that
        \begin{equation}
            \mathbf{F}(z) = \int_{\mathbb{R}^+} \frac{\mu^{\mathbf{F}}(\mathrm{d} \lambda)}{\lambda - z}, ~~\text{and}~~ \mu^{\mathbf{F}}(\mathbb{R}^+) = \mathbf{A}.
        \end{equation}
    \end{itemize}
\end{proposition}
\textit{Proof:} Items 1) - 3) can be found in \cite[Proposition 2.2]{hachem2007deterministic}. Items 4) and 5) can be proved by 3) directly. 6) is a direct corollary of \cite[Theorem 5.4]{gesztesy2000matrix}. \QED
\par
Directly proving the existence or uniqueness of solutions to the system of equations \eqref{DE_system_1} is challenging. Our strategy is to fix $\delta$ and treat $\underline{\omega}$ and $\gamma$ as functions of $\delta$ to study their properties. Recall the definition of $\mathcal{S}$, we introduce the following two propositions.
\begin{proposition} \label{Prop_S_omega_gamma}
Assuming that $\overline{s} \geq 0$, $\gamma_S^{(0)}, \underline{\omega}_S^{(0)} \in \mathcal{S}$ and denoting
\begin{equation}
\left\{ \begin{array}{lr}
        \underline{\omega}_S^{(1)}(y)  = \frac{1}{L} \mathrm{Tr} \left[\mathbf{T}_1^{\frac{1}{2}} \mathbf{R}_2 \mathbf{T}_1^{\frac{1}{2}} \mathbf{F}_{S, \omega}^{(0)}(y)\right], \\
        \gamma_S^{(1)}(y) = \frac{1}{M} \mathrm{Tr}\left[\mathbf{T}_2\mathbf{F}^{(0)}_{S, \gamma}(y)\right],
    \end{array} 
\right. \label{DE_Prop_1}
\end{equation}
where 
\begin{equation}
\left\{ \begin{array}{lr}
    \mathbf{F}^{(k)}_{S, \omega}(y) = \left( -y\left( \mathbf{I}_L - \frac{\overline{s}}{y} \mathbf{T}_1 +  {\gamma}^{(k)}_S(y) \mathbf{T}_1^{\frac{1}{2}} \mathbf{R}_2 \mathbf{T}_1^{\frac{1}{2}} \right) \right)^{-1}, \\ \mathbf{F}^{(k)}_{S, \gamma}(y) = \left( -y\left( \mathbf{I}_L  +  \frac{L}{M}{\underline{\omega}}^{(k)}_S(y) \mathbf{T}_2 \right) \right)^{-1},
    \end{array} 
\right. 
\end{equation}
then the following holds true
\begin{itemize}
    \item [1)] $\mathbf{F}^{(0)}_{S, \omega}(y)$ and $ \mathbf{F}^{(0)}_{S, \gamma}(y)$ are holomorphic over $\mathbb{C} - \mathbb{R}^+$.
    \item [2)] $\norm{ \mathbf{F}^{(0)}_{S, \omega}(y)} \leq \frac{1}{\mathrm{dist}(y, \mathbb{R}^+)}$, $\norm{ \mathbf{F}^{(0)}_{S, \gamma}(y)} \leq \frac{1}{\mathrm{dist}(y, \mathbb{R}^+)}$, $y \in \mathbb{C} - \mathbb{R}^+$.
    \item [3)] $\underline{\omega}_S^{(1)} \in \mathcal{S}$ and $\gamma_S^{(1)} \in \mathcal{S}$ with the corresponding total measure $\mu^{(1)}_{S, \underline{\omega}}(\mathbb{R}^+) = \frac{1}{L} \mathrm{Tr}\mathbf{R}_2 \mathbf{T}_1 $ and $\mu^{(1)}_{S, \gamma}(\mathbb{R}^+) = \frac{1}{M} \mathrm{Tr} \mathbf{T}_2$.
\end{itemize}
\end{proposition}
\textit{Proof:} 
For item 1), we check that $\underline{\mathbf{F}}^{(0)}_{S, \omega}(y) = -y( \mathbf{I}_L - \frac{\overline{s}}{y} \mathbf{T}_1 +  {\gamma}^{(0)}_S(y) \mathbf{T}_1^{\frac{1}{2}} \mathbf{R}_2 \mathbf{T}_1^{\frac{1}{2}})$ and $\underline{\mathbf{F}}^{(0)}_{S, \gamma}(y) = -y ( \mathbf{I}_L  +  \frac{L}{M}{\underline{\omega}}^{(0)}_S(y) \mathbf{T}_2)$ are invertible over $\mathbb{C}^+$, $\mathbb{C}_-$, and $(-\infty, 0)$. 
For any $ 
 \mathbf{v} \in \mathbb{C}^L$ such that $\norm{\mathbf{v}} = 1$ and $y \in \mathbb{C}^+$, we have
\begin{equation}
    \Im(\mathbf{v}^H \underline{\mathbf{F}}^{(0)}_{S, \omega}(y) \mathbf{v})  = -\Im(y) - \Im(y \gamma_S^{(0)}(y)) \mathbf{v}^H\mathbf{T}_1^{\frac{1}{2}} \mathbf{R}_2 \mathbf{T}_1^{\frac{1}{2}} \mathbf{v} \leq - \Im(y),
\end{equation}
so $\inf_{\norm{\mathbf{v}} = 1} \{ | \mathbf{v}^H \underline{\mathbf{F}}^{(0)}_{S, \omega}(y)\mathbf{v}| \} > 0$. A similar argument on $\underline{\mathbf{F}}^{(0)}_{S, \omega}(y) $ yields the same result. This implies the two matrices are invertible for $\mathbb{C}^+$. The argument for $\mathbb{C}^-$ and $(-\infty, 0)$ is similar and has been 
 omitted for brevity. As a result, $\mathbf{F}^{(0)}_{S, \omega}(y)  = (\underline{\mathbf{F}}^{(0)}_{S, \omega}(y) )^{-1}$ and $\mathbf{F}^{(0)}_{S, \gamma}(y)  = (\underline{\mathbf{F}}^{(0)}_{S, \gamma}(y) )^{-1}$  are holomorphic over $\mathbb{C} - \mathbb{R}^+$.
\par
For item 2), we need to check property 6) of Proposition \ref{Prop_Sset}. By using the identity $\mathbf{A} - \mathbf{B} = \mathbf{A}(\mathbf{B}^{-1} - \mathbf{A}^{-1})\mathbf{B}$, we have  
\begin{align}
    &\Im(\mathbf{F}^{(0)}_{S, \omega}(y)) = \frac{1}{2 \jmath }\left(\mathbf{F}^{(0)}_{S, \omega}(y) - (\mathbf{F}^{(0)}_{S, \omega}(y))^H  \right) =  \left( \mathbf{F}^{(0)}_{S, \omega}(y)\left( \Im(y) + \Im(y \gamma_S^{(0)}(y))\mathbf{T}_1^{\frac{1}{2}} \mathbf{R}_2 \mathbf{T}_1^{\frac{1}{2}} \right)(\mathbf{F}^{(0)}_{S, \omega}(y))^H\right) \notag \\
    &= \Im(y)\mathbf{F}^{(0)}_{S, \omega}(y)(\mathbf{F}^{(0)}_{S, \omega}(y))^H +  \Im(y \gamma_S^{(0)}(y))\mathbf{F}^{(0)}_{S, \omega}(y) \mathbf{T}_1^{\frac{1}{2}} \mathbf{R}_2 \mathbf{T}_1^{\frac{1}{2}}(\mathbf{F}^{(0)}_{S, \omega}(y))^H \succ 0,  \hspace*{1mm}y \in \mathbb{C}^+. \label{Im_F}
\end{align}
The relation $\Im(y\mathbf{F}^{(0)}_{S, \omega}(y)) \succ 0 $ can be obtained by using the similar method as in \eqref{Im_F}. Moreover,  
\begin{equation}
\lim_{y \rightarrow + \infty} -\jmath y \mathbf{F}^{(0)}_{S, \omega}(\jmath y)=\lim_{ y \rightarrow + \infty}  \left( \mathbf{I}_L - \frac{\overline{s}}{\jmath y} \mathbf{T}_1 +  {\gamma}^{(0)}_S(\jmath y) \mathbf{T}_1^{\frac{1}{2}} \mathbf{R}_2 \mathbf{T}_1^{\frac{1}{2}} \right)^{-1} = \mathbf{I}_L.  
\end{equation}
Thus, by properties 6) of Proposition \ref{Prop_Sset}, matrix ${\mathbf{F}}^{(0)}_{S, \omega}$ have integration representation ${\mathbf{F}}^{(0)}_{S, \omega}(y) = \int_{\mathbb{R}^+}\frac{\mu_{S, \omega}(\mathrm{d}\lambda)}{\lambda - y}$. Thus for any $y \in \mathbb{C}^+$,  we obtain
\begin{equation}
    \Im(y) \mathbf{F}^{(0)}_{S, \omega}(y)(\mathbf{F}^{(0)}_{S, \omega}(y))^H \overset{(a)}{\preceq} \Im(\mathbf{F}^{(0)}_{S, \omega}(y)) = \int_{\mathbb{R}^+}\frac{\Im(y) \mu_{S, \omega}(\mathrm{d}\lambda)}{|\lambda - y|^2} \overset{(b)}{\preceq } \frac{\Im(y)}{\dist(y, \mathbb{R}^+)^2} \mathbf{I}_L, 
\end{equation}
where $(a)$ follows by \eqref{Im_F} and $(b)$ follows from the fact that  $\mu_{S, \omega}(\mathbb{R}^+) = \mathbf{I}_L$. The continuity argument yields that for $y \in  \mathbb{C} - \mathbb{R}^+$, $\mathbf{F}^{(0)}_{S, \omega}(y)(\mathbf{F}^{(0)}_{S, \omega}(y))^H {\preceq } \frac{1}{\dist(y, \mathbb{R}^+)^2} \mathbf{I}_L$. The result for $ \mathbf{F}^{(0)}_{S, \gamma}(y)$ can be obtained using the same method, which completes the proof of 2).
\par
For item 3), we just need to check 3) in Proposition \ref{Prop_Sset} and is omitted here for brevity. Thus, we have proved Proposition \ref{Prop_S_omega_gamma}. \QED
\par
In fact, by matrix identity $\mathbf{A}(\mathbf{I} + \mathbf{B}\mathbf{A})^{-1} = (\mathbf{I} + \mathbf{A}\mathbf{B})^{-1}\mathbf{A}$ in Proposition \ref{Prop_S_omega_gamma}, the function $\underline{\omega}_{S}^{(1)}$  can be expressed as $\underline{\omega}_{S}^{(1)}(y) = \frac{1}{L}\Tr[\mathbf{R}_2\mathbf{T}_1( -y( \mathbf{I}_L - \frac{\overline{s}}{y} \mathbf{T}_1 +  {\gamma}^{(k)}_S(y) \mathbf{R}_2 \mathbf{T}_1) )^{-1}]$. For the sake of symmetry and convenience of analysis, we maintain the notation of Proposition \ref{Prop_S_omega_gamma}. To improve readability, we will omit the variable  $y$ in the appropriate places. The following proposition is related to the existence and uniqueness of $\underline{\omega}$ and $\gamma$ in \eqref{DE_system_1}.
\par
\begin{proposition} \label{Prop_fixed_point_omega_gamma}
There exist unique solutions $\underline{\omega}_S, \gamma_S \in \mathcal{S}$, such that 
\begin{equation}
\left\{ \begin{array}{lr}
        \underline{\omega}_S(y)  = \frac{1}{L} \mathrm{Tr} \left[\mathbf{T}_1^{\frac{1}{2}} \mathbf{R}_2 \mathbf{T}_1^{\frac{1}{2}} \mathbf{F}_{S, \omega}(y)\right], \\
        \gamma_S(y) = \frac{1}{M} \mathrm{Tr}\left[\mathbf{T}_2\mathbf{F}_{S, \gamma}(y)\right],
    \end{array} 
\right. \label{DE_omega_gamma}
\end{equation}
and the corresponding total measures are $\frac{1}{L} \Tr \mathbf{R}_2 \mathbf{T}_1$ and $\frac{1}{M} \Tr \mathbf{T}_2$. 
Moreover, for $y = -y_0 < 0$, the positive solution $\underline{\omega}_S(-y_0) > 0$ and $\gamma_S(-y_0) > 0$ is unique.
\end{proposition}
\textit{Proof:} We prove the existence by construction. Here, we consider the following iterative scheme
\begin{equation}
\left\{ \begin{array}{lr}
        \underline{\omega}_S^{(k+1)}(y)  = \frac{1}{L} \mathrm{Tr} \left[\mathbf{T}_1^{\frac{1}{2}} \mathbf{R}_2 \mathbf{T}_1^{\frac{1}{2}} \mathbf{F}_{S, \omega}^{(k)}(y)\right], \\
        \gamma_S^{(k+1)}(y) = \frac{1}{M} \mathrm{Tr}\left[\mathbf{T}_2\mathbf{F}^{(k)}_{S, \gamma}(y)\right],
    \end{array} 
\right. 
\end{equation}
with starting points $\gamma_S^{(0)}, \underline{\omega}_S^{(0)} \in \mathcal{S}$. According to Proposition \ref{Prop_S_omega_gamma} we know that $\underline{\omega}_S^{(k)}$ and $\gamma_S^{(k)}$ are elements of class $\mathcal{S}$ and the inequalities $\norm{ \mathbf{F}^{(0)}_{S, \omega}} \leq \frac{1}{\mathrm{dist}(y, \mathbb{R}^+)}$ and $\norm{ \mathbf{F}^{(0)}_{S, \gamma}} \leq \frac{1}{\mathrm{dist}(y, \mathbb{R}^+)}$ hold for all $k \geq 0$. By taking the difference, we obtain
\begin{equation}
\begin{split}
    &\abs{\underline{\omega}_S^{(k+1)} - \underline{\omega}_S^{(k)}} = \frac{1}{L}\abs{\mathrm{Tr}\mathbf{T}_1^{\frac{1}{2}} \mathbf{R}_2 \mathbf{T}_1^{\frac{1}{2}} \left(\mathbf{F}^{(k)}_{S, \omega} - \mathbf{F}^{(k-1)}_{S, \omega} \right) }  =  \abs{ \gamma_S^{(k)} - \gamma_S^{(k-1)}} \abs{\frac{y}{L} \mathrm{Tr} \mathbf{T}_1^{\frac{1}{2}} \mathbf{R}_2 \mathbf{T}_1^{\frac{1}{2}} \mathbf{F}^{(k)}_{S, \omega}\mathbf{T}_1^{\frac{1}{2}} \mathbf{R}_2 \mathbf{T}_1^{\frac{1}{2}} \mathbf{F}^{(k-1)}_{S, \omega}} \\
    & \overset{(a)}{\leq}  \abs{ \gamma_S^{(k)} - \gamma_S^{(k-1)} } \frac{\abs{y} r^4}{\dist(y, \mathbb{R}^+)^2}. \label{Diff_omega}
\end{split}
\end{equation}
Here, step $(a)$ comes from the inequality $\Tr \mathbf{A} \leq N \norm{\mathbf{A}}$ for $\mathbf{A} \in \mathbb{C}^{N \times N}$. Using the same method, we can obtain the upper bound for $\abs{\gamma_S^{(k+1)} - \gamma_S^{(k)}}$ as 
\begin{equation}
\begin{split}
    \abs{\gamma_S^{(k+1)} - \gamma_S^{(k)}} = \frac{1}{M}\abs{\mathrm{Tr}\mathbf{T}_2 \left(\mathbf{F}^{(k)}_{S, \gamma} - \mathbf{F}^{(k-1)}_{S, \gamma} \right) } {\leq}  \abs{\underline{\omega}_S^{(k)} - \underline{\omega}_S^{(k-1)}} \frac{L \abs{y} r^2}{M \dist(y, \mathbb{R}^+)^2}. \label{Diff_gamma}
\end{split}
\end{equation}
Define $\mathbf{M}^{(k)}(y) = \max(|\underline{\omega}_S^{(k+1)} - \underline{\omega}_S^{(k)}|(y), |\gamma_S^{(k+1)} - \gamma_S^{(k)}|(y))$ and $\mathcal{E}(y) = \frac{|y|}{\dist(y, \mathbb{R}^+)^2}\max(r^4, \frac{L}{M}r^2)$. Using \eqref{Diff_omega} and \eqref{Diff_gamma} yields that $\mathbf{M}^{(k)}(y) \leq \mathcal{E}(y) \mathbf{M}^{(k-1)}(y)$. Let $y \in \mathbb{C}^+$ satisfy that $\mathcal{E}(y) \leq 0.9$, then the two sequences $\{ \underline{\omega}_S^{(k)}(y) \}_{k \geq 0}$ and $\{ \gamma_S^{(k)}(y)\}_{k \geq 0}$ form Cauchy sequences. Denote by $\underline{\omega}_S(y)$ and $\gamma_S(y)$ the respective limits. By 2) in Proposition \ref{Prop_Sset}, $\underline{\omega}_S^{(k)}$ is  bounded on every compact set in $\mathbb{C} - \mathbb{R}^{+}$. Thus $\underline{\omega}_S^{(k)}$ forms a normal family \cite[Theorem 15.2.3]{hille2002analytic} over $\mathbb{C} - \mathbb{R}^{+}$. For each subsequence of $\underline{\omega}_S^{(k)}$, i.e. $ \underline{\omega}_S^{(k_n)}$, one can further choose a subsequence $ \underline{\omega}_S^{(k_{n_l})}$ that converges to a holomorphic function $\widetilde{\underline{\omega}}_S$.   Since $\widetilde{\underline{\omega}}_S$ is same as $\underline{\omega}_S$ in the domain $\{y:  \mathcal{E}(y) \leq 0.9 \}$ and the chosen of $k_n$ is arbitrary,  $\underline{\omega}_S^{(k)}$ converges to and holomorphic function which we still denote by $\underline{\omega}_S$ over $\mathbb{C} - \mathbb{R}^{+}$. Similarly, $\gamma_S^{(k)}$ will converge to a holomorphic function $\gamma_S$ over $\mathbb{C} - \mathbb{R}^{+}$.
\par
Next, we prove that $\underline{\omega}_S, \gamma_S \in \mathcal{S}$. The  convergence of $\underline{\omega}_S^{(k)}(y), \gamma_S^{(k)}(y)$ and  3) in Proposition \ref{Prop_S_omega_gamma} imply that 
\begin{subequations}
 \label{Limit_omega_gamma}
\begin{align}
     &\Im(\underline{\omega}_S(y)) \geq 0, \Im(y\underline{\omega}_S(y)) \geq 0, \abs{\underline{\omega}_S(y)} \leq \frac{\mathrm{Tr}\mathbf{R}_2 \mathbf{T}_1}{L \dist(y, \mathbb{R}^+)}, \\
     \text{ and }& \Im(\gamma_S(y)) \geq 0, \Im(y\gamma_S(y)) \geq 0, \abs{\gamma_S(y)} \leq \frac{\mathrm{Tr}\mathbf{T}_2}{M \dist(y, \mathbb{R}^+)}, ~~ y \in \mathbb{C}^+.
\end{align}
\end{subequations}
Therefore, from the convergence of $\mathbf{F}^{(k)}_{S, \omega} \xrightarrow[k \rightarrow + \infty]{ }\mathbf{F}_{S, \omega}$, we obtain
\begin{equation}
    \lim_{y \rightarrow + \infty} - \jmath y \frac{1}{L} \mathrm{Tr} \left[\mathbf{T}_1^{\frac{1}{2}} \mathbf{R}_2 \mathbf{T}_1^{\frac{1}{2}} \mathbf{F}_{S, \omega}(\jmath y)\right] = \lim_{y \rightarrow + \infty} \frac{1}{L}\mathrm{Tr} \mathbf{T}_1^{\frac{1}{2}} \mathbf{R}_2 \mathbf{T}_1^{\frac{1}{2}}\left( \mathbf{I}_L - \frac{\overline{s} \mathbf{T}_1}{\jmath y}  +  {\gamma}_S(\jmath y) \mathbf{T}_1^{\frac{1}{2}} \mathbf{R}_2 \mathbf{T}_1^{\frac{1}{2}} \right) ^{-1} = \frac{1}{L}\mathrm{Tr}  \mathbf{R}_2\mathbf{T}_1. \label{Limit_jy}
\end{equation}
 Suppose that there exists $y_0 \in \mathbb{C}^+$ such that $\Im(\underline{\omega}_S(y_0)) = 0$. Since $\underline{\omega}_S$ is holomorphic, $\Im(\underline{\omega}_S)$ is a harmonic function. According to the maximum principle \cite[Theorem 21]{ahlfors1979complex}, $\Im(\underline{\omega}_S)$ must be constant on every closed set $\mathcal{D}_t = \{ y \in \mathbb{C}^+: \Im(y) \geq \frac{1}{t}, |y - y_0| \leq t \}$. As $t \rightarrow + \infty$, we have $\Im(\underline{\omega}_S(y)) = 0$ for $y \in \mathbb{C}^+$, which is contradictive  with \eqref{Limit_jy}. Hence, it must hold that $\Im(\underline{\omega}_S(y)) > 0$ and $\Im(y\underline{\omega}_S(y)) > 0$ for $y \in \mathbb{C}^+$. The same argument can be applied to $\gamma_S$. Therefore, we complete the proof of the existence.
 \par
 Next, we prove the uniqueness, which consists of two parts.
  The first part shows that the positive scalar solutions $\underline{\omega}_S(-y_0)$ and $\gamma_S(-y_0)$ are unique for certain $y = -y_0 < 0$. 
 The second part shows that the function solutions $\underline{\omega}_S$ and $\gamma_S$ are unique in class $\mathcal{S}$. 
 For the first part, assume that there are two solutions $(\underline{\omega}_S(y), \gamma_S(y)) $ and $(\overline{\underline{\omega}}_S(y), \overline{\gamma}_S(y))$ both admit the relation \eqref{DE_omega_gamma}. We denote $(\mathbf{F}_{S, \omega}, \mathbf{F}_{S, \gamma})$ and $(\overline{\mathbf{F}}_{S, \omega}, \overline{\mathbf{F}}_{S, \gamma})$ as the associated matrices. Introducing $\boldsymbol{\epsilon} = (\underline{\omega}_S - \overline{\underline{\omega}}_S, \gamma_S - \overline{\gamma}_S)^T$, we have
 \begin{equation}
     \begin{bmatrix}
         1 &  -\frac{y}{L} \mathrm{Tr} \mathbf{T}_1^{\frac{1}{2}} \mathbf{R}_2 \mathbf{T}_1^{\frac{1}{2}} \mathbf{F}_{S, \omega}\mathbf{T}_1^{\frac{1}{2}} \mathbf{R}_2 \mathbf{T}_1^{\frac{1}{2}} \overline{\mathbf{F}}_{S, \omega} \\
         -\frac{L y}{M^2} \mathrm{Tr}\mathbf{T}_2\mathbf{F}_{S, \gamma}\mathbf{T}_2 \overline{\mathbf{F}}_{S, \gamma}  & 1
     \end{bmatrix} \boldsymbol{\epsilon} = \left(\mathbf{I}_2 + \mathbf{S}\right) \boldsymbol{\epsilon} = \mathbf{0}_2. \label{epsilon_omega_gamma}
 \end{equation}
To show $\boldsymbol{\epsilon} = \mathbf{0}_2$, we need to prove that $\mathbf{I}_2 + \mathbf{S}$ in \eqref{epsilon_omega_gamma} is invertible when $y = -y_0 < 0$. It is sufficient to show that the spectral radius of $\mathbf{S}$ is less than 1. Using \eqref{DE_omega_gamma}, we have $\underline{\omega}_S(- y_0) \leq \frac{ \Tr \mathbf{R}_1 \mathbf{T}_1}{L y_0} \leq \frac{r^2}{y_0}$ and $\gamma_S(- y_0) \leq \frac{\Tr \mathbf{T}_2}{M y_0} \leq \frac{r}{y_0} $. Define the components of $\widetilde{\mathbf{S}} \in \mathbb{R}^{2 \times 2}$ as $[\widetilde{\mathbf{S}}]_{11} = [\widetilde{\mathbf{S}}]_{22} = 0$, $[\widetilde{\mathbf{S}}]_{12} = \frac{y_0}{L} \mathrm{Tr} [\mathbf{T}_1^{\frac{1}{2}} \mathbf{R}_2 \mathbf{T}_1^{\frac{1}{2}} \mathbf{F}_{S, \omega}(-y_0)]^2 $, and $[\widetilde{\mathbf{S}}]_{21} = \frac{Ly_0}{M^2} \mathrm{Tr}[\mathbf{T}_2\mathbf{F}_{S, \gamma}(-y_0)]^2$, we have
  \begin{equation}
      \begin{bmatrix}
        \underline{\omega}_S(-y_0) \\
         \gamma_S(-y_0)
     \end{bmatrix} = \widetilde{\mathbf{S}}
     \begin{bmatrix}
        \underline{\omega}_S(-y_0) \\
         \gamma_S(-y_0)
     \end{bmatrix} + \begin{bmatrix}
        \frac{y_0}{L} \mathrm{Tr} \mathbf{T}_1^{\frac{1}{2}} \mathbf{R}_2 \mathbf{T}_1^{\frac{1}{2}} \mathbf{F}_{S, \omega}^2 + \frac{\overline{s}}{L} \mathrm{Tr} \mathbf{T}_1^{\frac{1}{2}} \mathbf{R}_2 \mathbf{T}_1^{\frac{1}{2}}\mathbf{F}_{S, \omega} \mathbf{T}_1 \mathbf{F}_{S, \omega} \\
          \frac{y_0}{M} \Tr \mathbf{T}_2 \mathbf{F}_{S, \gamma}^2
     \end{bmatrix}. \label{DE_identity}
 \end{equation} 
We rewrite \eqref{DE_identity} as $\mathbf{u} = \widetilde{\mathbf{S}}\mathbf{u} + \mathbf{v}$. Then, the following holds true
\begin{equation}
   \frac{1}{L}\mathrm{Tr} \mathbf{T}_1^{\frac{1}{2}} \mathbf{R}_2 \mathbf{T}_1^{\frac{1}{2}} \mathbf{F}_{S, \omega}^2 = \frac{1}{L}\sum_{i=1}^L \frac{\mathbf{w}_i^H \mathbf{T}_1^{\frac{1}{2}} \mathbf{R}_2 \mathbf{T}_1^{\frac{1}{2}} \mathbf{w}_i}{(y_0 + \lambda_i(\mathbf{G}))^2} \geq \frac{1}{L}\sum_{i=1}^L \frac{\mathbf{w}_i^H \mathbf{T}_1^{\frac{1}{2}} \mathbf{R}_2 \mathbf{T}_1^{\frac{1}{2}} \mathbf{w}_i}{(y_0 + \norm{\mathbf{G}})^2} =  \frac{\frac{1}{L} \Tr  \mathbf{R}_2 \mathbf{T}_1}{(y_0 + \norm{\mathbf{G}})^2} \overset{(a)}{\geq} \frac{l}{(y_0 + \overline{s} r + r^3)^2} > 0,
\end{equation}
where $\mathbf{G} = \overline{s} \mathbf{T}_1 + y_0 \gamma_S(-y_0)\mathbf{T}_1^{\frac{1}{2}} \mathbf{R}_2 \mathbf{T}_1^{\frac{1}{2}}$ with the  
eigenvalue decomposition $\mathbf{G} = \sum_{i=1}^L \lambda_i(\mathbf{G}) \mathbf{w}_i\mathbf{w}^H_i$. Step $(a)$ follows from $y_0\gamma_S(-y_0) \leq r$ and  
$\norm{\mathbf{G}} \leq \overline{s} \norm{\mathbf{T}_1} + y_0 \gamma_S(-y_0)\norm{\mathbf{T}_1^{\frac{1}{2}} \mathbf{R}_2 \mathbf{T}_1^{\frac{1}{2}}}\leq \overline{s}r + r^3$. Thus, we have $[\mathbf{u}]_1 > 0$. With the similar argument, we can check all the elements of $\mathbf{u}$, $\mathbf{v}$, and $\widetilde{\mathbf{S}}$ are positive. \cite[Lemma 5.2 (1)]{Hachem2008CLTVP} implies that $\rho(\widetilde{\mathbf{S}}) \leq 1 - \frac{\min([\mathbf{v}]_1, [\mathbf{v}]_2)}{\max([\mathbf{u}]_1, [\mathbf{u}]_2)} < 1$, where $\rho(\cdot)$ is the spectral radius operator. In the following, we bound $\rho (\mathbf{S})$. According to the Cauchy–Schwarz inequality $\Tr(\mathbf{A}\mathbf{B}) \leq \sqrt{\Tr(\mathbf{A}\mathbf{A}^H) \Tr(\mathbf{B}\mathbf{B}^H)}$ with $\mathbf{A} = [\mathbf{T}_1^{\frac{1}{2}} \mathbf{R}_2 \mathbf{T}_1^{\frac{1}{2}}]^{\frac{1}{2}} \mathbf{F}_{S, \omega}(y)[\mathbf{T}_1^{\frac{1}{2}} \mathbf{R}_2 \mathbf{T}_1^{\frac{1}{2}}]^{\frac{1}{2}} $, for $y = -y_0$, we have
\begin{equation}
    \abs{[\mathbf{S}]_{12}} \leq \sqrt{\frac{y_0}{L}\Tr[\mathbf{T}_1^{\frac{1}{2}} \mathbf{R}_2 \mathbf{T}_1^{\frac{1}{2}} \mathbf{F}_{S, \omega}]^2\frac{y_0}{L}\Tr[\mathbf{T}_1^{\frac{1}{2}} \mathbf{R}_2 \mathbf{T}_1^{\frac{1}{2}} \overline{\mathbf{F}}_{S, \omega}]^2 } = \sqrt{[\widetilde{\mathbf{S}}]_{12}[\overline{\widetilde{\mathbf{S}}}]_{12}},
    \label{Spectral_radius}
\end{equation}
where $\overline{\widetilde{\mathbf{S}}}$ is defined by replacing $(\underline{\omega}_S, \gamma_S) $ with $(\overline{\underline{\omega}}_S, \overline{\gamma}_S) $ in ${\widetilde{\mathbf{S}}}$. Similarly, we have $\abs{[\mathbf{S}]_{21}} \leq \sqrt{[\widetilde{\mathbf{S}}]_{21}[\overline{\widetilde{\mathbf{S}}}]_{21}}$. By \cite[Theorem 8.1.18]{horn2012matrix} and 
\cite[ Lemma 5.7.9]{roger1994topics}, it holds true that $ \rho(\mathbf{S}) \leq \sqrt{\rho(\widetilde{\mathbf{S}}) \rho(\overline{\widetilde{\mathbf{S}}})} < 1$.
Hence, we proved that $(\underline{\omega}_S(-y_0), \gamma_S(-y_0)) =(\overline{\underline{\omega}}_S(-y_0), \overline{\gamma}_S(-y_0))$. For the second part, we assume that there exist two function solutions $(\underline{\omega}_S, \gamma_S) \in \mathcal{S}^2$ and $(\overline{\underline{\omega}}_S, \overline{\gamma}_S) \in \mathcal{S}^2$. From the first part, we know that $(\underline{\omega}_S, \gamma_S)$ and $(\overline{\underline{\omega}}_S, \overline{\gamma}_S)$ must be same on $(- \infty, 0)$. Since $\underline{\omega}_S, \gamma_S$, $\overline{\underline{\omega}}_S$ and $\overline{\gamma}_S$ are holomorphic, it must hold that $(\underline{\omega}_S, \gamma_S) = (\overline{\underline{\omega}}_S, \overline{\gamma}_S)$ over $\mathbb{C} - \mathbb{R}^+$. Therefore, we proved the uniqueness.
\QED
\par
In the following, we prove a property similar to Proposition \ref{Prop_fixed_point_omega_gamma} for the fundamental equations \eqref{DE_system_1}, i.e., there exist unique solutions $(\delta, \overline{\omega}, \underline{\omega}, \gamma)$ in class $\mathcal{S}$ for \eqref{DE_system_1},
 and the positive solutions $(\delta, \overline{\omega}, \underline{\omega}, \gamma)(\overline{s}_0, z_0)$ are unique for $z = z_0 > 0$ and $s = \overline{s}_0 \geq 0$. We will prove the existence and uniqueness, respectively.
\subsection{Existence \label{App_S_existence}}
First, we fix $\overline{s} \geq 0$ by treating it as a constant. To study \eqref{DE_system_1}, it is useful to change the sign of the variable $z$ and extend the domain of $z$
 to the complex plane. Let $f_\delta(z) = \delta(\overline{s}, -z) $ in \eqref{DE_system_1}. Similar to the proof of Proposition \ref{Prop_fixed_point_omega_gamma}, we can construct the desired solution by induction. Assuming that $f_\delta^{(k)} \in \mathcal{S}$ for some $k \geq 0$, an immediate result is that $- \frac{1}{f_\delta^{(k)}(z)} \in \mathbb{C}^+$ for $z \in \mathbb{C}^+$.  Denote by $\overline{\omega}_S, \gamma_S \in \mathcal{S}$ the unique solutions of \eqref{DE_omega_gamma} which have the integration representations $
    \underline{\omega}_S(y) = \int_{\mathbb{R}^+} \frac{ \mu_{\underline{\omega}_S}(\mathrm{d} \lambda) }{\lambda - y}$ and $ \gamma_S(y) = \int_{\mathbb{R}^+} \frac{ \mu_{\gamma_S}(\mathrm{d} \lambda) }{\lambda - y}.
$
By plugging $y = - \frac{1}{f_\delta^{(k)}(z)}$ in the integration representation, 4) of Proposition \ref{Prop_Sset} implies that $f_{\underline{\omega}}^{(k)} = \underline{\omega}_S(- \frac{1}{f_\delta^{(k)}}) \in \mathcal{S}$ and $f_{\gamma}^{(k)}  = \gamma_S(- \frac{1}{f_\delta^{(k)}}) \in \mathcal{S}$. Moreover, by \eqref{DE_omega_gamma}, the following holds
\begin{equation}
\label{Eq_f_omega_gamma_delta}
\left\{ \begin{array}{lr}
       \frac{f_{\underline{\omega}}^{(k)}}{ f_\delta^{(k)}}  = \frac{1}{L} \mathrm{Tr} \left[\mathbf{T}_1^{\frac{1}{2}} \mathbf{R}_2 \mathbf{T}_1^{\frac{1}{2}} \left( \mathbf{I}_L +  \overline{s}f_\delta^{(k)} \mathbf{T}_1 +  f_{\gamma}^{(k)} \mathbf{T}_1^{\frac{1}{2}} \mathbf{R}_2 \mathbf{T}_1^{\frac{1}{2}}  \right)^{-1}\right], \\
        \frac{f_{\gamma}^{(k)}}{f_\delta^{(k)}} = \frac{1}{M} \mathrm{Tr}\left[\mathbf{T}_2\left( \mathbf{I}_L  +  \frac{L}{M}f_{\underline{\omega}}^{(k)} \mathbf{T}_2 \right)^{-1}\right].
    \end{array} 
\right. 
\end{equation}
Since $f_{\delta}^{(k)}$, $f_{\underline{\omega}}^{(k)}$, and $f_{\gamma}^{(k)}$ are of class $\mathcal{S}$, by 4) of Proposition \ref{Prop_Sset}, we have  $-\frac{f_{\underline{\omega}}^{(k)}}{z f_\delta^{(k)}}, -\frac{f_{\gamma}^{(k)}}{z f_\delta^{(k)}} \in \mathcal{S}$. Therefore, the following holds by 5) of Proposition \ref{Prop_Sset}
\begin{equation}
    -\frac{f_{\underline{\omega}}^{(k)}f_{\gamma}^{(k)}}{z (f_\delta^{(k)})^2} = -\frac{1}{M} \sum_{i=1}^M \frac{\lambda_i(\mathbf{T}_2)}{\frac{zf_{\delta}^{(k)}}{f_{\underline{\omega}}^{(k)}} + zf_{\delta}^{(k)}\frac{L}{M} \lambda_i(\mathbf{T}_2)} \in \mathcal{S},
    \label{prop_S_o_g_d}
\end{equation}
where $ \lambda_i(\mathbf{T}_2) $ is the eigenvalue of $\mathbf{T}_2$. Denoting  $f_{\overline{\omega}}^{(k)} = \frac{1}{L} \mathrm{Tr}\mathbf{T}_1( \mathbf{I}_L +  \overline{s}f_\delta^{(k)} \mathbf{T}_1 +  f_{\gamma}^{(k)} \mathbf{T}_1^{\frac{1}{2}} \mathbf{R}_2 \mathbf{T}_1^{\frac{1}{2}} )^{-1}$, we have $- \frac{f_{\overline{\omega}}^{(k)}}{z} \in \mathcal{S} $. Based on the previous argument, we can define an iterative scheme for $f_\delta$ as follows
\begin{equation}
    f_\delta^{(k + 1)} = \frac{1}{L} \Tr \mathbf{R}_1[-z( \mathbf{I}_N -  (\frac{\overline{s} f_{\overline{\omega}}^{(k)}}{z} +\frac{f_{\underline{\omega}}^{(k)}f_{\gamma}^{(k)}}{z (f_\delta^{(k)})^2}) \mathbf{R}_1]^{-1}.
    \label{Iterrative_f_delta}
\end{equation}
By \eqref{prop_S_o_g_d}, it holds true that $f_\delta^{(k + 1)} \in \mathcal{S}$. The convergence analysis of the iterative scheme is quite similar to Proposition \ref{Prop_fixed_point_omega_gamma}, we omit it here.
Denoting $(f_{\delta}, {f_{\overline{\omega}}}, f_{\underline{\omega}},  f_{\gamma})$ the corresponding limits and by taking $(\delta, \overline{\omega}, \underline{\omega}, \gamma)(\overline{s}, z) = 
 (f_{\delta}, {f_{\overline{\omega}}}, \frac{f_{\underline{\omega}}}{f_{\delta}},  \frac{f_{\gamma}}{f_{\delta}})(-z)$, 
 we prove the existence of the solution for \eqref{DE_system_1}. Moreover, treating $\overline{s} \geq 0$ as a constant,
  functions $\delta(\overline{s}, -z)$, $-\frac{1}{z}\overline{\omega}(\overline{s}, -z)$, $-\frac{1}{z}\underline{\omega}(\overline{s}, -z)$, and $-\frac{1}{z}\gamma(\overline{s}, -z)$
   are in class $\mathcal{S}$ and have the integration representations with total measures $\frac{1}{L} \Tr \mathbf{R}_1, \frac{1}{L} \Tr \mathbf{T}_1, \frac{1}{L} \Tr \mathbf{R}_2\mathbf{T}_1$, and $\frac{1}{M} \Tr \mathbf{T}_2$. We summarize the iterative scheme \eqref{Iterrative_f_delta} in \textbf{Algorithm \ref{DE_algorithm}}.
\begin{algorithm}[t]
\caption{Algorithm for obtaining solutions of fundamental equations \eqref{DE_system_1}}
\label{DE_algorithm}
\begin{algorithmic}[1]
\small
\STATE Set iteration indices $t_1, t_2=0$, iteration steps $N_1, N_2$, correlation matrices $\mathbf{R}_i$, $\mathbf{T}_i$, and $ \overline{s} \geq 0, z > 0$.
\STATE Initialize $\delta^{(0)} = \frac{1}{z}$, $\overline{\omega}^{(0)} = 1$, $\underline{\omega}^{(0, N_2)} = 1$, $\gamma^{(0, N_2)} = 1$, and $t_1 = 1$.
\REPEAT
\STATE  Set $\underline{\omega}^{(t_1, 0)} = \underline{\omega}^{(t_1 -1, N_2)}$, $\gamma^{(t_1, 0)} = \gamma^{(t_1 -1, N_2)}$, $t_2 = 1$ .
\REPEAT
\STATE $\underline{\omega}^{(t_1, t_2)} = \frac{1}{L} \Tr \mathbf{R}_2\mathbf{T}_1(\mathbf{I}_L + \overline{s} \delta^{(t_1-1)} \mathbf{T}_1 + \delta^{(t_1-1)}\gamma^{(t_1, t_2 - 1)}  \mathbf{R}_2\mathbf{T}_1 )^{-1}$.
\STATE ${\gamma}^{(t_1, t_2)} = \frac{1}{M} \Tr \mathbf{T}_2(\mathbf{I}_M + \frac{L}{M}\delta^{(t_1-1)}\underline{\omega}^{(t_1, t_2 - 1)}\mathbf{T}_2 )^{-1}$.
\STATE $t_2 = t_2 + 1$.
\UNTIL $t_2 > N_2$
\STATE $\overline{\omega}^{(t_1)} = \frac{1}{L}\Tr \mathbf{T}_1(\mathbf{I}_L + \overline{s} \delta^{(t_1)}\mathbf{T}_1 + \delta^{(t_1)} \gamma^{(t_1, N_2)}\mathbf{R}_2\mathbf{T}_1)^{-1}$.
\STATE $\delta^{(t_1 )} = \frac{1}{L} \Tr \mathbf{R}_1(z \mathbf{I}_N + (\overline{s} \overline{\omega}^{(t_1)} + \gamma^{(t_1, N_2)} \underline{\omega}^{(t_1, N_2)})\mathbf{R}_1)^{-1}$.
\STATE $t_1 = t_1 + 1$.
\UNTIL $t_1 > N_1$.
\STATE \textbf{Output:}  $\delta^{(N_1)}$, $\overline{\omega}^{(N_1)}$, $\underline{\omega}^{(N_1, N_2)}$, ${\gamma}^{(N_1, N_2)}$.
\end{algorithmic}
\end{algorithm}
\subsection{Uniqueness}
 In the previous section, we showed that \textbf{Algorithm} \ref{DE_algorithm} converges to the positive solutions for the initial values $\overline{s} \geq 0$ and $z > 0$.
Here, we will demonstrate that these solutions are unique.
The proof of the uniqueness for the solutions of \eqref{DE_system_1} is quite similar to that of Proposition \ref{Prop_fixed_point_omega_gamma} and we first prove the uniqueness of the scalar solutions for certain $\overline{s} = \overline{s}_0 \geq 0$ and $z = z_0 > 0$. For that purpose, we assume that there exist two positive solutions pairs $(\delta, \overline{\omega}, \underline{\omega}, \gamma)$ and $(\overline{\delta}, \overline{\overline{\omega}}, \overline{\underline{\omega}}, \overline{\gamma})$ satisfying \eqref{DE_system_1}.
Moreover, we define the vectors $\mathbf{u} = (\delta, \overline{\omega}, \underline{\omega}, \gamma, \gamma\underline{\omega}, \delta\gamma, \delta \underline{\omega})^T$, $\overline{\mathbf{u}} = (\overline{\delta}, \overline{\overline{\omega}}, \overline{\underline{\omega}}, \overline{\gamma}, \overline{\gamma\underline{\omega}}, \overline{\delta\gamma}, \overline{\delta} \overline{\underline{\omega}})^T$, and $\boldsymbol{\epsilon} = \mathbf{u} - \overline{\mathbf{u}}$. Similar to \eqref{epsilon_omega_gamma}, we can get $(\mathbf{I}_7 + \mathbf{S}) \boldsymbol{\epsilon} = \mathbf{0}_7$, where the entries of $\mathbf{S}$ can be obtained by using the identity $\mathbf{A} - \mathbf{B} = \mathbf{A}(\mathbf{B}^{-1} - \mathbf{A}^{-1})\mathbf{B}$. The relation $\rho(\mathbf{S}) < 1$ can be proved by the same argument in \eqref{DE_identity}-\eqref{Spectral_radius}. Therefore, it must hold that $\mathbf{u} = \overline{\mathbf{u}}$ when $z > 0$ and $\overline{s} \geq 0$. Secondly, if there exist two function solutions  $(\delta, \overline{\omega}, \underline{\omega}, \gamma)$ and $(\overline{\delta}, \overline{\overline{\omega}}, \overline{\underline{\omega}}, \overline{\gamma})$ satisfying that $(\delta(\overline{s}, -z), -\frac{1}{z}\overline{\omega}(\overline{s}, -z), -\frac{1}{z}\underline{\omega}(\overline{s}, -z), -\frac{1}{z}\gamma(\overline{s}, -z)) \in \mathcal{S}^4$ and $(\overline{\delta}(\overline{s}, -z), -\frac{1}{z}\overline{\overline{\omega}}(\overline{s}, -z), -\frac{1}{z}\overline{\underline{\omega}}(\overline{s}, -z), -\frac{1}{z}\overline{\gamma}(\overline{s}, -z)) \in \mathcal{S}^4$ ($\overline{s}$ is treated as a constant), it holds true that $(\delta, \overline{\omega}, \underline{\omega}, \gamma)(\overline{s}, -z) = (\overline{\delta}, \overline{\overline{\omega}}, \overline{\underline{\omega}}, \overline{\gamma})(\overline{s}, -z)$ when $z \in (-\infty, 0)$. Due to the property of holomorphic functions, they are also equal on the domain $\mathbb{C} - \mathbb{R}^+$. Therefore, we complete the proof of Proposition \ref{Prop_Stieltjes}. \QED
\section{Proof of Proposition \ref{Prop_tightness}} \label{App_Prop_tightness}
In this section, we will prove that the matrix $\mathbf{V}(\overline{s}, \underline{s}, z)$ satisfies the regularity condition \eqref{regularity_cond}. Here, we first show that the diagonal elements of $\mathbf{V}$ are uniformly bounded, i.e.,
\begin{equation}
    0 < m_{i}(\overline{s}, \underline{s}, z) \leq \inf_{N \geq 1}[\mathbf{V}]_{ii} \leq \sup_{N \geq 1} [\mathbf{V}]_{ii} \leq M_{i}(\overline{s}, \underline{s}, z) < + \infty, i = 1, 2. \label{Bound_of_lamda}
\end{equation}
Then, we show that the bounded property also holds for $\det(\mathbf{V})$ as following
\begin{equation}
    0 < m_{\det}(\overline{s}, \underline{s}, z) \leq  \inf_{N \geq 1} \det(\mathbf{V}) \leq \sup_{N \geq 1} \det(\mathbf{V}) \leq M_{\det}(\overline{s}, \underline{s}, z)< + \infty. \label{Bound_of_det}
\end{equation}
Assuming  \eqref{Bound_of_lamda} and \eqref{Bound_of_det} hold, the inequalities $\det(\mathbf{V}) > 0$ and $[\mathbf{V}]_{11} > 0$ imply that $\mathbf{V}$ is positive definite and $\norm{\mathbf{V}}
\leq [\mathbf{V}]_{11} + [\mathbf{V}]_{22}$. Thus we have $\norm{\mathbf{V}}$ is uniformly bounded, 
which means there exists $M_{V} (\overline{s}, \underline{s}, z) > 0$ such that $\sup_N \norm{\mathbf{V}}  \leq M_V(\overline{s}, \underline{s}, z)< +\infty$.
 According to the identity $\lambda_{max}(\mathbf{V})\lambda_{min}(\mathbf{V}) = \det(\mathbf{V})$, the following holds true
\begin{equation}
    \inf_{N \geq 1} \lambda_{min}(\mathbf{V}) = \inf_{N \geq 1} \frac{\det(\mathbf{V})}{\norm{\mathbf{V}}} \geq \frac{m_{\det}(\overline{s}, \underline{s}, z)}{M_V(\overline{s}, \underline{s}, z)}
     = m_{V}(\overline{s}, \underline{s}, z) > 0.
\end{equation}
Therefore, if equations \eqref{Bound_of_lamda} and \eqref{Bound_of_det} hold, then Proposition \ref{Prop_tightness} holds and it is sufficient to prove them.
\subsection{The Boundedness of $[\mathbf{V}]_{ii}$}
In order to prove \eqref{Bound_of_lamda}, we first estimate the upper and lower bouds for each term of the fundamental equations \eqref{DE_system_1} in the following lemma.
\begin{lemma} \label{Lemma_bound_of_DE}
    Under the assumptions \textbf{\ref{A-1}}-\textbf{\ref{A-3}}, for $\overline{s} \geq 0$, $z > 0$, denoting $(\delta, \overline{\omega}, \underline{\omega}, \gamma)$ the positive solutions of \eqref{DE_system_1}, then the following holds 
\begin{equation}
    \delta \in \left[\frac{Nl}{L(z + \overline{s}r^2 + r^3)}, \frac{N r}{L z}\right],  \overline{\omega} \in \left[\frac{l}{1 + \frac{N (\overline{s} r^2 + r^4)}{L z}}, r \right],  \underline{\omega} \in \left[ \frac{l}{1 + \frac{N (\overline{s} r^2 + r^4)}{L z}}, r^2 \right],  \gamma \in \left[ \frac{l}{1 + \frac{Nr^4}{Mz}}, r \right].
\end{equation}
\end{lemma}
\textit{Proof:}  The upper bound of $(\delta, \overline{\omega}, \underline{\omega}, \gamma)$ can be obtained using the inequality 
$\mathrm{Tr}[\mathbf{A}] \leq N\norm{\mathbf{A}}$ for $ \mathbf{A} \in \mathbb{C}^{N \times N}$. Taking $\delta$ as an example, 
the upper bound of $\delta$ can be obtained as follows
\begin{equation}
    \delta = \frac{1}{L} \mathrm{Tr}[\mathbf{R}_1 \mathbf{F}_{\delta}] \leq \frac{N}{L} \norm{\mathbf{R}_1} \norm{(z \mathbf{I}_N + (\overline{s} \overline{\omega} + \gamma \underline{\omega}) \mathbf{R}_1)^{-1}} \leq \frac{N r}{L z}.
\end{equation}
The upper bound of $\overline{\omega}, \underline{\omega}$ and $\gamma$ can be derived similarly.
\par
For the lower bound, we prove the case of $\overline{\omega}$. The other cases are similar and therefore omitted.
To continue, we change the form of $\overline{\omega}$. 
Denoting $\mathbf{W} = \overline{s} \mathbf{T}_1 + \gamma \mathbf{T}_1^{\frac{1}{2}}\mathbf{R}_2\mathbf{T}_1^{\frac{1}{2}}$ and ${\mathbf{F}}_{\omega, H} = (\mathbf{I}_L +  \delta \mathbf{W} )^{-1}$, then ${\mathbf{F}}_{\omega, H}$ is a Hermitian matrix and the following holds
\begin{equation}
    \overline{\omega} = \frac{1}{L} \mathrm{Tr}[\mathbf{T}_1 \mathbf{F}_{\omega}] \overset{(a)}{=} \frac{1}{L} \mathrm{Tr}[\mathbf{T}_1^{\frac{1}{2}} \mathbf{F}_{\omega, H}\mathbf{T}_1^{\frac{1}{2}}]
    =\frac{1}{L} \mathrm{Tr}[\mathbf{T}_1\mathbf{F}_{\omega, H}],
\end{equation}
where $(a)$ follows from the matrix identity $\mathbf{A}(\mathbf{I} + \mathbf{B}\mathbf{A})^{-1} = (\mathbf{I} + \mathbf{A}\mathbf{B})^{-1}\mathbf{A}$.
Next, we introduce the eigenvalue decomposition $\mathbf{W} = \sum_{i=1}^L \lambda_i(\mathbf{W}) \mathbf{v}_i \mathbf{v}_i^H$.  Using this decomposition, we can get
\begin{equation}
    \overline{\omega} = \frac{1}{L} \sum_{i=1}^{L} \frac{ \mathbf{v}_i^H \mathbf{T}_1 \mathbf{v}_i}{1 + \delta \lambda_i(\mathbf{W})} \geq \frac{\sum_{i=1}^{L} 
     \mathbf{v}_i^H \mathbf{T}_1 \mathbf{v}_i}{L(1 + \frac{Nr}{Lz} \norm{\mathbf{W}})} \overset{(a)}{\geq}
       \frac{ \frac{1}{L} \mathrm{Tr}[\mathbf{T}_1 (\sum_{i=1}^{L}\mathbf{v}_i\mathbf{v}_i^H)]}{1
        + \frac{Nr}{Lz} ( \overline{s} r + r^3)} \geq \frac{l}{1 + \frac{N(\overline{s}r^2 + r^4)}{Lz} },
\end{equation}
where $(a)$ follows from $\norm{\mathbf{W}} \leq \overline{s} \norm{\mathbf{T}_1} + \gamma \norm{\mathbf{T}_1^{\frac{1}{2}}\mathbf{R}_2\mathbf{T}_1^{\frac{1}{2}}} \leq \overline{s}r + r^3$. This completes the proof of Lemma \ref{Lemma_bound_of_DE}. \QED 
\par
By assumptions \textbf{\ref{A-1}}-\textbf{\ref{A-3}} , $l > 0$ and $r < + \infty$. So the terms $(\delta, \overline{\omega}, \underline{\omega}, \gamma)$ are 
all bounded away from $0$ uniformly in $N$. Next, we will estimate the lower bound of $\delta_{2, I}$. 
Denoting the eigenvalue decomposition by $\mathbf{R}_1 = \sum_{i=1}^N \lambda_i(\mathbf{R}_1) \mathbf{u}_i \mathbf{u}_i^H$, then
\begin{equation}
    \delta_{2,I} = \frac{1}{L}\mathrm{Tr}[\mathbf{R}_1 \mathbf{F}_{\delta}^2] 
    =\frac{1}{L}\sum_{i=1}^N \frac{\lambda_i(\mathbf{R}_1)}{(z + (\overline{s} \overline{\omega} + \gamma \underline{\omega})\lambda_i(\mathbf{R}_1))^2} 
     \geq \frac{ Nl}{L(z + \overline{s}r^2 + r^3)^2}.
\end{equation}
To this end, $\delta_{2,I}$ is bounded away from $0$ uniformly in $N$. 
By $\frac{1}{L} \mathrm{Tr}[\mathbf{R}_2 \mathbf{T}_1 \mathbf{F}_{\omega}] 
= \frac{1}{L} \mathrm{Tr}[\mathbf{R}_2 \mathbf{T}_1 \mathbf{F}_{\omega}\mathbf{F}_{\omega}^{-1}\mathbf{F}_{\omega}]$, 
we have $\underline{\omega} = \underline{\omega}_{2,I} + \overline{s} \delta \overline{\underline{\omega}}_{1,1} + \delta \gamma \underline{\omega}_2 = B_{\underline{\omega}} + \delta \gamma \underline{\omega}_2$, where 
\begin{equation}
    B_{\underline{\omega}} = \underline{\omega}_{2,I} + \overline{s} \delta \overline{\underline{\omega}}_{1,1} 
    \overset{(a)}{\geq} \underline{\omega}_{2,I} = \frac{1}{L} \mathrm{Tr}[\mathbf{T}_1^{\frac{1}{2}}\mathbf{R}_2 \mathbf{T}_1^{\frac{1}{2}} \mathbf{F}_{\omega, H}^2] 
    = \frac{1}{L} \sum_{i=1}^{L} \frac{ \mathbf{v}_i^H \mathbf{T}_1^{\frac{1}{2}}\mathbf{R}_2 \mathbf{T}_1^{\frac{1}{2}} \mathbf{v}_i}{(1 + \delta \lambda_i(\mathbf{W}))^2} 
    \geq \frac{l}{(1 + \frac{N(\overline{s}r^2 + r^4)}{Lz})^2 },
\end{equation}
where $(a)$ follows from $\overline{\underline{\omega}}_{1,1} = \frac{1}{L} \mathrm{Tr}[\mathbf{T}_1^{\frac{1}{2}}\mathbf{F}_{\omega, H}\mathbf{T}_1^{\frac{1}{2}}\mathbf{R}_2 \mathbf{T}_1^{\frac{1}{2}} \mathbf{F}_{\omega, H}\mathbf{T}_1^{\frac{1}{2}} ] \geq 0$.
Meanwhile, the term $\gamma = \frac{1}{M}\Tr [\mathbf{T}_2 \mathbf{F}_{\gamma}^2] + \frac{L}{M} \delta \underline{\omega} \gamma_2= B_{\gamma} + \frac{L}{M} \delta \underline{\omega} \gamma_2$. The lower bound of $B_{\gamma} \geq \frac{l}{(1 + \frac{Nr^4}{Mz})^2}$ can be obtained in a similar manner. Multiplying the above two equations together we have
\begin{equation}
    \underline{\omega} \gamma = (B_{\underline{\omega}} + \delta \gamma \underline{\omega}_2)(B_{\gamma} + \frac{L}{M} \delta \underline{\omega} \gamma_2) \geq B_{\underline{\omega}}B_{\gamma} + (1 - \Delta) \underline{\omega} \gamma.
\end{equation}
Hence, the following holds
\begin{equation}
    \Delta \geq \frac{B_{\underline{\omega}}B_{\gamma}}{\underline{\omega} \gamma} \geq \frac{l^2}{(1 + \frac{N(\overline{s}r^2 + r^4)}{Lz})^2(1 + \frac{Nr^4}{Mz})^2r^3}.
\end{equation}
Therefore, $\Delta$ is bounded away from $0$ uniformly in $N$. On the other hand, 
\begin{align}
    \underline{\omega}_2 &= \frac{1}{L}\mathrm{Tr}[(\mathbf{R}_2 \mathbf{T}_1 \mathbf{F}_{\omega})^2 ] = \frac{1}{L}\mathrm{Tr}[(\mathbf{R}_2 \mathbf{T}_1^{\frac{1}{2}} \mathbf{F}_{\omega, H}\mathbf{T}_1^{\frac{1}{2}})^2] = \frac{1}{L}\mathrm{Tr}[(\mathbf{F}_{\omega, H}^{\frac{1}{2}}\mathbf{T}_1^{\frac{1}{2}}\mathbf{R}_2 \mathbf{T}_1^{\frac{1}{2}} \mathbf{F}_{\omega, H}^{\frac{1}{2}})^2]\notag \\
    &\overset{(a)}{\geq} (\frac{1}{L}\mathrm{Tr}[\mathbf{F}_{\omega, H}^{\frac{1}{2}}\mathbf{T}_1^{\frac{1}{2}}\mathbf{R}_2 \mathbf{T}_1^{\frac{1}{2}} \mathbf{F}_{\omega, H}^{\frac{1}{2}}])^2 = \underline{\omega}^2,
\end{align}
where $(a)$ follows from the inequality $\mathrm{Tr}[\mathbf{A}^2] \geq \frac{1}{N} (\mathrm{Tr}\mathbf{A})^2$ for Hermitian matrix $\mathbf{A} \in \mathbb{C}^{N \times N}$. Similarly, we have $\gamma_2 \geq \gamma^2$. Hence, the upper bound of $\Delta$ is 
given by $\Delta \leq 1 - \frac{L}{M} \gamma^2 \underline{\omega}^2 \delta^2$.
 From Lemma \ref{Lemma_bound_of_DE} we have $ \underset{N \geq 1}{\inf}(\gamma^2 \underline{\omega}^2 \delta^2) > 0$, so there exists $M_{\Delta_{V_1}}(\overline{s}, \underline{s}, z) < 1$ 
 such that ${\sup}_{N \geq 1} \Delta \leq M_{\Delta_{V_1}}(\overline{s}, \underline{s}, z)$.  Note that this bound is actually independent of $\underline{s}$, but we write it this way for the sake of uniformity.
\par
By the proof in Appendix \ref{App_Prop_Stiltjes}, we have $\delta(\overline{s}, -z) \in \mathcal{S}$, and  $\delta(\overline{s}, z)$ is calculated based on the following integration representation
\begin{equation}
    \delta(\overline{s}, z) = \int \frac{\mu_{\overline{s}, \delta}(\mathrm{d}\lambda)}{\lambda + z},
\end{equation}
where $\mu_{\overline{s}, \delta}(\cdot)$ is a positive measure supported over $\mathbb{R}^+$ and $\mu_{\overline{s}, \delta}(\mathbb{R}^+) = \frac{1}{L} \mathrm{Tr}\mathbf{R}_1 \leq \frac{N r}{L}$. Then, by the dominated convergence theorem, we have 
\begin{equation}
    -\delta' = \int_{\mathbb{R}^+} \frac{\mu_{\overline{s}, \delta}(\mathrm{d}\lambda)}{(\lambda + z)^2} \leq \int_{\mathbb{R}^+} \frac{\mu_{\overline{s}, \delta}(\mathrm{d}\lambda)}{z^2} \leq \frac{N r}{L z^2},
\end{equation}
where the derivative $\delta'$ is taken with respect to $z$. According to the relation \eqref{Diff_EQ_deltas}, we have
\begin{equation}
    \Delta_{V_1} = -\frac{ \delta_{2,I} \Delta}{\delta'} \geq \frac{ z^2l^3}{ (z + \overline{s}r^2 + r^3)^2(1 + \frac{N(\overline{s}r^2 + r^4)}{Lz})^2(1 + \frac{Nr^4}{Mz})^2r^4}.
    \label{lower_bounds_Delta_V1}
\end{equation}
Therefore, there exits $m_{\Delta_{V_1}}(\overline{s}, \underline{s}, z) > 0$ such that ${\inf}_{N \geq 1} \Delta_{V_1} \geq m_{\Delta_{V_1}}(\overline{s}, \underline{s}, z)$. Recalling that $[\mathbf{V}]_{11} = -\log(\Delta_{V_1})$, 
there exist positive numbers $0 < m_{1}(\overline{s}, \underline{s}, z) \leq M_{1}(\overline{s}, \underline{s}, z) < + \infty$ satisfying
\begin{equation}
    0 < m_{1}(\overline{s}, \underline{s}, z) \leq \inf_{N \geq 1}[\mathbf{V}]_{11} \leq \sup_{N \geq 1} [\mathbf{V}]_{11} \leq M_{1}(\overline{s}, \underline{s}, z) < + \infty.
\end{equation}
The lower bound $\Delta_{V_2} \geq \frac{ lz^2}{(\underline{s}r^2 + z)^2} $ can be derived in a similar way thus is omitted here for brevity.
\subsection{The Boundedness of $\det({\mathbf{V}})$}
The upper bound of $\det(\mathbf{V})$ can be obtained by 
\begin{equation}
    \sup_{N \geq 1} \det(\mathbf{V}) \leq \sup_{N \geq 1}  [\mathbf{V}]_{11} \sup_{N \geq 1}[\mathbf{V}]_{22} \leq M_1(\overline{s}, \underline{s}, z)M_2(\overline{s}, \underline{s}, z) := M_{\det}(\overline{s}, \underline{s}, z).
\end{equation}
To derive the lower bound of $\det(\mathbf{V})$ in \eqref{Bound_of_det}, we first introduce a useful inequality in the following lemma.
\begin{lemma} \label{lemma_cauchy}
    If $a$, $b$ $\in (0, 1)$, then $\log(1-a) \log(1-b) \geq [\log(1 - \sqrt{ab})]^2$. 
\end{lemma}
\textit{Proof:} For a given integer $N > 0$, by using Cauchy–Schwarz inequality, we have
\begin{equation}
    \left(\sum_{n=1}^{N} \frac{a^n}{n}\right)\left(\sum_{n=1}^{N} \frac{b^n}{n}\right) \geq \left(\sum_{n=1}^{N} \frac{(ab)^\frac{n}{2}}{n}\right)^2. \label{Taylor_series}
\end{equation}
By taking $N$ to infinity and using Taylor's theorem $\log(1 - x) = - \sum_{n=1}^{+\infty} \frac{x^n}{n}, \forall x \in [-1, 1)$, we complete the proof of Lemma \ref{lemma_cauchy}. \QED
\par
Next, we prove that  $\varsigma \delta_2 < 1$ holds uniformly. By the identity $\Tr [(\overline{s} \mathbf{I}_L + \gamma \mathbf{R}_1)\mathbf{T}_1 \mathbf{F}_{\omega}] = \Tr [(\overline{s} \mathbf{I}_L + \gamma \mathbf{R}_1)\mathbf{T}_1 \mathbf{F}_{\omega}\mathbf{F}_{\omega}^{-1}\mathbf{F}_{\omega}]$, we have
    $ \overline{s} \overline{\omega} + \gamma \underline{\omega} = \overline{s} \frac{1}{L} \Tr \mathbf{T}_1 \mathbf{F}_{\omega}^2 + \gamma \underline{\omega}_{2, I} + \delta \varsigma = B_{\omega} + \delta \varsigma.$
Similarly, we have $\delta = \delta_{2, I} + (\overline{s}  \overline{\omega} + \gamma \underline{\omega}) \delta_2 
= B_{\delta} + (\overline{s}  \overline{\omega} + \gamma \underline{\omega}) \delta_2$. Then, the following holds
\begin{equation}
    (\overline{s}  \overline{\omega} + \gamma \underline{\omega}) \delta = (B_{\omega} + \delta \varsigma)( B_{\delta} 
    + (\overline{s}  \overline{\omega} + \gamma \underline{\omega}) \delta_2) \geq B_{\omega}B_{\delta} +  \delta_2\varsigma \delta(s \overline{\omega} + \gamma \underline{\omega}).
\end{equation}
Thus, we have $1 - \varsigma \delta_2 \geq  \frac{B_{\omega} B_{\delta}}{\overline{s}  \overline{\omega} + \gamma \underline{\omega}} > 0$.
\par
Recall the term $\Delta_C = 1 - \overline{s}  \vartheta (\gamma \underline{\phi} + s \overline{\phi})$ in Table \ref{tabel_of_notations}. First we derive the upper bound of $\vartheta$. By Cauchy–Schwarz inequality, we have
\begin{equation}
    \vartheta = \frac{1}{L} \mathrm{Tr}[\mathbf{R}_1 \mathbf{F}_{\delta} \mathbf{R}_1 \mathbf{G}_{\tau}] \leq \frac{1}{L} \sqrt{\mathrm{Tr}[\mathbf{R}_1 \mathbf{F}_{\delta}(\mathbf{R}_1 \mathbf{F}_{\delta})^H] \mathrm{Tr}[\mathbf{R}_1 \mathbf{G}_{\tau}(\mathbf{R}_1 \mathbf{G}_{\tau})^H]} = \sqrt{\delta_2 \tau_2}.
\end{equation}
The upper bound of $\gamma \underline{\phi} + \overline{s}  \overline{\phi}$ can be derived in a similar manner using Cauchy–Schwarz inequality as follows
\begin{equation}
    \gamma \underline{\phi} + \overline{s} \overline{\phi} = \frac{1}{L} \mathrm{Tr} [(\overline{s} \mathbf{I}_L + \gamma \mathbf{R}_2) \mathbf{T}_1 \mathbf{G}_{\overline{\tau}}\mathbf{T}_1 \mathbf{F}_{\omega}] 
    =\frac{1}{L} \mathrm{Tr} [ (\mathbf{T}_1^{\frac{1}{2}} \mathbf{G}_{\overline{\tau}}\mathbf{T}_1^{\frac{1}{2}})(\mathbf{T}_1^{\frac{1}{2}} \mathbf{F}_{\omega}(\overline{s} \mathbf{I}_L + \gamma \mathbf{R}_2)\mathbf{T}_1^{\frac{1}{2}})] \leq \sqrt{\varsigma \overline{\tau}_2}.
\end{equation}
Since $1 - \varsigma \delta_2 > 0$ and $\Delta_{V_2} > 0$, according to Lemma \ref{lemma_cauchy}, we have
\begin{equation}
    [\log(\Delta_C)]^2 \leq [\log(1 - \sqrt{\varsigma \delta_2 \underline{s}^2 \tau_2 \overline{\tau}_2})]^2 \leq \log(1 - \varsigma \delta_2) \log(\Delta_{V_2}),
\end{equation}
which further gives 
\begin{equation}
\begin{split}
    \det(\mathbf{V}) &= \log(\Delta_{V_1})\log(\Delta_{V_2}) - [\log(\Delta_C)]^2 \geq \log((1 - \varsigma \delta_2)\Delta)\log(\Delta_{V_2}) - [\log(\Delta_C)]^2  \geq \log(\Delta)\log(\Delta_{V_2}).
\end{split}
\end{equation}
The previous analysis shows that $\Delta$ and $\Delta_{V_1}$ are bounded away from $0$ and $1$. As a result, there exists $m_{\det}(\overline{s}, \underline{s}, z) > 0$ such that $\underset{N \geq 1}{\inf} \det(\mathbf{V}) \geq m_{\det}(\overline{s}, \underline{s}, z) > 0$. Thus, we have proved the regularity condition of $\mathbf{V}$. \QED
\section{Proof of Lemma \ref{Lemma_DE_Q1_1}} \label{App_Lemma_DE_Q1_1}
To prove Lemma \ref{Lemma_DE_Q1_1}, we first introduce the main mathematical tools used in the proof. The derivation of Lemma \ref{Lemma_DE_Q1_1} is mainly based on the integral by parts formula and Poincaré-Nash inequality. These two formulas together are called Gaussian tools, which make good use of the Gaussian properties of r.v.s and have wide applications in RMT. In this section, we will use the definitions of random and deterministic terms from Section \ref{Sec_First-order}. For convenience of calculation, we further define 
\begin{equation}
    \overline{\mathbf{\Theta}}_{\omega} = \mathbf{T}_1 \mathbf{\Theta}_{\omega}(\overline{s} \mathbf{I}_L + \alpha_{\gamma} \mathbf{R}_2). \label{overline_Theta_omega}
\end{equation}
Then, there holds by Woodbury matrix identity
\begin{equation}
    {\mathbf{\Theta}}_{\omega}^H = \mathbf{I}_L - \alpha_{\delta} \mathbf{T}_1 \mathbf{\Theta}_{\omega} (\overline{s}\mathbf{I}_L + \alpha_\gamma \mathbf{R}_2) = \mathbf{I}_L - \alpha_{\delta} \overline{\mathbf{\Theta}}_{\omega}.
    \label{Wood_b_Theta_omega}
\end{equation}
\subsection{Mathematical Tools}
\subsubsection{Gaussian Distribution}
For matrix $\mathbf{H} = \mathbf{R}^{\frac{1}{2}} \mathbf{X} \mathbf{T}^{\frac{1}{2}} \in \mathbb{C}^{N \times M}$, where the entries of $\mathbf{X} \in \mathbb{C}^{N \times M}$ are i.i.d r.v.s with distribution $\mathcal{CN}(0, 1)$, then the vector $\mathrm{Vec}(\mathbf{H}) = (\mathbf{T}^{\frac{T}{2}}  \otimes \mathbf{R}^{\frac{1}{2}}) \mathrm{Vec}(\mathbf{X})$ is a Gaussian random vector with zero mean and correlation 
\begin{equation}
    \mathbb{E}
    [[\mathbf{H}]_{ij}[\mathbf{H}]_{ab}^*] = \mathbb{E}[\mathbf{e}_i^T \mathbf{R}^{\frac{1}{2}} \mathbf{X} \mathbf{T}^{\frac{1}{2}} \mathbf{e}_j \mathbf{e}_b^T \mathbf{T}^{\frac{1}{2}}\mathbf{X}^H \mathbf{R}^{\frac{1}{2}} \mathbf{e}_a] = [\mathbf{R}]_{ia} [\mathbf{T}]_{bj}. \label{Gaussian_Corr}
\end{equation}
\subsubsection{Integration by Parts Formula} Let $\mathbf{x} \in \mathbb{C}^N$ be a complex Gaussian random vector with $\mathbb{E} [\mathbf{x}] = \mathbf{0}_N $, $\mathbb{E} [\mathbf{x}\mathbf{x}^T] = \mathbf{0}_{N \times N}$, and $\mathbb{E} [\mathbf{x}\mathbf{x}^H] = \mathbf{\Xi}$. Let $f = f(\mathbf{x}, \mathbf{x}^*)$ be a $\mathcal{C}^1$ complex function polynomial bounded together with
its partial derivatives. Then, the following equality holds true \cite[Eq. (17)]{Hachem2008ANewApproach}, \cite[Eq. (2.17)]{pastur2005simple}
\begin{equation}
    \mathbb{E}\left[[\mathbf{x}]_i f(\mathbf{x}, \mathbf{x}^*)\right] = \sum_{j=1}^N [\mathbf{\Xi}]_{ij} \mathbb{E} \left[\frac{\partial f(\mathbf{x}, \mathbf{x}^*)}{\partial{[\mathbf{x}]_j^*}}\right], \label{Integration_by_parts}
\end{equation}
where the differential operators are defined as $\frac{\partial}{\partial z} = \frac{1}{2} \left( \frac{\partial}{\partial \Re(z) } - \jmath \frac{\partial}{\partial \Im(z)}\right)$ and $\frac{\partial}{\partial z^*} = \frac{1}{2} \left( \frac{\partial}{\partial \Re(z) } + \jmath \frac{\partial}{\partial \Im(z)}\right)$.
\par
\subsubsection{The Poincaré-Nash Inequality} With $\mathbf{x}$ and $f$ given
above, we have \cite[Eq. (18)]{Hachem2008ANewApproach}, \cite[Proposition 2.5]{pastur2005simple}
\begin{equation}
    \mathrm{Var}[f(\mathbf{x}, {\mathbf{x}}^*)] \leq \mathbb{E}\Big[\nabla_{\mathbf{x}} f(\mathbf{x}, {\mathbf{x}^*})^T \mathbf{\Xi} (\nabla_{\mathbf{x}} f(\mathbf{x}, \mathbf{x}^*))^* \Big] +  \mathbb{E}\Big[(\nabla_{{\mathbf{x}}^*}f(\mathbf{x}, {\mathbf{x}^*}))^H \mathbf{\Xi} \nabla_{\mathbf{x}^*} f(\mathbf{x}, {\mathbf{x}^*}) \Big],
\label{Poincare_Nash_ineq}
\end{equation}
where $\nabla_\mathbf{x}f = [\frac{\partial f}{\partial [\mathbf{x}]_1}, \ldots, \frac{\partial f}{\partial [\mathbf{x}]_N}]^T $ and $\nabla_{\mathbf{x}^*}f = [\frac{\partial f}{\partial [\mathbf{x}]_1^*}, \ldots, \frac{\partial f}{\partial [\mathbf{x}]_N^*}]^T $. Assuming the variables of $f$ are the matrices $\mathbf{H}_1$, $\mathbf{H}_2$, $\mathbf{H}_1^H$, and  $\mathbf{H}_2^H$ defined in \eqref{Eq_Channel_Model}, the upper bound of the variance can be calculated as
\begin{equation}
\begin{split}
     \mathrm{Var}(f) &\leq \mathbb{E} \left\{ \sum_{i,j,a,b} \frac{1}{L}[\frac{\partial f}{\partial \mathbf{H}_1}]_{ij} [\mathbf{R}_1]_{ia}[\mathbf{T}_1]_{bj}[(\frac{\partial f}{\partial \mathbf{H}_1})^H]_{ba} +   \sum_{i,j,a,b} \frac{1}{M}[\frac{\partial f}{\partial \mathbf{H}_2}]_{ij} [\mathbf{R}_2]_{ia}[\mathbf{T}_2]_{bj}[(\frac{\partial f}{\partial \mathbf{H}_2})^H]_{ba} \right\}  \\
    & + \mathbb{E} \left\{ \sum_{i,j,a,b} \frac{1}{L}[(\frac{\partial f}{\partial \mathbf{H}_1^H})^H]_{ij} [\mathbf{R}_1]_{ia}[\mathbf{T}_1]_{bj}[\frac{\partial f}{\partial \mathbf{H}_1^H}]_{ba} +   \sum_{i,j,a,b} \frac{1}{M}[(\frac{\partial f}{\partial \mathbf{H}_2^H})^H]_{ij} [\mathbf{R}_2]_{ia}[\mathbf{T}_2]_{bj}[\frac{\partial f}{\partial \mathbf{H}_2^H}]_{ba} \right\}  \\
    & \leq \mathbb{E} \left\{ \frac{1}{L} \mathrm{Tr}\left[\mathbf{R}_1^T\frac{\partial f}{\partial \mathbf{H}_1}\mathbf{T}_1^T(\frac{\partial f}{\partial \mathbf{H}_1})^H\right] + \frac{1}{M} \mathrm{Tr}\left[\mathbf{R}_2^T\frac{\partial f}{\partial \mathbf{H}_2}\mathbf{T}_2^T(\frac{\partial f}{\partial \mathbf{H}_2})^H\right]\right\} \\
    & + \mathbb{E} \left\{ \frac{1}{L} \mathrm{Tr}\left[\mathbf{R}_1^T(\frac{\partial f}{\partial \mathbf{H}_1^H})^H\mathbf{T}_1^T\frac{\partial f}{\partial \mathbf{H}_1^H}\right] + \frac{1}{M} \mathrm{Tr}\left[\mathbf{R}_2^T (\frac{\partial f}{\partial \mathbf{H}_2^H})^H \mathbf{T}_2^T \frac{\partial f}{\partial \mathbf{H}_2^H}\right]\right\}. 
    \label{Variance_f} 
\end{split}
\end{equation}
\subsubsection{The Boundedness of Empirical Spectral Norm} Assume that the entries of random matrix $\mathbf{X} \in \mathbb{C}^{N \times M}$ are i.i.d. zero mean and unit variance Gaussian r.v.s. Moreover, $N$ and $M$ grow in the same rate, i.e., $0 < \liminf_{N}  \frac{N}{M} \leq \limsup_{N } \frac{N}{M} < + \infty$. For $p \geq 0$, there exists a constant $K_p < + \infty$ that is independent of $N$ and $M$ such that \cite[Lemma 2]{CLTMVDRRubio2012}
\begin{equation}
    \sup_{N \geq 1} \mathbb{E}\Big[ \norm{\frac{\mathbf{X}}{\sqrt{N}}}^{p} \Big] \leq K_{p}. \label{Bound_norm_}
\end{equation}
For the channel matrices defined in \eqref{Eq_Channel_Model} and given numbers $p_1$, $p_2 > 0$, we have the following result
\begin{equation}
    \sup_{N \geq 1} \mathbb{E}  \norm{\mathbf{H}_1}^{p_1} \norm{\mathbf{H}_2}^{p_2}  =  \sup_{N \geq 1} \mathbb{E}  \norm{\mathbf{R}_1^{\frac{1}{2}}\mathbf{X}_1\mathbf{T}_1^{\frac{1}{2}}}^{p_1} \norm{\mathbf{R}_2^{\frac{1}{2}}\mathbf{X}_2\mathbf{T}_2^{\frac{1}{2}}}^{p_2}  \overset{(a)}{\leq} r^{p_1 + p_2} \sup_{N \geq 1}  \sqrt{\mathbb{E}\norm{\mathbf{X}_1}^{2p_1} \mathbb{E}\norm{\mathbf{X}_1}^{2p_2}} < +\infty, \label{Bounded_norm}
\end{equation}
where $(a)$ follows from the sub-multiplicative property of the spectral norm and Cauchy–Schwarz inequality. 
\subsubsection{Useful Matrix Derivatives}
Assume $\mathbf{A}$ and $\mathbf{B}$ are deterministic matrices with appropriate dimensions. By matrix derivative formula $\partial \mathbf{X}^{-1} = - \mathbf{X}^{-1}(\partial \mathbf{X}) \mathbf{X}^{-1}$, we have
\begin{subequations}
\label{diff_Q_1_H12}
\begin{align}
\frac{\partial{[\mathbf{A}\mathbf{Q}_1\mathbf{B}]_{ij}}}{\partial{{[\mathbf{H}_1]_{kl}^*}}} 
&= -[\mathbf{A}\mathbf{Q}_1\mathbf{H}_1 \mathbf{H}_2 \mathbf{H}_2^H]_{il}[\mathbf{Q}_1\mathbf{B}]_{kj}- \overline{s} [\mathbf{A}\mathbf{Q}_1\mathbf{H}_1]_{il}[\mathbf{Q}_1\mathbf{B}]_{kj}, \label{diff_Q_1_H1}\\
\frac{\partial{[\mathbf{A}\mathbf{Q}_1\mathbf{B}]_{ij}}}{\partial[\mathbf{H}_2]_{kl}^*} &= -[\mathbf{A}\mathbf{Q}_1\mathbf{H}_1 \mathbf{H}_2]_{il}[\mathbf{H}_1^H \mathbf{Q}_1\mathbf{B}]_{kj}.\label{diff_Q_1_H2}%
\end{align}
\end{subequations}
where $\mathbf{Q}_1$ is defined in \eqref{Res_Q1}.
\subsection{Proof of Lemma \ref{Lemma_DE_Q1_1}}
In this section, we will utilize Gaussian tools to show that the term $\mathbb{E}\mathbf{Q}_1$ can be approximated by $\mathbf{\Theta}_{\delta}$.  Before proving the theorem, we first use Gaussian tools to give the following property of variance control, which will be used in estimating the error.
\begin{proposition} \label{Prop_variance}
    Assume that $\mathbf{H}_1$, $\mathbf{H}_2$ are defined in \eqref{Eq_Channel_Model} and assumptions \textbf{\ref{A-1}} - \textbf{\ref{A-3}} holds. Given the deterministic matrices $\mathbf{A}, \mathbf{B}, \mathbf{C}$, $ \mathbf{D}$ with uniformly bounded spectral norm, i.e.,  $\sup_{N \geq 1} \{ \norm{\mathbf{A}},\norm{\mathbf{B}}, \norm{\mathbf{C}}, \norm{\mathbf{D}} \} < + \infty$, and deterministic vectors $\mathbf{a}, \mathbf{b}$ with uniformly bounded Euclidean norm, i.e., $\sup_{N \geq 1} \{ \norm{\mathbf{a}},\norm{\mathbf{b}}\} < + \infty$, then there holds for the trace forms
\begin{subequations}
\label{Prop_variance_3_Trace}
\begin{align}
     &\mathrm{Var}[\mathrm{Tr}\mathbf{A} \mathbf{Q}_1] = \mathcal{O}_z^s(\frac{1}{z^2}),  \label{Prop_variance_3_Trace_1} \\
      &\mathrm{Var}[{\mathrm{Tr}} \mathbf{A} \mathbf{Q}_1 \mathbf{H}_1 \mathbf{B} \mathbf{H}_1^H] = \mathcal{O}_z^s(1), \label{Prop_variance_3_Trace_2} \\
       &\mathrm{Var}[\mathrm{Tr} \mathbf{A} \mathbf{Q}_1 \mathbf{H}_1 \mathbf{B} \mathbf{H}_2 \mathbf{C} \mathbf{H}_2^H \mathbf{D} \mathbf{H}_1^H] = \mathcal{O}_z^s(1), \label{Prop_variance_3_Trace_3} 
\end{align}
\end{subequations}
and for the bilinear forms
\begin{subequations}
    \label{Prop_variance_3_bilinear}
    \begin{align}
    &\mathrm{Var}[\mathbf{a}^H \mathbf{Q}_1 \mathbf{H}_1 \mathbf{B} \mathbf{H}_1^H \mathbf{b}] = \mathcal{O}_z^s(\frac{1}{N}),   \label{Prop_variance_3_bilinear_1}\\
    &\mathrm{Var}[\mathbf{a}^H \mathbf{Q}_1 \mathbf{H}_1 \mathbf{B} \mathbf{H}_2 \mathbf{C} \mathbf{H}_2^H \mathbf{D} \mathbf{H}_1^H \mathbf{b}] 
    = \mathcal{O}_z^s(\frac{1}{N}).
    \end{align}
\end{subequations}
\end{proposition}
\textit{Proof:} The proof of Proposition \ref{Prop_variance} is given in Appendix \ref{App_Prop_variance}. \QED
\par
Using the mathematical tools listed above and the variance control proposition, we now prove Lemma \ref{Lemma_DE_Q1_1}. We first study the entries of $\mathbb{E} \mathbf{Q}_1$. According to the resolvent identity \eqref{Resolvent_identity_Q1}, we have $ z [\mathbf{Q}_1]_{ij} = [\mathbf{I}_N]_{ij}-[\mathbf{Q}_1\mathbf{H}_1 \mathbf{H}_2\mathbf{H}_2^H\mathbf{H}_1^H]_{ij} - \overline{s} [\mathbf{Q}_1\mathbf{H}_1\mathbf{H}_1^H]_{ij}$. We will use Gaussian tools to estimate $[\mathbf{Q}_1\mathbf{H}_1 \mathbf{H}_2\mathbf{H}_2^H\mathbf{H}_1^H]_{ij}$ and $[\mathbf{Q}_1\mathbf{H}_1\mathbf{H}_1^H]_{ij}$, respectively.
According to \eqref{Gaussian_Corr}, we know that for random matrices $\mathbf{H}_1$ and $\mathbf{H}_2$ defined in \eqref{Eq_Channel_Model}, 
the joint distribution of their entries is Gaussian. 
Thus, we apply the integration by parts formula \eqref{Integration_by_parts} on variable $[\mathbf{H}_2]_{kl}$ and  \eqref{diff_Q_1_H2} to obtain
\begin{equation}
\begin{split}
    & \sum_{k}\mathbb{E}[\mathbf{Q}_1 \mathbf{H}_1]_{ik}[\mathbf{H}_2]_{kl}[\mathbf{H}_2]_{ms}^* [\mathbf{H}_1]_{jp}^* 
    =  \sum_{k, a, b} \frac{[\mathbf{R}_2]_{ka} [\mathbf{T}_2]_{bl}}{M} \mathbb{E}\frac{\partial[\mathbf{Q}_1 \mathbf{H}_1]_{ik} [\mathbf{H}_2]_{ms}^*[\mathbf{H}_1]_{jp}^*}{\partial  [\mathbf{H}_2]_{ab}^*} \\
    & =  \sum_{k, a, b} \frac{[\mathbf{R}_2]_{ka} [\mathbf{T}_2]_{bl}}{M} \mathbb{E}\Big\{ -[\mathbf{Q}_1\mathbf{H}_1 \mathbf{H}_2]_{ib}[\mathbf{H}_1^H \mathbf{Q}_1\mathbf{H}_1]_{ak}[\mathbf{H}_2]_{ms}^*[\mathbf{H}_1]_{jp}^* +   [\mathbf{Q}_1 \mathbf{H}_1]_{ik}\delta(m-a) \delta(s-b) [\mathbf{H}_1]_{jp}^* \Big\} . 
\end{split} \label{App_Lemma_DE_Q1_1_eq_InteByPart}
\end{equation}
Consequently, the following is obtained by \eqref{App_Lemma_DE_Q1_1_eq_InteByPart}
\begin{equation}
\mathbb{E}[\mathbf{Q}_1 \mathbf{H}_1 \mathbf{H}_2\mathbf{\Theta}_{\gamma}^{-1}]_{il}[\mathbf{H}_2]_{ms}^*[\mathbf{H}_1]_{jp}^*  \overset{(a)}{=}  \mathbb{E}\Big\{ - \mathring{\kappa}[\mathbf{Q}_1\mathbf{H}_1 \mathbf{H}_2 \mathbf{T}_2]_{il}[\mathbf{H}_2]_{ms}^*[\mathbf{H}_1]_{jp}^* + \frac{[\mathbf{T}_2]_{sl}}{M} [\mathbf{Q}_1 \mathbf{H}_1 \mathbf{R}_2]_{im} [\mathbf{H}_1]_{jp}^*\Big\}, \label{APP_DE_Q1_eq1} 
\end{equation}
where $(a)$ follows from $\widehat{\kappa} = \kappa + \mathring{\kappa}$. By multiplying $[\mathbf{\Theta}_{\gamma}]_{ls}$, $[\mathbf{T}_1]_{mq}$ on both sides of \eqref{APP_DE_Q1_eq1} and summing over the subscripts $l, s, m$, and $ q$, the following holds
\begin{equation}
\mathbb{E}[\mathbf{Q}_1 \mathbf{H}_1 \mathbf{H}_2\mathbf{H}_2^H \mathbf{T}_1]_{iq}[\mathbf{H}_1]_{jp}^* = -\mathbb{E} \mathring{\kappa}[\mathbf{Q}_1\mathbf{H}_1 \mathbf{H}_2 \mathbf{T}_2\mathbf{\Theta}_{\gamma}\mathbf{H}_2^H\mathbf{T}_1]_{iq}[\mathbf{H}_1]_{jp}^* + \alpha_{\gamma} \mathbb{E}[\mathbf{Q}_1 \mathbf{H}_1 \mathbf{R}_2 \mathbf{T}_1]_{iq} [\mathbf{H}_1]_{jp}^*. \label{App_DE_Q1_eq1_sum_1} 
\end{equation}
By the similar calculation in \eqref{App_Lemma_DE_Q1_1_eq_InteByPart} and the integration by parts formula, the evaluation for $\mathbb{E}  [\mathbf{Q}_1 \mathbf{H}_1]_{iq} [\mathbf{H}_1]_{jp}^*$ is given by
\begin{equation}
    \mathbb{E}  [\mathbf{Q}_1 \mathbf{H}_1]_{iq} [\mathbf{H}_1]_{jp}^* = \mathbb{E}\Big\{ -\widehat{\alpha}_{\delta} [\mathbf{Q}_1\mathbf{H}_1 \mathbf{H}_2 \mathbf{H}_2^H \mathbf{T}_1]_{iq}[\mathbf{H}_1]_{jp}^*  - \overline{s} \widehat{\alpha}_{\delta} [\mathbf{Q}_1\mathbf{H}_1\mathbf{T}_1]_{iq}[\mathbf{H}_1]_{jp}^* + \frac{[\mathbf{T}_1]_{pq}}{L}[\mathbf{Q}_1\mathbf{R}_1]_{ij} \Big\} . \label{App_DE_Q1_eq2}
\end{equation}
Multiplying both sides of \eqref{App_DE_Q1_eq1_sum_1} by $- \alpha_{\delta}$ and then adding it to \eqref{App_DE_Q1_eq2}, the following can be obtained by writing $\widehat{\alpha}_{\delta} = \alpha_{\delta} + \mathring{\alpha}_{\delta}$
\begin{align}
    & \mathbb{E} [\mathbf{Q}_1 \mathbf{H}_1 \mathbf{\Theta}_{\omega}^{-1}]_{iq} [\mathbf{H}_1]_{jp}^* = \mathbb{E} \Big\{ -\mathring{\alpha_{\delta}}[\mathbf{Q}_1\mathbf{H}_1 \mathbf{H}_2 \mathbf{H}_2^H \mathbf{T}_1]_{iq}[\mathbf{H}_1]_{jp}^* - \overline{s}\mathring{\alpha}_{\delta}[\mathbf{Q}_1\mathbf{H}_1\mathbf{T}_1]_{iq}[\mathbf{H}_1]_{jp}^*  \notag \\
     &+ \alpha_{\delta}  \mathring{\kappa}[\mathbf{Q}_1\mathbf{H}_1 \mathbf{H}_2 \mathbf{T}_2\mathbf{\Theta}_{\gamma}\mathbf{H}_2^H \mathbf{T}_1]_{iq}[\mathbf{H}_1]_{jp}^*  + \frac{[\mathbf{T}_1]_{pq}}{L}[\mathbf{Q}_1\mathbf{R}_1]_{ij} \Big\}. \label{App_DE_Q1_eq3}
\end{align}
Multiply both sides of \eqref{App_DE_Q1_eq3} by $[\mathbf{\Theta}_{\omega}(\overline{s} \mathbf{I}_L + \alpha_{\gamma}\mathbf{R}_2)]_{qp}$ and sum over the subscripts $q$ and $ p$. Multiply $[\mathbf{\Theta}_{\gamma}]_{ls}$ and $\delta(m-p)$ on both sides of \eqref{APP_DE_Q1_eq1} and sum over the subscripts $l, s, m$ and $ p$. Then, we add the two results together and use \eqref{Wood_b_Theta_omega} to obtain the following
\begin{equation}
\label{Eq_DE_pre_resolve_E_delta_I}
\begin{split}
     & \mathbb{E}\Big\{[\mathbf{Q}_1 \mathbf{H}_1 \mathbf{H}_2\mathbf{H}_2^H\mathbf{H}_1^H]_{ij} + \overline{s}  [\mathbf{Q}_1 \mathbf{H}_1\mathbf{H}_1^H]_{ij}\Big\} =  (\overline{s}  \alpha_{\overline{\omega}} + \alpha_\gamma \alpha_{\underline{\omega}} )[\mathbb{E}\mathbf{Q}_1\mathbf{R}_1]_{ij} + \mathbb{E} \Big\{ - \overline{s}\mathring{\alpha_{\delta}}[\mathbf{Q}_1\mathbf{H}_1\overline{\mathbf{\Theta}}_{\omega}\mathbf{H}_1^H]_{ij}  \\
     & - \mathring{\kappa}[\mathbf{Q}_1\mathbf{H}_1 \mathbf{H}_2 \mathbf{T}_2\mathbf{\Theta}_{\gamma}\mathbf{H}_2^H \mathbf{\Theta}_{\omega}^H \mathbf{H}_1^H]_{ij}  - \mathring{\alpha_{\delta}}[\mathbf{Q}_1\mathbf{H}_1 \mathbf{H}_2 \mathbf{H}_2^H \overline{\mathbf{\Theta}}_{\omega} \mathbf{H}_1^H]_{ij} \Big\} =  (\overline{s}\alpha_{\overline{\omega}} + \alpha_\gamma \alpha_{\underline{\omega}} )[\mathbb{E}\mathbf{Q}_1\mathbf{R}_1]_{ij} - [\boldsymbol{\mathcal{E}}_{\delta, I}]_{ij}. 
\end{split}
\end{equation}
By using the resolvent identity of $\mathbf{Q}_1$, i.e., \eqref{Resolvent_identity_Q1} at the LHS of the first line of the above equation, we obtain
\begin{equation}
   \mathbb{E}[\mathbf{Q}_1 \mathbf{\Theta}_{\delta}^{-1}]_{ij} = \delta(i-j) + [\boldsymbol{\mathcal{E}}_{\delta, I}]_{ij}. \label{Q_1_E}
\end{equation}
By \eqref{Approx_of_Q_1_eq1}, we have $\boldsymbol{\mathcal{E}}_{\delta} = \boldsymbol{\mathcal{E}}_{\delta, I}\mathbf{\Theta}_{\delta}$. Each entry of $\boldsymbol{\mathcal{E}}_{\delta}$  can be calculated as
\begin{equation}
\begin{split}
    & [\boldsymbol{\mathcal{E}}_{\delta}]_{ij} = \mathbb{E}\{ \overline{s} \mathring{\alpha_{\delta}}\mathbf{e}_i^T\mathbf{Q}_1\mathbf{H}_1\overline{\mathbf{\Theta}}_{\omega}\mathbf{H}_1^H\mathbf{\Theta}_{\delta}\mathbf{e}_j + \mathring{\kappa}\mathbf{e}_i^T\mathbf{Q}_1\mathbf{H}_1 \mathbf{H}_2 \mathbf{T}_2\mathbf{\Theta}_{\gamma}\mathbf{H}_2^H \mathbf{\Theta}_{\omega}^H \mathbf{H}_1^H\mathbf{\Theta}_{\delta}\mathbf{e}_j \notag \\
    & +\mathring{\alpha_{\delta}}\mathbf{e}_i^T\mathbf{Q}_1\mathbf{H}_1 \mathbf{H}_2 \mathbf{H}_2^H \overline{\mathbf{\Theta}}_{\omega} \mathbf{H}_1^H\mathbf{\Theta}_{\delta}\mathbf{e}_j \} := [\boldsymbol{\mathcal{E}}_{\delta, 1} + \boldsymbol{\mathcal{E}}_{\delta, 2} + \boldsymbol{\mathcal{E}}_{\delta, 3}]_{ij}.
\end{split}
\end{equation}
Next, we will estimate the order of $[\boldsymbol{\mathcal{E}}_{\delta, k}]_{ij}$, $k=1,2,3$. Based on the variance control in Proposition \ref{Prop_variance}, we can obtain the variance of $\widehat{\alpha}_{\delta}$ is of order $\mathcal{O}(\frac{1}{N^2z^4})$. Noting that $\norm{\mathbf{\Theta}_{\delta}} \leq \frac{1}{z}$ and using Cauchy–Schwarz inequality, we have
\begin{equation}
    \sup_{i,j} |[\boldsymbol{\mathcal{E}}_{\delta, 1}]_{ij}| \leq  \frac{\overline{s}}{z} \sqrt{\Var(\mathring{\alpha}_{\delta})\sup_{ij} \left\{\Var[z\mathbf{e}_i^T\mathbf{Q}_1\mathbf{H}_1\overline{\mathbf{\Theta}}_{\omega}\mathbf{H}_1^H\mathbf{\Theta}_{\delta}\mathbf{e}_j]\right\} } = \mathcal{O}_z^s({N^{-\frac{3}{2}}z^{-2}}). 
    \label{Bound_of_E_delta_1}
\end{equation}
Using the same approach as in \eqref{Bound_of_E_delta_1}, we obtain $\sup_{i,j}|[\boldsymbol{\mathcal{E}}_{\delta, 2}]_{ij}| = \mathcal{O}_z^s({N^{-\frac{3}{2}}z^{-1}})$ and $\sup_{i, j}|[\boldsymbol{\mathcal{E}}_{\delta, 3}]_{ij}| = \mathcal{O}_z^s({N^{-\frac{3}{2}}z^{-2}})$.
 Therefore, we have proven that the order of magnitude of the elements in matrix $\boldsymbol{\mathcal{E}}_{\delta}$ is $\mathcal{O}_z^s({N^{-\frac{3}{2}}}z^{-1})$.
\par
By multiplying $[\mathbf{\Theta}_{\delta}\mathbf{A}]_{ji}$, for any deterministic matrix $\mathbf{A}$ with bounded spectral norm, on both sides of \eqref{Q_1_E} and summing over the subscripts $i$ and $j$, we have
\begin{equation}
   \mathrm{Tr}\mathbf{A}\mathbb{E}\mathbf{Q}_1 = \mathrm{Tr}\mathbf{A}\mathbf{\Theta}_{\delta} + \Tr (\mathbf{A}\boldsymbol{\mathcal{E}}_{\delta}) = \mathrm{Tr}\mathbf{A}\mathbf{\Theta}_{\delta} + \varepsilon_{\delta}(\mathbf{A}),
\end{equation}
where 
\begin{equation}
    \begin{split}
    & \varepsilon_{\delta}(\mathbf{A}) = \mathbb{E} \mathring{\kappa}\mathrm{Tr}[\mathbf{\Theta}_{\delta}\mathbf{A}\mathbf{Q}_1\mathbf{H}_1 \mathbf{H}_2 \mathbf{T}_2\mathbf{\Theta}_{\gamma}\mathbf{H}_2^H \mathbf{\Theta}_{\omega}^H \mathbf{H}_1^H]  + \mathbb{E} \mathring{\alpha_{\delta}}\mathrm{Tr}[\mathbf{\Theta}_{\delta}\mathbf{A}\mathbf{Q}_1\mathbf{H}_1 \mathbf{H}_2 \mathbf{H}_2^H \overline{\mathbf{\Theta}}_{\omega}\mathbf{H}_1^H] \notag \\
    & + \overline{s}  \mathbb{E} \mathring{\alpha_{\delta}}\mathrm{Tr}[\mathbf{\Theta}_{\delta}\mathbf{A}\mathbf{Q}_1\mathbf{H}_1\overline{\mathbf{\Theta}}_{\omega}\mathbf{H}_1^H]  :=  \varepsilon_{\delta,1}(\mathbf{A})+\varepsilon_{\delta,2}(\mathbf{A})+\varepsilon_{\delta,3}(\mathbf{A}).
\end{split}
\end{equation}
Next, we will estimate $|\varepsilon_{\delta, k}(\mathbf{A})|$, $k=1,2,3$. 
By Cauchy–Schwarz inequality, the boundedness property of the norm $\norm{\mathbf{\Theta_{\delta}}} \leq \frac{1}{z}$, and the Proposition \ref{Prop_variance}, the following upper bound holds true
\begin{equation}
    |\varepsilon_{\delta,1}(\mathbf{A})| \leq \frac{1}{Nz}\sqrt{\Var(N \widehat\kappa) \Var [\mathrm{Tr}z\mathbf{\Theta}_{\delta}\mathbf{A}\mathbf{Q}_1\mathbf{H}_1 \mathbf{H}_2 \mathbf{T}_2\mathbf{\Theta}_{\gamma}\mathbf{H}_2^H \mathbf{\Theta}_{\omega} \mathbf{H}_1^H]} = \mathcal{O}_z^s(\frac{1}{Nz}).
\label{Bound_error_Tr_AQ_1}
\end{equation}
Similar to the evaluation in \eqref{Bound_error_Tr_AQ_1}, we can obtain that $|\varepsilon_{\delta,2}(\mathbf{A})| = \mathcal{O}_z^s(\frac{1}{Nz^2})$ and $|\varepsilon_{\delta,3}(\mathbf{A})| = \mathcal{O}_z^s(\frac{1}{Nz^2})$. Consequently, $ \Tr (\mathbf{A}\boldsymbol{\mathcal{E}}_{\delta}) = \varepsilon_{\delta}(\mathbf{A}) = \mathcal{O}_z^s(\frac{1}{Nz})$.  Thus, we complete the proof of Lemma \ref{Lemma_DE_Q1_1}. \QED
\section{Proof of Proposition \ref{Prop_variance}}
\label{App_Prop_variance}
We will prove the bounds for the variances of trace forms and bilinear forms, respectively, by utilizing the Poincaré-Nash inequality \eqref{Poincare_Nash_ineq}. Before we start, we introduce two useful inequalities in the following lemma.
\begin{lemma}
\label{Lemma_Matrix_Norm_Trace_ineq}
For matrix sequence $\{\mathbf{A}_k\}_{k=1}^K$, matrix $\mathbf{C}$ and Hermitian non-negative matrix $\mathbf{D}$ with appropriate dimensions, there hold
\begin{align}
    \norm{\sum_{k=1}^K \mathbf{A}_k}^2 &\leq K \sum_{k=1}^K \norm{\mathbf{A}_k}^2, \label{Matrix_norm_ineq} \\
    \Tr\left[\mathbf{C}\left( \sum_{k=1}^K \mathbf{A}_k  \right)\mathbf{D}\left( \sum_{k=1}^K \mathbf{A}_k^H  \right)\mathbf{C}^H\right]& \leq K \sum_{k=1}^K \Tr\left[\mathbf{C}\mathbf{A}_k\mathbf{D}\mathbf{A}_k^H\mathbf{C}^H\right]. \label{Matrix_Trace_ineq} 
\end{align}
\end{lemma}
\textit{Proof:} According to the subadditivity property of the norm, we have
\begin{equation}
    \norm{\sum_{k=1}^K \mathbf{A}_k}^2 \leq \left( \sum_{k=1}^K \norm{\mathbf{A}_k} \right)^2 \leq K \sum_{k=1}^K \norm{\mathbf{A}_k}^2,
\end{equation}
which proves \eqref{Matrix_norm_ineq}. Defining $\mathbf{M}_k = \mathbf{C} \mathbf{A}_k  \mathbf{D}^{\frac{1}{2}}$, then the LHS of \eqref{Matrix_Trace_ineq} is equal to $\Tr (\sum_{k=1}^K\mathbf{M}_k) ( \sum_{k=1}^K\mathbf{M}_k )^H$ and 
\begin{equation}
 \Tr \left(\sum_{k=1}^K\mathbf{M}_k\right) \left( \sum_{k=1}^K\mathbf{M}_k \right)^H =  \norm{\sum_{k=1}^K\mathbf{M}_k}_F^2 \leq \left( \sum_{k=1}^K \norm{\mathbf{M}_k}_F \right)^2 \leq  K \sum_{k=1}^K \norm{\mathbf{M}_k}^2_F = K \sum_{k=1}^K \Tr\left[\mathbf{C}\mathbf{A}_k\mathbf{D}\mathbf{A}_k^H\mathbf{C}^H\right],
\end{equation}
where $\norm{\cdot}_F$ is the Frobenius norm. Hence, Lemma \ref{Lemma_Matrix_Norm_Trace_ineq} is proved. \QED
\subsection{Variance Control for Trace Forms}
By using inequality $|\Tr \mathbf{A}| \leq N \norm{\mathbf{A}}$ for $\mathbf{A} \in \mathbb{C}^{N \times N}$ and the identity $\norm{\mathbf{A}} = \norm{\mathbf{A}^H} = \norm{\mathbf{A}^T} $, 
we can continue to bound \eqref{Variance_f} as
 \begin{align}
    \mathrm{Var}(f) \leq \frac{N r^2}{L} \mathbb{E}  \norm {\frac{\partial f}{\partial \mathbf{H}_1}}^2 + \frac{N r^2}{L} \mathbb{E}  \norm {\frac{\partial f}{\partial \mathbf{H}_1^H}}^2  + \frac{ L r^2}{M} \mathbb{E}\norm{ \frac{\partial f}{\partial \mathbf{H}_2}}^2 + \frac{ L r^2}{M} \mathbb{E}\norm{ \frac{\partial f}{\partial \mathbf{H}_2^H}}^2. 
    \label{variance_f_Tr}
 \end{align}
 Next, we start to prove equation \eqref{Prop_variance_3_Trace}.  Define $f_{\Tr}(\mathbf{H}_1, \mathbf{H}_2, \mathbf{H}_1^H, \mathbf{H}_2^H) = \mathrm{Tr}[\mathbf{A} \mathbf{Q}_1 \mathbf{H}_1 \mathbf{B} \mathbf{H}_1^H]$. By taking the derivative and using formula $\partial \mathbf{X}^{-1} = - \mathbf{X}^{-1}(\partial \mathbf{X}) \mathbf{X}^{-1}$ and \eqref{diff_Q_1_H12}, we have
\begin{subequations}
\label{Diff_f_Tr}
\begin{align}
    & \frac{\partial f_{\mathrm{Tr}}}{ \partial [\mathbf{H}_1]_{ab}} 
    = -[\mathbf{H}_2 \mathbf{H}_2^H \mathbf{H}_1^H\mathbf{Q}_1\mathbf{H}_1 \mathbf{B} \mathbf{H}_1^H \mathbf{A} \mathbf{Q}_1]_{ba} -\overline{s}[\mathbf{H}_1^H\mathbf{Q}_1\mathbf{H}_1 \mathbf{B} \mathbf{H}_1^H \mathbf{A} \mathbf{Q}_1]_{ba} + [\mathbf{B}  \mathbf{H}_1^H\mathbf{A} \mathbf{Q}_1]_{ba}, \\
    &\frac{\partial f_{\mathrm{Tr}}}{ \partial [\mathbf{H}_1^H]_{ab}} = 
    -[\mathbf{Q}_1\mathbf{H}_1 \mathbf{B}   \mathbf{H}_1^H\mathbf{A}\mathbf{Q}_1 \mathbf{H}_{1} \mathbf{H}_2 \mathbf{H}_2^H]_{ba} - \overline{s}[\mathbf{Q}_1\mathbf{H}_1 \mathbf{B} \mathbf{H}_1^H\mathbf{A}\mathbf{Q}_1 \mathbf{H}_{1}]_{ba} + [ \mathbf{A} \mathbf{Q}_1 \mathbf{H}_{1} \mathbf{B}]_{ba}, \\
     & \frac{\partial f_{\Tr}}{ \partial [\mathbf{H}_2]_{ab}} 
    =  -[\mathbf{H}_2^H \mathbf{H}_1^H\mathbf{Q}_1\mathbf{H}_1 \mathbf{B} \mathbf{H}_1^H\mathbf{A}\mathbf{Q}_1\mathbf{H}_{1}]_{ba}, \\
    & \frac{\partial f_{\Tr}}{ \partial [\mathbf{H}_2^H]_{ab}} 
    =  -[\mathbf{H}_1^H\mathbf{Q}_1\mathbf{H}_1 \mathbf{B}  \mathbf{H}_1^H \mathbf{A}\mathbf{Q}_1\mathbf{H}_{1} \mathbf{H}_{2}]_{ba}.
\end{align} 
\end{subequations}
According to the variance upper bound \eqref{variance_f_Tr} and inequality \eqref{Matrix_norm_ineq}, we obtain
\begin{equation}
     \mathrm{Var}(f_{\Tr}) \leq U_{1,1} + U_{1,2} + U_{1,3} + U_{1,4},
\end{equation}
where 
\begin{subequations}
\begin{align}
    U_{1,1} &= \frac{3 Nr^2}{L}\mathbb{E} \norm{\mathbf{H}_2 \mathbf{H}_2^H \mathbf{H}_1^H\mathbf{Q}_1\mathbf{H}_1 \mathbf{B} \mathbf{H}_1^H \mathbf{A} \mathbf{Q}_1}^2 + \frac{3 Nr^2}{L}\mathbb{E} \norm{\mathbf{Q}_1\mathbf{H}_1 \mathbf{B}   \mathbf{H}_1^H\mathbf{A}\mathbf{Q}_1 \mathbf{H}_{1} \mathbf{H}_2 \mathbf{H}_2^H}^2 \notag \\
    &\leq \frac{6 Nr^2\norm{\mathbf{A}}^2\norm{\mathbf{B}}^2}{L}\mathbb{E}\norm{\mathbf{Q}_1}^4\norm{\mathbf{H}_1}^6\norm{\mathbf{H}_2}^4, \\
    U_{1,2} &=  \frac{3Nr^2 \overline{s}^2}{L} \mathbb{E}\norm{\mathbf{H}_1^H\mathbf{Q}_1\mathbf{H}_1 \mathbf{B} \mathbf{H}_1^H \mathbf{A} \mathbf{Q}_1}^2 + \frac{3Nr^2 \overline{s}^2}{L} \mathbb{E}\norm{\mathbf{Q}_1\mathbf{H}_1 \mathbf{B} \mathbf{H}_1^H\mathbf{A}\mathbf{Q}_1 \mathbf{H}_{1}}^2 \notag \\
    &\leq \frac{6 Nr^2 \overline{s}^2\norm{\mathbf{A}}^2 \norm{\mathbf{B}}^2}{L} \mathbb{E}  \norm{\mathbf{Q}_1}^4 \norm{\mathbf{H}_1}^6, \\
     U_{1,3} &= \frac{3 Nr^2}{L} \mathbb{E} \norm{\mathbf{B}  \mathbf{H}_1^H\mathbf{A} \mathbf{Q}_1}^2 + \frac{3 Nr^2}{L} \mathbb{E} \norm{\mathbf{A} \mathbf{Q}_1 \mathbf{H}_{1} \mathbf{B}}^2 \leq \frac{6 Nr^2\norm{\mathbf{A}}^2 \norm{\mathbf{B}}^2}{L} \mathbb{E} \norm{\mathbf{Q}_1}^2 \norm{\mathbf{H}_1}^2,  \\
     U_{1,4} &= \frac{ L r^2}{M} \mathbb{E} \norm{\mathbf{H}_2^H \mathbf{H}_1^H\mathbf{Q}_1\mathbf{H}_1 \mathbf{B} \mathbf{H}_1^H\mathbf{A}\mathbf{Q}_1\mathbf{H}_{1}}^2 + \frac{ L r^2}{M} \mathbb{E} \norm{\mathbf{H}_1^H\mathbf{Q}_1\mathbf{H}_1 \mathbf{B}  \mathbf{H}_1^H \mathbf{A}\mathbf{Q}_1\mathbf{H}_{1} \mathbf{H}_{2}}^2 \notag \\
     &\leq \frac{2 L r^2\norm{\mathbf{A}}^2 \norm{\mathbf{B}}^2}{M}  \mathbb{E}  \norm{\mathbf{Q}_1}^4 \norm{\mathbf{H}_1}^8 \norm{\mathbf{H}_2}^2.
\end{align} \label{U_1_s} %
\end{subequations}
By applying the inequality \eqref{Bounded_norm} to the above bounds and noticing that $\norm{\mathbf{Q}_1} \leq \frac{1}{z}$, we obtain
\begin{subequations}
\begin{align}
    & \sup_{N \geq 1} U_{1, 1}  \leq \sup_{N \geq 1} \frac{6 Nr^2\norm{\mathbf{A}}^2\norm{\mathbf{B}}^2}{L z^4} \sup_{N \geq 1} \mathbb{E}\norm{\mathbf{H}_1}^6\norm{\mathbf{H}_2}^4 = \mathcal{O}(\frac{1}{z^4}), \\
    & \sup_{N \geq 1} U_{1, 2}  \leq \sup_{N \geq 1} \frac{6 Nr^2 \overline{s}^2 \norm{\mathbf{A}}^2\norm{\mathbf{B}}^2}{L z^4} \sup_{N \geq 1} \mathbb{E}\norm{\mathbf{H}_1}^6 = \mathcal{O}(\frac{\overline{s}^2}{z^4}), \\
    & \sup_{N \geq 1} U_{1, 3} \leq \sup_{N \geq 1} \frac{6 Nr^2\norm{\mathbf{A}}^2 \norm{\mathbf{B}}^2}{Lz^2} \sup_{N \geq 1} \mathbb{E} \norm{\mathbf{H}_1}^2 = \mathcal{O}(\frac{1}{z^2}), \\
    & \sup_{N \geq 1} U_{1, 4} \leq \sup_{N \geq 1} \frac{2 Lr^2\norm{\mathbf{A}}^2 \norm{\mathbf{B}}^2}{Mz^4} \sup_{N \geq 1} \mathbb{E} \norm{\mathbf{H}_1}^8 \norm{\mathbf{H}_2}^2 = \mathcal{O}(\frac{1}{z^4}).
\end{align} \label{Bounding_U}%
\end{subequations}
Thus, we conclude that $\mathrm{Var}(f_{\Tr}) = \mathcal{O}_z^s(1)$. The proof for \eqref{Prop_variance_3_Trace_1} and \eqref{Prop_variance_3_Trace_3} are quite similar and thus are omitted here for brevity.
\subsection{Variance Control for Bilinear Forms}
Now, we focus on the proof of \eqref{Prop_variance_3_bilinear}. We first define $f_{\mathrm{Bi}} = \mathbf{a}^H \mathbf{Q}_1 \mathbf{H}_1 \mathbf{B} \mathbf{H}_1^H \mathbf{b}$. We will only prove $f_{\mathrm{Bi}}$ here, as the other terms can be obtained using the same method. In fact, $f_{\mathrm{Bi}}$ can be written as $\Tr[\mathbf{b}\mathbf{a}^H \mathbf{Q}_1 \mathbf{H}_1 \mathbf{B} \mathbf{H}_1^H ]$, and the terms for differentiation are very similar to \eqref{Diff_f_Tr} by replacing $\mathbf{A}$ with $\mathbf{b}\mathbf{a}^H$. Using formula \eqref{Variance_f} and \eqref{Matrix_Trace_ineq}, we observe that the upper bound on the variance of $f_{\mathrm{Bi}}$ can be written as
\begin{equation}
    \Var(f_{\mathrm{Bi}}) \leq \overline{U}_{1,1} + \overline{U}_{1,2} + \overline{U}_{1,3} + \overline{U}_{1,4},
\end{equation}
where 
\begin{subequations}
\begin{align}
    \overline{U}_{1,1} &= \frac{3}{L}\mathbb{E} \mathbf{a}^H \mathbf{Q}_1 \mathbf{R}_1 \mathbf{Q}_1\mathbf{a} \mathbf{b}^H \mathbf{H}_1 \mathbf{B}^H \mathbf{H}_1^H \mathbf{Q}_1\mathbf{H}_1\mathbf{H}_2\mathbf{H}_2^H \mathbf{T}_1\mathbf{H}_2\mathbf{H}_2^H\mathbf{H}_1^H \mathbf{Q}_1\mathbf{H}_1 \mathbf{B} \mathbf{H}_1^H\mathbf{b} \notag \\
    &+ \frac{3}{L} \mathbb{E} \mathbf{a}^H \mathbf{Q}_1 \mathbf{H}_1 \mathbf{H}_2 \mathbf{H}_2^H \mathbf{T}_1 \mathbf{H}_2 \mathbf{H}_2^H \mathbf{H}_1^H \mathbf{Q}_1 \mathbf{a} \mathbf{b}^H \mathbf{H}_1 \mathbf{B}^H \mathbf{H}_1^H \mathbf{Q}_1 \mathbf{R}_1 \mathbf{Q}_1\mathbf{H}_1 \mathbf{B} \mathbf{H}_1^H \mathbf{b} \leq \frac{6 r^2C_{a, b, B}}{ Lz^4}\mathbb{E}\norm{\mathbf{H}_1}^6\norm{\mathbf{H}_2}^4, \\
    \overline{U}_{1,2} &=  \frac{3 \overline{s}^2}{L} \mathbb{E} \mathbf{a}^H \mathbf{Q}_1 \mathbf{R}_1 \mathbf{Q}_1\mathbf{a} \mathbf{b}^H \mathbf{H}_1 \mathbf{B}^H \mathbf{H}_1^H \mathbf{Q}_1\mathbf{H}_1\mathbf{T}_1\mathbf{H}_1^H \mathbf{Q}_1\mathbf{H}_1 \mathbf{B} \mathbf{H}_1^H\mathbf{b}  \notag \\
    &+ \frac{3\overline{s}^2}{L} \mathbb{E}\mathbf{a}^H \mathbf{Q}_1 \mathbf{H}_1  \mathbf{T}_1 \mathbf{H}_1^H \mathbf{Q}_1 \mathbf{a} \mathbf{b}^H \mathbf{H}_1 \mathbf{B}^H \mathbf{H}_1^H \mathbf{Q}_1 \mathbf{R}_1 \mathbf{Q}_1\mathbf{H}_1 \mathbf{B} \mathbf{H}_1^H \mathbf{b} \leq \frac{6r^2 \overline{s}^2C_{a, b, B}}{Lz^4} \mathbb{E} \norm{\mathbf{H}_1}^6, \\
     \overline{U}_{1,3} &= \frac{3}{L} \mathbb{E}\mathbf{a}^H \mathbf{Q}_1 \mathbf{R}_1 \mathbf{Q}_1\mathbf{a}\mathbf{b}^H \mathbf{H}_1 \mathbf{B}^H \mathbf{T}_1\mathbf{B}\mathbf{H}_1^H\mathbf{b} + \frac{3}{L} \mathbb{E} \mathbf{a}^H \mathbf{Q}_1 \mathbf{H}_1 \mathbf{B}  \mathbf{T}_1 \mathbf{B} ^H \mathbf{H}_1^H \mathbf{Q}_1 \mathbf{a} \norm{\mathbf{b}}^2 \leq \frac{6r^2C_{a, b, B}}{L z^2} \mathbb{E}  \norm{\mathbf{H}_1}^2,  \\
     \overline{U}_{1,4} &= \frac{ 1}{M} \mathbb{E} \mathbf{a}^H \mathbf{Q}_1  \mathbf{H}_1 \mathbf{R}_2 \mathbf{H}_1^H\mathbf{Q}_1\mathbf{a} \mathbf{b}^H \mathbf{H}_1 \mathbf{B}^H \mathbf{H}_1^H \mathbf{Q}_1 \mathbf{H}_1 \mathbf{H}_2 \mathbf{R}_2\mathbf{H}_2^H\mathbf{H}_1^H \mathbf{Q}_1 \mathbf{H}_1 \mathbf{B}\mathbf{H}_1^H\mathbf{b}  \notag \\
     &+ \frac{1}{M} \mathbb{E} \mathbf{a}^H\mathbf{Q}_1\mathbf{H}_{1} \mathbf{H}_{2} \mathbf{T}_2\mathbf{H}_{2}^H \mathbf{H}_{1}^H \mathbf{Q}_1 \mathbf{a}\mathbf{b}^H \mathbf{H}_1\mathbf{B}^H \mathbf{H}_1^H \mathbf{Q}_1 \mathbf{H}_1 \mathbf{R}_2 \mathbf{H}_1^H\mathbf{Q}_1\mathbf{H}_1 \mathbf{B}  \mathbf{H}_1^H \mathbf{b} \leq \frac{2 r^2 C_{a, b, B}}{M z^4}  \mathbb{E}  \norm{\mathbf{H}_1}^8 \norm{\mathbf{H}_2}^2,
\end{align}
with $C_{a, b, B} = \norm{\mathbf{a}}^2 \norm{\mathbf{b}}^2\norm{\mathbf{B}}^2$. Using the same argument as \eqref{Bounding_U}, we can prove that $\overline{U}_{1,1} = \mathcal{O}(\frac{1}{N z^4})$, $\overline{U}_{1,2} = \mathcal{O}(\frac{\overline{s}^2}{N z^4})$, $\overline{U}_{1,3} = \mathcal{O}(\frac{1}{N z^2})$ and $\overline{U}_{1,4} = \mathcal{O}(\frac{1}{N z^4})$. This completes the proof of Proposition \ref{Prop_variance}. \QED
\end{subequations}
\section{Proof of Lemma \ref{Lemma_DE_Q1_2}}
\label{App_Lemma_DE_Q1_2}
We will divide the proof into two parts. In the first part, we will use the normal family theorem to prove a weaker result, i.e., $ \alpha_{\delta}-\delta$, $  \alpha_{\overline{\omega}}-\overline{\omega}$, $\alpha_{\underline{\omega}} - \underline{\omega}$, and $\alpha_{\gamma}-\gamma$  are of order $o(1)$. In the second part, we will strengthen the result $o(1)$ to $\mathcal{O}(\frac{1}{N^2})$ by studying the difference between $\mathbf{\Theta}_{\delta}$ and $\mathbf{F}_{\delta}$. To improve the readability of the proof, we list the notations used in this section in Table \ref{Temp_Notations}.
\begin{table}[t]
\centering
\caption{Table of Intermediate Notations}
 \label{Temp_Notations}
 \begin{tabular}{|c c| c c| cc |} 
 \toprule[1pt]
\midrule
 Notation & Expression & Notation & Expression & Notation & Expression \\
\midrule
 $\widetilde{\alpha}_{\delta}$ & $\frac{1}{L} \mathrm{Tr}[\mathbf{R}_1 \mathbf{\Theta}_{\delta}] $ & $\widetilde{\delta}_2$ & $\frac{1}{L} \mathrm{Tr} [\mathbf{R}_1 \mathbf{\Theta}_\delta \mathbf{R}_1 \mathbf{F}_\delta]$ &  $\widetilde{\overline{\omega}}_2$ & $\frac{1}{L}\mathrm{Tr} [ \mathbf{T}_1 \mathbf{\Theta}_\omega \mathbf{T}_1 \mathbf{F}_\omega]$ \\ 
 $\widetilde{\underline{\omega}}_2$ & $\frac{1}{L}  \mathrm{Tr} [\mathbf{R}_2 \mathbf{T}_1 \mathbf{\Theta}_\omega \mathbf{R}_2 \mathbf{T}_1 \mathbf{F}_\omega] $ & $\widetilde{\overline{\underline{\omega}}}_{1,1}$ & $\frac{1}{L}\mathrm{Tr} [\mathbf{R}_2 \mathbf{T}_1 \mathbf{\Theta}_\omega \mathbf{T}_1 \mathbf{F}_\omega]$ & $\widetilde{\gamma}_2$ & $\frac{1}{M} \mathrm{Tr} [\mathbf{T}_2 \mathbf{\Theta}_\gamma \mathbf{T}_2 \mathbf{F}_\gamma]$   \\ 
\bottomrule[1pt]
\end{tabular}
\end{table}
First, we establish the integration representation and bounded properties of terms $\alpha_{\delta}, \widetilde{\alpha}_{\delta}, \alpha_{\overline{\omega}}, \alpha_{\underline{\omega}}, \alpha_{\gamma}$, and $ \kappa$ in the following proposition. Similar to the discussion in Appendix \ref{App_Prop_Stiltjes}, we need to extend the domain of $z$ to the complex plane.
\begin{proposition} \label{Prop_Stieltjes_Eterms}
Assuming that assumptions \textbf{\ref{A-1}}-\textbf{\ref{A-3}} hold and fixing $\overline{s} \geq 0$, we have 
\begin{equation}
\alpha_{\delta}(\overline{s}, -z), \widetilde{\alpha}_{\delta}(\overline{s} , -z) , -\frac{\alpha_{\overline{\omega}}(\overline{s} , -z)}{z} , -\frac{\alpha_{\underline{\omega}}(s, -z)}{z} , -\frac{\alpha_{\gamma}(\overline{s} , -z)}{z} , {\kappa}(\overline{s} , -z) \in \mathcal{S}.
\end{equation} 
\end{proposition}
\textit{Proof:} The proof only requires verifying item 3) of Proposition \ref{Prop_Sset}, and is similar to the discussion in Proposition \ref{Prop_S_omega_gamma}. Therefore, we omit it here. According to item 2) of Proposition \ref{Prop_Sset},  
we know that for $z$ in any compact subset of $\mathbb{C} - \mathbb{R}_{-}$, 
functions $\alpha_{\delta}(\overline{s} , z), \widetilde{\alpha}_{\delta}(\overline{s} , z), \alpha_{\overline{\omega}}(\overline{s} , z), 
\alpha_{\underline{\omega}}(\overline{s} , z), \alpha_{\gamma}(\overline{s} , z)$, and $ \kappa(\overline{s} , z)$ 
are bounded. Note that here we changed the sign of $z$, which is equivalent to changing the sign of $z$ in the integration 
representation \eqref{Integration_rep}, without affecting the conclusion.
\subsection{The Rate of $o(1)$ Convergence} 
\label{weak_result_convergence}
By multiplying $[\mathbf{A}]_{ji}$, $[\mathbf{\Theta}_{\omega}\mathbf{B}]_{qp}$ on both sides of \eqref{App_DE_Q1_eq3} for any deterministic matrix $\mathbf{A}$, $\mathbf{B}$ with bounded spectral norm and summing over $i, j, p$ and $q$, we obtain
\begin{equation}
    \mathbb{E}  \mathrm{Tr}[\mathbf{A}\mathbf{Q}_1 \mathbf{H}_1\mathbf{B} \mathbf{H}_1^H] 
    = \frac{1}{L}\mathrm{Tr} [\mathbf{B}\mathbf{T}_1 \mathbf{\Theta}_{\omega}]\mathrm{Tr}[\mathbf{A}\mathbb{E} \mathbf{Q}_1\mathbf{R}_1] + \varepsilon_{\Phi_1}(\mathbf{A}, \mathbf{B}), \label{DE_Phi_1_pre}
\end{equation} 
where 
\begin{equation}
\begin{split}
    \varepsilon_{\Phi_1}(\mathbf{A}, \mathbf{B}) &= \alpha_{\delta}  \mathring{\kappa}\mathrm{Tr}[\mathbf{A}\mathbf{Q}_1\mathbf{H}_1 \mathbf{H}_2 \mathbf{T}_2\mathbf{\Theta}_{\gamma}\mathbf{H}_2^H\mathbf{T}_1\mathbf{\Theta}_{\omega} \mathbf{B} \mathbf{H}_1^H]  - \mathbb{E}\mathring{\alpha_{\delta}}\mathrm{Tr}[\mathbf{A}\mathbf{Q}_1\mathbf{H}_1 \mathbf{H}_2 \mathbf{H}_2^H \mathbf{T}_1\mathbf{\Theta}_{\omega} \mathbf{B} \mathbf{H}_1^H] \\
    &-  \overline{s}  \mathbb{E}\mathring{\alpha_{\delta}}\mathrm{Tr}[\mathbf{A}\mathbf{Q}_1\mathbf{H}_1\mathbf{T}_1\mathbf{\Theta}_{\omega} \mathbf{B} \mathbf{H}_1^H] := \varepsilon_{\Phi_1,1}(\mathbf{A}, \mathbf{B}) + \varepsilon_{\Phi_1,2}(\mathbf{A}, \mathbf{B}) + \varepsilon_{\Phi_1,3}(\mathbf{A}, \mathbf{B}).
\end{split}
\end{equation}
For $\overline{s} \geq 0$ and $z > 0$, by Proposition \ref{Prop_Stieltjes_Eterms} and Proposition \ref{Prop_Sset}, 
we have $|\alpha_{\delta}(\overline{s} , z)| \leq \frac{\Tr \mathbf{R}_1 }{L z} \leq \frac{N r}{L z}$, $ \norm{\mathbf{\Theta}_{\gamma}} \leq 1$,  and $ \norm{\mathbf{\Theta}_{\omega}} \leq 1$. 
Using the variance control in Proposition \ref{Prop_variance}, we have $|\varepsilon_{\Phi_1,i}(\mathbf{A}, \mathbf{B})| = \mathcal{O}_z^s(\frac{1}{N z})$, $ i= 1,2,3$. Letting $\mathbf{A} = \mathbf{I}_N$, $\mathbf{B} = \mathbf{R}_2$ in \eqref{DE_Phi_1_pre} and applying Lemma \ref{Lemma_DE_Q1_1} to approximate $\mathrm{Tr}[\mathbb{E} \mathbf{Q}_1\mathbf{R}_1]$, we obtain
\begin{equation}
    \kappa = \frac{L}{M} \alpha_{\underline{\omega}} \alpha_{\delta} + \mathcal{O}_z^s(\frac{1}{N^2z}). \label{DE_kappa_approx}
\end{equation}
By using the identity $\mathbf{A} - \mathbf{B} = \mathbf{A}(\mathbf{B}^{-1} - \mathbf{A}^{-1})\mathbf{B}$, we have 
\begin{equation}
\begin{split}
    \alpha_{\delta} - \delta &= \alpha_{\delta} - \widetilde{\alpha}_{\delta} + \widetilde{\alpha}_{\delta} - \delta = \alpha_{\delta} - \widetilde{\alpha}_{\delta} - \frac{1}{L}\Tr \mathbf{R}_1 \mathbf{\Theta_{\delta}} \left( (\overline{s} \alpha_{\overline{\omega}}- \overline{s} {\overline{\omega}}) + (\alpha_\gamma \alpha_{\underline{\omega}}  - \gamma {\underline{\omega}}  )\right) \mathbf{R}_1\mathbf{F_{\delta}}\\
    &= \alpha_{\delta} - \widetilde{\alpha}_{\delta} - \overline{s} (\alpha_{\overline{\omega}}- {\overline{\omega}})\widetilde{\delta}_2 - (\alpha_\gamma \alpha_{\underline{\omega}}  - \alpha_\gamma {\underline{\omega}} + \alpha_\gamma {\underline{\omega}} - \gamma {\underline{\omega}}  )\widetilde{\delta}_2 \\
    &= \alpha_{\delta} - \widetilde{\alpha}_{\delta} -  \overline{s}\widetilde{\delta}_2 (\alpha_{\overline{\omega}}- {\overline{\omega}}) - \alpha_\gamma \widetilde{\delta}_2(\alpha_{\underline{\omega}}  - {\underline{\omega}}) - {\underline{\omega}}\widetilde{\delta}_2(\alpha_\gamma  - \gamma   ).
\end{split}
\end{equation}
Similarly, we obtain
\begin{equation}
    \begin{bmatrix}
        1& \overline{s}\widetilde{\delta}_2& \alpha_{\gamma} \widetilde{\delta}_2& \underline{\omega}\widetilde{\delta}_2 \\
        \overline{s} \widetilde{\overline{\omega}}_2 + \gamma \widetilde{\overline{\underline{\omega}}}_{1,1} &1 & 0& \alpha_{\delta}\widetilde{\overline{\underline{\omega}}}_{1,1} \\
        \overline{s}\widetilde{\overline{\underline{\omega}}}_{1,1} + \gamma \widetilde{\underline{\omega}}_2 & 0 & 1 & \alpha_{\delta} \widetilde{\underline{\omega}}_2  \\
        \frac{L}{M} \underline{\omega} \widetilde{\gamma}_2 & 0& \frac{L}{M} \alpha_{\delta} \widetilde{\gamma}_2 & 1
    \end{bmatrix} \begin{bmatrix}
        \alpha_\delta - \delta \\
        \alpha_{\overline{\omega}} - \overline{\omega} \\
        \alpha_{\underline{\omega}} - \underline{\omega} \\
        \alpha_{\gamma} - \gamma
    \end{bmatrix} =  \begin{bmatrix}
        \alpha_\delta - \widetilde{\alpha}_{\delta} \\
        0 \\
        0 \\
        \left( \frac{L}{M}\alpha_{\delta} \alpha_{\underline{\omega}} - \kappa \right) \widetilde{\gamma}_2
    \end{bmatrix}. \label{Matrix_Eq}
\end{equation}
By Propositions \ref{Prop_Stieltjes}  and  \ref{Prop_Stieltjes_Eterms}, and inequality $\Tr \mathbf{A} \leq N \norm{\mathbf{A}}$ for $\mathbf{A} \in \mathbb{C}^{N \times N}$, we can obtain the upper bound of terms defined in Table \ref{Temp_Notations}. We present the result in the following without details
\begin{equation}
 \abs{\widetilde{\delta}_2} \leq \frac{Nr^2}{Lz^2},\hspace*{2mm}  \abs{\widetilde{\overline{\omega}}_2} \leq r^2,  \hspace*{2mm}\abs{\widetilde{\underline{\omega}}_2} \leq r^4, \hspace*{2mm}  \abs{\widetilde{\overline{\underline{\omega}}}_{1,1}} \leq r^3, \hspace*{2mm}\abs{\tilde{\gamma}_2} \leq r^2.
\end{equation}
We write \eqref{Matrix_Eq} in the form of $\widetilde{\mathbf{K}} \mathbf{d} = \mathbf{r}$ and obtain the following
\begin{equation}
\begin{split}
    \det(\widetilde{\mathbf{K}}) &= 1 - \frac{L}{M} \widetilde{\delta}_2 \underline{\omega}^2 \widetilde{\gamma}_2 - \frac{L}{M} \alpha_{\delta}^2 \widetilde{\underline{\omega}}_2 \widetilde{\gamma}_2 - \overline{s}^2 \widetilde{\delta}_2 \widetilde{\overline{\omega}}_2 - \overline{s}\widetilde{\delta}_2\widetilde{\underline{\overline{\omega}}}_2 \gamma - \widetilde{\delta}_2 \widetilde{\underline{\omega}}_2 \gamma \alpha_{\gamma} - 
    \overline{s}\widetilde{\delta}_2 \widetilde{\overline{\underline{\omega}}}_{1,1}\alpha_{\gamma} - \overline{s}^2\frac{L}{M} \widetilde{\delta}_2 \widetilde{\overline{\underline{\omega}}}_{1,1}^2\alpha_{\delta}^2\widetilde{\overline{\omega}}_2 \\
    &+ \frac{L}{M}\widetilde{\delta}_2\underline{\omega} \alpha_{\delta}\widetilde{\underline{\omega}}_2 + 2\overline{s} \frac{L}{M}\widetilde{\delta}_2 \underline{\omega}\widetilde{\overline{\underline{\omega}}}_{1,1} \alpha_{\delta}\widetilde{\overline{\omega}}_2 + \frac{L}{M} \widetilde{\delta}_2\underline{\omega}\alpha_{\delta}\widetilde{\underline{\omega}}_2 \widetilde{\overline{\omega}}_2\alpha_{\gamma} 
    + \overline{s}^2 \frac{L}{M} \widetilde{\delta}_2\widetilde{\overline{\omega}}_2\alpha_{\delta}^2\widetilde{\underline{\omega}}_2\widetilde{\overline{\omega}}_2  \geq 1 - \sum_{i=2}^4C_{i} z^{-i},
\end{split}
\end{equation}
where $C_i, i= 2,3,4$, are positive constants that are related to $\overline{s}$ and independent of $N$. Hence, there exists a sufficiently large constant $z_0$ such that for $z > z_0$,  
 $ {\inf}_{N \geq 1} \det( \widetilde{\mathbf{K}}) > 0$. This indicates that $\widetilde{\mathbf{K}}$ is uniformly invertible. 
Since every entry of $\widetilde{\mathbf{K}}$ is uniformly upper and lower bounded, which can be proved by the similar method in Appendix \ref{App_Prop_tightness},  
the spectral norm of $\widetilde{\mathbf{K}}^{-1} = \det( \widetilde{\mathbf{K}})^{-1} \mathrm{adj}(\widetilde{\mathbf{\mathbf{K}}})$ is uniformly bounded, where $\mathrm{adj}({\mathbf{M}})$ represents the adjugate matrix of $\mathbf{M}$. 
By letting $\mathbf{A} = \mathbf{R}_1$ in Lemma \ref{Lemma_DE_Q1_1} and \eqref{DE_kappa_approx}, we have $\norm{\mathbf{r}} = \mathcal{O}_z^s(\frac{1}{N^2 z})$. Therefore, for given $\overline{s} \geq 0$, 
we have
\begin{equation}
    \norm{ \mathbf{d} } \leq \sup_{N \geq 1} \norm{ \widetilde{\mathbf{K}}^{-1} } \norm{\mathbf{r}} = \mathcal{O}_z^s(\frac{1}{N^2 z}),  z > z_0. \label{Matrix_ineq_1}
\end{equation} 
In order to obtain more convenient analytical properties, we extend the domain of $z$ to the complex plane. Similar to Proposition \ref{Prop_S_omega_gamma} 2), we can obtain $\norm{\mathbf{\Theta}}_{\delta} \leq  \frac{1}{ \dist(z, \mathbb{R}_{-})}$, $ \norm{\mathbf{\Theta}_{\omega}} \leq \frac{|z|}{ \dist(z, \mathbb{R}_{-})}$, and $ \norm{\mathbf{\Theta}_{\gamma}} \leq \frac{|z|}{ \dist(z, \mathbb{R}_{-})}$. Thus, we can slightly modify the Proposition \ref{Prop_variance} to conclude that each element of $\mathbf{d}$ is bounded on any compact subset of $\mathbb{C} - \mathbb{R}_{-}$. Since each element in $\mathbf{d}$ is holomorphic when $\overline{s} \geq 0$ is fixed and $\lim_{N \rightarrow + \infty}\mathbf{d} = \mathbf{0}_4$ when $z \in (z_0, + \infty)$, by normal family theorem,  we have $  \lim_{N \rightarrow + \infty} \mathbf{d}  = \mathbf{0}_4 $ for $z \in \mathbb{C} - \mathbb{R}_-$. This indicates that $ \alpha_{\delta}-\delta$, $  \alpha_{\overline{\omega}}-\overline{\omega}$ , $\alpha_{\underline{\omega}} - \underline{\omega}$, and $\alpha_{\gamma}-\gamma$  are all of order $o(1)$.
\subsection{The Rate of $\mathcal{O}(\frac{1}{N^2})$ Convergence}
The previous result implies that the $\alpha_{sym}$ is a good approximation of $sym$, where $sym$ can be any 
symbol of $\delta$, $\overline{\omega}$, $\underline{\omega}$, $\gamma$. However, the convergence rate is still $o(1)$ for $z$ in the interval $(0, z_0)$. In this section, 
 we will prove a stronger result that the convergence rate $\mathcal{O}_s^z(\frac{1}{zN^2})$ holds everywhere for $z > 0$. 
 The analysis in Appendix \ref{weak_result_convergence} further indicates 
 that $\widetilde{\delta}_2 - \delta_2$, $\widetilde{\overline{\omega}}_2 - \overline{\omega}_2$, $\widetilde{\underline{\omega}}_2 - 
 \underline{\omega}_2$, $\widetilde{\overline{\underline{\omega}}}_{1,1} - \overline{\underline{\omega}}_{1,1}$,
  and $\widetilde{\gamma}_2 - \gamma_2$ are of order $o(1)$. In the following discussion, we assume $\overline{s} \geq 0$ is fixed. 
  Given $\mathbf{K}_1$ defined in \eqref{Diff_funcamental_equations}, we have
\begin{equation}
    \norm{\mathbf{K}_1 - \widetilde{\mathbf{K}}} = o(1), \forall z > 0.
\end{equation}
According to Proposition \ref{Prop_tightness}, we further have
\begin{equation}
    \liminf_{N \rightarrow + \infty } \det(\widetilde{\mathbf{K}}) = \liminf_{N \rightarrow + \infty } \det(\mathbf{K}_1) + o(1) \geq  \inf_{N \geq 1} \Delta_{V_1} > 0, \forall z > 0.
\end{equation}
Thus, $\widetilde{\mathbf{K}}$ is asymptotically invertible and 
\begin{equation}
    \underset{N \rightarrow + \infty  }{\limsup} \norm{\widetilde{\mathbf{K}}^{-1}} \leq  \frac{\underset{N \rightarrow + \infty  }{\limsup} \norm{\mathrm{adj}(\widetilde{\mathbf{K}})}}{\underset{N \rightarrow + \infty  }{\liminf} \det(\widetilde{\mathbf{K}})} = \mathcal{O}(\mathcal{G}_s(z^{-1})).
\end{equation}
Therefore, the following holds
\begin{equation}
    \norm {\mathbf{d}} \leq \norm{ \widetilde{\mathbf{K}}^{-1} } \norm{\mathbf{r}} = \mathcal{O}_z^s(\frac{1}{N^2 z}). \label{App_strong_bound}
\end{equation}
Moreover, by identity $\mathbf{A} - \mathbf{B} = \mathbf{A}(\mathbf{B}^{-1} - \mathbf{A}^{-1})\mathbf{B} $, following the definition of $\widetilde{\boldsymbol{\mathcal{E}}}_{\delta}$ in \eqref{Lemma_DE_Q1_2_Eq}, we have  
\begin{equation}
\begin{split}
    \widetilde{\boldsymbol{\mathcal{E}}}_{\delta} &= \mathbf{\Theta}_{\delta} - \mathbf{F}_{\delta} = \mathbf{\Theta}_{\delta}(\mathbf{F}_{\delta}^{-1} - \mathbf{\Theta}_{\delta}^{-1})\mathbf{F}_{\delta} 
    = \left(\overline{s} \overline{\omega} + \gamma \underline{\omega} - (\overline{s} \alpha_{\overline{\omega}} + \alpha_{\gamma} \alpha_{\underline{\omega}}) \right) \mathbf{\Theta}_{\delta}\mathbf{R}_1 \mathbf{F}_{\delta} \\
    &= \left[ \overline{s} ({\overline{\omega}} - \alpha_{\overline{\omega}}) + \alpha_\gamma ({\underline{\omega}} - \alpha_{\underline{\omega}}) + {\underline{\omega}}(\gamma - \alpha_\gamma) \right] \mathbf{\Theta}_{\delta}\mathbf{R}_1 \mathbf{F}_{\delta} \overset{(a)}{=} \mathcal{O}_z^s(\frac{1}{N^2 z})\mathbf{\Theta}_{\delta}\mathbf{R}_1 \mathbf{F}_{\delta}. \label{App_diff_trace_Q1}
\end{split}
\end{equation}
Here, step $(a)$ follows from \eqref{App_strong_bound} and $\abs{[\mathbf{d}]_i} \leq \norm{\mathbf{d}} $. Thus, we have 
\begin{equation}
    \sup_{i,j}\abs{[\widetilde{\boldsymbol{\mathcal{E}}}_{\delta}]_{ij}} = \sup_{i,j}\abs{\mathbf{e}_i^T \widetilde{\boldsymbol{\mathcal{E}}}_{\delta} \mathbf{e}_j} =  \mathcal{O}_z^s(\frac{1}{N^2 z})\sup_{i,j}\abs{\mathbf{e}_i^T\mathbf{\Theta}_{\delta}\mathbf{R}_1 \mathbf{F}_{\delta} \mathbf{e}_j}
     \leq \mathcal{O}_z^s(\frac{1}{N^2 z}) \norm{\mathbf{\Theta}_{\delta}} \norm{\mathbf{R}_1} \norm{ \mathbf{F}_{\delta} } 
     = \mathcal{O}_z^s(\frac{1}{N^2 z^3}), \\
\end{equation} 
and for matrix$\mathbf{A}$ with uniformly bounded spectral norm we have
\begin{equation}
    \abs{ \Tr \mathbf{A} \widetilde{\boldsymbol{\mathcal{E}}}_{\delta}} =  \mathcal{O}_z^s(\frac{1}{N^2 z}) 
    \abs{ \Tr \mathbf{A} \mathbf{\Theta}_{\delta}\mathbf{R}_1 \mathbf{F}_{\delta}} 
    \leq \mathcal{O}_z^s(\frac{1}{N z}) \norm{\mathbf{A}} \norm{\mathbf{\Theta}_{\delta}} \norm{\mathbf{R}_1} \norm{ \mathbf{F}_{\delta} } = \mathcal{O}_z^s(\frac{1}{N z^3}). \\
\end{equation} 
Similarly, \eqref{App_strong_bound} also implies that
\begin{subequations}
\begin{align}
    \sup_{i,j}\abs{ [\mathbf{\Theta_{\omega}} - \mathbf{F}_{\omega}]_{ij}} = \mathcal{O}_z^s(\frac{1}{N^2z}), \hspace*{2mm} \abs{\Tr[\mathbf{A}(\mathbf{\Theta_{\omega}} - \mathbf{F}_{\omega})]} = \mathcal{O}_z^s(\frac{1}{Nz}),\\
    \sup_{i,j}\abs{ [\mathbf{\Theta_{\gamma}} - \mathbf{F}_{\gamma}]_{ij}} = \mathcal{O}_z^s(\frac{1}{N^2z}), \hspace*{2mm} \abs{\Tr[\mathbf{A}(\mathbf{\Theta_{\gamma}} - \mathbf{F}_{\gamma})]} = \mathcal{O}_z^s(\frac{1}{Nz}).
\end{align}
\label{Approx_omega_gamma}%
\end{subequations}
\QED
\section{Proof of Lemma \ref{Lemma_tight_CLT}}
\label{App_Lemma_tight_CLT}
We first claim that $\mathbf{V}_n^{-\frac{1}{2}} \mathbf{x}_n$ is tight. For any $\epsilon > 0$, there exists a sufficiently small $a > 0$ such that
\begin{equation}
\begin{split}
    &\frac{1}{a^p} \int_{[-a, a]^p} \Big( 1 -  \exp[- \frac{1}{2}\mathbf{t}^T \mathbf{V}_n\mathbf{t}] \Big) \mathrm{d} \mathbf{t} \overset{(a)}{=} \frac{1}{a^p} \int_{\mathbf{U}_n^{-1}([-a, a]^p)} \Big( 1 -  \exp[- \frac{1}{2}\mathbf{x}^T \mathbf{\Lambda}_n\mathbf{x}] \Big) \mathrm{d} \mathbf{x}  \\
    & \overset{(b)}{\leq} \frac{1}{a^p} \int_{[-\sqrt{p}a, \sqrt{p}a]^p} \Big( 1 - \exp[-\frac{M_Vp^2 a^2}{2}] \Big) \mathrm{d} \mathbf{x} \leq 2^p p^{\frac{p}{2}} \Big( 1 - \exp[-\frac{M_Vp^2 a^2}{2}] \Big) \leq \epsilon,
\end{split} 
\end{equation}
where $(a)$ follows from the orthogonal transformation $\mathbf{x} = \mathbf{U}_n^{-1} \mathbf{t}$, where the columns of $\mathbf{U}_n$ are all the eigenvectors of $\mathbf{V}_n$. $\mathbf{\Lambda}_n = \mathrm{diag}\{\lambda_1(\mathbf{V}_n), \lambda_2(\mathbf{V}_n), \cdots , \lambda_p(\mathbf{V}_n) \}$ is a diagonal matrix with $\mathbf{V}_n = \mathbf{U}_n \mathbf{\Lambda}_n \mathbf{U}_n^T $. $(b)$ follows from the fact that $\mathbf{U}_n^{-1}([-a, a]^p) \subset [-\sqrt{p}a, \sqrt{p}a]^p$ and $\mathbf{x}^T \mathbf{\Lambda}_n \mathbf{x} \leq M_V \norm{\mathbf{x}}^2 \leq M_V p^2 a^2$. Based on \eqref{CLT_Lemma_Cond1}, due to the dominated convergence theorem, for sufficiently large $n$, i.e., $n > N_0$, where $N_0$ is a positive integer, we have
\begin{equation}
    \abs{ \frac{1}{a^p} \int_{[-a, a]^p} \Big( 1 -  \varphi_n(\mathbf{t}) \Big) \mathrm{d} \mathbf{t} - \frac{1}{a^p} \int_{[-a, a]^p} \Big( 1 -  \exp[- \frac{1}{2}\mathbf{t}^T \mathbf{V}_n\mathbf{t}] \Big) \mathrm{d} \mathbf{t} } \leq \epsilon. \label{Eq_Converge_chara}
\end{equation}
Next, assuming $\varphi(\mathbf{t}) = \mathbb{E} \exp[\jmath \mathbf{t}^T \mathbf{x}]$ is the  characteristic function of random vector $\mathbf{x} \in \mathbb{R}^p$ and exploiting Fubini’s theorem, we have
\begin{equation}
\begin{split}
&\frac{1}{a^p}\int_{[-a, a]^p}\Big( 1 -  \varphi(\mathbf{t}) \Big) \mathrm{d} \mathbf{t} = \frac{1}{a^p}\int_{[-a, a]^p}\mathrm{d} \mathbf{t}  \int_{\mathbb{R}^p} \Big( 1 -  \exp[ \jmath \sum_{i=1}^p t_i [\mathbf{x}]_i] \Big) F_{\mathbf{x}}(\mathrm{d}\mathbf{x})  \\
&=\frac{1}{a^p} \int_{\mathbb{R}^p}   F_{\mathbf{x}}(\mathrm{d}\mathbf{x})  \int_{[-a, a]^p} \Big( 1 -  \prod_{i=1}^p \exp[\jmath t_i [\mathbf{x}]_i] \Big)\mathrm{d} \mathbf{t} = 2^p \int_{\mathbb{R}^p}  \Big( 1 -  \prod_{i=1}^p\frac{\sin (a[\mathbf{x}]_i)}{a[\mathbf{x}]_i} \Big)  F_{\mathbf{x}}(\mathrm{d}\mathbf{x}) \\
&\geq 2^p \int_{\{\mathbf{x}: |[\mathbf{x}]_i| \geq \frac{2}{a}\}}  \Big( 1 -  \prod_{i=1}^p\frac{\sin (a[\mathbf{x}]_i)}{a[\mathbf{x}]_i} \Big)  F_{\mathbf{x}}(\mathrm{d}\mathbf{x}) \geq 2^p \int_{\{\mathbf{x}: |[\mathbf{x}]_i| \geq \frac{2}{a}\}}  \Big( 1 -  \prod_{i=1}^p\frac{1}{a|[\mathbf{x}]_i|} \Big)  F_{\mathbf{x}}(\mathrm{d}\mathbf{x}) \\
&\geq (2^p - 1) \mathbb{P}\{\mathbf{x}: |[\mathbf{x}]_i| \geq \frac{2}{a}\}. \label{Eq_multi_dim_chara}
\end{split}
\end{equation}
Applying  \eqref{Eq_multi_dim_chara} to $\mathbf{x}_n$ in \eqref{Eq_Converge_chara}, there exists a sufficiently small $a$ such that
\begin{equation}
    \mathbb{P}\{\mathbf{x}_n: |[\mathbf{x}_n]_i| \geq \frac{2}{a}\} \leq \frac{2 \epsilon}{2^p - 1} = \epsilon',  n > N_0.
\end{equation}
Thus, the sequence $\mathbf{x}_n$ is tight. Moreover, according to \eqref{Regu_lemma_Gaussian}, we have $\forall \mathbf{y} \in \mathbb{R}^p$, such that  $ \norm{\mathbf{y}} = R$ for some $R > 0$, then $ \norm{ \mathbf{V}_n^{\frac{1}{2}} \mathbf{y} } \geq  \sqrt{\lambda_{min}(\mathbf{V}_n)}R \geq \sqrt{m_V}R > 0$. It thus follows that 
\begin{equation}
\begin{split}
     1 - \mathbb{P}\{\mathbf{V}_n^{-\frac{1}{2}} \mathbf{x}_n \in B_p(R) \} \leq 1 -  \mathbb{P}\{ \mathbf{x}_n \in  B_p(\sqrt{m_V}R) \} \leq 1 -  \mathbb{P}\{ \mathbf{x}_n : |[\mathbf{x}_{n}]_{i}| \leq \sqrt{\frac{m_V}{p}}R \}, \label{tight_1}
\end{split}
\end{equation}
where $B_p(R) = \{\mathbf{x}\in \mathbb{R}^p: \norm{\mathbf{x}} \leq R \}$. Therefore, for sufficiently large $R$, the LHS of \eqref{tight_1} is sufficiently small, which implies the sequence $\mathbf{V}_n^{-\frac{1}{2}} \mathbf{x}_n$ is tight. Due to the tightness, we can find a subsequence $\mathbf{V}_{n_m}^{-\frac{1}{2}} \mathbf{x}_{n_m}$ that converges in distribution. Since the inequality $-\lambda_{max}(\mathbf{V}_n) \leq [\mathbf{V}_n]_{ij} \leq \lambda_{max}(\mathbf{V}_n)$ implies that each element of $\mathbf{V}_{n_m}$ belongs to a compact set, one can further find a subsequence of $n_{m}$, namely $n_{m_k}$, such that $\mathbf{V}_{n_{m_k}} \xrightarrow[k \rightarrow + \infty]{} \mathbf{V}^{\infty}$. Hence, we have $\mathbf{V}_{n_{m_k}}^{-\frac{1}{2}} \mathbf{x}_{n_{m_k}} \xrightarrow[k \rightarrow +\infty]{d} \mathcal{N}(\mathbf{0}_p, \mathbf{I}_p)$. Therefore, $\mathbf{V}_{n_m}^{-\frac{1}{2}} \mathbf{x}_{n_m} \xrightarrow[m \rightarrow +\infty]{d} \mathcal{N}(\mathbf{0}_p, \mathbf{I}_p)$ must hold. From the above discussion, we can conclude that any subsequence of $\mathbf{V}_n^{-\frac{1}{2}} \mathbf{x}_n$ that converges in distribution converges to $\mathcal{N}(\mathbf{0}_p, \mathbf{I}_p)$. Therefore, $\mathbf{V}_n^{-\frac{1}{2}} \mathbf{x}_n$ converges to $\mathcal{N}(\mathbf{0}_p, \mathbf{I}_p)$, which can be derived from the scalar case \cite[Theorem 3.2.15]{durrett2019probability}. \QED
\section{Proof of Lemma \ref{Lemma_DE_terms_Approx_Psi}}
\label{App_Lemma_DE_terms_Approx_Psi}
In this section, we will use the definitions of variables in 
\eqref{Random_quantities}-\eqref{Deterministic_quantities} and \eqref{overline_Theta_omega}. We divide the proof into four parts. 
First, we consider $\Psi_1$ and $\Psi_2$ where the degree of the resolvent $\mathbf{Q}_1$ is 1. Then we analysis $\Psi_3 $ and $\Psi_4$ where the degree of $\mathbf{Q}_1$ is 2.
In $\Psi_5 $ and $\Psi_6$, the degree of the resolvent is 2, but they contain random matrices $\mathbf{H}_1\mathbf{R}_2\mathbf{H}_1^H$ between the resolvents. Finally we analysis $\Psi_7$ which contains the correlation between $\mathbf{Q}_1$ and $\mathbf{Q}_2$.  According to the bounds in \eqref{Bound_norm_} and \eqref{Bounded_norm}, all $\Psi_j$s are $\mathcal{O}(N)$. 
\subsection{Evaluation of $\Psi_1$ and $\Psi_2$}
By applying Lemma \ref{Lemma_DE_Q1_1} and \ref{Lemma_DE_Q1_2} to \eqref{DE_Phi_1_pre} and using $\mathbf{F}_{\omega}$ to approximate $\mathbf{\Theta}_{\omega}$ in \eqref{App_strong_bound}, relation \eqref{DE_Psi_1} holds.
By multiplying $[\mathbf{\Theta}_{\gamma}\mathbf{B}]_{ls}$, $[\mathbf{C}]_{mp}$ and $[\mathbf{A}]_{ji}$ on both sides of \eqref{APP_DE_Q1_eq1} 
and summing over  all the subscripts $j, i, s, l,  m$, and $p$, we have the following equation
\begin{equation}
    \Psi_2(\mathbf{A}, \mathbf{B}, \mathbf{C})=\frac{1}{M} \mathrm{Tr}[\mathbf{T}_2\mathbf{\Theta}_{\gamma} \mathbf{B}]  \Psi_1(\mathbf{A}, \mathbf{R}_2\mathbf{C}) + \varepsilon_{\Psi_2}(\mathbf{A}, \mathbf{B}, \mathbf{C}),
\end{equation}
where the term $\varepsilon_{\Psi_2}(\mathbf{A}, \mathbf{B}, \mathbf{C})$ is defined as 
$ \mathbb{E} -\mathring{\kappa}\mathrm{Tr}[\mathbf{A}\mathbf{Q}_1\mathbf{H}_1 \mathbf{H}_2 \mathbf{T}_2\mathbf{\Theta}_{\gamma} \mathbf{B} \mathbf{H}_2\mathbf{C} \mathbf{H}_1^H ] $. 
According to the variance control of the trace forms in \eqref{Prop_variance_3_Trace} and $\norm{\mathbf{\Theta}_{\gamma}} \leq 1$,  
we have $\Var(\mathring{\kappa}) = \mathcal{O}_z^s(\frac{1}{N^2})$ and $\Var(\mathrm{Tr}[\mathbf{A}\mathbf{Q}_1\mathbf{H}_1 \mathbf{H}_2 \mathbf{T}_2\mathbf{\Theta}_{\gamma} \mathbf{B} \mathbf{H}_2\mathbf{C} \mathbf{H}_1^H ]) = \mathcal{O}_z^s{(1)}$. Based on the Cauchy–Schwarz inequality $\mathbb{E}|xy| \leq \sqrt{\mathbb{E}|x|^2\mathbb{E}|y|^2} $, we have $|\varepsilon_{\Psi_2}(\mathbf{A}, \mathbf{B}, \mathbf{C})| = \mathcal{O}_z^s(\frac{1}{N})$. By applying the strong convergence result \eqref{App_strong_bound}, replacing $\mathbf{\Theta}_{\gamma}$ with $\mathbf{F}_{\gamma}$, and 
using the approximation of \eqref{DE_Psi_1} on $\Psi_1(\mathbf{A}, \mathbf{R}_2\mathbf{C})$, we have
\begin{equation}
    \Psi_2(\mathbf{A}, \mathbf{B}, \mathbf{C}) = \frac{1}{M}\mathrm{Tr}[\mathbf{B}\mathbf{T}_2 \mathbf{F}_{\gamma}]\frac{1}{L}\mathrm{Tr}[\mathbf{R}_2 \mathbf{C} \mathbf{T}_1 \mathbf{F}_{\omega}]\mathrm{Tr}[\mathbf{A}\mathbf{F}_{\delta} \mathbf{R}_1 ] 
    + \mathcal{O}_z^s(\frac{1}{N}).
\end{equation}
This completes the proof of \eqref{DE_Psi_2}. 
\subsection{Evaluation of $\Psi_3$ and $\Psi_4$}
In order to calculate $\Psi_3$ and $\Psi_4$, we define the random quantity $\widehat\Gamma(\mathbf{A}, \mathbf{B}) = \mathrm{Tr}[\mathbf{A} \mathbf{Q}_1 \mathbf{B} \mathbf{Q}_1]$. 
It is obvious that $\Gamma(\mathbf{A}, \mathbf{B}) = \mathbb{E}\widehat\Gamma(\mathbf{A}, \mathbf{B}) \leq N \norm{\mathbf{A}} \norm{\mathbf{B}} z^{-2} $,  which is $\mathcal{O}(N)$ and the approximation result is given by the following lemma.
\begin{lemma} \label{DE_Gamma}
    Assuming that $\mathbf{A}$ and $\mathbf{B}$ are deterministic matrices with bounded spectral norms, and the same settings as Lemma \ref{Lemma_DE_Q1_1} hold, then
\begin{equation}
    \Gamma(\mathbf{A}, \mathbf{B}) = \frac{\Delta \varsigma + \frac{L}{M} \underline{\omega}_{2, I}^2 \gamma_2}{\Delta_{V_1}} \frac{1}{L} \mathrm{Tr}[\mathbf{B}\mathbf{F}_{\delta}\mathbf{R}_1 \mathbf{F}_{\delta}] \mathrm{Tr}[\mathbf{A}\mathbf{F}_{\delta}\mathbf{R}_1 \mathbf{F}_{\delta}]
    + \mathrm{Tr}[\mathbf{A}\mathbf{F}_{\delta}\mathbf{B} \mathbf{F}_{\delta}]  + \mathcal{O}_{z}^s(\frac{1}{N}).
\end{equation}
\end{lemma}
\textit{Proof:} By taking $f = [\mathbf{Q}_1 \mathbf{B} \mathbf{Q}_1 \mathbf{H}_1]_{ik}  [\mathbf{H}_2]_{ms}^* [\mathbf{H}_1]_{jp}^* $  and  $[\mathbf{H}_2]_{kl}$ as the variable in
integration by parts formula \eqref{Integration_by_parts}, we have
\begin{equation}
\begin{split}
     &\sum_{k} \mathbb{E} [\mathbf{Q}_1 \mathbf{B} \mathbf{Q}_1 \mathbf{H}_1]_{ik} [\mathbf{H}_2]_{kl} [\mathbf{H}_2]_{ms}^* [\mathbf{H}_1]_{jp}^* 
     = \sum_{k, a, b}\frac{[\mathbf{R}_2]_{ka}[\mathbf{T}_2]_{bl}}{M} \mathbb{E} \Big\{ -[\mathbf{Q}_1 \mathbf{H}_1\mathbf{H}_2]_{ib} [\mathbf{H}_1^H\mathbf{Q}_1\mathbf{B} \mathbf{Q}_1 \mathbf{H}_1]_{ak}  [\mathbf{H}_2]_{ms}^* [\mathbf{H}_1]_{jp}^* \\
    & - [\mathbf{Q}_1\mathbf{B}\mathbf{Q}_1 \mathbf{H}_1\mathbf{H}_2]_{ib} [\mathbf{H}_1^H\mathbf{Q}_1\mathbf{H}_1 ]_{ak}  [\mathbf{H}_2]_{ms}^* [\mathbf{H}_1]_{jp}^* +  [\mathbf{Q}_1 \mathbf{B} \mathbf{Q}_1 \mathbf{H}_1]_{ik}  \delta(m-a) \delta(s-b) [\mathbf{H}_1]_{jp}^*  \Big\}.
\end{split}
\end{equation}
By writing $\widehat{\kappa} = \kappa + \mathring \kappa$, we have
\begin{equation}
\begin{split}
 \mathbb{E} [\mathbf{Q}_1 \mathbf{B} \mathbf{Q}_1 \mathbf{H}_1 \mathbf{H}_2 \mathbf{\Theta}_{\gamma}^{-1}]_{il} [\mathbf{H}_2]_{ms}^* [\mathbf{H}_1]_{jp}^* =  \mathbb{E} \Big\{ - \frac{\widehat{\Psi}_{3}(\mathbf{I}_N, \mathbf{B}, \mathbf{R}_2)}{M}[\mathbf{Q}_1 \mathbf{H}_1\mathbf{H}_2 \mathbf{T}_2]_{il}  [\mathbf{H}_2]_{ms}^* [\mathbf{H}_1]_{jp}^*\\
 -  \mathring{\kappa} [\mathbf{Q}_1\mathbf{B}\mathbf{Q}_1 \mathbf{H}_1\mathbf{H}_2 \mathbf{T}_2 ]_{il} [\mathbf{H}_2]_{ms}^* [\mathbf{H}_1]_{jp}^*+ \frac{[\mathbf{T}_2]_{sl}}{M} [\mathbf{Q}_1 \mathbf{B} \mathbf{Q}_1 \mathbf{H}_1 \mathbf{R}_2 ]_{im}  [\mathbf{H}_1]_{jp}^* \Big\}. \label{DE_Psi_3_eq_1}
\end{split}
\end{equation}
Similarly, we have
\begin{equation}
\begin{split}
    &\mathbb{E} [\mathbf{Q}_1 \mathbf{B} \mathbf{Q}_1 \mathbf{H}_1]_{iq} [\mathbf{H}_1]_{jp}^* 
    =  \mathbb{E} \Big\{ -\frac{\widehat\Gamma(\mathbf{R}_1,\mathbf{B})}{L}\left( [\mathbf{Q}_1 \mathbf{H}_1\mathbf{H}_2\mathbf{H}_2^H \mathbf{T}_1]_{iq} + \overline{s} [\mathbf{Q}_1 \mathbf{H}_1\mathbf{T}_1]_{iq} \right) [\mathbf{H}_1]_{jp}^*  \\
    & -  \widehat{\alpha}_{\delta}[\mathbf{Q}_1 \mathbf{B}\mathbf{Q}_1 \mathbf{H}_1\mathbf{H}_2\mathbf{H}_2^H \mathbf{T}_1]_{iq}   [\mathbf{H}_1]_{jp}^* - \overline{s} \widehat{\alpha}_{\delta}[\mathbf{Q}_1 \mathbf{B}\mathbf{Q}_1 \mathbf{H}_1\mathbf{T}_1]_{iq} [\mathbf{H}_1]_{jp}^* + \frac{[\mathbf{T}_1]_{pq}}{L} [\mathbf{Q}_1 \mathbf{B} \mathbf{Q}_1 \mathbf{R}_1]_{ij} \Big\}. 
    \label{DE_Psi_3_eq_2}
\end{split}
\end{equation}
By multiplying both sides of equation \eqref{DE_Psi_3_eq_1} with $[\mathbf{\Theta}_{\gamma}]_{ls}$ and $[\mathbf{T}_1]_{mq}$, and then summing over subscripts $l$, $s$, and $m$, we obtain the following equation
 \begin{equation}
\begin{split}
     \mathbb{E} [\mathbf{Q}_1 \mathbf{B} \mathbf{Q}_1 \mathbf{H}_1 \mathbf{H}_2\mathbf{H}_2^H \mathbf{T}_1]_{iq} [\mathbf{H}_1]_{jp}^* =  \mathbb{E} \Big\{ - \frac{\widehat{\Psi}_{3}(\mathbf{I}_N, \mathbf{B}, \mathbf{R}_2)}{M}[\mathbf{Q}_1 \mathbf{H}_1\mathbf{H}_2 \mathbf{T}_2 \mathbf{\Theta}_{\gamma}\mathbf{H}_2^H \mathbf{T}_1]_{iq} [\mathbf{H}_1]_{jp}^*\\
     -  \mathring{\kappa} [\mathbf{Q}_1\mathbf{B}\mathbf{Q}_1 \mathbf{H}_1\mathbf{H}_2 \mathbf{T}_2\mathbf{\Theta}_{\gamma}\mathbf{H}_2^H\mathbf{T}_1]_{iq}[\mathbf{H}_1]_{jp}^* 
     + {\alpha}_{\gamma} [\mathbf{Q}_1 \mathbf{B} \mathbf{Q}_1 \mathbf{H}_1 \mathbf{R}_2 \mathbf{T}_1]_{iq}[\mathbf{H}_1]_{jp}^* \Big\}.
\label{DE_Psi_Eq_pq}%
\end{split}
\end{equation}
By multiplying both sides of the above equation with $-\alpha_{\delta}$ and then adding the result to both sides of \eqref{DE_Psi_3_eq_2}, the following can be obtained by
expressing $\widehat{\alpha}_{\delta} = \alpha_{\delta} + \mathring{\alpha}_{\delta}$ 
\begin{equation}
\begin{split}
    & \mathbb{E} [\mathbf{Q}_1 \mathbf{B} \mathbf{Q}_1 \mathbf{H}_1\mathbf{\Theta}_{\omega}^{-1}]_{iq} [\mathbf{H}_1]_{jp}^* = \mathbb{E} \Big\{ -\frac{\widehat\Gamma(\mathbf{R}_1, \mathbf{B})}{L} ([\mathbf{Q}_1 \mathbf{H}_1\mathbf{H}_2\mathbf{H}_2^H \mathbf{T}_1]_{iq} +  \overline{s} [\mathbf{Q}_1 \mathbf{H}_1\mathbf{T}_1]_{iq})[\mathbf{H}_1]_{jp}^* \\
    &-   \mathring{\alpha}_{\delta}[\mathbf{Q}_1 \mathbf{B} \mathbf{Q}_1 \mathbf{H}_1\mathbf{H}_2\mathbf{H}_2^H \mathbf{T}_1]_{iq}   [\mathbf{H}_1]_{jp}^* + \frac{\alpha_{\delta}}{M} \widehat{\Psi}_{3}(\mathbf{I}_N, \mathbf{B},\mathbf{R}_2)[\mathbf{Q}_1 \mathbf{H}_1\mathbf{H}_2 \mathbf{T}_2\mathbf{\Theta}_{\gamma} \mathbf{H}_2^H \mathbf{T}_1]_{iq} [\mathbf{H}_1]_{jp}^* \\
    &+  \alpha_{\delta}\mathring \kappa[\mathbf{Q}_1\mathbf{B}\mathbf{Q}_1 \mathbf{H}_1\mathbf{H}_2 \mathbf{T}_2 \mathbf{\Theta}_{\gamma} \mathbf{H}_2^H \mathbf{T}_1]_{iq} [\mathbf{H}_1]_{jp}^*- \overline{s} \mathring{\alpha}_{\delta}[\mathbf{Q}_1 \mathbf{B}\mathbf{Q}_1 \mathbf{H}_1\mathbf{T}_1]_{iq} [\mathbf{H}_1]_{jp}^* + \frac{[\mathbf{T}_{1}]_{pq}}{L} [\mathbf{Q}_1 \mathbf{B} \mathbf{Q}_1 \mathbf{R}_1]_{ij} \Big\} 
    \label{DE_Psi_3_eq_3}.
\end{split}
\end{equation}
By multiplying $[\mathbf{\Theta}_{\omega}(\overline{s} \mathbf{I}_L + \alpha_{\gamma} \mathbf{R}_2)]_{qp}$, $[\mathbf{\Theta}_{\delta}\mathbf{A}]_{ji}$ 
on both sides of \eqref{DE_Psi_3_eq_3} and summing over the subscripts $q$, $p$, $j$, and $i$ , we get one equation. By multiplying both sides of \eqref{DE_Psi_3_eq_1} with 
$[\mathbf{\Theta}_{\gamma}]_{ls}$, $[\mathbf{\Theta}_{\delta}\mathbf{A}]_{ji}$ and $\delta(m-p)$ on both sides of \eqref{DE_Psi_3_eq_1} and summing over $l, s, m$ , $p$, $j$, and $i$, 
we get another equation. Combining these two equations, we obtain the following by using the resolvent identity of $\mathbf{Q}_2$ \eqref{Resolvent_identity_Q2} to eliminate 
the term $\mathbb{E}\Tr \mathbf{Q}_1 \mathbf{B} \mathbf{Q}_1 \mathbf{H}_1 \mathbf{H}_2 \mathbf{H}_2^H \mathbf{H}_1^H\mathbf{\Theta}_{\delta}\mathbf{A} + \overline{s}\Tr \mathbf{Q}_1 \mathbf{B} \mathbf{Q}_1 \mathbf{H}_1 \mathbf{H}_1^H\mathbf{\Theta}_{\delta}\mathbf{A}$
\begin{equation}
\begin{split}
\Gamma(\mathbf{A}, \mathbf{B}) &= \mathbb{E} \Big\{ \frac{\widehat\Gamma(\mathbf{R}_1,\mathbf{B})}{L} [ \widehat{\Psi}_2(\mathbf{\Theta}_{\delta} \mathbf{A}, \mathbf{I}_M, \overline{\mathbf{\Theta}}_{\omega})  + \overline{s} \widehat{\Psi}_1(\mathbf{\Theta}_{\delta} \mathbf{A}, \overline{\mathbf{\Theta}}_{\omega})] 
+ \mathring{\alpha}_{\delta}\widehat{\Psi}_4(\mathbf{\Theta}_{\delta} \mathbf{A},\mathbf{B}, \mathbf{I}_M,\overline{\mathbf{\Theta}}_{\omega}) + \mathrm{Tr}[\mathbf{Q}_1 \mathbf{B}\mathbf{\Theta}_{\delta} \mathbf{A}] \\
&+ \overline{s} \mathring{\alpha}_{\delta}\widehat{\Psi}_3(\mathbf{\Theta}_{\delta} \mathbf{A}, \mathbf{B},\overline{\mathbf{\Theta}}_{\omega})  + \frac{1}{M} \widehat{\Psi}_{3}(\mathbf{I}_N, \mathbf{B},\mathbf{R}_2)\widehat{\Psi}_2(\mathbf{\Theta}_{\delta} \mathbf{A}, \mathbf{T}_2 \mathbf{\Theta}_{\gamma},\mathbf{\Theta}_{\omega}^H)+ \mathring \kappa \widehat{\Psi}_4(\mathbf{\Theta}_{\delta} \mathbf{A}, \mathbf{B},\mathbf{T}_2 \mathbf{\Theta}_{\gamma},\mathbf{\Theta}_{\omega}^H) \Big\}.
\label{Gamma_AB_pre}
\end{split}
\end{equation}
Writing all the r.v.s  $\widehat x = x + \mathring x$ in the above equation, using the approximation rules of \eqref{DE_Psi_1} and \eqref{DE_Psi_2} to the terms $\Psi_1$ and $\Psi_2$ and replacing $\mathbf{\Theta}_{sym}$ with $\mathbf{F}_{sym}$, where $sym$ is one of symbol $\delta, \omega$ and 
$\gamma$, we obtain
\begin{equation}
     \Gamma(\mathbf{A}, \mathbf{B})  =  \frac{\Gamma(\mathbf{R}_1,\mathbf{B}) }{L} \varsigma \mathrm{Tr}[\mathbf{A} \mathbf{F}_{\delta} \mathbf{R}_1\mathbf{F}_{\delta}] +  \mathrm{Tr}[\mathbf{A} \mathbf{F}_{\delta} \mathbf{B}\mathbf{F}_{\delta}] + \frac{\Psi_{3}(\mathbf{I}_N, \mathbf{B}, \mathbf{R}_2)}{M}  \gamma_2 \underline{\omega}_{2, I} \mathrm{Tr}[\mathbf{A} \mathbf{F}_{\delta} \mathbf{R}_1\mathbf{F}_{\delta}] +\varepsilon_\Gamma(\mathbf{A}, \mathbf{B}) + \mathcal{O}_{z}^s(\frac{1}{N}), 
     \label{DE_Gamma_eq_1}
\end{equation}
where the error term is given by
\begin{equation}
    \varepsilon_{\Gamma}(\mathbf{A}, \mathbf{B}) =   
    \mathbb{E} \Big\{ \frac{\mathring\Gamma}{L}\mathring{\Psi}_2 + \overline{s}\frac{\mathring\Gamma}{L}\mathring{\Psi}_1+ 
    \mathring{\alpha}_{\delta}\mathring{\Psi}_4 + \overline{s}\mathring{\alpha}_{\delta}\mathring{\Psi}_3 
    + \frac{1}{M} \mathring{\Psi}_{3}\mathring{\Psi}_2 + \mathring \kappa \mathring{\Psi}_4\Big \}.
\end{equation}
Here, variables are omitted in the RHS without causing ambiguity. By observing the identity that $z\widehat{\Gamma}(\mathbf{A}, \mathbf{B}) + \overline{s}\widehat{\Psi}_3(\mathbf{A}, \mathbf{B}, \mathbf{I}_L) + \widehat{\Psi}_4(\mathbf{A}, \mathbf{B}, \mathbf{I}_L, \mathbf{I}_M) = \mathrm{Tr}\mathbf{A} \mathbf{Q}_1 \mathbf{B}$ and using  the variance control in \eqref{Prop_variance_3_Trace} and \eqref{Var_control}, 
we have the variance control $\Var(\widehat{\Gamma}(\mathbf{A}, \mathbf{B})) = \mathcal{O}_s^z(1)$. Note that a more detailed discussion similar to the proof of Proposition \ref{Prop_variance} shows
$\Var(\widehat{\Gamma}(\mathbf{A}, \mathbf{B})) = \mathcal{O}_s^z(z^{-4})$. However, $\mathcal{O}_s^z(1)$ is enough here. Cauchy–Schwarz inequality and  variance control on $\mathring{\kappa}$, 
$\mathring{\alpha}_{\delta}$, $\mathring{\Gamma}$, and $\mathring{\Psi}_j$ together imply that
$\varepsilon_{\Gamma}(\mathbf{A}, \mathbf{B}) = \mathcal{O}_{z}^s(\frac{1}{N})$.
\par
Setting $\mathbf{A} = \mathbf{R}_1$ in \eqref{DE_Gamma_eq_1}, we have
\begin{equation}
    \Gamma(\mathbf{R}_1, \mathbf{B}) =
    \Gamma(\mathbf{R}_1, \mathbf{B}) \varsigma \delta_2 
    + \mathrm{Tr}[\mathbf{R}_1 \mathbf{F}_{\delta} \mathbf{B}\mathbf{F}_{\delta}] 
    + \frac{L}{M} \Psi_{3}(\mathbf{I}_N, \mathbf{B}, \mathbf{R}_2) \gamma_2 \underline{\omega}_{2, I} \delta_2 + \mathcal{O}_{z}^s(\frac{1}{N}). \label{DE_Gamma_R1_B}
\end{equation}
On the other hand, by multiplying $[\mathbf{A}]_{ji}$ and $[\mathbf{\Theta}_{\omega}\mathbf{C}]_{qp}$ on both sides of \eqref{DE_Psi_3_eq_3} and summing over the subscripts $j$, $i$, $q$, and $p$, the following holds true
\begin{equation}
\begin{split}
 \Psi_3(\mathbf{A}, \mathbf{B}, \mathbf{C}) &= \mathbb{E} \Big\{-\frac{\widehat\Gamma(\mathbf{R}_1, \mathbf{B})}{L} \widehat{\Psi}_2(\mathbf{A}, \mathbf{I}_M, \mathbf{T}_1\mathbf{\Theta}_{\omega}\mathbf{C}) - \frac{\overline{s} \widehat\Gamma(\mathbf{R}_1, \mathbf{B})}{L}\widehat{\Psi}_1(\mathbf{A},  \mathbf{T}_1\mathbf{\Theta}_{\omega}\mathbf{C})  \\
    &-  \mathring{\alpha}_{\delta}\widehat{\Psi}_4(\mathbf{A}, \mathbf{B},  \mathbf{T}_1\mathbf{\Theta}_{\omega}\mathbf{C}) +  \frac{\alpha_{\delta}}{M} \widehat{\Psi}_{3}(\mathbf{I}_N, \mathbf{B}, \mathbf{R}_2)\widehat{\Psi}_2(\mathbf{A}, \mathbf{T}_2 \mathbf{\Theta}_{\gamma}, \mathbf{T}_1\mathbf{\Theta}_{\omega}\mathbf{C})  \\
    &+  \alpha_{\delta}\mathring \kappa \widehat{\Psi}_4(\mathbf{A},\mathbf{B}, \mathbf{T}_2 \mathbf{\Theta}_{\gamma}, \mathbf{T}_1\mathbf{\Theta}_{\omega}\mathbf{C}) - \overline{s}\mathring{\alpha}_{\delta}\widehat{\Psi}_3(\mathbf{A},\mathbf{B},  \mathbf{T}_1\mathbf{\Theta}_{\omega}\mathbf{C})\Big\} 
    + \frac{\mathrm{Tr}[\mathbf{C}\mathbf{T}_{1}\mathbf{\Theta}_{\omega}]}{L} \Gamma(\mathbf{R}_1 \mathbf{A}, \mathbf{B}).
\label{Psi_3_ABC_pre}
\end{split}
\end{equation}
Using the same argument as in \eqref{DE_Gamma_eq_1}, the following holds true
\begin{equation}
\begin{split}
    \Psi_3(\mathbf{A}, \mathbf{B}, \mathbf{C}) &= - \frac{\Gamma(\mathbf{R}_1, \mathbf{B})}{L} \frac{1}{L} \mathrm{Tr}[(\overline{s} \mathbf{I}_L + \gamma \mathbf{R}_2) \mathbf{T}_1\mathbf{F}_{\omega}\mathbf{C}\mathbf{T}_1 \mathbf{F}_{\omega}]\mathrm{Tr}[\mathbf{A} \mathbf{F}_{\delta} \mathbf{R}_1] \\
    &+  \delta \gamma_{2} \frac{{\Psi}_{3}(\mathbf{I}_N, \mathbf{B}, \mathbf{R}_2)}{M} \frac{1}{L} \mathrm{Tr}[\mathbf{R}_2\mathbf{T}_1\mathbf{F}_{\omega}\mathbf{C}\mathbf{T}_1 \mathbf{F}_{\omega}]\mathrm{Tr}[\mathbf{A} \mathbf{F}_{\delta} \mathbf{R}_1] + \frac{\mathrm{Tr}[\mathbf{C}\mathbf{T}_{1}\mathbf{F}_{\omega}]}{L}  \Gamma(\mathbf{R}_1 \mathbf{A}, \mathbf{B}) + \mathcal{O}_{z}^s(\frac{1}{N}). \label{DE_Psi_3_A}%
\end{split}
\end{equation}

To study the relation between $\Gamma(\mathbf{R}_1, \mathbf{B})$ and $\Psi_{3}(\mathbf{I}_N, \mathbf{B}, \mathbf{R}_2)$, 
we set $\mathbf{A} = \mathbf{I}_N$ and $\mathbf{C} = \mathbf{R}_2$ in \eqref{DE_Psi_3_A} 
and exploit the relation $\underline{\omega} = \underline{\omega}_{2, I} + \overline{s} \delta \overline{\underline{\omega}}_{1,1} + \delta \gamma \underline{\omega}_2$, resulting in the following equation
\begin{equation}
 \Psi_3(\mathbf{I}_N, \mathbf{B}, \mathbf{R}_2) = \underline{\omega}_{2, I} \Gamma(\mathbf{R}_1,\mathbf{B}) 
 +  \frac{L}{M}\gamma_{2}  \underline{\omega}_2 \delta^2 {\Psi}_{3}(\mathbf{I}_N,\mathbf{B}, \mathbf{R}_2) + \mathcal{O}_{z}^s(\frac{1}{N}).
\end{equation}
Combining the above equation with \eqref{DE_Gamma_R1_B}, we obtain the following two terms
\begin{subequations}
\begin{align}
    \Gamma(\mathbf{R}_1, \mathbf{B}) = \frac{\Delta}{\Delta_{V_1}}\mathrm{Tr}[\mathbf{B}\mathbf{F}_{\delta}\mathbf{R}_1 \mathbf{F}_{\delta}]+ \mathcal{O}_{z}^s(\frac{1}{N}), \\
    \Psi_3(\mathbf{I}_N, \mathbf{B}, \mathbf{R}_2) = \frac{\underline{\omega}_{2, I}}{\Delta_{V_1}}\mathrm{Tr}[\mathbf{B}\mathbf{F}_{\delta}\mathbf{R}_1 \mathbf{F}_{\delta}]+ \mathcal{O}_{z}^s(\frac{1}{N}). \label{Psi_3_I_B_R2}%
\end{align}
\end{subequations}
Plugging the above relations into \eqref{DE_Gamma_eq_1}, we complete the proof of Lemma \ref{DE_Gamma}. \QED
\par
By applying Lemma \ref{DE_Gamma} to \eqref{DE_Psi_3_A} , we obtain
\begin{equation}
\begin{split}
    \Psi_3(\mathbf{A}, \mathbf{B}, \mathbf{C})
    &= -\frac{\Delta}{\Delta_{V_1}} \frac{1}{L}\mathrm{Tr}[\mathbf{B}\mathbf{F}_{\delta} \mathbf{R}_1 \mathbf{F}_\delta]\frac{1}{L} \mathrm{Tr}[( \overline{s} \mathbf{I}_L + \gamma \mathbf{R}_2)\mathbf{T}_1\mathbf{F}_{\omega}\mathbf{C}\mathbf{T}_1 \mathbf{F}_{\omega}]\mathrm{Tr}[\mathbf{A} \mathbf{F}_{\delta} \mathbf{R}_1]  \\
    &+ \frac{\frac{L}{M}  \underline{\omega}_{2, I} \gamma_{2}\delta}{\Delta_{V_1}}\frac{1}{L}\mathrm{Tr}[\mathbf{B}\mathbf{F}_{\delta} \mathbf{R}_1 \mathbf{F}_\delta]\frac{1}{L} \mathrm{Tr}[\mathbf{R}_2\mathbf{T}_1\mathbf{F}_{\omega}\mathbf{C}\mathbf{T}_1 \mathbf{F}_{\omega}]\mathrm{Tr}[\mathbf{A} \mathbf{F}_{\delta} \mathbf{R}_1]  \\
    &+ \frac{\Delta \varsigma + \frac{L}{M} \underline{\omega}_{2, I}^2 \gamma_2}{\Delta_{V_1}} \frac{1}{L}\mathrm{Tr}[\mathbf{C}\mathbf{T}_{1}\mathbf{F}_{\omega}]\frac{1}{L} \mathrm{Tr}[\mathbf{B}\mathbf{F}_{\delta}\mathbf{R}_1 \mathbf{F}_{\delta}] \mathrm{Tr}[\mathbf{A}\mathbf{F}_{\delta}\mathbf{R}_1 \mathbf{F}_{\delta}\mathbf{R}_1]  \\
    &+ \frac{1}{L}\mathrm{Tr}[\mathbf{C}\mathbf{T}_{1}\mathbf{F}_{\omega}]\mathrm{Tr}[\mathbf{A}\mathbf{F}_{\delta}\mathbf{B} \mathbf{F}_{\delta}\mathbf{R}_1] + \mathcal{O}_{z}^s(\frac{1}{N}),
\end{split}
\end{equation}
which concludes \eqref{DE_Psi_3}. 
By multiplying $[\mathbf{A}]_{ji}$, $[\mathbf{\Theta}_{\gamma}\mathbf{C}]_{ls}$ and $[\mathbf{D}]_{mp}$ on both sides of \eqref{DE_Psi_3_eq_1} and summing over all subscripts,
the following holds by applying the similar argument in \eqref{DE_Gamma_eq_1}
\begin{equation}
\begin{split}
     \Psi_4(\mathbf{A}, \mathbf{B}, \mathbf{C}, \mathbf{D}) &= -\frac{ \underline{\omega}_{2, I}}{\Delta_{V_1}} \frac{1}{M} \mathrm{Tr}[\mathbf{B} \mathbf{F}_{\delta} \mathbf{R}_1 \mathbf{F}_{\delta}]\frac{1}{M} \mathrm{Tr}[\mathbf{C} \mathbf{T}_2 \mathbf{F}_{\gamma} \mathbf{T}_2 \mathbf{F}_{\gamma}] \frac{1}{L} \mathrm{Tr}[\mathbf{R}_2 \mathbf{D} \mathbf{T}_1 \mathbf{F}_{\omega}] \mathrm{Tr}[\mathbf{A} \mathbf{F}_{\delta} \mathbf{R}_1] \\
    &+ \frac{1}{M}\mathrm{Tr} [\mathbf{C} \mathbf{T}_{2}\mathbf{\Theta}_{\gamma}] \Psi_3(\mathbf{A}, \mathbf{B}, \mathbf{R}_2 \mathbf{D}) +  \varepsilon_{\Psi_4}(\mathbf{A}, \mathbf{B}, \mathbf{C}, \mathbf{D}) + \mathcal{O}_{z}^s(\frac{1}{N}),
\end{split}
\end{equation}
where 
$
    \varepsilon_{\Psi_4}(\mathbf{A}, \mathbf{B}, \mathbf{C}, \mathbf{D}) =   \mathbb{E} (-\frac{1}{M}\mathring{\Psi}_{3}\mathring\Psi_2 - \mathring \kappa \mathring{\Psi}_4) $ is proved to have the order $\mathcal{O}_{z}^s(\frac{1}{N}).
$
Thus, we obtain \eqref{DE_Psi_3} and \eqref{DE_Psi_4}.
\subsection{Evaluation of $\Psi_5$ and $\Psi_6$}
Now, we focus on calculating $\Psi_5$ and $\Psi_6$. First, by applying integration by parts formula \eqref{Integration_by_parts} to $[\mathbf{H}_{2}]_{kl}$, we obtain
\begin{equation}
\label{DE_Psi_5_eq_0}
\begin{split}
    &\sum_k \mathbb{E}[\mathbf{Q}_1 \mathbf{H}_1 \mathbf{R}_2 \mathbf{H}_1^H \mathbf{Q}_1 \mathbf{H}_1 ]_{ik} [\mathbf{H}_2]_{kl} [\mathbf{H}_2]_{ms}^* [\mathbf{H}_1]_{jp}^* = \sum_{k, a, b} \mathbb{E}[\mathbf{H}_2]_{kl}[\mathbf{H}_2]_{ab}^* \mathbb{E}\Big[\frac{\partial[\mathbf{Q}_1 \mathbf{H}_1 \mathbf{R}_2 \mathbf{H}_1^H \mathbf{Q}_1\mathbf{H}_1 ]_{ik} [\mathbf{H}_2]_{ms}^* [\mathbf{H}_1]_{jp}^*}{\partial [\mathbf{H}_2]_{ab}^*}\Big] \\
    &= \sum_{k, a, b}\frac{[\mathbf{R}_2]_{ka}[\mathbf{T}_2]_{bl}}{M} \mathbb{E}\Big\{ -[\mathbf{Q}_1 \mathbf{H}_1 \mathbf{H}_2]_{ib} [\mathbf{H}_1^H \mathbf{Q}_1 \mathbf{H}_1 \mathbf{R}_2 \mathbf{H}_1^H \mathbf{Q}_1 \mathbf{H}_1]_{ak} [\mathbf{H}_2]_{ms}^* [\mathbf{H}_1]_{jp}^* \\
    &-  [ \mathbf{Q}_1 \mathbf{H}_1 \mathbf{R}_2 \mathbf{H}_1^H\mathbf{Q}_1 \mathbf{H}_1 \mathbf{H}_2]_{ib} [ \mathbf{H}_1^H \mathbf{Q}_1 \mathbf{H}_1]_{ak} [\mathbf{H}_2]_{ms}^* [\mathbf{H}_1]_{jp}^* +  [\mathbf{Q}_1 \mathbf{H}_1 \mathbf{R}_2 \mathbf{H}_1^H \mathbf{Q}_1 \mathbf{H}_1 ]_{ik} \delta(m-a) \delta(s-b) [\mathbf{H}_1]_{jp}^* \Big\}.
\end{split}
\end{equation}
As a result, the following is true
\begin{equation}
\begin{split}
    & \mathbb{E}[\mathbf{Q}_1 \mathbf{H}_1 \mathbf{R}_2 \mathbf{H}_1^H \mathbf{Q}_1 \mathbf{H}_1 \mathbf{H}_2 \mathbf{\Theta}_{\gamma}^{-1}]_{il} [\mathbf{H}_2]_{ms}^* [\mathbf{H}_1]_{jp}^*= \mathbb{E}\Big\{-\frac{\widehat{\Psi}_5(\mathbf{I}_N, \mathbf{R}_2)}{M}[\mathbf{Q}_1 \mathbf{H}_1 \mathbf{H}_2 \mathbf{T}_2 ]_{il}  [\mathbf{H}_2]_{ms}^* [\mathbf{H}_1]_{jp}^*  \\
    & - \mathring{\kappa}[ \mathbf{Q}_1 \mathbf{H}_1 \mathbf{R}_2 \mathbf{H}_1^H\mathbf{Q}_1 \mathbf{H}_1 \mathbf{H}_2 \mathbf{T}_2]_{il} [\mathbf{H}_2]_{ms}^* [\mathbf{H}_1]_{jp}^* + \frac{[\mathbf{T}_2]_{sl}}{M} [\mathbf{Q}_1 \mathbf{H}_1 \mathbf{R}_2 \mathbf{H}_1^H \mathbf{Q}_1 \mathbf{H}_1 \mathbf{R}_2 ]_{im} [\mathbf{H}_1]_{jp}^*\Big\}. \label{DE_Psi_5_eq_1}
\end{split}
\end{equation}
Similar to the evaluation in \eqref{DE_Psi_5_eq_0} and  \eqref{DE_Psi_5_eq_1}, we can obtain
\begin{equation}
    \begin{split}
    & \mathbb{E} [\mathbf{Q}_1 \mathbf{H}_1 \mathbf{R}_2 \mathbf{H}_1^H \mathbf{Q}_1 \mathbf{H}_1 ]_{iq}  [\mathbf{H}_1]_{jp}^*
     =  \mathbb{E}\Big\{ -\frac{1}{L}\widehat{\Psi}_3(\mathbf{I}_N, \mathbf{R}_1, \mathbf{R}_2)[\mathbf{Q}_1 \mathbf{H}_1 \mathbf{H}_2 \mathbf{H}_2^H 
     \mathbf{T}_1 ]_{iq}  [\mathbf{H}_1]_{jp}^* \\
    &- \overline{s}  \frac{1}{L} \widehat{\Psi}_3(\mathbf{I}_N, \mathbf{R}_1,\mathbf{R}_2) [\mathbf{Q}_1 \mathbf{H}_1 \mathbf{T}_1 ]_{iq} [\mathbf{H}_1]_{jp}^* - \widehat{\alpha}_{\delta} [\mathbf{Q}_1 \mathbf{H}_1 \mathbf{R}_2 \mathbf{H}_1^H\mathbf{Q}_1 \mathbf{H}_1 \mathbf{H}_2 \mathbf{H}_2^H \mathbf{T}_1 ]_{iq} [\mathbf{H}_1]_{jp}^* \\
    &-  \overline{s} \widehat{\alpha}_{\delta} [\mathbf{Q}_1 \mathbf{H}_1 \mathbf{R}_2 \mathbf{H}_1^H\mathbf{Q}_1 \mathbf{H}_1 \mathbf{T}_1 ]_{iq}[\mathbf{H}_1]_{jp}^* + \frac{[\mathbf{T}_1]_{pq}}{L} [\mathbf{Q}_1 \mathbf{H}_1 \mathbf{R}_2 \mathbf{H}_1^H \mathbf{Q}_1 \mathbf{R}_1 ]_{ij} + \widehat{\alpha}_{\delta} [\mathbf{Q}_1 \mathbf{H}_1 \mathbf{R}_2 \mathbf{T}_1 ]_{iq} [\mathbf{H}_1]_{jp}^* \Big \}. 
    \label{DE_Psi_5_eq_2}
\end{split}
\end{equation}
Combining equations \eqref{DE_Psi_5_eq_1} and \eqref{DE_Psi_5_eq_2} similarly as \eqref{DE_Psi_3_eq_1}-\eqref{DE_Psi_3_eq_3}, we have
\begin{equation}
\begin{split}
    &\mathbb{E} [\mathbf{Q}_1 \mathbf{H}_1 \mathbf{R}_2 \mathbf{H}_1^H \mathbf{Q}_1 \mathbf{H}_1 \mathbf{\Theta}_{\omega}^{-1}]_{iq}  [\mathbf{H}_1]_{jp}^* =\mathbb{E}\Big\{  -\frac{\widehat{\Psi}_3(\mathbf{I}_N, \mathbf{R}_1,\mathbf{R}_2)}{L}[\mathbf{Q}_1 \mathbf{H}_1 \mathbf{H}_2 \mathbf{H}_2^H 
     \mathbf{T}_1 ]_{iq}  [\mathbf{H}_1]_{jp}^*  \\
    &- \frac{\overline{s}}{L}\widehat{\Psi}_3(\mathbf{I}_N, \mathbf{R}_1,\mathbf{R}_2) [\mathbf{Q}_1 \mathbf{H}_1 \mathbf{T}_1 ]_{iq} [\mathbf{H}_1]_{jp}^* - \mathring{\alpha}_{\delta} [\mathbf{Q}_1 \mathbf{H}_1 \mathbf{R}_2 \mathbf{H}_1^H\mathbf{Q}_1 \mathbf{H}_1 \mathbf{H}_2 \mathbf{H}_2^H \mathbf{T}_1 ]_{iq} [\mathbf{H}_1]_{jp}^* \\
    &-  \overline{s}  \mathring{\alpha}_{\delta} [\mathbf{Q}_1 \mathbf{H}_1 \mathbf{R}_2 \mathbf{H}_1^H\mathbf{Q}_1 \mathbf{H}_1 \mathbf{T}_1 ]_{iq}[\mathbf{H}_1]_{jp}^* + \frac{\alpha_{\delta}}{M} \widehat{\Psi}_5(\mathbf{I}_N, \mathbf{R}_2)[\mathbf{Q}_1 \mathbf{H}_1 \mathbf{H}_2 \mathbf{T}_2 \mathbf{\Theta}_{\gamma} \mathbf{H}_2^H\mathbf{T}_1]_{iq} [\mathbf{H}_1]_{jp}^* \\
    &+ \alpha_{\delta} \mathring\kappa [ \mathbf{Q}_1 \mathbf{H}_1 \mathbf{R}_2 \mathbf{H}_1^H\mathbf{Q}_1 \mathbf{H}_1 \mathbf{H}_2 \mathbf{T}_2 \mathbf{\Theta}_{\gamma}\mathbf{H}_2^H\mathbf{T}_1]_{iq} [\mathbf{H}_1]_{jp}^* +  \frac{[\mathbf{T}_1]_{pq}}{L} [\mathbf{Q}_1 \mathbf{H}_1 \mathbf{R}_2 \mathbf{H}_1^H \mathbf{Q}_1 \mathbf{R}_1 ]_{ij}  + \widehat{\alpha}_{\delta} [\mathbf{Q}_1 \mathbf{H}_1 \mathbf{R}_2 \mathbf{T}_1 ]_{iq} [\mathbf{H}_1]_{jp}^* \Big\}.
\end{split}
\end{equation}
Multiplying $[\mathbf{\Theta}_{\omega} \mathbf{B}]_{qp}$ and $[\mathbf{A}]_{ij}$ to both sides of the above equation and summing over $i, j, p$ and $q$, we have 
\begin{equation}
    \begin{split}
    &\Psi_5(\mathbf{A}, \mathbf{B}) =  \mathbb{E}\Big\{ -\frac{\widehat{\Psi}_3(\mathbf{I}_N, \mathbf{R}_1,\mathbf{R}_2)}{L}\widehat{\Psi}_2(\mathbf{A}, \mathbf{I}_M,  \mathbf{T}_1  \mathbf{\Theta}_{\omega}\mathbf{B}) - \frac{\overline{s}}{L}\widehat{\Psi}_3(\mathbf{I}_N, \mathbf{R}_1,\mathbf{R}_2) \widehat{\Psi}_1(\mathbf{A}, \mathbf{T}_1  \mathbf{\Theta}_{\omega}\mathbf{B})  \\
    & -  \mathring{\alpha}_{\delta} \widehat{\Psi}_6(\mathbf{A}, \mathbf{I}_M, \mathbf{T}_1  \mathbf{\Theta}_{\omega}\mathbf{B}) -  \overline{s} \mathring{\alpha}_{\delta} \widehat{\Psi}_5(\mathbf{A},  \mathbf{T}_1  \mathbf{\Theta}_{\omega}\mathbf{B}) + \frac{\alpha_{\delta}}{M} \widehat{\Psi}_5(\mathbf{I}_N, \mathbf{R}_2)\widehat{\Psi}_2 (\mathbf{A}, \mathbf{T}_2 \mathbf{\Theta}_{\gamma}, \mathbf{T}_1\mathbf{\Theta}_{\omega}\mathbf{B}) \\
    &+ \alpha_{\delta} \mathring\kappa \widehat{\Psi}_{6}(\mathbf{A}, \mathbf{T}_2 \mathbf{\Theta}_{\gamma}, \mathbf{T}_1\mathbf{\Theta}_{\omega}\mathbf{B}) + \widehat{\alpha}_{\delta} \widehat{\Psi}_1(\mathbf{A}, \mathbf{R}_2 \mathbf{T}_1 \mathbf{\Theta}_{\omega} \mathbf{B})\Big \} +  \frac{\mathrm{Tr}[\mathbf{T}_{1} \mathbf{\Theta}_{\omega}\mathbf{B}] }{L}\Psi_3(\mathbf{I}_N, \mathbf{R}_1 \mathbf{A}, \mathbf{R}_2).
\end{split}
\end{equation}
Next, similar to the derivation of \eqref{Gamma_AB_pre}-\eqref{DE_Gamma_R1_B}, we write all the r.v.s as  $\widehat{x} = x + \mathring{x}$ and obtain
\begin{equation}
\begin{split}
    \Psi_5(\mathbf{A}, \mathbf{B}) 
    &=  -\frac{\delta_2 \underline{\omega}_{2, I}}{\Delta_{V_1}} \frac{1}{L} \mathrm{Tr}[(\overline{s}\mathbf{I}_L + \gamma \mathbf{R}_2) \mathbf{T}_1 \mathbf{F}_{\omega} \mathbf{B} \mathbf{T}_1 \mathbf{F}_{\omega}] \mathrm{Tr}[\mathbf{A} \mathbf{F}_{\delta} \mathbf{R}_1]+ \delta \frac{1}{L} \mathrm{Tr}[\mathbf{R}_2 \mathbf{T}_1 \mathbf{F}_{\omega} \mathbf{B} \mathbf{T}_1 \mathbf{F}_{\omega}] \mathrm{Tr}[\mathbf{A} \mathbf{F}_{\delta} \mathbf{R}_1]  \\
    & +   {\delta}\gamma_2 \frac{{\Psi}_5(\mathbf{I}_N, \mathbf{R}_2)}{M} \frac{1}{L} \mathrm{Tr}[\mathbf{R}_2 \mathbf{T}_1 \mathbf{F}_{\omega} \mathbf{B} \mathbf{T}_1 \mathbf{F}_{\omega}]\mathrm{Tr}[\mathbf{A} \mathbf{F}_{\delta} \mathbf{R}_1] +  \frac{\underline{\omega}_{2, I}}{\Delta_{V_1}}\frac{1}{L}\mathrm{Tr}[\mathbf{T}_{1} \mathbf{F}_{\omega}\mathbf{B}]\mathrm{Tr}[\mathbf{A}\mathbf{F}_{\delta} \mathbf{R}_1 \mathbf{F}_{\delta} \mathbf{R}_1 ]  + \mathcal{O}_{z}^s(\frac{1}{N}).
     \label{DE_Psi_5_eq_4}
\end{split}
\end{equation}
Setting $\mathbf{A} = \mathbf{I}$ and $\mathbf{B} = \mathbf{R}_2$ in \eqref{DE_Psi_5_eq_4}, we have
\begin{equation}
    \Psi_5(\mathbf{I}_N, \mathbf{R}_2) = \frac{L \underline{\omega}_{2, I}^2 \delta_2}{\Delta_{V_1} \Delta} + \frac{L\delta^2 \underline{\omega}_2}{\Delta}  + \mathcal{O}_{z}^s(\frac{1}{N}) = \frac{M(1 - \varsigma \delta_2 - \Delta_{V_1})}{\Delta_{V_1} \gamma_2} + \mathcal{O}_{z}^s(\frac{1}{N}). \label{DE_Psi_5_I_R2}
\end{equation}
Plugging \eqref{DE_Psi_5_I_R2} into \eqref{DE_Psi_5_eq_4}, we have
\begin{equation}
\begin{split}
    &\Psi_5(\mathbf{A}, \mathbf{B}) 
  =  -\frac{\delta_2 \underline{\omega}_{2, I}}{\Delta_{V_1}} \frac{1}{L} \mathrm{Tr}[(\overline{s}\mathbf{I}_L + \gamma \mathbf{R}_2) \mathbf{T}_1 \mathbf{F}_{\omega} \mathbf{B} \mathbf{T}_1 \mathbf{F}_{\omega}] \mathrm{Tr}[\mathbf{A} \mathbf{F}_{\delta} \mathbf{R}_1] \\
    &+  \frac{\delta(1 - \varsigma \delta_2)}{\Delta_{V_1}} \frac{1}{L} \mathrm{Tr}[\mathbf{R}_2 \mathbf{T}_1 \mathbf{F}_{\omega} \mathbf{B} \mathbf{T}_1 \mathbf{F}_{\omega}]\mathrm{Tr}[\mathbf{A} \mathbf{F}_{\delta} \mathbf{R}_1] 
   +  \frac{\underline{\omega}_{2, I}}{\Delta_{V_1}}\frac{1}{L}\mathrm{Tr}[\mathbf{T}_{1} \mathbf{F}_{\omega}\mathbf{B}]\mathrm{Tr}[\mathbf{A}\mathbf{F}_{\delta} \mathbf{R}_1 \mathbf{F}_{\delta} \mathbf{R}_1 ] + \mathcal{O}_{z}^s(\frac{1}{N}).
\end{split}
\end{equation}
Hence, the proof of \eqref{DE_Psi_5} is completed. On the other hand, \eqref{DE_Psi_6} can be proved by multiplying $[\mathbf{A}]_{ji}$, $[\mathbf{\Theta}_{\gamma}\mathbf{B}]_{ls}$, and $[\mathbf{C}]_{mp}$ on both sides of \eqref{DE_Psi_5_eq_1} and summing over all the subscripts.
\subsection{Evaluation of $\Psi_7$}
In order to obtain an approximation for $\Psi_7$, we first define some quantities following the procedures as in \eqref{Random_quantities} and \eqref{Deterministic_quantities} as following
\begin{align}
    \widehat{\beta}_{\tau} = \frac{\Tr \mathbf{R}_1 \mathbf{Q}_2}{L},   \mathbf{\Omega}_{\overline{\tau}} = (\mathbf{I}_L + \underline{s} \beta_{\tau} \mathbf{T}_1)^{-1},  \beta_{\overline{\tau}} = \frac{1}{L} \Tr[\mathbf{T}_1\mathbf{\Omega}_{\overline{\tau}}],  \mathbf{\Omega}_{\tau} = (z\mathbf{I}_N + \underline{s} \beta_{\overline{\tau}}\mathbf{R}_1)^{-1}. \label{Q_2_quantities}
\end{align}
Similar to Lemma \ref{Lemma_DE_Q1_1}, Lemma \ref{Lemma_DE_Q1_2}, and the proof steps in Appendices \ref{App_Lemma_DE_Q1_1} and \ref{App_Lemma_DE_Q1_2}, we obtain the the following the variance control results
 and the approximation rules
\begin{subequations}
 \label{Approx_rules_Omega}
\begin{align}
    & \Var(\widehat{\beta}_{\tau}) = \mathcal{O}_z^s(\frac{1}{z^2 N^2}), \label{variance_control_beta_tau} \\
    &\sup_{i,j} \abs{ [\mathbb{E}\mathbf{Q}_2 - \mathbf{\Omega}_{\tau}]_{ij}} = \mathcal{O}_z^s(\frac{1}{N^{\frac{3}{2}}z }), \Tr \mathbf{A}[\mathbb{E}\mathbf{Q}_2 -  \mathbf{\Omega}_{\tau}] = \mathcal{O}_z^s(\frac{1}{N z}), \\
    &\sup_{i,j} \abs{ [\mathbf{\Omega}_{\tau} - \mathbf{G}_{\tau} ]_{ij}} = \mathcal{O}_z^s(\frac{1}{N^{2}z^3 }), \Tr \mathbf{A}[\mathbf{\Omega}_{\tau} - \mathbf{G}_{\tau}] = \mathcal{O}_z^s(\frac{1}{N z^3}), \\
    &\sup_{i,j} \abs{ [\mathbf{\Omega}_{\overline{\tau}} - \mathbf{G}_{\overline{\tau}} ]_{ij}} = \mathcal{O}_z^s(\frac{1}{N^{2}z }), \Tr \mathbf{A}[\mathbf{\Omega}_{\overline{\tau}} - \mathbf{G}_{\overline{\tau}} ]= \mathcal{O}_z^s(\frac{1}{N z}),
\end{align}
\end{subequations}
where $\mathbf{A}$ is a deterministic matrix with bounded norm. As the proof is very similar to Lemmas \ref{Lemma_DE_Q1_1} and \ref{Lemma_DE_Q1_2}, we omit it here for brevity. If both 
$\mathbf{R}_1$ and  $\mathbf{T}_1$ are diagonal matrices, part of the above result could be degenerated to \cite[Eq. (35)]{Hachem2008ANewApproach}.
To simplify the calculation, we define $\widehat{\Upsilon}(\mathbf{A}, \mathbf{B}) = \mathrm{Tr}[\mathbf{A} \mathbf{Q}_1 \mathbf{B} \mathbf{Q}_2] $ and estimate its mean $\Upsilon(\mathbf{A}, \mathbf{B})$ first. The approximation result is given by the following lemma.
\begin{lemma} \label{Lemma_DE_Upsilon}
    Assuming that $\mathbf{A}$ and $\mathbf{B}$ are  deterministic matrices with bounded spectral norms  and the same settings as Lemma \ref{Lemma_DE_Q1_1} hold, then
\begin{equation}
\Upsilon(\mathbf{A}, \mathbf{B}) = \frac{ \underline{s} \gamma \underline{\phi} + \underline{s}\overline{s} \overline{\phi}}{\Delta_{C}}\mathrm{Tr}[\mathbf{A} \mathbf{F}_{\delta} \mathbf{R}_1 \mathbf{G}_{\tau}] \mathrm{Tr} [\mathbf{R}_1\mathbf{F}_{\delta} \mathbf{B}\mathbf{G}_{\tau}] 
+ \mathrm{Tr}[\mathbf{A} \mathbf{F}_{\delta} \mathbf{B} \mathbf{G}_{\tau}] + \mathcal{O}_{z}^s(\frac{1}{N}). \label{Lemma_Eq_Upsilon}
\end{equation}
\end{lemma}
\textit{Proof:} Applying the integration by parts formula \eqref{Integration_by_parts} on variable $[\mathbf{H}_1]_{kl}$, we obtain
\begin{equation}
\begin{split}
\sum_{k}\mathbb{E} [\mathbf{Q}_1 \mathbf{B} \mathbf{Q}_2]_{ik} [\mathbf{H}_1]_{kl} [\mathbf{H}_1]_{jm}^* &= \sum_{k,a, b} \mathbb{E} [[\mathbf{H}_1]_{kl} [\mathbf{H}_1]_{ab}^* ] \mathbb{E} \Big[\frac{\partial [\mathbf{Q}_1 \mathbf{B} \mathbf{Q}_2]_{ik} [\mathbf{H}_1]_{jm}^*}{\partial [\mathbf{H}_1]_{ab}^*} \Big] \\
& = \sum_{k,a, b}\frac{[\mathbf{R}_1]_{ka}[\mathbf{T}_1]_{bl}}{L} \mathbb{E}\Big\{ -([\mathbf{Q}_1 \mathbf{H}_1 \mathbf{H}_2 \mathbf{H}_2^H]_{ib} + \overline{s} [\mathbf{Q}_1 \mathbf{H}_1]_{ib}) [\mathbf{Q}_1 \mathbf{B} \mathbf{Q}_2]_{ak} [\mathbf{H}_1]_{jm}^*  \\
& - \underline{s} [\mathbf{Q}_1 \mathbf{B} \mathbf{Q}_2 \mathbf{H}_1]_{ib} [\mathbf{Q}_{2}]_{ak} [\mathbf{H}_1]_{jm}^* + [\mathbf{Q}_1 \mathbf{B} \mathbf{Q}_2]_{ik} \delta(j - a) \delta(m - b) \Big\}. \\
\end{split}
\end{equation} 
Therefore, by writing $\widehat{\beta}_{\tau} = {\beta}_{\tau} + \mathring{\beta}_{\tau}$, we derive the following equation
\begin{equation}
    \begin{split}
 \mathbb{E} [\mathbf{Q}_1 \mathbf{B} \mathbf{Q}_2 \mathbf{H}_1 \mathbf{\Omega}_{\overline{\tau}}^{-1}]_{il}  [\mathbf{H}_1]_{jm}^*  &=  \mathbb{E}\Big\{ -\frac{1}{L}\widehat{\Upsilon}(\mathbf{R}_1, \mathbf{B})([\mathbf{Q}_1 \mathbf{H}_1 \mathbf{H}_2 \mathbf{H}_2^H \mathbf{T}_1]_{il} + \overline{s} [\mathbf{Q}_1 \mathbf{H}_1 \mathbf{T}_1]_{il})  [\mathbf{H}_1]_{jm}^* \\
& - \underline{s} \mathring{\beta}_{\tau}[\mathbf{Q}_1 \mathbf{B} \mathbf{Q}_2 \mathbf{H}_1 \mathbf{T}_1 ]_{il} [\mathbf{H}_1]_{jm}^* + \frac{ [\mathbf{T}_{1}]_{ml}}{L}[\mathbf{Q}_1 \mathbf{B} \mathbf{Q}_2 \mathbf{R}_1]_{ij} \Big\}. \label{DE_Psi_7_eq_1}
\end{split}
\end{equation} 
By multiplying $[\mathbf{\Omega}_{\overline{\tau}}]_{lm}$ on both sides of above equation and summing over subscripts $l$ and $m$, we have
\begin{equation}
    \begin{split}
 \mathbb{E} [\mathbf{Q}_1 \mathbf{B} \mathbf{Q}_2 \mathbf{H}_1 \mathbf{H}_1^H]_{ij} &=  \mathbb{E}\Big\{ -\frac{1}{L}\widehat{\Upsilon}(\mathbf{R}_1, \mathbf{B})([\mathbf{Q}_1 \mathbf{H}_1 \mathbf{H}_2 \mathbf{H}_2^H \mathbf{T}_1\mathbf{\Omega}_{\overline{\tau}}\mathbf{H}_1^H]_{ij} + \overline{s} [\mathbf{Q}_1 \mathbf{H}_1 \mathbf{T}_1\mathbf{\Omega}_{\overline{\tau}}\mathbf{H}_1^H]_{ij}) \\
& - \underline{s} \mathring{\beta}_{\tau}[\mathbf{Q}_1 \mathbf{B} \mathbf{Q}_2 \mathbf{H}_1 \mathbf{T}_1\mathbf{\Omega}_{\overline{\tau}}\mathbf{H}_1^H]_{ij}  +  \beta_{\overline{\tau}}[\mathbf{Q}_1 \mathbf{B} \mathbf{Q}_2 \mathbf{R}_1]_{ij} \Big\}. 
\end{split}
\end{equation} 
Then, we use the resolvent identity of $\mathbf{Q}_2$ \eqref{Resolvent_identity_Q2}
on the term $\mathbb{E} [\mathbf{Q}_1 \mathbf{B} \mathbf{Q}_2 \mathbf{H}_1 \mathbf{H}_1^H]_{ij} $ and multiply  $[\mathbf{\Omega}_{\tau}\mathbf{A}]_{ji}$ on both sides of the above equation. Then, we obtain the following equation by summing over subscripts $j$ and $i$
\begin{equation}
    \begin{split}
& \Upsilon(\mathbf{A}, \mathbf{B}) = \mathbb{E}\Big\{ \frac{\underline{s} \widehat{\Upsilon}(\mathbf{R}_1,\mathbf{B}) }{L} \big( \widehat{\Psi}_2(\mathbf{\Omega}_{\tau}\mathbf{A},\mathbf{I}_M, \mathbf{T}_1 \mathbf{\Omega}_{\overline{\tau}})  +  \overline{s}\widehat{\Psi}_1(\mathbf{\Omega}_{\tau}\mathbf{A}, \mathbf{T}_1 \mathbf{\Omega}_{\overline{\tau}}) \big) + \underline{s}^2 \mathring{\beta}_{\tau}\widehat{\Psi}_7(\mathbf{\Omega}_{\tau}\mathbf{A}, \mathbf{T}_1 \mathbf{\Omega}_{\overline{\tau}}) + \mathrm{Tr} [\mathbf{Q}_1 \mathbf{B}\mathbf{\Omega}_{\tau}\mathbf{A}] \Big\} \\
&\overset{(a)}{=} \frac{\underline{s}{\Upsilon}(\mathbf{R}_1, \mathbf{B})}{L} \gamma \underline{\phi} \mathrm{Tr}[\mathbf{A}\mathbf{F}_{\delta} \mathbf{R}_1 \mathbf{G}_{\tau}] +\frac{\overline{s}\underline{s}{\Upsilon}(\mathbf{R}_1,\mathbf{B})}{L} \overline{\phi}\mathrm{Tr}[\mathbf{A}\mathbf{F}_{\delta} \mathbf{R}_1 \mathbf{G}_{\tau}] +  \mathrm{Tr}[\mathbf{A} \mathbf{F}_{\delta} \mathbf{B} \mathbf{G}_{\tau}] + \varepsilon_{\Upsilon}(\mathbf{A}, \mathbf{B}) + \mathcal{O}_{z}^s(\frac{1}{N}), \label{DE_Psi_7_eq_2}
\end{split}
\end{equation} 
where in step $(a)$, we write all the r.v.s $\widehat{\Psi}_j$, $\widehat{\Upsilon}$ as the sum of their means and centered versions.  Furthermore, we apply the approximation rules \eqref{DE_Psi_1} to the term ${\Psi}_1(\mathbf{\Omega}_{\tau}\mathbf{A}, \mathbf{T}_1 \mathbf{\Omega}_{\overline{\tau}})$, \eqref{DE_Psi_2} to the term  ${\Psi}_2(\mathbf{\Omega}_{\tau}\mathbf{A},\mathbf{I}_M, \mathbf{T}_1 \mathbf{\Omega}_{\overline{\tau}})$, and Lemma \ref{Lemma_DE_Q1_1} and Lemma \ref{Lemma_DE_Q1_2} to the term $ \mathrm{Tr}[\mathbf{A} \mathbb{E}\mathbf{Q}_{1} \mathbf{B} \mathbf{\Omega}_{\tau}]$. We also use the approximation rules \eqref{Approx_rules_Omega} to replace $\mathbf{\Omega}_{sym}$ with $\mathbf{G}_{sym}$, where $sym$ can be either $\tau$ or $\overline{\tau}$.
The error term is given by
\begin{equation}
    \varepsilon_{\Upsilon}(\mathbf{A}, \mathbf{B})
    = \mathbb{E}\Big\{ \frac{\underline{s}}{L}  \mathring{\Upsilon}\mathring{\Psi}_2 + \frac{\underline{s}\overline{s}}{L}  \mathring{\Upsilon} \mathring{\Psi}_1 +  \underline{s}^2 \mathring{\beta}_{\tau}\mathring{\Psi}_7\Big\}.
\end{equation}
We omit variables in functions $\mathring{\Psi}_j$ and $\mathring{\Upsilon}$ when it does not cause ambiguity. 
Following the same steps as in Proposition \ref{Prop_variance}, we can prove that $\Var(\widehat{\Upsilon}(\mathbf{A}, \mathbf{B})) = \mathcal{O}_z^s(z^{-4})$. According to variance control result \eqref{Var_control} and \eqref{Prop_variance_3_Trace}
and Cauchy–Schwarz inequality $\mathbb{E}{|xy|} \leq \sqrt{\mathbb{E}{|x|^2}\mathbb{E}{|y|^2}}$,
we have $\varepsilon_{\Upsilon}(\mathbf{A}, \mathbf{B}) = \mathcal{O}_{z}^s(\frac{1}{N})$.  Letting $\mathbf{A} = \mathbf{R}_1$, 
we have 
\begin{equation}
\Upsilon(\mathbf{R}_1, \mathbf{B}) = \frac{\mathrm{Tr} [\mathbf{R}_1\mathbf{F}_{\delta} \mathbf{B}\mathbf{G}_{\tau}] }{\Delta_{C}}+ \mathcal{O}_{z}^s(\frac{1}{N}).
\end{equation}
Then, by substituting the above equation back into \eqref{DE_Psi_7_eq_2}, the equation \eqref{Lemma_Eq_Upsilon} holds.
\QED
\par
Let $\mathbf{B} = \mathbf{R}_1$ in \eqref{DE_Psi_7_eq_1}. Then, we multiply $[\mathbf{A}]_{ji}$ and $[\mathbf{\Omega}_{\overline{\tau}}\mathbf{B}]_{lm}$ on both sides of \eqref{DE_Psi_7_eq_1} and sum over
the subscripts $i, j, l$, and $ m$, resulting in the following equation 
\begin{equation}
\begin{split}
    & \Psi_7(\mathbf{A}, \mathbf{B}) =  \mathbb{E}\Big\{-\frac{1}{L}\widehat{\Upsilon}(\mathbf{R}_1, \mathbf{R}_1)[\widehat\Psi_2(\mathbf{A}, \mathbf{I}_{M}, \mathbf{T}_1 \mathbf{\Omega}_{\overline{\tau}} \mathbf{B}) +  \overline{s} \widehat\Psi_1(\mathbf{A},  \mathbf{T}_1 \mathbf{\Omega}_{\overline{\tau}} \mathbf{B})]  - \underline{s} \mathring{\beta}_{\tau} \widehat\Psi_7(\mathbf{A},  \mathbf{T}_1 \mathbf{\Omega}_{\overline{\tau}} \mathbf{B})\Big\} \notag \\
    & + \frac{\mathrm{Tr} [\mathbf{B}\mathbf{T}_{1}\mathbf{\Omega}_{\overline{\tau}}]}{L} {\Upsilon}(\mathbf{R}_1 \mathbf{A},\mathbf{R}_1) \overset{(a)}{=} - \frac{\vartheta}{\Delta_C}   \frac{1}{L} \mathrm{Tr}[(\overline{s} \mathbf{I}_L + \gamma\mathbf{R}_2) \mathbf{T}_1 \mathbf{G}_{\overline{\tau}} \mathbf{B} \mathbf{T}_1 \mathbf{F}_{\omega}] \mathrm{Tr}[\mathbf{A} \mathbf{F}_{\delta} \mathbf{R}_1]\notag \\
   & + \frac{1}{\Delta_C}\frac{1}{L}\mathrm{Tr} [\mathbf{B}\mathbf{T}_{1}\mathbf{G}_{\overline{\tau}}] \mathrm{Tr}[\mathbf{R}_1 \mathbf{A} \mathbf{F}_{\delta} \mathbf{R}_1 \mathbf{G}_{\tau}] + \varepsilon_{\Upsilon}(\mathbf{A}, \mathbf{B}) + \mathcal{O}_{z}^s(\frac{1}{N}).
\end{split}
\end{equation}
Here, step $(a)$ is similar to step $(a)$ in \eqref{DE_Psi_7_eq_2}, where we rewrite the r.v.s $\widehat{\Psi}_j = \Psi_j + \mathring{\Psi}_j$, $\widehat{\Upsilon} = \Upsilon + \mathring{\Upsilon}$ and calculate $\Upsilon(\mathbf{R}_1, \mathbf{R}_1)$ and $\Upsilon(\mathbf{R}_1\mathbf{A}, \mathbf{R}_1)$ by Lemma \ref{Lemma_DE_Upsilon}. 
Moreover, the term $\varepsilon_{\Upsilon}(\mathbf{A}, \mathbf{B}) =  \mathbb{E}[-\frac{1}{L} \mathring{\Upsilon}\mathring\Psi_2 - \frac{s}{L} \mathring{\Upsilon} \mathring \Psi_1 - s \mathring{\beta}_{\tau} \mathring\Psi_7 ]$ can be proved to have the order $\mathcal{O}_{z}^s(\frac{1}{N})$.
This completes the proof of \eqref{DE_Psi_7}. \QED
\section{Proof of Lemma \ref{Lemma_DE_terms_Approx_Phi}}
\label{App_Lemma_DE_terms_Approx_Phi}
In this section, we will use the definitions of some intermediate quantities in \eqref{Random_quantities}, \eqref{Deterministic_quantities}, and \eqref{Q_2_quantities}.
Similar to the discussion in Appendix \ref{App_Lemma_DE_terms_Approx_Psi}, we calculate the deterministic approximations of $\Phi_1$, $\Phi_2$ and 
$\Phi_3$, $\Phi_4$ separately.
\subsection{Evaluation of $\Phi_1$}
According to the integration by part formula \eqref{Integration_by_parts}, we take $[\mathbf{H}_1]_{kl}$ as variable and obtain
\begin{equation}
\begin{split}
    \sum_{k} \mathbb{E}  [\mathbf{Q}_{2}]_{ik} [\mathbf{H}_1]_{kl} [\mathbf{H}_1]_{jm}^*  &= \sum_{k, a,b} \mathbb{E}[\mathbf{H}_1]_{kl} [\mathbf{H}_1]_{ab}^* \mathbb{E}\frac{\partial [\mathbf{Q}_{2}]_{ik}  [\mathbf{H}_1]_{jm}^*}{\partial [\mathbf{H}_1]_{ab}^*} \\
    &=  \sum_{k,a,b} \frac{[\mathbf{R}_1]_{ka}[\mathbf{T}_1]_{bl}}{L}  \mathbb{E}\Big\{ -\underline{s}[\mathbf{Q}_2 \mathbf{H}_1]_{ib} [\mathbf{Q}_{2}]_{ak} [\mathbf{H}_1]_{jm}^* +  [\mathbf{Q}_{2}]_{ik} \delta(j-a) \delta(m - b) \Big\}.
\end{split}
\end{equation}
Therefore, the following holds
\begin{equation}
    \mathbb{E}  [\mathbf{Q}_2 \mathbf{H}_1]_{il} [\mathbf{H}_1]_{jm}^* =   -\mathbb{E} \underline{s} \widehat{\beta}_{\tau}[\mathbf{Q}_2 \mathbf{H}_1 \mathbf{T}_1]_{il}  [\mathbf{H}_1]_{jm}^* +   \frac{1}{L} [\mathbf{T}_1]_{ml} \mathbb{E}[\mathbf{Q}_2 \mathbf{R}_1]_{ij}.
\end{equation}
By splitting $\widehat{\beta}_{\tau}$ into the sum of ${\beta}_{\tau}$ and $\mathring{\beta}_{\tau}$, we have
\begin{equation}
    \mathbb{E}  [\mathbf{Q}_2 \mathbf{H}_1\mathbf{\Omega}_{\overline{\tau}}^{-1}]_{il} [\mathbf{H}_1]_{jm}^* =  
     -\mathbb{E} \underline{s} \mathring{\beta}_{\tau}[\mathbf{Q}_2 \mathbf{H}_1 \mathbf{T}_1]_{il}  [\mathbf{H}_1]_{jm}^* +  
      \frac{1}{L} [\mathbf{T}_1]_{ml} \mathbb{E}[\mathbf{Q}_2 \mathbf{R}_1]_{ij}. 
\end{equation}
By multiplying $[\mathbf{\Omega}_{\overline{\tau}} \mathbf{B}]_{lm}$ and $[\mathbf{A}]_{ji}$ on both sides of the above equation and summing over $i$, $j$, $l$, and $m$, the following equation is obtained
\begin{equation}
    \begin{split}
    \Phi_1(\mathbf{A}, \mathbf{B}) 
    &=     -\mathbb{E} \underline{s} \mathring{\beta}_{\tau}\widehat{\Phi}_1(\mathbf{A},  \mathbf{T}_1 \mathbf{\Omega}_{\overline{\tau}} \mathbf{B}) +   \frac{1}{L} \mathrm{Tr} [\mathbf{T}_1\mathbf{\Omega}_{\overline{\tau}} \mathbf{B}]   \mathrm{Tr}[\mathbb{E}\mathbf{Q}_2 \mathbf{R}_1\mathbf{A}] \\
    &=   \frac{1}{L} \mathrm{Tr} [\mathbf{T}_1\mathbf{G}_{\overline{\tau}} \mathbf{B}] 
     \mathrm{Tr}[\mathbf{A}\mathbf{G}_{\tau} \mathbf{R}_1] + \varepsilon_{\Phi_1}(\mathbf{A}, \mathbf{B}) + \mathcal{O}_{z}^s(\frac{1}{N}),
    \label{Phi_1_error_term_1}
\end{split}
\end{equation}
where we use the approximation rules of $\mathbf{\Omega}_{\overline{\tau}}$ and $\mathbf{\Omega}_{\tau}$ in \eqref{Approx_rules_Omega}. 
The term $\varepsilon_{\Phi_1}(\mathbf{A}, \mathbf{B})$ is defined as 
$-\mathbb{E} \underline{s} \mathring{\beta}_{\tau}\widehat{\Phi}_1(\mathbf{A},  \mathbf{T}_1 \mathbf{\Omega}_{\overline{\tau}} \mathbf{B}) = -\mathbb{E} \underline{s} \mathring{\beta}_{\tau}\mathring{\Phi}_1(\mathbf{A},  \mathbf{T}_1 \mathbf{\Omega}_{\overline{\tau}} \mathbf{B})$ which can be proved to have the order of $\mathcal{O}_{z}^s(\frac{1}{N})$ by using the 
Cauchy–Schwarz inequality and variance control results \eqref{variance_control_beta_tau} and \eqref{Var_control}.
\subsection{Evaluation of $\Phi_2$}
To evaluate $\Phi_2$, we first define the random quantity $\widehat{\eta}(\mathbf{A}) =  \mathrm{Tr}[\mathbf{A}\mathbf{Q}_{2}\mathbf{R}_1 \mathbf{Q}_{2} ]$ and determine the deterministic approximation of 
its mean ${\eta}(\mathbf{A}) = \mathbb{E} \widehat{{\eta}}(\mathbf{A})$. We provide the result in the following lemma.
\begin{lemma}
    Assuming that $\mathbf{A}$ is a deterministic matrix with bounded spectral norm, and the same settings as Lemma \ref{Lemma_DE_Q1_1} hold, then we have
\begin{equation}
\eta(\mathbf{A}) = \frac{1}{\Delta_{V_2}}\mathrm{Tr}[\mathbf{A}\mathbf{G}_{\tau}\mathbf{R}_1\mathbf{G}_{\tau} ] + \mathcal{O}_{z}^s(\frac{1}{N}). \label{DE_eta}
\end{equation}
\end{lemma}
\textit{Proof:} By applying the integration by parts formula \eqref{Integration_by_parts} on $[\mathbf{H}_1]_{kl}$ and taking
 $f =  [\mathbf{Q}_{2}\mathbf{R}_1 \mathbf{Q}_{2}]_{ik} [\mathbf{H}_1]_{jm}^*$, we obtain
\begin{equation}
\begin{split}
    &\mathbb{E}  [\mathbf{Q}_{2}\mathbf{R}_1 \mathbf{Q}_{2} \mathbf{H}_1]_{il} [\mathbf{H}_1]_{jm}^* = \sum_{k} \mathbb{E}  [\mathbf{Q}_{2}\mathbf{R}_1 \mathbf{Q}_{2}]_{ik} [\mathbf{H}_1]_{kl} [\mathbf{H}_1]_{jm}^* =  \sum_{k, a,b} \frac{[\mathbf{R}_1]_{ka}[\mathbf{T}_1]_{bl}}{L}  \mathbb{E}\Big\{ [\mathbf{Q}_{2}\mathbf{R}_1 \mathbf{Q}_{2}]_{ik}\delta(j-a)\delta(m-b) \\
     &\hspace*{5mm}- \underline{s}[\mathbf{Q}_2 \mathbf{H}_1]_{ib} [\mathbf{Q}_2 \mathbf{R}_1 \mathbf{Q}_2]_{ak} [\mathbf{H}_1]_{jm}^*   -\underline{s}[\mathbf{Q}_{2}\mathbf{R}_1\mathbf{Q}_2 \mathbf{H}_1]_{ib} [\mathbf{Q}_2]_{ak} [\mathbf{H}_1]_{jm}^*\Big\}.\\
    &\hspace*{5mm}= \mathbb{E}\Big\{ - \frac{\underline{s}}{L} \widehat{\eta}(\mathbf{R}_1)[\mathbf{Q}_2 \mathbf{H}_1 \mathbf{T}_1]_{il}  [\mathbf{H}_1]_{jm}^* 
    - \underline{s} \widehat{\beta}_{\tau} [\mathbf{Q}_{2}\mathbf{R}_1\mathbf{Q}_2 \mathbf{H}_1 \mathbf{T}_1]_{il}[\mathbf{H}_1]_{jm}^* +  \frac{1}{L} [\mathbf{T}_1]_{ml} [\mathbf{Q}_{2}\mathbf{R}_1 \mathbf{Q}_{2} \mathbf{R}_1 ]_{ij} \Big\}. \label{Phi_2_ij_Eq_1}
\end{split}
\end{equation}
By multiplying $[\mathbf{\Omega}_{\overline{\tau}}]_{lm}$ on both sides of the above equation and writing $\widehat{\beta}_{\tau} = \beta_{\tau} + \mathring{\beta}_{\tau}$, 
the following is obtained by summing over $l$ and $m$
\begin{equation}
\begin{split}
    \mathbb{E}  [\mathbf{Q}_{2}\mathbf{R}_1 \mathbf{Q}_{2} \mathbf{H}_1  \mathbf{H}_1^H]_{ij} &= \mathbb{E}\Big\{ - \frac{\underline{s}}{L} \widehat{\eta}(\mathbf{R}_1)[\mathbf{Q}_2 \mathbf{H}_1 \mathbf{T}_1 \mathbf{\Omega}_{\overline{\tau}} \mathbf{H}_1^H]_{ij}  \\
    & - \underline{s} \mathring{\beta}_{\tau} [\mathbf{Q}_{2}\mathbf{R}_1\mathbf{Q}_2 \mathbf{H}_1 \mathbf{T}_1\mathbf{\Omega}_{\overline{\tau}}\mathbf{H}_1^H]_{ij}  +  \beta_{\overline{\tau}} [\mathbf{Q}_{2}\mathbf{R}_1 \mathbf{Q}_{2} \mathbf{R}_1 ]_{ij} \Big\}.
\label{App_DE_Phi_2_Eq_1}
\end{split}
\end{equation}
By using the resolvent identity of $\mathbf{Q}_2$ \eqref{Resolvent_identity_Q2} on the term $\mathbb{E}  [\mathbf{Q}_{2}\mathbf{R}_1 \mathbf{Q}_{2} \mathbf{H}_1  \mathbf{H}_1^H]_{ij}$ 
of \eqref{App_DE_Phi_2_Eq_1}, multiplying both sides by $[\mathbf{\Omega}_{\tau} \mathbf{A}]_{ji}$, and then summing over subscripts $i$ and $j$, 
the following holds
\begin{equation}
\begin{split}
\eta(\mathbf{A}) &= \mathbb{E}\Big\{ \frac{\underline{s}^2}{L} \widehat{\eta}(\mathbf{R}_1) \widehat{\Phi}_1(\mathbf{\Omega}_{\tau}\mathbf{A}, \mathbf{T}_1 \mathbf{\Omega}_{\overline{\tau}})  + \underline{s}^2 \mathring{\beta}_{\tau} \widehat{\Phi}_2(\mathbf{\Omega}_{\tau}\mathbf{A}, \mathbf{T}_1 \mathbf{\Omega}_{\overline{\tau}})  +  \mathrm{Tr}[\mathbf{Q}_{2}\mathbf{R}_1\mathbf{\Omega}_{\tau} \mathbf{A}]\Big\} \\
& \overset{(a)}{=} \frac{\underline{s}^2}{L} {\eta}(\mathbf{R}_1) {\Phi}_1(\mathbf{\Omega}_{\tau}\mathbf{A}, \mathbf{T}_1 \mathbf{\Omega}_{\overline{\tau}})  +  \mathrm{Tr}[\mathbb{E}\mathbf{Q}_{2}\mathbf{R}_1\mathbf{\Omega}_{\tau} \mathbf{A}] +  \mathbb{E}\Big\{\frac{\underline{s}^2}{L} \mathring{\eta}(\mathbf{R}_1) \mathring{\Phi}_1(\mathbf{\Omega}_{\tau}\mathbf{A}, \mathbf{T}_1 \mathbf{\Omega}_{\overline{\tau}}) + \underline{s}^2 \mathring{\beta}_{\tau} \mathring{\Phi}_2(\mathbf{\Omega}_{\tau}\mathbf{A}, \mathbf{T}_1 \mathbf{\Omega}_{\overline{\tau}}) \Big\} \\    
& \overset{(b)}{=} \frac{\underline{s}^2}{L} {\eta}(\mathbf{R}_1) \overline{\tau}_2 \mathrm{Tr}[\mathbf{A} \mathbf{G}_{\tau} \mathbf{R}_1 \mathbf{G}_{\tau}]  +   \mathrm{Tr}[\mathbf{A}\mathbf{G}_{\tau}\mathbf{R}_1\mathbf{G}_{\tau} ] + \varepsilon_{\eta}(\mathbf{A}) + \mathcal{O}_{z}^s(\frac{1}{N}). \label{eta_Eq_1}
\end{split}
\end{equation}
Here, step $(a)$ follows by writing the r.v.s
$\widehat{\eta}$, $\widehat{\Phi}_1$, and $\widehat{\Phi}_2 $ as $\eta + \mathring{\eta}$, $ \Phi_1 + \mathring{\Phi}_1$, and $ \Phi_2 + \mathring{\Phi}_2$,
respectively. In step $(b)$, we use the approximation rules \eqref{DE_Phi_1} 
and \eqref{Approx_rules_Omega}. 
Using the Poincaré-Nash inequality as in Proposition \ref{Prop_variance}, we can prove that the variance of $\widehat{\eta}(\mathbf{A})$ is $\mathcal{O}_z^s(z^{-4})$.
Following the same discussion as for  $\varepsilon_{\Phi_1}$ in  \eqref{Phi_1_error_term_1}, we can obtain that $\varepsilon_{\eta}(\mathbf{A}) = \mathcal{O}_z^s(\frac{1}{N})$.
\par
By setting $\mathbf{A} = \mathbf{R}_1$ in \eqref{eta_Eq_1}, we have $\eta(\mathbf{R}_1) = \frac{L \tau_2}{\Delta_{V_2}} + \mathcal{O}_z^s(\frac{1}{N})$. Finally, \eqref{DE_eta} can be obtained by substituting this result back into the RHS of \eqref{eta_Eq_1}. \QED
\par
By multiplying $[\mathbf{\Omega}_{\overline{\tau}}\mathbf{B}]_{lm}$, $[\mathbf{A}]_{ji}$ on both sides of \eqref{Phi_2_ij_Eq_1} and writing $\widehat{\beta}_{\tau} = \beta_{\tau} + \mathring{\beta}_{\tau}$, 
the following is obtained by summing over all subscripts
\begin{equation}
\begin{split}
    \Phi_2(\mathbf{A}, \mathbf{B}) &= \mathbb{E}\Big\{- \frac{\underline{s}}{L} \widehat{\eta}(\mathbf{R}_1) \widehat{\Phi}_1(\mathbf{A}, \mathbf{T}_1 \mathbf{\Omega}_{\overline{\tau}} \mathbf{B}) - \underline{s} \mathring{\beta}_{\tau} \widehat{\Phi}_2(\mathbf{A}, \mathbf{T}_1 \mathbf{\Omega}_{\overline{\tau}} \mathbf{B}) \Big\}
     +  \frac{1}{L} \mathrm{Tr}[\mathbf{T}_1\mathbf{\Omega}_{\overline{\tau}} \mathbf{B}] \eta(\mathbf{R}_1 \mathbf{A}) \\
    &\overset{(a)}{=} -\frac{ \underline{s} \tau_2}{\Delta_{V_2}} \frac{1}{L} \mathrm{Tr}[\mathbf{T}_1 \mathbf{G}_{\overline{\tau}}\mathbf{T}_1
     \mathbf{G}_{\overline{\tau}}\mathbf{B}] \mathrm{Tr}[\mathbf{A} \mathbf{G}_{\tau} \mathbf{R}_1] 
     + \frac{1}{\Delta_{V_2}}\mathrm{Tr}[\mathbf{T}_1 \mathbf{G}_{\overline{\tau}}\mathbf{B}] \mathrm{Tr}[\mathbf{A} \mathbf{G}_{\tau} \mathbf{R}_1\mathbf{G}_{\tau} \mathbf{R}_1]
     + \mathcal{O}_{z}^s(\frac{1}{N}),
\end{split}
\end{equation}
where step $(a)$ follows the similar argument in \eqref{eta_Eq_1}. This completes the proof of \eqref{DE_Phi_2}. 
\subsection{Evaluation of $\Phi_3$ and $\Phi_4$}
By applying the integration by parts formula \eqref{Integration_by_parts} on variable $[\mathbf{H}_2]_{kl}$, we have 
\begin{align}
    \sum_{k} \mathbb{E} [\mathbf{Q}_2 \mathbf{R}_1 \mathbf{Q}_1 \mathbf{H}_1]_{ik} [\mathbf{H}_2]_{kl} [\mathbf{H}_{2}]_{ms}^{*}[\mathbf{H}_{1}]_{jp}^{*} 
    &= \sum_{k, a, b}\frac{[\mathbf{R}_2]_{ka}[\mathbf{T}_2]_{bl}}{M} \mathbb{E}\Big\{ -[\mathbf{Q}_2 \mathbf{R}_1\mathbf{Q}_1\mathbf{H}_1\mathbf{H}_2]_{ib} [\mathbf{H}_1^H \mathbf{Q}_1 \mathbf{H}_1]_{ak}[\mathbf{H}_{2}]_{ms}^{*}[\mathbf{H}_{1}]_{jp}^{*}\notag \\
    & + [\mathbf{Q}_2 \mathbf{R}_1 \mathbf{Q}_1\mathbf{H}_1]_{ik} \delta(m-a) \delta(s-b) [\mathbf{H}_{1}]_{jp}^{*}\Big\}. \label{Phi_34_Eq_t_1}
\end{align}
Consequently, the following holds
\begin{equation}
\begin{split}
    \mathbb{E} [\mathbf{Q}_2 \mathbf{R}_1 \mathbf{Q}_1 \mathbf{H}_1 \mathbf{H}_2 \mathbf{\Theta}_{\gamma}^{-1}]_{il}[\mathbf{H}_{2}]_{ms}^{*} [\mathbf{H}_{1}]_{jp}^{*}  &=  \mathbb{E}\Big\{- \mathring{\kappa}[\mathbf{Q}_2 \mathbf{R}_1\mathbf{Q}_1\mathbf{H}_1\mathbf{H}_2 \mathbf{T}_2]_{il} [\mathbf{H}_{2}]_{ms}^{*}[\mathbf{H}_{1}]_{jp}^{*} \\ 
    &+ \frac{[\mathbf{T}_2]_{sl}}{M} [\mathbf{Q}_2 \mathbf{R}_1 \mathbf{Q}_1 \mathbf{H}_1\mathbf{R}_2 ]_{im} [\mathbf{H}_{1}]_{jp}^{*}\Big\}. \label{Phi_34_Eq_t_3}
\end{split}
\end{equation}
By multiplying $[\mathbf{\Theta}_{\gamma}]_{ls}$ and $[\mathbf{T}_1]_{mq}$ to both sides of the above equation and summing over $l$, $s$, and $m$, we have
\begin{equation}
\begin{split}
    \mathbb{E} [\mathbf{Q}_2 \mathbf{R}_1 \mathbf{Q}_1 \mathbf{H}_1 \mathbf{H}_2 \mathbf{H}_{2}^H \mathbf{T}_1]_{iq} [\mathbf{H}_{1}]_{jp}^{*} & =  \mathbb{E}\Big\{- \mathring{\kappa}[\mathbf{Q}_2 \mathbf{R}_1\mathbf{Q}_1\mathbf{H}_1\mathbf{H}_2 \mathbf{T}_2\mathbf{\Theta}_{\gamma} \mathbf{H}_{2}^H \mathbf{T}_1]_{iq}[\mathbf{H}_{1}]_{jp}^{*} \\
    & + \alpha_{\gamma} [\mathbf{Q}_2 \mathbf{R}_1 \mathbf{Q}_1 \mathbf{H}_1\mathbf{R}_2 \mathbf{T}_1]_{iq} [\mathbf{H}_{1}]_{jp}^{*}\Big\}. \label{Phi_34_Eq_t_2}
\end{split}
\end{equation}
Using the same steps as in \eqref{Phi_34_Eq_t_1}-\eqref{Phi_34_Eq_t_2}, we obtain
\begin{equation}
    \begin{split}
    & \mathbb{E} [\mathbf{Q}_2 \mathbf{R}_1 \mathbf{Q}_1 \mathbf{H}_1]_{iq} [\mathbf{H}_1]_{jp}^* = 
    \sum_{k}\mathbb{E} [\mathbf{Q}_2 \mathbf{R}_1 \mathbf{Q}_1]_{ik} [\mathbf{H}_1]_{kq} [\mathbf{H}_1]_{jp}^* 
    = \mathbb{E}\Big\{ -\frac{\underline{s}}{L}\widehat{\Upsilon}(\mathbf{R}_1,\mathbf{R}_1)[\mathbf{Q}_2 \mathbf{H}_1 \mathbf{T}_1]_{iq}  [\mathbf{H}_1]_{jp}^* \\
    &-  \widehat{\alpha}_{\delta}[\mathbf{Q}_2 \mathbf{R}_1 \mathbf{Q}_1 \mathbf{H}_1 \mathbf{H}_2\mathbf{H}_2^H \mathbf{T}_1]_{iq} [\mathbf{H}_1]_{jp}^* -   
    \overline{s}\widehat{\alpha}_{\delta}[\mathbf{Q}_2 \mathbf{R}_1 \mathbf{Q}_1 \mathbf{H}_1 \mathbf{T}_1]_{iq} [\mathbf{H}_1]_{jp}^* + \frac{[\mathbf{T}_1]_{pq}}{L} [\mathbf{Q}_2 \mathbf{R}_1 \mathbf{Q}_1 \mathbf{R}_1]_{ij} \Big\}.
\end{split}
\end{equation}
Expressing 
 $\widehat{\alpha}_{\delta} = \alpha_{\delta} + \mathring{\alpha}_{\delta}$ and substituting \eqref{Phi_34_Eq_t_2} into the 
 term $ -{\alpha}_{\delta}\mathbb{E} [\mathbf{Q}_2 \mathbf{R}_1 \mathbf{Q}_1 \mathbf{H}_1 \mathbf{H}_2\mathbf{H}_2^H \mathbf{T}_1]_{iq} [\mathbf{H}_1]_{jp}^*$ in the second line of above equation, we have
 \begin{equation}
    \begin{split}
    & \mathbb{E} [\mathbf{Q}_2 \mathbf{R}_1 \mathbf{Q}_1 \mathbf{H}_1 \mathbf{\Theta}_{\omega}^{-1}]_{iq} [\mathbf{H}_1]_{jp}^*
    =  \mathbb{E}\Big\{-\frac{\underline{s}}{L}\widehat{\Upsilon}(\mathbf{R}_1,\mathbf{R}_1)[\mathbf{Q}_2 \mathbf{H}_1 \mathbf{T}_1]_{iq}  [\mathbf{H}_1]_{jp}^*  - \mathring{\alpha}_{\delta}[\mathbf{Q}_2 \mathbf{R}_1 \mathbf{Q}_1 \mathbf{H}_1 \mathbf{H}_2\mathbf{H}_2^H \mathbf{T}_1]_{iq} [\mathbf{H}_1]_{jp}^* \\
    &+  \alpha_{\delta}\mathring{\kappa}[\mathbf{Q}_2 \mathbf{R}_1\mathbf{Q}_1\mathbf{H}_1\mathbf{H}_2 \mathbf{T}_2\mathbf{\Theta}_{\gamma} \mathbf{H}_2^H \mathbf{T}_1]_{iq} [\mathbf{H}_1]_{jp}^* -   \overline{s} \mathring{\alpha}_{\delta}[\mathbf{Q}_2 \mathbf{R}_1 \mathbf{Q}_1 \mathbf{H}_1 \mathbf{T}_1]_{iq} [\mathbf{H}_1]_{jp}^* +\frac{[\mathbf{T}_1]_{pq}}{L} [\mathbf{Q}_2 \mathbf{R}_1 \mathbf{Q}_1 \mathbf{R}_1]_{ij} \Big\}.
\end{split}
\end{equation}
By multiplying $[\mathbf{\Theta}_{\omega}\mathbf{B}]_{qp}$, $[\mathbf{A}]_{ji}$  on both sides of the above equation and summing over the subscripts $p$, $q$, $i$, and $j$, the following holds
\begin{equation}
    \begin{split}
     \Phi_3(\mathbf{A}, \mathbf{B})
    & = \mathbb{E}\Big\{ -\frac{\underline{s}}{L}\widehat{\Upsilon}(\mathbf{R}_1,\mathbf{R}_1)\widehat{\Phi}_1(\mathbf{A}, \mathbf{T}_1\mathbf{\Theta}_{\omega}\mathbf{B}) -  \mathring{\alpha}_{\delta}\widehat{\Phi}_4(\mathbf{A}, \mathbf{I}_M, \mathbf{T}_1\mathbf{\Theta}_{\omega}\mathbf{B}) +  \alpha_{\delta}  \mathring{\kappa}\widehat{\Phi}_4(\mathbf{A}, \mathbf{T}_2 \mathbf{\Theta}_{\gamma}, \mathbf{T}_1\mathbf{\Theta}_{\omega}\mathbf{B})\\
    & - \overline{s} \mathring{\alpha}_{\delta}\widehat{\Phi}_3(\mathbf{A}, \mathbf{T}_1 \mathbf{\Theta}_{\omega} \mathbf{B}) \Big\} + \frac{1}{L}\mathrm{Tr}[\mathbf{T}_1 \mathbf{\Theta}_{\omega}\mathbf{B}] \Upsilon(\mathbf{R}_1, \mathbf{R}_1 \mathbf{A})
     \overset{(a)}{=}
     -\frac{\underline{s} \vartheta}{\Delta_C}\frac{1}{L} \mathrm{Tr}[\mathbf{B} \mathbf{T}_1 \mathbf{G}_{\overline{\tau}} \mathbf{T}_1 \mathbf{F}_{\omega} ] \mathrm{Tr}[\mathbf{A}\mathbf{G}_{\tau} \mathbf{R}_1] \\
    &+ \frac{1}{\Delta_C}\frac{1}{L}\mathrm{Tr}[\mathbf{T}_1 \mathbf{F}_{\omega}\mathbf{B}] \mathrm{Tr}[\mathbf{R}_1 \mathbf{F}_{\delta}\mathbf{R}_1\mathbf{A} \mathbf{G}_{\tau} ]
    + \varepsilon_{\Phi_3}(\mathbf{A}, \mathbf{B}) + \mathcal{O}_{z}^s(\frac{1}{N}),
\end{split}
\end{equation}
where step $(a)$ is obtained by expressing the r.v.s $\widehat{\Phi}_j$ and $\widehat{\Upsilon}$ as the sum of their 
means ${\Phi}_j$ and ${\Upsilon}$ and the centralized versions $\mathring{\Phi}_j$ and $\mathring{\Upsilon}$, applying the approximation rule \eqref{DE_Phi_1} to $\Phi_1(\mathbf{A}, \mathbf{T}_1\mathbf{\Theta}_{\omega}\mathbf{B})$, Lemma \ref{Lemma_DE_Upsilon} to 
$\Upsilon(\mathbf{R}_1, \mathbf{R}_1)$ and $\Upsilon(\mathbf{R}_1, \mathbf{R}_1\mathbf{A})$, and replacing $\mathbf{\Theta}_{\omega}$
with $\mathbf{F}_{\omega}$.  Without causing ambiguity, we express 
$
    \varepsilon_{\Phi_3}(\mathbf{A}, \mathbf{B})   
    = \mathbb{E} \{ -\frac{\underline{s}}{L}\mathring{\Upsilon}\mathring{\Phi}_1 -  \mathring{\alpha}_{\delta}\mathring{\Phi}_4 +  \alpha_{\delta} \mathring{\kappa}\mathring{\Phi}_4-
    \overline{s}\mathring{\alpha}_{\delta}\mathring{\Phi}_3 \}
$, which can be proved to have the order  of $\mathcal{O}_z^s(\frac{1}{N})$ by Cauchy–Schwarz inequality. Thus, we obtain \eqref{DE_Phi_3}.
\par
On the other hand, by multiplying $[\mathbf{\Theta}_{\gamma}\mathbf{B}]_{ls}$, $[\mathbf{C}]_{mp}$, $[\mathbf{A}]_{ji}$ on both sides of \eqref{Phi_34_Eq_t_3} and summing over all the subscripts $l$, $s$, $m$, $p$, $j$, and $i$, we obtain
\begin{equation}
    \begin{split}
    \Phi_4(\mathbf{A}, \mathbf{B}, \mathbf{C}) 
    &=  -\mathbb{E}\mathring{\kappa}\widehat{\Phi}_4(\mathbf{A}, \mathbf{T}_2 \mathbf{\Theta}_{\gamma} \mathbf{B}, \mathbf{C}) + \frac{\mathrm{Tr}[\mathbf{T}_{2} \mathbf{\Theta}_{\gamma}\mathbf{B}]}{M} \Phi_3(\mathbf{A}, \mathbf{R}_2 \mathbf{C}) \\
    &\overset{(a)}{=}  \frac{\mathrm{Tr}[\mathbf{T}_{2} \mathbf{F}_{\gamma}\mathbf{B}]}{M} \Phi_3(\mathbf{A}, \mathbf{R}_2 \mathbf{C})  +  \varepsilon_{\Phi_4}(\mathbf{A}, \mathbf{B}) + \mathcal{O}_{z}^s(\frac{1}{N}),
\end{split}
\end{equation}
where $(a)$ follows by replacing $\mathbf{\Theta}_{\gamma}$ with $\mathbf{F}_{\gamma}$. The term $
    \varepsilon_{\Phi_4}(\mathbf{A}, \mathbf{B}) 
    =  -\mathbb{E}\mathring{\kappa}\mathring{\Phi}_4(\mathbf{A}, \mathbf{T}_2 \mathbf{\Theta}_{\gamma} \mathbf{B}, \mathbf{C}) = \mathcal{O}_{z}^s(\frac{1}{N})
$ can be obtained by Cauchy–Schwarz inequality. Hence, the proof of \eqref{DE_Phi_4} is completed. \QED
\section{Proof of Lemma \ref{Lemma_Z_alpha_kappa}}
\label{App_Lemma_Z_alpha_kappa}
We start the proof from the equation \eqref{CLT_Hole}. Applying the resolvent identity \eqref{Resolvent_identity_Q1} at the LHS of the first line of \eqref{CLT_Hole}, we obtain
\begin{equation}
\begin{split}
    \mathbb{E} \delta(i-j) \widehat{\psi} &= \mathbb{E} \widehat{\psi} \Big\{ [\mathbf{Q}_{1} \mathbf{\Theta}_{\delta}^{-1}]_{ij}   - \mathring \alpha_{\delta} [\mathbf{Q}_1 \mathbf{H}_1 \mathbf{H}_2 \mathbf{H}_2^H  \overline{\mathbf{\Theta}}_{\omega} \mathbf{H}_1^H]_{ij} - \overline{s}  \mathring \alpha_{\delta} [\mathbf{Q}_1 \mathbf{H}_1\overline{\mathbf{\Theta}}_{\omega} \mathbf{H}_1^H]_{ij} \\
    &+ \frac{\jmath t_1}{L}   [ \mathbf{Q}_1\mathbf{R}_1 \mathbf{Q}_1 \mathbf{H}_1 \mathbf{H}_2 \mathbf{H}_2^H \overline{\mathbf{\Theta}}_{\omega} \mathbf{H}_1^H]_{ij} + \frac{\jmath \overline{s} t_1}{L}  [\mathbf{Q}_1 \mathbf{R}_1 \mathbf{Q}_1\mathbf{H}_1\overline{\mathbf{\Theta}}_{\omega} \mathbf{H}_1^H]_{ij} + \frac{\jmath \underline{s} t_2}{L} [\mathbf{Q}_1 \mathbf{R}_1 \mathbf{Q}_2\mathbf{H}_1\overline{\mathbf{\Theta}}_{\omega} \mathbf{H}_1^H]_{ij} \\
    &-  \mathring \kappa[\mathbf{Q}_1\mathbf{H}_1 \mathbf{H}_2 \mathbf{T}_2\mathbf{\Theta}_{\gamma} \mathbf{H}_2^H\mathbf{\Theta}_{\omega}^H \mathbf{H}_1^H]_{ij} +  \frac{\jmath t_1}{M} [\mathbf{Q}_1\mathbf{H}_1\mathbf{R}_2 \mathbf{H}_1^H \mathbf{Q}_1 \mathbf{H}_1 \mathbf{H}_2 \mathbf{T}_2\mathbf{\Theta}_{\gamma} \mathbf{H}_2^H \mathbf{\Theta}_{\omega}^H\mathbf{H}_1^H]_{ij} \Big\}.
\end{split}
\end{equation}
By multiplying $[\mathbf{\Theta}_{\delta}\mathbf{R}_1]_{ji}$ on both sides of the above equation and summing over the subscripts $i$ and $j $, we obtain
\begin{equation}
\begin{split}
    & [-L + {\Psi}_2(\mathbf{\Theta}_{\delta}\mathbf{R}_1, \mathbf{I}_M, \overline{\mathbf{\Theta}}_{\omega}) + \overline{s}{\Psi}_1(\mathbf{\Theta}_{\delta}\mathbf{R}_1,  \overline{\mathbf{\Theta}}_{\omega}) ] \mathcal{Z}_{\alpha_{\delta}} 
    + {\Psi}_2(\mathbf{\Theta}_{\delta}\mathbf{R}_1, \mathbf{T}_2\mathbf{\Theta}_{\gamma}, \mathbf{\Theta}_{\omega}^H)\mathcal{Z}_{\kappa} = \Big[ \frac{\jmath t_1}{L} {\Psi}_4(\mathbf{\Theta}_{\delta}\mathbf{R}_1, \mathbf{R}_1, \mathbf{I}_M,\overline{\mathbf{\Theta}}_{\omega}) \\
    & + \frac{\jmath \overline{s} t_1}{L} {\Psi}_3(\mathbf{\Theta}_{\delta}\mathbf{R}_1, \mathbf{R}_1, \overline{\mathbf{\Theta}}_{\omega})+ \frac{\jmath \underline{s} t_2}{L} {\Psi}_7(\mathbf{\Theta}_{\delta}\mathbf{R}_1,  \overline{\mathbf{\Theta}}_{\omega})  +  \frac{\jmath t_1}{M} {\Psi}_6(\mathbf{\Theta}_{\delta}\mathbf{R}_1, \mathbf{T}_2\mathbf{\Theta}_{\gamma}, \mathbf{\Theta}_{\omega}^H)\Big]{\psi}  + \varepsilon_{\mathcal{Z},1}, \label{Z_eq_1}
\end{split}
\end{equation}
where 
\begin{equation}
\begin{split}
    \varepsilon_{\mathcal{Z},1} &= \mathbb{E}\widehat{\psi} \Big\{ {\mathrm{Tr}}[\mathbf{R}_1(\mathbb{E}\mathbf{Q}_1 - \mathbf{\Theta}_{\delta})] - \mathring \alpha_{\delta} \mathring{\Psi}_2(\mathbf{\Theta}_{\delta}\mathbf{R}_1, \mathbf{I}_M, \overline{\mathbf{\Theta}}_{\omega})  - \overline{s} \mathring \alpha_{\delta} \mathring{\Psi}_1(\mathbf{\Theta}_{\delta}\mathbf{R}_1,  \overline{\mathbf{\Theta}}_{\omega}) 
     + \frac{\jmath t_1}{L}   \mathring{\Psi}_4(\mathbf{\Theta}_{\delta}\mathbf{R}_1, \mathbf{R}_1, \mathbf{I}_M,\overline{\mathbf{\Theta}}_{\omega}) \\
     & + \frac{\jmath \overline{s} t_1}{L} \mathring{\Psi}_3(\mathbf{\Theta}_{\delta}\mathbf{R}_1, \mathbf{R}_1, \overline{\mathbf{\Theta}}_{\omega}) + \frac{\jmath \underline{s} t_2}{L} \mathring{\Psi}_7(\mathbf{\Theta}_{\delta}\mathbf{R}_1,  \overline{\mathbf{\Theta}}_{\omega}) 
     -  \mathring \kappa\mathring{\Psi}_2(\mathbf{\Theta}_{\delta}\mathbf{R}_1, \mathbf{T}_2\mathbf{\Theta}_{\gamma}, \mathbf{\Theta}_{\omega}^H)  +  \frac{\jmath t_1}{M} \mathring{\Psi}_6(\mathbf{\Theta}_{\delta}\mathbf{R}_1, \mathbf{T}_2\mathbf{\Theta}_{\gamma}, \mathbf{\Theta}_{\omega}^H)  \Big\} \\
     &:= \mathbb{E}\widehat{\psi} \varpi_1.
\end{split}
\end{equation}
Similar to the evaluation of $\varepsilon_{\mathcal{X}}$ in \eqref{CLT_res_X}, it can be proved that $\mathbb{E}|\varpi_1| = \mathcal{O}_{z}^s(\frac{1}{zN})$.  Therefore, we have $|\varepsilon_{\mathcal{Z},1}| = \mathcal{O}_{z}^s(\frac{1}{zN})$. Then, the LHS of \eqref{Z_eq_1} is rewritten as follows 
\begin{equation}
    (-L + L \varsigma \delta_2) \mathcal{Z}_{\alpha_{\delta}} + L \gamma_{2}\underline{\omega}_{2,I}\delta_2 \mathcal{Z}_{\kappa} + \mathcal{O}_{z}^s(\frac{1}{N}).
\end{equation}
The calculation of the RHS of \eqref{Z_eq_1} is quite similar to \eqref{CLT_X_terms} since the variable $\mathbf{\Theta}_{\delta}$ in \eqref{X_to_Psi_terms} 
becomes $\mathbf{\Theta}_{\delta}\mathbf{R}_1$ in \eqref{Z_eq_1}, so we omit the detailed steps here for brevity. The approximation result for the RHS of \eqref{Z_eq_1}  is equal to $-X_{1,V_1} - X_{1,C} + \mathcal{O}_{z}^s(\frac{1}{zN})$ as defined in \eqref{Lemma_Z_alpha_kappa_X_terms}. 
Similarly, by multiplying  $[\mathbf{\Theta}_{\omega}\mathbf{R}_2]_{qp}$ on both sides of \eqref{CLT_2} and summing over all subscripts $i$, $j$, $q$, and $p$, we have
\begin{equation}
\begin{split}
     & [M-\alpha_{\delta}   \Psi_2(\mathbf{I}_N, \mathbf{T}_2\mathbf{\Theta}_{\gamma}, \mathbf{T}_1\mathbf{\Theta}_{\omega}\mathbf{R}_2)] \mathcal{Z}_{\kappa}   + [\Psi_2(\mathbf{I}_N, \mathbf{I}_M, \mathbf{T}_1\mathbf{\Theta}_{\omega}\mathbf{R}_2) + \overline{s}  \Psi_1(\mathbf{I}_N, \mathbf{T}_1\mathbf{\Theta}_{\omega}\mathbf{R}_2) -  L\alpha_{\underline{\omega}}]\mathcal{Z}_{\alpha_\delta}  \\
     & = \Big[ - \frac{\jmath t_1 \alpha_{\delta} }{M} \Psi_6(\mathbf{I}_N, \mathbf{T}_2\mathbf{\Theta}_{\gamma}, \mathbf{T}_1\mathbf{\Theta}_{\omega}\mathbf{R}_2) + \frac{\jmath \underline{s} t_2}{L}\Psi_7(\mathbf{I}_N, \mathbf{T}_1\mathbf{\Theta}_{\omega}\mathbf{R}_2)    + \frac{\jmath t_1}{L} \Psi_4(\mathbf{I}_N, \mathbf{R}_1, \mathbf{I}_M,  \mathbf{T}_1\mathbf{\Theta}_{\omega}\mathbf{R}_2) \\
     & + \frac{\jmath \overline{s} t_1}{L}\Psi_3(\mathbf{I}_N, \mathbf{R}_1,  \mathbf{T}_1\mathbf{\Theta}_{\omega}\mathbf{R}_2)\Big] {\psi}+ \mathcal{O}_z^s(\frac{1}{N}). \label{Z_eq_2}
\end{split}
\end{equation}
The LHS of the equation \eqref{Z_eq_2} can be evaluated as follows
\begin{equation}
    (M - L\delta^2 \gamma_2\underline{\omega}_2) \mathcal{Z}_{\kappa} - L (\underline{\omega} - \overline{s} \delta \overline{\underline{\omega}}_{1,1} -\gamma\delta\underline{\omega}_2 ) \mathcal{Z}_{\alpha_{\delta}} + \mathcal{O}_{z}^s(\frac{1}{N}) \overset{(a)}{=} M\Delta \mathcal{Z}_{\kappa} - L \underline{\omega}_{2,I} \mathcal{Z}_{\alpha_{\delta}} + \mathcal{O}_{z}^s(\frac{1}{N}), 
\end{equation}
where $(a)$ follow sfrom tha fact that $\underline{\omega} = \frac{1}{L}\mathrm{Tr}[\mathbf{R}_2\mathbf{T}_2\mathbf{F}_{\omega}(\mathbf{I}_L + \overline{s}\delta\mathbf{T}_1 + \delta\gamma \mathbf{R}_2\mathbf{T}_1)\mathbf{F}_{\omega}] = \underline{\omega}_{2,I}+ \overline{s}\delta \overline{\underline{\omega}}_{1,1} + \gamma\delta\underline{\omega}_2$. The RHS of \eqref{Z_eq_2}
 can be calculated in a similar manner, and is equal to $X_{2, V_1} + X_{1, C} + \mathcal{O}_{z}^s(\frac{1}{N})$. Combining this result with \eqref{Z_eq_1} and \eqref{Z_eq_2}, we have
\begin{equation}
    \begin{bmatrix}
        L(1 - \varsigma \delta_2) & -L \gamma_2 \underline{\omega}_{2,I} \delta_2 \\
        -L \underline{\omega}_{2,I} & M \Delta
    \end{bmatrix} \begin{bmatrix}
        \mathcal{Z}_{\alpha_{\delta}} \\
        \mathcal{Z}_{\kappa}
    \end{bmatrix} = \begin{bmatrix}
            X_{1,V_1} + X_{1, C} + \mathcal{O}_{z}^s(\frac{1}{N})\\
            X_{2,V_1} + X_{2, C} + \mathcal{O}_{z}^s(\frac{1}{N})
        \end{bmatrix} .
\end{equation}
By solving the above system of equations, we complete the proof of Lemma \ref{Lemma_Z_alpha_kappa}. \QED

\bibliographystyle{IEEEtran}
\bibliography{Reference_List}
\end{document}